\documentclass[11pt,a4paper,twoside]{book}
\usepackage[paper=a4paper,left=2cm,right=2cm,top=2cm,bottom=2cm]{geometry}%margins and page size (margins only for titlepage)
\usepackage[breaklinks=true,colorlinks=true,linkcolor=blue,urlcolor=blue,citecolor=blue]{hyperref}%hyperlinks
\usepackage{bookmark}%Correct Hirachy of Chapter/Section links in pdf
\usepackage[utf8]{inputenc}%umlauts and coding
\usepackage[T1]{fontenc}%west-european coding
\usepackage{lmodern}%smooth font style for pdf
\usepackage[english]{babel}%language
\usepackage{comment}%comments in code
\usepackage{appendix}%Appendix
\usepackage{epigraph}%Quotes
\usepackage{fancybox,framed}%Boxes around Text
\usepackage{datetime}%month and year
\newdateformat{monthyeardate}{\monthname[\THEMONTH] \THEYEAR}
\usepackage{xcolor}
\usepackage{tikz-cd}%commutative diagrams
\graphicspath{{Figures/}}
\usepackage[Lenny]{fncychap}
\usepackage{fancyhdr}
\usepackage{titletoc}
\usepackage{chngcntr}
\counterwithin*{section}{chapter}
\titlecontents{chapter}[0pt]{\bfseries\protect\addvspace{15pt}\titlerule\addvspace{1.5ex}}
{\contentslabel[\thecontentslabel]{0pt}\qquad}{\chaptername~}{\hfill\contentspage}[\addvspace{0.7ex}\titlerule\addvspace{1.5ex}]
\addto\captionsenglish{}
\usepackage{graphicx}%Figures
\usepackage{float}%Fixing Figures with [H]
\usepackage[font=normalsize]{subfig} %Subfigures
\usepackage[export]{adjustbox}
\usepackage{dsfont}%for mathds
\usepackage{mathrsfs}%for \mathscr
\usepackage[tbtags]{amsmath}%mathematic einvoronment: align
\usepackage{amssymb}%mathematical symbols
\usepackage{braket}%braket notation
\usepackage{amsthm,bm}%includes \newtheorem
\usepackage[framemethod=TikZ]{mdframed}%framed environments
\theoremstyle{definition}%not italic inside environment
\newmdtheoremenv[everyline=true,linewidth=1.1pt,innertopmargin=-2pt, innerbottommargin=7pt,splittopskip=17pt]{Definition}{Definition}[chapter]
\newmdtheoremenv[everyline=true,linewidth=1.1pt,innertopmargin=-2pt, innerbottommargin=7pt,splittopskip=17pt]{Theorem}[Definition]{Theorem}
\newmdtheoremenv[everyline=true,linewidth=1.1pt,innertopmargin=-2pt, innerbottommargin=7pt,splittopskip=17pt]{Proposition}[Definition]{Proposition}
\newmdtheoremenv[everyline=true,linewidth=1.1pt,innertopmargin=-2pt, innerbottommargin=7pt,splittopskip=17pt]{Lemma}[Definition]{Lemma}
\newmdtheoremenv[everyline=true,linewidth=1.1pt,innertopmargin=-2pt, innerbottommargin=7pt,splittopskip=17pt]{Definition and Theorem}[Definition]{Definition and Theorem}
\newmdtheoremenv[everyline=true,linewidth=1.1pt,innertopmargin=-2pt, innerbottommargin=7pt,splittopskip=17pt]{Definition and Lemma}[Definition]{Definition and Lemma}
\newmdtheoremenv[everyline=true,linewidth=1.1pt,innertopmargin=-2pt, innerbottommargin=7pt,splittopskip=17pt]{Definition and Proposition}[Definition]{Definition and Proposition}
\newmdtheoremenv[everyline=true,linewidth=1.1pt,innertopmargin=-2pt, innerbottommargin=7pt,splittopskip=17pt]{Corollary}[Definition]{Corollary}
\newmdtheoremenv[everyline=true,linewidth=1.1pt,innertopmargin=-2pt, innerbottommargin=7pt,splittopskip=17pt]{Axiom}{Axiom}
\newtheorem{Remarkp}[Definition]{Remark}
\newenvironment{Remark}
  {\pushQED{\qed}   
  \Remarkp}{\popQED\endRemarkp}
\newtheorem{Examplesp}[Definition]{Examples}
\newenvironment{Examples}
  {\pushQED{\qed}
  \Examplesp}{\popQED\endExamplesp}
\newtheorem{Remarksp}[Definition]{Remarks}
\newenvironment{Remarks}
  {\pushQED{\qed}   
  \Remarksp}{\popQED\endRemarksp}
\newtheorem{Remarkspq}[Definition]{Remarks and Examples}
\newenvironment{Remarks and Examples}
  {\pushQED{\qed}   
  \Remarkspq}{\popQED\endRemarkspq}
\newcommand{\KN}{\mathbin{\bigcirc\mspace{-15mu}\wedge\mspace{3mu}}}
\usepackage[hang]{footmisc}
\usepackage{lipsum}
\counterwithout{footnote}{chapter}
\let\oldbibliography\thebibliography
\renewcommand{\thebibliography}[1]{\oldbibliography{#1}
\setlength{\itemsep}{0\baselineskip}}
\begin{document}
%
%
%
%
%
%Title page:
\begin{titlepage}

\begin{figure}[H]
\centering
\subfloat{\includegraphics[width=0.3\textwidth]{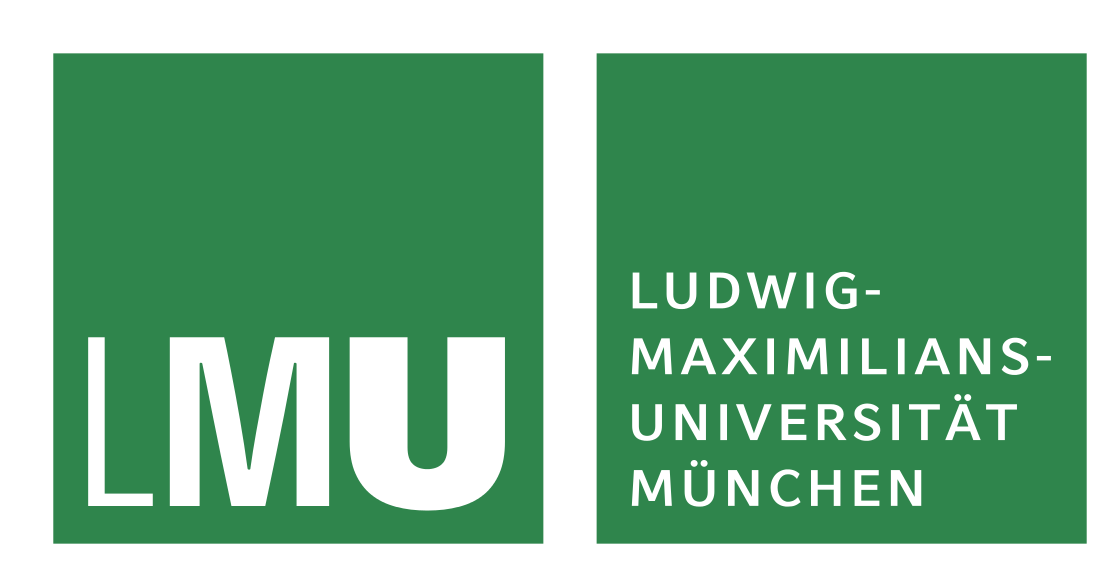}}\hfill
\subfloat{\includegraphics[width=0.28\textwidth]{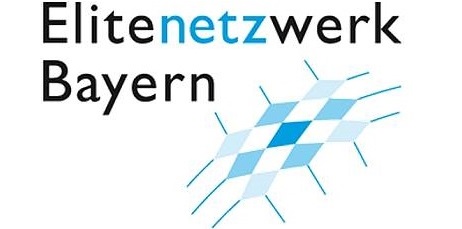}}\hfill
\subfloat{\includegraphics[clip,trim=10.35cm 14.15cm 5.5cm 13.5cm,scale=1.35]{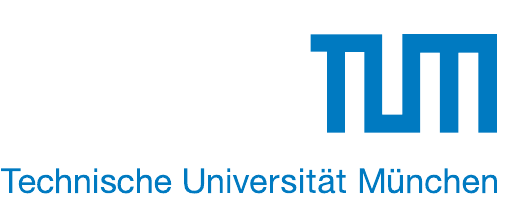}}
\end{figure}

\vspace*{1cm}

\begin{itemize}
\centering
\item[]\begin{LARGE}\textsc{Master's Thesis}\end{LARGE}
\end{itemize}
\hrulefill
\begin{itemize}
\centering
%\item[]Title
\item[]\begin{huge}\textbf{On 3-Dimensional Quantum Gravity and}\end{huge}
\item[]\begin{huge}\textbf{Quasi-Local Holography in Spin Foam}\end{huge}
\item[]\begin{huge}\textbf{Models and Group Field Theory}\end{huge}
%\item[]\begin{Large}Subtitle?\end{Large}
\end{itemize}
\hrulefill

\vspace*{1cm}

\begin{Large}
\begin{tabular}[t]{@{}l} 
\textit{Author:}\\ \textsc{Gabriel Schmid}
\end{tabular}
\hfill
\begin{tabular}[t]{l@{}}
\textit{Supervisor:}\\\textsc{Dr. Daniele Oriti}\\ 
\\
\textit{Co-Supervisor:}\\\textsc{Dr. Christophe Goeller}
\end{tabular}
\end{Large}

\vspace*{2cm}
%\vfill

{\begin{itemize}
\centering
\item[]\begin{Large}\textit{A thesis submitted for the degree of}\end{Large}
\item[]\begin{Large}\textsc{Master of Science}\end{Large}
\item[]\begin{Large}\textit{in}\end{Large}
\item[]\begin{Large}\textsc{Theoretical and Mathematical Physics}\end{Large}
\end{itemize}

%\vspace*{1cm}
\vfill

\begin{figure}[H]
\centering
\includegraphics[width=0.5\textwidth]{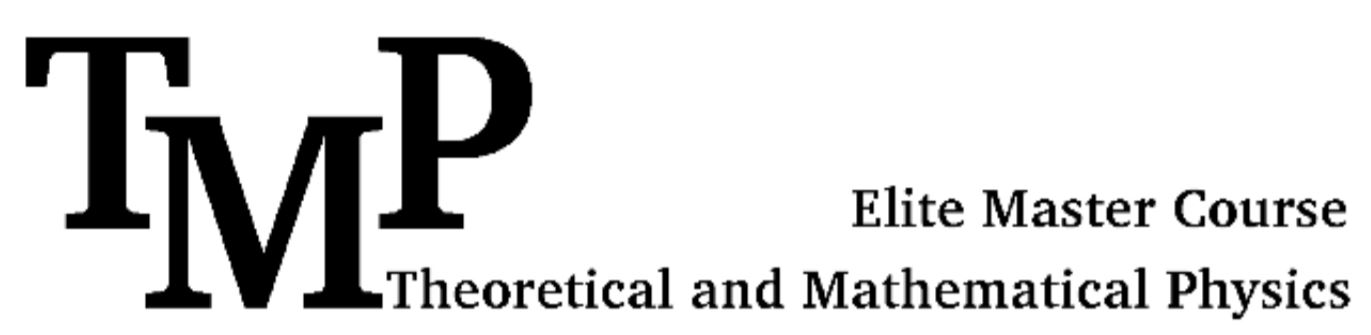}
\end{figure}}

%\vspace*{1cm}
\vfill

\begin{itemize}
\centering
\item[]\begin{Large}Munich, the 5$^{\mathrm{th}}$ of March 2022\end{Large}
\end{itemize}

\end{titlepage}
%
%
%
%
%
%Change Margins (1 inch on each side) and create headers:
\newgeometry{top=1in,bottom=1in,outer=1in,inner=1in}
\pagestyle{fancy}
\fancyhead[RO]{\nouppercase{\leftmark}}
\fancyhead[LE]{\nouppercase{\rightmark}}
\fancyhead[RE,LO]{\thepage}
\fancyfoot[C]{}
\newpage
\thispagestyle{empty}
\begin{itemize}\item[]\end{itemize}
\vfill
\begin{Large}
\begin{tabular}[t]{@{}l} 
\\ 
\end{tabular}
\hfill
\begin{tabular}[t]{l@{}}
\underline{\textsc{Disputation:}}\\Munich, the 21$^{\mathrm{st}}$ of March 2022 (10:00 a.m.)\\
\\
\underline{\textsc{Examination Committee:}}\\
Dr. Daniele Oriti (Supervisor)\\
Dr. Christophe Goeller (2$^{\mathrm{nd}}$ Referee)
\end{tabular}
\end{Large}
%
%
%
%
%
%\newpage\null\thispagestyle{empty}\newpage
\newpage
\thispagestyle{empty}
\section*{Declaration of Authorship}
I hereby declare that the thesis submitted is my own unaided work and that all direct or indirect sources used are acknowledged as references. This paper was not previously presented to another examination board and has not been published. 

\vspace*{1cm}

\begin{otherlanguage}{ngerman}
\section*{Eidesstattliche Erklärung}
Ich erkläre hiermit ehrenwörtlich, dass ich die vorliegende Arbeit selbständig angefertigt habe und dass die aus fremden Quellen direkt und indirekt übernommenen Gedanken als solche kenntlich gemacht sind. Die Arbeit wurde weder einer anderen Prüfungsbehörde vorgelegt noch veröffentlicht. 
\end{otherlanguage}
 
\vspace*{1cm}
%\vspace*{4cm}

\begin{figure}[H]\includegraphics[width=0.3\textwidth,right]{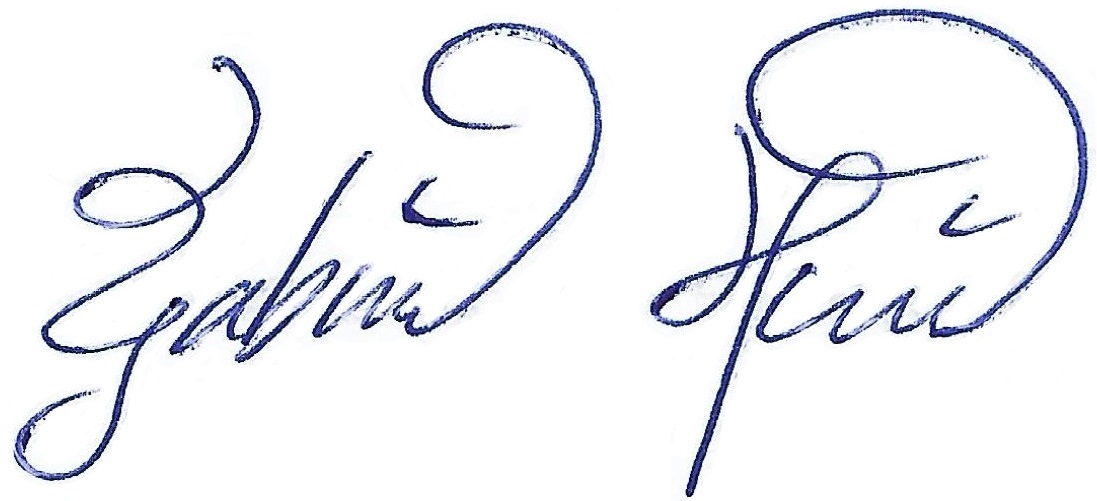} \end{figure}
Munich, March 2022\hfill Gabriel Schmid\hspace*{10mm}
%
%
%
%
%
%Dedication and Quote
\newpage\null\thispagestyle{empty}\newpage
\thispagestyle{empty}
\vspace*{3cm}
\begin{itemize}
\setlength\itemsep{-5pt}
\centering
\item[]\textit{To my family:}
\item[]\textit{My mother Renate with Markus,}
\item[]\textit{my father Bernhard}
\item[]\textit{and my brother Benjamin,}
\item[]\textit{for all their overwhelming support over the last years.}
\end{itemize}
\vfill
\setlength{\epigraphwidth}{2.6in}
\setlength{\epigraphrule}{0pt} 
\begin{minipage}[t]{0.5\textwidth}\raggedleft
\epigraph{\textit{``To those who do not know mathematics it is difficult to get across a real feeling as to the beauty, the deepest beauty, of nature. [${\dots}$]
If you want to learn about nature, to appreciate nature, it is necessary to understand the language that she speaks in.''}}{-- Richard P. Feynman  \cite[p.58]{FeynmanQuote}}
\end{minipage}
\begin{minipage}[t]{0.5\textwidth}\raggedleft
\epigraph{\textit{``Young man, in mathematics you don't understand things. You just get used to them.''}\\
\\Reply to a young physicist who said\newline\textit{``I'm afraid I don't understand the method of characteristics.''}}{-- John von Neumann \cite[p.208]{Zukav}}
\end{minipage}

\hfill

\begin{minipage}[t]{0.5\textwidth}\raggedleft
\epigraph{\textit{``Whenever a theory appears to you as the only possible one, take this as a sign that you have neither understood the theory nor the problem which it was intended to solve.''}}{-- Karl R. Popper \cite[p.266]{PopperQuote}}
\end{minipage}
\begin{minipage}[t]{0.5\textwidth}\raggedleft
\epigraph{\textit{``The scientist does not study nature because it is useful to do so. He studies it because he takes pleasure in it, and he takes pleasure in it because it is beautiful.''}}{-- J. Henri Poincaré \cite[p.22]{PoincareQuote}}
\end{minipage}
%
%
%
%
%
%Abstract:
\newpage\null\thispagestyle{empty}\newpage
\newpage
\thispagestyle{empty}

\section*{Abstract}\vspace{-9pt}
This thesis is devoted to the study of $3$-dimensional quantum gravity as a spin foam model and group field theory. In the first part of this thesis, we review some general physical and mathematical aspects of $3$-dimensional gravity, focusing on its topological nature. Afterwards, we review some important aspects of the Ponzano-Regge spin foam model for $3$-dimensional Riemannian quantum gravity and explain in some details how it is related to the discretized path integral of general relativity in its first-order formulation. Furthermore, we discuss briefly some related spin foam models and review the notion of spin network states in order to properly define transition amplitudes of these models.\newline
\hspace*{2cm}The main results of this thesis are contained in the second part. We start by reviewing the Boulatov group field theory and explain how it is related to the Ponzano-Regge model and some advantages of introducing colouring. Afterwards, we give a very detailed review of the topology of coloured graphs with non-empty boundary and review techniques, which are devolved in crystallization theory, a branch of geometric topology. In the last part of this chapter, we apply these techniques in order to define suitable boundary observables and transition amplitudes of this model and in order to set up a formalism for dealing with transition amplitudes in the coloured Boulatov model in a more systematic way by writing them as topological expansions. We also apply these techniques to the simplest possible boundary state representing a $2$-sphere.\newline
\hspace*{2cm}Last but not least, we review some results regarding quasi-local holography in the Ponzano-Regge model, construct some explicit examples of coloured graphs representing manifolds with torus boundary  and discuss the transition amplitude of some fixed boundary graph representing a $2$-torus.

\vfill

\begin{otherlanguage}{ngerman}
\section*{Abstrakt}\vspace{-9pt}
Die vorliegende Arbeit ist dem Studium $3$-dimensionaler Quantengravitation als Spin-Schaum Modell und Gruppenfeldtheorie gewidmet. Im erstem Teil dieser Arbeit geben wir einen Überblick über einige allgemeine physikalische und mathematische Aspekte von $3$-dimesionaler Gravitation, wobei wir uns auf deren topologische Natur fokussieren. Anschließend beschreiben wir einige wichtige Aspekte des Ponzano-Regge Spin-Schaum Modells für $3$-dimensionale Riemann'sche Quantengravitation und erklären detailliert, wie dieses mit der Diskretisierung des Pfadintegral der allgemeinen Relativitätstheorie im Erste-Ordnung Formalismus zusammenhängt. Des Weiteren diskutieren wir kurz einige verwandte Spin-Schaum Modelle und geben einen kurzen Überblick über Spin-Netzwerke, um die Übergangsamplituden der oben genannten Modelle zu definieren.\newline
\hspace*{2cm}Die Hauptresultate dieser Arbeit finden sich im zweiten Teil. Zu Beginn geben wir einen Überblick über die Boulatov'sche Gruppenfeldtheorie, erklären wie diese mit dem Ponzano-Regge Modell zusammhängt und welche Vorteile eine Färbung des Modells mit sich bringt. Anschließend geben wir einen detaillierten Überblick über die Topologie gefärbter Grafen mit nicht-leerem Rand und diskutieren Techniken, welche in der Krystallisierungstheorie, einem Teilgebiet der geometrischen Topologie, entwickelt wurden. Im letzen Teil dieses Kapitels wenden wir diese Techniken auf das Boulatov Modell an, um geeignete Randobservablen und Übergangsamplituden zu definieren und um einen Formalismus einzuführen, der es erlaubt, die Übergangsamplitudes dieses Modells in eine systematischere Weise als topologische Entwicklung zu schreiben. Schlussendlich wenden wir diesen Formalismus auf den einfachsten Randzustand, der eine $2$-Sphäre beschreibt, an.\newline
\hspace*{2cm}Abschließend geben wir einen Überblick über einige Resultate zur quasi-lokalen Holografie im Ponzano-Regge Modell, konstruieren einige explizite gefärbte Grafen, die Mannigfaltigeiten mit toroidalem Rand beschreiben und disuktieren die Übergangsamplitude bezüglich eines fixierten Randgrafen mit toroidaler Topologie.
\end{otherlanguage}
\vfill
%
%
%
%
%
%Acknowledgements:
\newpage\null\thispagestyle{empty}\newpage
\newpage
\thispagestyle{empty}
\textcolor{white}{ }
\vfill
\section*{Acknowledgements}\vspace{-9pt}
First and foremost I would like to express my deepest gratitude to my co-supervisor Dr. Christophe Goeller for his patience and the huge amount of time he spent in our weekly meetings to explain to me so many interesting physics and mathematics, starting from the very basics of classical and quantum gravity, to discussing all the conceptual and technical problems popping up throughout the work of this project. There was almost no question he couldn't answer and it has always been a pleasure to work with him. Thank you very much for all your patience, encouragement and support! Secondly, I would like to equally thank my supervisor Dr. Daniele Oriti for all his support and encouragement during this work, especially for sharing his very broad knowledge and expertise in order to help us out whenever we were stuck due to some conceptual or technical misunderstandings, as well as for all his valuable advices and encouragements regarding the thesis, but also regarding my academic career towards the end of my project. Thank you very much! In this light, I would also like to thank him for giving me the opportunity to work in the ``Quantum Gravity and Foundations of Physics'' research group of the Arnold Sommerfeld Center for Theoretical Physics (ASC) at the Ludwig-Maximilians Universität in Munich, which provided me with such a friendly and productive environment to work on my thesis project. Having said this, I would also like to express my gratitude to all the members of the group for all their support and all the interesting presentations and discussions which we had during our weekly group meetings.\newline\hspace*{2cm}
Secondly, I am very grateful to PD Dr. Michael Haack, my mentor in the TMP-mentoring program, for all his great support and his valuable advices regarding my academic career, starting from the choice of my master thesis topic, to the problem of finding a future PhD supervisor. I would also like to thank Dr. Robert Helling for making TMP such an unforgettable time. Despite the unfortunate circumstances over the last years due to the COVID-19 pandemic, I greatly enjoyed my time in Munich and I cannot express in words how many things I have learnt throughout my studies here.\newline\hspace*{2cm}
Last, but certainly not least, I want to thank my family, to whom I dedicate this work, as well as all my friends in Tyrol and around the world for all their unwavering support during the last years of my academic career and in my life in general. Thank you all!
%
%
%
%
%Table of Contents (with hyperlinks, but in black colour)
\newpage
{\hypersetup{linkcolor=black}
\tableofcontents}
\newpage
\phantomsection
\chapter*{Introduction}
\addcontentsline{toc}{section}{\hspace{-16pt}\textbf{Introduction}}
\markboth{Introduction}{Introduction}
The search for a consistent and non-perturbative theory of quantum gravity is one of the most striking open problems of theoretical physics within the last decades. It aims at finding a description of gravity from the smallest to the largest scale, combining the two cornerstones of modern theoretical physics, which are \textsc{Einstein's General Theory of Relativity}, the basic theory underlying the \textsc{Standard Model of Cosmology}, and \textsc{Quantum Field Theory}, which provides the foundation of the \textsc{Standard Model of Particle Physics}. Many experiments have shown the great success of these two ``\textit{pillars}'' of modern physics, like the discovery of the Higgs particle in 2012 (Nobel Prize 2013), the detection of gravitational waves in 2015 (Nobel Prize 2017) as well as the discovery of indirect and direct evidences for the existence of black holes (Nobel Prize 2020), just to name a few recent results. However, despite their enormous success, there are still many open problems, like the nature of singularities and black holes and the origin and nature of the cosmological constant, which hints towards the need of a unified treatment of these two theories and hence also to a consistent and non-perturbative theory of quantum gravity. Although the problem of finding a complete formulation of quantum gravity is still far from being solved, many results have been obtained in the last decades and many different approaches have led to various striking results. The problem of formulating such a theory already starts at the conceptual level: Quantum field theories, like quantum electrodynamics and the standard model of particle physics, are usually defined on some fixed background, but quantum gravity requires the quantization of spacetime itself, since general relativity is a theory of dynamical spacetime with no preferred reference frame. In other words, classical gravity is a background-independent theory, whereas quantum field theory is a local theory defined on a fixed background with a splitting into space and time. It is important to note at this point that a theory of quantum gravity does not need to be the result of some kind of quantization procedure of classical gravity. Quantum gravity is about the understanding of the microscopic structure of spacetime itself and the only requirement is that it should reduce to the theory of classical gravity in an appropriate limit and regime of the theory.
\newline\hspace*{2cm}
In this thesis, we will focus on the case of Riemannian quantum gravity in three dimensions. ``\textit{Riemannian}'' in this context means that we treat all three dimensions in the same way, i.e. we use a metric with definite signature. $3$-dimensional general relativity is a topological field theory and has no local degrees of freedom and hence, $3$-dimensional gravity can be used as a simple toy model of studying various different aspects and phenomena, which we expect to be there at the $4$-dimensional case, like holographic dualities. We will mainly focus on manifestly background-independent approaches to quantum gravity, such as \textsc{Spin Foam Models}, also known as \textsc{Covariant Loop Quantum Gravity}, as well as the \textsc{Group Field Theory} approach. Manifestly background-independent in this case means that we do not use any spacetime manifold or some kind of preexisting metric in order to formulate our theory, but rather purely combinatorial and algebraic data. The notion of a spacetime should then emerge in some classical and continuum limit. A very important feature of these approaches is that we are working in a discrete setting. Indeed, the notion of a smooth spacetime and of a point in spacetime has to be questioned in general relativity, since by background-independence of classical gravity, the theory does not depend on the chosen background. In manifestly background-independent approaches, we are hence aiming at a quantum theory of the very ``atoms'' of spacetime itself. 

\section*{Plan of this Thesis}
This thesis is organized into four chapters containing a total number of eleven sections.\newline\hspace*{2cm}
In Chapter \ref{Chap1} we review Einstein's general theory of relativity, focusing both on its mathematical structure and on the peculiarities in the $3$-dimensional case. In particular, we count the physical degrees of freedom, derive the Newtonian limit, analyse the curvature tensor and discuss the solutions of Einstein's field equations in vacuum. Each of these discussions will lead us to the same conclusion: Gravity in three dimensions is a topological field theory. There are no local degrees of freedom and hence no propagation in form of gravitational waves. We will then introduce the triadic first-order formalism à la Palatini and will see that it can be understood as a particular case of topological $BF$-theory and Chern-Simons theory, which are two well-studied topological field theories of Schwarz type.\newline\hspace*{2cm}
Chapter \ref{Chap2} is devoted to a short discussion of the Ponzano-Regge spin foam model for $3$-dimensional non-perturbative Riemannian quantum gravity without a cosmological constant. This is a manifestly background-independent approach and in fact the first model for $3$-dimensional quantum gravity ever proposed. We will comment on its historical origin and will see that it can be understood as the discretization of the quantum partition function of gravity in the triadic first-order formalism. Afterwards, we will discuss how to gauge-fix this model in a suitable way and will argue that the Ponzano-Regge model is a well-defined topological invariant. Finally, this chapter ends with a discussion of all the necessary ingredients needed in order to define and calculate transition amplitudes of this and related models.
\newline\hspace*{2cm}
The third part of this thesis is dedicated to yet another approach to $3$-dimensional Riemannian quantum gravity without a cosmological constant, namely the Boulatov group field theory. At the beginning of Chapter \ref{Chap3}, we start be discussing some general features of this model, which is by definition a proper field theory defined on three copies of the Lie group $\mathrm{SU}(2)$. Most importantly, we will see that the Feynman graphs are dual to $3$-dimensional simplicial complexes and that the corresponding amplitudes are nothing else than the Ponzano-Regge partition functions. In general, the perturbative expansion of observables in this model is quite hard to control, since it generically involves a sum over all topologies including highly singular topologies. We will see that the situation is drastically improved by introducing a colouring of the model, which sorts out all topologies which are more singular than pseudomanifolds, i.e. topologies, which fail to be manifolds at a finite number of isolated points. The Feynman graphs of the coloured Boulatov model can be understood as proper edge-coloured graphs and it turns out that these type of graphs are well known in the mathematical literature in a branch of geometric topology known as crystallization theory. We will then review many important definitions and results obtained in crystallization theory for the general case of graphs with non-empty boundary. This extensive discussion will provide is will all the tools needed in order to define suitable boundary observables and transition amplitudes in the coloured Boulatov model. Furthermore, we will be able to write the transition amplitudes in a more systematic way in terms of a topological expansion.\newline\hspace*{2cm}
The final part of this thesis, Chapter \ref{Chap4}, is devoted to a short overview of results obtained in the study of (quasi-local) holographic dualities in the Ponzano-Regge model. In particular, we will review a work devoted to the study of holographic dualities in the Ponzano-Regge model on the solid torus. We will end this chapter by discussing what will change when doing the same analysis in the coloured Boulatov model, in which we only fix the boundary graph and sum over all bulk topologies matching the given boundary topology.
\chapter{General Relativity in 3D and Topological Field Theory}\label{Chap1}
\fancyhead[LE]{\nouppercase{\rightmark}}
General relativity is one of the cornerstones of modern theoretical physics. Since it was firstly published by A. Einstein around 1915 \cite{Einstein}, many experiments have shown the great success of this theory in many of its aspects, two of the most recent ones being the detection of gravitational waves in September 2015 \cite{Ligo} as well as the first direct observation of a black hole \cite{BlackHole} in 2019. In dimension three it turns out that general relativity is rather peculiar, since it has no local degrees of freedom. This is often stated by saying that $3$-dimensional general relativity is a ``\textit{topological field theory}'', since it only depends on the topology of spacetime. In the first chapter of this thesis, we focus on general relativity in three dimensions. We start by discussion some general features of the theory and some of its peculiarities in the $3$-dimensional case. Afterwards, we introduce a first-order formulation of general relativity, namely the triadic Palatini formalism, which is often used as a starting point for theories of quantum gravity. Furthermore, this formalism will allow us to compare general relativity to other well-known topological field theories used in physics, such as topological $BF$ and Chern-Simons theory.

\section{General Relativity in Three Dimensions}
The first section is devoted to a short overview of the mathematical structure of Einstein's theory of general relativity. Afterwards, we will discuss various aspects of the $3$-dimensional case. More precisely, we will look at the physical degrees of freedom, the Newtonian limit, the curvature tensor in vacuum as well as at the vacuum solutions. Each of these discussions will lead us to the same conclusion: General relativity in three dimensions is a topological field theory, i.e. there are no local degrees of freedom. Last but not least, we introduce the triadic Palatini formalism and explain how it is related to the Einstein-Hilbert action of gravity.

\subsection{A Glance of Einstein's General Theory of Relativity}
In Einstein's general theory of relativity spacetime is modelled by a smooth $4$-dimensional Lorentzian manifold $(\mathcal{M},g)$. More precisely, this means that $\mathcal{M}$ is a $4$-dimensional smooth manifold with or without boundary and that $g\in\Gamma^{\infty}(T^{\ast}\mathcal{M}^{\otimes 2})$ is a smooth tensor field of rank $(0,2)$, such that $g_{p}:T_{p}\mathcal{M}\times T_{p}\mathcal{M}\to\mathbb{R}$ is for all $p\in\mathcal{M}$ a symmetric and non-degenerate bilinear form of signature $(+,-,-,-)$. In addition, we normally assume some further topological and geometrical properties. First of all, usually we assume $\mathcal{M}$ to be connected, because it would not be possible to experience anything from some disconnected component of spacetime. Furthermore, we want our manifold to be oriented, because it should be possible to integrate over it. Last but not least, we normally assume our spacetime manifold also to be time-orientable, which means that we assume that there exist an overall non-zero timelike smooth vector field on $\mathcal{M}$, i.e. a vector field $X\in\mathfrak{X}(\mathcal{M}):=\Gamma^{\infty}(T\mathcal{M})$ satisfying $g_{p}(X_{p},X_{p})>0$ for all $p\in\mathcal{M}$. Such a vector field allows us to define the past $\mathcal{J}^{-}(p)$ and the future $\mathcal{J}^{+}(p)$ with respect to some spacetime point $p\in\mathcal{M}$ via
\begin{align}\mathcal{J}^{\pm}(p):=\{q\in\mathcal{M}\mid \exists\text{ path }\gamma\in C^{\infty}([0,1],\mathcal{M}):\gamma(0)=p,\gamma(1)=q\text{ and } g(X,\dot{\gamma})\gtrless 0\}.\end{align}
This assumption seems plausible, since it should be possible to distinguish between a past and future, for example using thermodynamical quantities like the entropy. Furthermore, it allows us to incorporate the ``\textit{principle of causality}'', which states that an event at $p$ can only be influenced by events in $\mathcal{J}^{-}(p)$ and the event itself can only influence events in $\mathcal{J}^{+}(p)$. In addition, in some situations it is also useful to have even more restrictive properties. An important concept to mention here at this point are manifolds which are ``\textit{globally hyperbolic}'', which means that they admit a foliation into Cauchy surfaces\footnote{For a precise definition and discussion of global hyperbolicity as well as causality in Lorentzian manifolds in general the reader is guided to the excellent text book \cite{Oneill}.}. Such manifolds are particularly useful for constructing quantum field theoretic models in curved spacetime, because it can be shown that these type of manifolds do not admit closed timelike curves \cite[p.23]{Baer}, which would lead to several conceptual and technical difficulties. Up to now, we have explained how to model spacetime in Einstein's general theory of relativity. The metric $g$ takes the role of the gravitational field on the spacetime manifold $\mathcal{M}$. One of the key properties of general relativity is ``\textit{diffeomorphism invariance}'', which mathematically means that two metrics $g_{1}$ and $g_{2}$ on some given spacetime manifold $\mathcal{M}$ describe the same physical situation, if there exists a smooth isometry $\Phi:(\mathcal{M},g_{1})\to (\mathcal{M},g_{2})$, i.e. a smooth diffeomorphism from $\mathcal{M}$ to itself with the property $g_{1}=\Phi^{\ast}g_{2}$, or equivalently,
\begin{align}(g_{1})_{p}(v,w)=(\Phi^{\ast}g_{2})_{p}(v,w)=(g_{2})_{\Phi(p)}(\Phi_{\ast}v,\Phi_{\ast}w)\end{align}
for all $p\in\mathcal{M}$ and $v,w\in T_{p}\mathcal{M}$. The gravitational field is therefore modelled by an equivalence class of metrics on $\mathcal{M}$, where two metric are equivalent if they are related by an orientation and time-orientation preserving isometry. The dynamics of general relativity is described by Einstein's field equations, which are invariant under such transformations, and which can be written as the following equation of tensor fields:
\begin{align}\label{EFE}\mathrm{Ric}_{g}-\frac{1}{2}R_{g}g+\Lambda g=\kappa T,\end{align}
where $\Lambda$ denotes the cosmological constant, $\kappa:=\frac{8\pi G}{c^{4}}$ is ``\textit{Einstein's gravitational constant}'', $\mathrm{Ric}_{g}$ denotes the Ricci tensor and $R_{g}$ the Ricci scalar of the Lorentzian manifold $(\mathcal{M},g)$ and where $T$ is a smooth, symmetric and divergence-free $(0,2)$-tensor field, called the ``\textit{energy-momentum tensor}'', describing the matter content. It is well known that Einstein's field equations can also be derived from an action functional, called the``\textit{Einstein-Hilbert action}'', which is given by
\begin{align}\mathcal{S}_{\mathrm{EH}}[g]:=\frac{1}{2\kappa}\int_{\mathcal{M}}\!R_{g}\,\mathrm{d}\mu_{g},\end{align}
where $\mu_{g}$ is the Radon measure on $\mathcal{M}$ induced by the volume form of $g$. Varying with respect to the metric $g$ yields Einstein's field equations in the case $\Lambda=0$ and in vacuum, i.e. $T=0$. More generally, the Einstein-Hilbert action is given by
\begin{align}\mathcal{S}_{\mathrm{EH},\Lambda}[g,\{\varphi_{i}\}_{i\in I}]:=\int_{\mathcal{M}}\!\bigg (\frac{1}{2\kappa}(R_{g}-2\Lambda)+\mathcal{L}_{\mathrm{Matter}}[g,\{\varphi_{i}\}_{i\in I}]\,\bigg )\mathrm{d}\mu_{g},\end{align}
where $\mathcal{L}_{\mathrm{Matter}}[g,\{\varphi_{i}\}_{i\in I}]$ describes the matter content modelled by fields $\{\varphi_{i}\}_{i\in I}$, which are labelled by an index set $I$. Such a field could for example be a scalar field $\varphi\in C^{\infty}(\mathcal{M},\mathbb{R})$ (=a section of a trivial line bundle), a fermionic field (=a section of a spinor bundle over $\mathcal{M}$), or a gauge field (=a connection $1$-forms of a principal fibre bundle over $\mathcal{M}$). Varying with respect to $g$ yields Einstein's field equations (\ref{EFE}) with $T_{\mu\nu}=\frac{-2}{\sqrt{-g}}\frac{\delta(\sqrt{-g}\mathcal{L}_{\mathrm{Matter}})}{\delta g^{\mu\nu}}$. \cite{HawkingEllis,Kriele,Sachs}

\subsection{Physical Degrees of Freedom and Newtonian Limit}
Before going into more details, let us discuss some properties of general relativity in three dimensions using simple physical arguments. For this, we consider a general $d$-dimensional spacetime manifold together with a Lorentzian metric. What are the physical degrees of freedom of this theory? To answer this question, we have to look at the phase space, which is given by the spatial metric on some hypersurface with constant time. On such a hypersurface, the metric has $d(d-1)/2$ components, since it is a symmetric rank $(0,2)$-tensor field. Together with their time derivatives (=conjugate momenta), we get in total $d(d-1)$ degrees of freedom per point in spacetime. However, there are also a number of constraints. First of all, the Einstein tensor is divergence-free and hence, we get $d$ additional constraints. Furthermore, we also have the freedom of choosing a different coordinate system, which eliminates another $d$ degrees of freedom. In total, we are left with $d(d-3)$ degrees of freedom per spacetime point. As a consistency check, let us look at the $4$-dimensional case: Using $d=4$, our calculation yields $4$ physical degrees of freedom, which coincides with the two polarizations for gravitational waves together with their conjugate momenta. Looking at this result, we directly see that $d=3$ is somehow special and we conclude that there are \textit{no physical degrees of freedom in $3$-dimensional gravity}. In other words, $3$-dimensional gravity is not propagating. In the next section, we will derive this result more rigorously by analysing the curvature tensor of the theory. \cite[p.4]{CarlipQG}\\
\\
As a consequence of the above considerations, we also expect that the Newtonian limit of the $3$-dimensional theory should have a rather unusual form. Let $(\mathcal{M},g)$ be a $d$-dimensional Lorentzian manifold modelling spacetime. In order to derive the Newtonian limit of Einstein's field equations (\ref{EFE}), we have to make the following three assumptions:
\begin{itemize}
\item[(a)]The \textbf{gravitational field is weak}, i.e. we are able to expend the metric tensor $g$ in a local chart $(U,\varphi)$ of $\mathcal{M}$, where $U\subset\mathcal{M}$ is some open set, as $g_{\mu\nu}=\eta_{\mu\nu}+h_{\mu\nu}$, where $\eta=\operatorname{diag}(1,-1,\dots,-1)$ denotes the $d$-dimensional Minkowski metric and where the perturbations $h_{\mu\nu}:U\to\mathbb{R}$ are ``small'' compared to $g_{\mu\nu}:=g(\partial_{\mu},\partial_{\nu})$.
\item[(b)]The \textbf{metric is approximately stationary}, i.e. $\partial_{0}g_{\mu\nu}\approx 0$.
\item[(c)]\textbf{Objects are moving slowly compared to the speed of light}: More precisely, consider a space-like trajectory $\gamma:I\to\mathcal{M}$, where $I\subset\mathbb{R}$ denotes a closed interval, which in local coordinates $(U,\varphi)$ is given by $x^{\mu}:=(\varphi\circ\gamma)^{\mu}$. Then the velocity vectors $v^{i}=\mathrm{d}x^{i}/\mathrm{d}t$ should be small compared to the speed of light $c$, where $t:=\frac{1}{c}x^{0}$. Equivalently, we may write
\begin{align}\frac{\mathrm{d}x^{i}}{\mathrm{d}\tau}\ll\frac{\mathrm{d}x^{0}}{\mathrm{d}\tau},\end{align}
where $\tau$ denotes the ``proper time'', which parametrizes the curve $\gamma:I\to\mathcal{M}$.
\end{itemize}

The equations of motion for some spacelike trajectory $\gamma:I\to\mathcal{M}$ are given by the geodesic equations, which in local coordinates have the form
\begin{align}\label{geo}\frac{\mathrm{d}^{2}x^{\mu}}{\mathrm{d}\tau^{2}}+\Gamma_{\alpha\beta}^{\mu}\frac{\mathrm{d}x^{\alpha}}{\mathrm{d}\tau}\frac{\mathrm{d}x^{\beta}}{\mathrm{d}\tau}=0,\end{align}
where we use Einstein's summation convention. Using assumption $(c)$, we approximate this as
\begin{align}\label{appgeo}\frac{\mathrm{d}^{2}x^{\mu}}{\mathrm{d}\tau^{2}}+\Gamma_{\alpha\beta}^{\mu}\frac{\mathrm{d}x^{\alpha}}{\mathrm{d}\tau}\frac{\mathrm{d}x^{\beta}}{\mathrm{d}\tau}\approx\frac{\mathrm{d}^{2}x^{\mu}}{\mathrm{d}\tau^{2}}+\Gamma_{00}^{\mu}\bigg (\frac{\mathrm{d}x^{0}}{\mathrm{d}\tau}\bigg )^{2}\end{align}
and hence, we only have to consider the Christoffel symbols $\{\Gamma_{00}^{\mu}\}_{\mu\in\{0,\dots,d-1\}}$. First of all, lets look at the case $\mu=0$, which yields
\begin{align}\Gamma_{00}^{0}=\frac{1}{2}g^{0\delta}(\underbrace{\partial_{0}g_{0\delta}}_{\approx 0}+\underbrace{\partial_{0}g_{\delta 0}}_{\approx 0}-\partial_{\delta}g_{00})\approx -\frac{1}{2}g^{00}\partial_{0}g_{00}\approx 0,\end{align}
where we used that our metric is assumed to be (approximately) stationary, i.e. property (b) in the list above. Secondly, we have to look at the case, where the coordinate $\mu$ is spacelike. Let us denote spatial coordinates by $i\in\{1,\dots,d\}$, as usual. Then
\begin{align}\Gamma_{00}^{i}=\frac{1}{2}g^{i\delta}(\underbrace{\partial_{0}g_{0\delta}}_{\approx 0}+\underbrace{\partial_{0}g_{\delta 0}}_{\approx 0}-\partial_{\delta}g_{00})\approx -\frac{1}{2}(\eta^{ij}+h^{ij})\partial_{j}(\eta_{00}+h_{00})\approx\frac{1}{2}\delta^{ij}\partial_{j}h_{00}=\frac{1}{2}\partial_{i}h_{00},\end{align}
where we used assumptions (a) and (b). Plugging this back into our original Equation (\ref{appgeo}), we get the following set of equations:
\begin{align}
&\text{(1) }\frac{\mathrm{d}^{2}t}{\mathrm{d}\tau^{2}}=0\\
&\text{(2) }\frac{\mathrm{d}^{2}x^{i}}{\mathrm{d}\tau^{2}}+\frac{1}{2}c^{2}\bigg(\frac{\mathrm{d}t}{\mathrm{d}\tau}\bigg)^{2}\partial_{i}h_{00}=0
\end{align}
Combining both of them finally yields the following equation for the Newtonian limit:
\begin{align}\label{edf}\frac{\mathrm{d}^{2}x^{i}}{\mathrm{d}t^{2}}+\frac{1}{2}c^{2}\partial_{i}h_{00}=0\end{align}
In order to compare this equation with the equations of motion of Newton's law of gravity, we have to find a relation between the Newtonian potential $\Phi$ and the metric component $h_{00}$. For this, we use the general fact that it is always possible to choose a particular gauge, usually called ``Lorentz gauge'', or ``harmonic gauge'', such that Einstein's field equations reduce to 
\begin{align}\label{eq2}-\frac{1}{2}\eta^{\mu\nu}\partial_{\mu}\partial_{\nu}\overline{h}_{\alpha\beta}\approx \kappa T_{\mu\nu},\end{align}
where the ``trace-reverse'' $\overline{h}_{\alpha\beta}$ fulfils the constraint $\eta^{\mu\nu}\partial_{\mu}\overline{h}_{\nu\alpha}=\partial^{\nu}\overline{h}_{\nu\alpha}=0$ and is defined via the equation
\begin{align}\label{1}\overline{h}_{\alpha\beta}:=h_{\alpha\beta}-\frac{1}{2}\eta_{\alpha\beta}\eta^{\mu\nu}h_{\mu\nu}.\end{align}
The derivation of Equation (\ref{eq2}) is straightforward and can be found in most textbooks discussing linearized gravity and gravitational waves (e.g. \cite[p.470ff.]{Hobson}). Inverting the above equation yields 
\begin{align}h_{\alpha\beta}=\overline{h}_{\alpha\beta}-\frac{1}{d-2}\eta_{\alpha\beta}\eta^{\mu\nu}\overline{h}_{\mu\nu}.\end{align}
Using $T_{00}\approx\rho c^{2}$, where $\rho$ denotes the matter density, as well as assumption (a), Equation (\ref{eq2}) reduces in the case $\alpha=\beta=0$ to 
\begin{align}-\frac{1}{2}\eta^{ij}\partial_{i}\partial_{j}\overline{h}_{00}\approx\frac{8\pi G}{c^{2}}\rho.\end{align}
Comparing this equation with the ``\textit{Poisson equation}'' $\Delta\Phi=4\pi G\rho$ for classical gravity and using the fact that $\Delta=-\eta^{ij}\partial_{i}\partial_{j}=\delta^{ij}\partial_{i}\partial_{j}$ we finally arrive at the identification 
\begin{align}\overline{h}_{00}\approx \frac{4}{c^{2}}\Phi.\end{align}
Combing this identification with the definition of the trace-reverse (\ref{1}) and our expression of the equations of motion (\ref{edf}), we arrive at the following result for the Newtonian limit of $d$-dimensional general relativity:
\begin{align}\frac{\mathrm{d}^{2}x^{i}}{\mathrm{d}t^{2}}+\frac{2(d-3)}{d-2}\partial_{i}\Phi=0\end{align}
In the case $d=4$, this expression just reduces to the usual Newtonian equation for gravity for a test particle with unit mass $m$, i.e. $\frac{\mathrm{d}^{2}\vec{x}}{\mathrm{d}t^{2}}=-\frac{1}{m}\vec{\nabla}\Phi=\frac{1}{m}\vec{F}_{\mathrm{grav.}}$. In the case $d=3$, we again encounter something special. The Newtonian limit of $3$-dimensional gravity shows that \textit{test particles do not experience any force}. \cite[p.4f.]{CarlipQG}, \cite[p.3ff.]{GarciaExactSolutions3DGravity}

\subsection{Weyl Tensor and Vacuum Solutions}
In the last section, we have seen that $3$-dimensional general relativity is quite special. In the following, we will have a closer look at the vacuum solutions of the theory. To start with, let $(\mathcal{M},g)$ be a $d$-dimensional pseudo-Riemannian manifold with corresponding Levi-Civita connection $\nabla:\mathfrak{X}(\mathcal{M})\times\mathfrak{X}(\mathcal{M})\to\mathfrak{X}(\mathcal{M})$. Furthermore, let us denote by
\begin{align}\mathfrak{X}(\mathcal{M})^{4}\ni(X,Y,Z,W)\mapsto\operatorname{R}_{g}^{\mathrm{(0,4)}}(X,Y,Z,W):=g(\nabla_{X}\nabla_{Y}Z-\nabla_{Y}\nabla_{X}Z-\nabla_{[X,Y]}Z,W)\end{align}
the totally covariant Riemann curvature tensor, which is a $(0,4)$-tensor field on $\mathcal{M}$. In the case of vacuum and $\Lambda=0$, Einstein's field equations (\ref{EFE}) tell us that the Ricci tensor is zero. Since the Ricci tensor is by definition the trace of the curvature tensor, the gravitational field in vacuum (and $\Lambda=0$) is described by the trace-less part of the Riemann tensor. Mathematically, this is characterized by the so-called ``\textit{Weyl curvature tensor}''. Let us assume that $d\neq 1,2$, for definiteness. The Weyl tensor $C\in\Gamma^{\infty}(T^{\ast}\mathcal{M}^{\otimes 4})$ is globally given by \cite[p.215]{LeeRiemannianManifolds}
\begin{align}C:=\operatorname{R}_{g}^{\mathrm{(0,4)}}-\frac{1}{d
-2}\mathrm{Ric}_{g}\KN g+\frac{R_{g}}{2(d-2)(d-1)}g\KN g,\end{align}
where $\KN$ denotes the ``\textit{Kulkarni–Nomizu product}'', which is for two rank $(0,2)$-tensor fields $T,S\in\Gamma^{\infty}(T^{\ast}\mathcal{M}^{\otimes 2})$ defined by
\begin{equation}\begin{aligned}(T\KN S)(X,Y,Z,W):=& T(X,Z)S(Y,W)+T(Y,W)S(X,Z)\\&-T(X,W)S(Y,Z)-T(Y,Z)S(X,W)\end{aligned}\end{equation}
for all vector fields $X,Y,Z,W\in\mathfrak{X}(\mathcal{M})$. A straightforward calculation shows that the Weyl tensor is trace-free. Furthermore, the Weyl tensor $C$ has all the symmetry properties of the Riemann curvature tensor, e.g. antisymmetry in the first and second pair of indices, symmetry in interchanging the first with the second pair of indices and it satisfies the first Bianchi identity. Choosing a local coordinate chart $(U,\varphi=\{x^{\mu}\})$ of $\mathcal{M}$ with $U\subset\mathcal{M}$ open, it is straightforward to show that the coordinates $C_{\alpha\beta\gamma\delta}:=C(\partial_{\alpha},\partial_{\beta},\partial_{\gamma},\partial_{\delta})$ of $C$ are given by\footnote{We use the convention that the anti-symmetrization symbol $[\dots]$ comes with a normalization, i.e. we write $T_{[\alpha\beta]}:=\frac{1}{2!}(T_{\alpha\beta}-T_{\beta\alpha})$ for some $T\in\Gamma^{\infty}(T^{\ast}\mathcal{M}^{\otimes 2})$. Hence, $(T\KN S)_{\alpha\beta}=2(T_{\alpha[\gamma}S_{\delta]\beta}+T_{\beta[\delta}S_{\gamma]\alpha})$.}
\begin{align}\label{Weyl}C_{\alpha\beta\gamma\delta}=R_{\alpha\beta\gamma\delta}-\frac{2}{d-2}\big(g_{\gamma [\alpha}R_{\beta]\delta}+R_{\gamma [\alpha}g_{\beta]\delta}\big)+\frac{2}{(d-1)(d-2)}R_{g}g_{\gamma [\alpha}g_{\beta]\delta}.\end{align}
As a next step, let us count the number of independent components of the Riemann curvature tensor in arbitrary dimensions: In total, the curvature tensor has $d^{4}$ components. However, we also have to take the symmetry properties of the Riemann tensor into account:
\begin{itemize}
\item[(1)]The first pair of indices can take $d(d-1)/2$ independent configurations, since the Riemann tensor is antisymmetric in interchanging the first two indices. Similarly, antisymmetry in the second pair of indices again leads to $d(d-1)/2$ independent choices. Let us use the notation $n:=d(d-1)/2$. Using symmetry under changing the first and second pair of indices, we are left with a total of $n(n+1)/2$ independent components.
\item[(2)]Secondly, we have to take into account the first Bianchi identity, i.e. $R_{\alpha(\beta\gamma\delta)}=0$. Note that whenever two indices are equivalent, the first Bianchi identity is trivially fulfilled. Furthermore, reshuffling the indices obviously yields the same constraint. Therefore, we have to subtract the number of components by $\binom{d}{4}=\frac{d!}{4!(d-4)!}$.
\end{itemize}
In total, the number of independent components of the Riemann curvature tensor is hence
\begin{align}\frac{n(n+1)}{2}-\frac{d!}{4!(d-4)!}=\frac{d^{2}(d^{2}-1)}{12}.\end{align}
The Ricci tensor is a symmetric $(0,2)$-tensor field and as such it has a total number of $d(d+1)/2$ independent components. Comparing the number of independent components of the Riemann tensor and the Ricci tensor, we see that they coincide in the case $d=3$. As a consequence, the Weyl tensor is identically zero in the $3$-dimensional case. Since the Weyl tensor exactly describes the gravitational field in vacuum, once again we come to the conclusion that there are \textit{no gravitational degrees of freedom in three dimensions}. In other words, \textit{gravity is not propagating and there are no gravitational waves in three dimensions}.\\
\\
The fact that the Weyl tensor is zero in three dimensions is a generic fact for 3-dimensional pseudo-Riemannian manifolds. As a next step, we apply the equations of motion of general relativity (\ref{EFE}), in order to analyse the Riemann curvature tensor on-shell. Using Formula (\ref{Weyl}), the Riemann curvature tensor in the case $d=3$ can be written as
\begin{align}\label{riemann}R_{\alpha\beta\gamma\delta}=g_{\alpha\gamma}R_{\beta\delta}+g_{\beta\delta}R_{\alpha\gamma}-g_{\beta\gamma}R_{\alpha\delta}-g_{\alpha\delta}R_{\beta\gamma}+\frac{1}{2}R_{g}\big (g_{\beta\gamma}g_{\alpha\delta}-g_{\alpha\gamma}g_{\beta\delta}\big ).\end{align}
Using Einstein's field equations for $3$d gravity and their trace, i.e. the formulas $R_{\mu\nu}=2\Lambda g_{\mu\nu}$ and $R_{g}=6\Lambda$, we can rewrite Formula (\ref{riemann}) as
\begin{align}R_{\alpha\beta\gamma\delta}=\Lambda\big (g_{\alpha\gamma}g_{\beta\delta}-g_{\alpha\delta}g_{\beta\gamma}\big ),\end{align}
or in coordinate-free notation, $R_{g}^{(0,4)}=\frac{\Lambda}{2}g\KN g$. This is exactly the generic form of the curvature tensor for ``\textit{constant curvature spaces}''. Therefore, \textit{$3$-dimensional general relativity is a theory of constant curvature}, which means that it has the following solutions in vacuum:
\begin{itemize}
\item[(1)]For $\Lambda>0$ the solutions of Einstein's equations are locally de Sitter space $\mathrm{dS}_{3}$.
\item[(2)]For $\Lambda=0$ the solutions of Einstein's equations are locally flat.
\item[(3)]For $\Lambda<0$ the solutions of Einstein's equations are locally anti-de Sitter space $\mathrm{AdS}_{3}$.
\end{itemize}
However, it is important to note that $3$-dimensional gravity is not \textit{globally} trivial, as there are for example local defects (``particles''), black hole solutions (``BTZ black holes'' \cite{BTZ}) in $3$-dimensional anti-de Sitter space as well as asymptotic symmetries \cite{Brown}. \cite[p.36ff.]{CompereBook}

\subsection{The Triadic Palatini-Formalism of General Relativity}
Previously, we reviewed the Einstein-Hilbert action of general relativity, firstly proposed by D. Hilbert in 1915 \cite{HilbertAction}. However, there are also other variational approaches to gravity, using different types of variables. One of them is the so-called ``\textit{Palatini formalism}'', in which one uses both the metric and the affine connection as independent variables in the action principle. This method is usually credited to A. Palatini \cite{PalatiniFormalism}, although it seems that the historical development was non-linear\footnote{As discussed in \cite{PalatiniHistory}, the original paper by A. Palatini from 1919 is quite far from what is usually meant by ``Palatini formalism'' nowadays and it seems that A. Einstein's work from 1925 \cite{EinsteinPalatini1925} is more suitable to regard as the origin of this variational method.}. It is a first-order formalism, meaning that the objects to vary only contain up to first derivatives. Yet another and related first-order formalism for gravity is the ``\textit{triadic Palatini formalism}'' in three dimensions, or ``\textit{tetradic Palatini formalism}'' in four dimensions, in which one uses ``frame fields'' and the ``spin connection'' as independent variables. This approach is for example necessary if we want to couple fermions to gravity and in that sense it seems to be more natural. Furthermore, it is also the starting point for many background-independent approaches to quantum gravity, as well as for Einstein-Cartan-Sciama-Kibble-type theories of gravity with torsion and gauge-theoretic formulations of general relativity. In the following, we review the main ideas and the mathematics of this formalism in more details.\\
\\
In the following, we will use the language of mathematical gauge theory, which is reviewed in Appendix \ref{gaugetheory}. General references for the following discussion are \cite{BaezKnots,Cat2,Cat1,MathCoFrame,WiseThesis}. Let $\mathcal{M}$ be some orientable $d$-dimensional manifold, which admits Lorentzian metrics, however, we a priori do not equip it with such a metric. Now, let $\mathcal{T}$ be an oriented vector bundle, which is isomorphic to the tangent bundle $T\mathcal{M}$ via a vector bundle isomorphism
\begin{align}e:T\mathcal{M}\to\mathcal{T}\in\Omega^{1}(\mathcal{M},\mathcal{T}),\end{align}
called ``\textit{co-frame field}'' and which is equipped with some fixed Lorentzian bundle metric. Next, let us denote by $P:=\mathcal{F}_{\mathrm{Ort}}(T\mathcal{M})$ the bundle of orthonormal frames, which is the principal $\mathrm{SO}(1,d-1)$-bundle whose fibres at some point $p\in\mathcal{M}$ are given by
\begin{align}P_{p}:=\{\{v_{\mu}\}_{\mu=0,\dots,d-1}\subset T_{p}\mathcal{M}\mid\text{ basis of }T_{p}\mathcal{M}\text{ and }g_{p}(v_{\mu},v_{\nu})=\eta_{\mu\nu}\},\end{align}
where the metric $g$ on $\mathcal{M}$ is induced via the isomorphism $e$ by the Lorentzian metric defined on $\mathcal{T}$ and where $\eta=\mathrm{diag}(1,-1,\dots,-1)$ denotes the $d$-dimensional Minkowski metric. Using the isomorphism $e$, we find an isomorphism of vector bundles
\begin{align}\mathcal{T}\cong P\times_{\rho}\mathbb{M}^{d},\end{align}
where $(\mathbb{M}^{d},\eta)$ denotes the $d$-dimensional Minkowski space and where $\rho:\mathrm{SO}(1,d-1)\to\mathrm{Aut}(\mathbb{M}^{d})$ denotes the fundamental representation. Now, the main object of interest in the following is the form
\begin{align}\underbrace{e\wedge\dots\wedge e}_{(d-2)-\text{times}}\in\Omega^{d-2}\bigg(\mathcal{M},{\bigwedge}^{d-2}\mathcal{T}\bigg).\end{align}
Using the explicit description of $\mathcal{T}$ above, we can view this form equivalently as a form in $\Omega^{d-2}(\mathcal{M},\mathrm{Ad}(P))$, where $\mathrm{Ad}(P):=P\times_{\mathrm{Ad}}\mathfrak{so}(1,d-1)$ denotes the adjoint bundle \cite[p.61f.]{WiseThesis}. As a second ingredient, we choose a connection $1$-form $\omega\in\Omega^{1}(P,\mathfrak{so}(1,d-1))$, called the ``\textit{spin connection}''. The corresponding curvature $F[\omega]:=\mathrm{d}\omega+\frac{1}{2}[\omega\wedge\omega]\in\Omega^{2}(P,\mathfrak{so}(1,d-1))$ can as usual equivalently be viewed as an element in $\Omega^{2}(\mathcal{M},\mathrm{Ad}(P))$. The action of gravity in the first-order formalism (without a cosmological constant) is then defined to be
\begin{align}\label{ActionTriad}\mathcal{S}[e,\omega]:=\frac{1}{2\kappa}\int_{\mathcal{M}}\mathrm{tr}(\underbrace{(e\wedge\dots\wedge e)}_{(d-2)-\text{times}}\wedge F[\omega]),\end{align}
where $\mathrm{tr}(\cdot\wedge\cdot):\Omega^{d-2}(\mathcal{M},\mathrm{Ad}(P))\times\Omega^{2}(\mathcal{M},\mathrm{Ad}(P))\to\Omega^{d}(\mathcal{M})$ denotes the wedge product, which is induced by the bundle metric on $\mathrm{Ad}(P)$, which in turn is induced by the Killing form on $\mathrm{SO}(1,d-1)$, which is also the reason for the notation ``$\mathrm{tr}$''\footnote{The Killing form of the matrix Lie algebra $\mathfrak{so}(p,q)$ is given by $B(X,Y):=\mathrm{tr}(\mathrm{ad}_{X}\circ\mathrm{ad}_{Y})=(p+q-2)\mathrm{tr}(XY)$ for all $X,Y\in\mathfrak{so}(p,q)$, e.g. see \cite[p.192]{Baum}}. The domain of the action is $\Omega^{1}_{\mathrm{nd}}(T\mathcal{M},\mathcal{T})\times\mathcal{C}(P)$, where $\Omega^{1}_{\mathrm{nd}}(T\mathcal{M},\mathcal{T})$ denotes the set of non-degenerate $1$-forms, or equivalently, the set of bundle isomorphism from $T\mathcal{M}$ to $\mathcal{T}$ and where $\mathcal{C}(P)$ denotes the affine space of connection $1$-forms of $P$. However, a priori, we can extend the action also to non-degenerate forms, i.e. to maps $e:T\mathcal{M}\to\mathcal{T}$ which are non necessarily bundle isomorphism, but just homomorphism. This provides a generalization of the Einstein-Hilbert action, as we will explain below.\\
\\
In order to explain the relation between the first-order formalism discussed above with the Einstein-Hilbert action, it is convenient to work in coordinates. Furthermore, let us restrict now to dimension $d=3$. If $\mathcal{M}$ is parallelizable, we can identify a co-frame field with $e^{a}v_{a}\in\Omega^{1}(\mathcal{M},\mathbb{M}^{3})$ such that $e^{a}=\Omega^{1}(\mathcal{M})$ and such that $\{v_{a}\}$ is an orthonormal basis of $\mathbb{M}^{3}$. As a consequence, the elements $e^{a}$ can locally in some chart $(U,\varphi=\{x^{\mu}\})$ of $\mathcal{M}$ be written as $e^{a}=e^{a}_{\mu}\mathrm{d}x^{\mu}$. Furthermore, by definition of the induced metric $g$, they fulfil the relation:
\begin{align}\eta_{ab}e_{\mu}^{a}e^{b}_{\nu}=g_{\mu\nu}.\end{align}
Next, we can also write the curvature in coordinates, i.e. $F^{ab}[\omega]:=\mathrm{d}\omega^{ab}+{\omega^{a}}_{c}\wedge\omega^{cb}$, where $\omega^{ab}$ are the local connection forms. A straightforward calculation shows that the curvature is directly related to the Ricci scalar $R_{g}$ of the metric $g_{\mu\nu}=\eta_{ab}e_{\mu}^{a}e^{b}_{\nu}$ via
\begin{align}R_{g}=e_{a}^{\mu}e_{b}^{\nu}F^{ab}_{\mu\nu},\end{align}
where $F^{ab}=F^{ab}_{\mu\nu}\mathrm{d}x^{\mu}\wedge\mathrm{d}x^{\nu}$. Last but not least, observe that 
\begin{align}\mathrm{tr}(e\wedge F[\omega])=\varepsilon_{abc}e^{a}\wedge F^{bc}[\omega]=e_{a}^{\mu}e_{b}^{\nu}F^{ab}_{\mu\nu}\vert e\vert\mathrm{d}^{4}x=R_{g}\sqrt{-g}\mathrm{d}^{4}x,\end{align}
where $\varepsilon_{abc}$ denotes the Levi-Civita pseudotensor density and where $\vert e\vert$ denotes the determinant $\sqrt{-g}$ written in terms of the cotriads. Hence, we have established the relation between the first-order formalism and the Einstein-Hilbert action. To sum up, the triadic action in the $3$-dimensional case can be written as
\begin{align}\label{triad3d}\mathcal{S}[e,\omega]=\frac{1}{2\kappa}\int_{\mathcal{M}}\!\operatorname{tr}(e\wedge F[\omega])=\frac{1}{2\kappa}\int_{\mathcal{M}}\!\varepsilon_{abc}e^{a}\wedge F^{bc}[\omega].\end{align}
Since we are working in dimension three, we can also introduce a (local) Lorentz vector via $\omega^{a}:=\frac{1}{2}\varepsilon_{abc}\omega^{bc}$ in order to enforce antisymmetry of the spin-connection. A straightforward calculation yields $\frac{1}{2}\varepsilon_{abc}F^{ab}[\omega]=\mathrm{d}\omega_{a}+\frac{1}{2}\varepsilon_{abc}\omega^{b}\wedge\omega^{c}$ and hence, we can equivalently write
\begin{align}\label{triad3d2}\mathcal{S}[e,\omega]=\frac{1}{2\kappa}\int_{\mathcal{M}}2e^{a}\wedge \bigg(\mathrm{d}\omega_{a}+\frac{1}{2}\varepsilon_{abc}\omega^{b}\wedge\omega^{c}\bigg).\end{align} 
The Euler-Lagrange equations of the triadic Palatini action in three dimensions are given by
\begin{align}\mathrm{d}_{\omega}e=0\hspace{1cm}\text{and}\hspace{1cm}F[\omega]=0.\end{align}
If $e$ is non-degenerate, then the first equation tells us that the connection $\nabla$ on $T\mathcal{M}$ induced by $\omega$ via the isomorphism $e$ is torsion-free and hence given by the unique Levi-Civita connection of the induced metric $g$. The second equation is then equivalent to Einstein's field equation in vacuum with $\Lambda=0$ on-shell of the first equation. In order to get also the cosmological constant term we have to add a term proportional to $\mathrm{tr}(e\wedge e\wedge e)$ in the triadic Palatini action (\ref{ActionTriad}), or in coordinates, $\varepsilon_{abc}e^{a}\wedge e^{b}\wedge e^{c}$ to the actions (\ref{triad3d}) and (\ref{triad3d2}). As already mentioned, in the discussion of the triadic Palatini action it is natural to allow also for degenerate triads. In this case, we get some additional field configurations, which are not present in the Einstein-Hilbert approach. At the classical level, these additional configurations might not matter, but they do at the quantum level. Since the triadic Palatini formalism is necessary to couple fermions to gravity, one often argues that it is the more natural one to describe gravity \cite[p.34]{RovelliQG}.

\section{Topological Field Theory and 3D Gravity}
As discussed in the last section, $3$-dimensional gravity has no local degrees of freedom and hence is an example of a ``topological field theory''. These are certain types of (quantum) field theories, where observables do not depend on a chosen spacetime metric and hence do compute topological invariants. They play an important role in both physics, for example in condensed matter physics and string theory, as well as in pure mathematics, like in knot theory and in the theory of $4$-manifold. Furthermore, M. Atiyah formulated, based on previous axiomatic approaches to conformal field theory by G. Segal, a mathematical precise axiomatic formulation in the language of category theory \cite{AtiyahTFT}. Since then, TFT has become a very rich research direction with many results and applications. In this section, we discuss two important and well-known classes of such field theories and show how they are related to gravity in three dimensions. Early works on the topological nature of $3$-dimensional gravity are \cite{Deser1,Deser2}.

\subsection{Topological BF-Theories and 3D Gravity}\label{BF}
A very important class of topological field theories are $BF$-theories. They are free of any choice of particular background, i.e. we do not need to use any preexisting metric on our spacetime manifold. As a consequence, they are an example of TFTs of ``\textit{Schwarz type}'', in which the action functional is inherently metric-independent. $BF$-theories are also the only known examples of topological field theories, which can be defined in any dimension. Furthermore, they can be viewed as the simplest possible gauge theories in some sense. As we will see in this section, $3$-dimensional general relativity is in fact an example of a $BF$-theory. Furthermore, also $4$-dimensional general relativity can be viewed as a $BF$-theory, when adding additional constraints. Historically, $BF$-theories were firstly discussed by G. Horowitz \cite{HorowitzBFTheory} as a generalization of E. Witten's work on $3$-dimensional gravity \cite{WittenChernSimons}. General references for this section are \cite{BaezBFTheory,BFTheory,WiseThesis}. For a short discussion about some relevant definitions of mathematical gauge theory, see Appendix \ref{gaugetheory}.\\
\\
Let $G$ be a Lie group with Lie algebra $\mathfrak{g}$ and let $\pi:P\to\mathcal{M}$ be a principal $G$-bundle over a smooth orientable manifold $\mathcal{M}$. Furthermore, let $\langle\cdot,\cdot\rangle_{\mathfrak{g}}$ be an $\mathrm{Ad}$-invariant, non-degenerate and symmetric bilinear form on $\mathfrak{g}$. A generic $d$-dimensional $BF$-theory is defined using the following two types of fields:
\begin{itemize}
\item[(1)]A connection 1-form $A\in\Omega^{1}(P,\mathfrak{g})$.
\item[(2)]An $\mathrm{Ad}(P)$-valued $(d-2)$-form $B\in\Omega^{d-2}(\mathcal{M},\mathrm{Ad}(P))$, where $\mathrm{Ad}(P):=P\times_{\mathrm{Ad}}\mathfrak{g}$ denotes the adjoint bundle.
\end{itemize}
Recall that the curvature $F[A]:=\mathrm{d}A+\frac{1}{2}[A\wedge A]\in\Omega^{2}(P,\mathfrak{g})$ is horizontal and of type $\mathrm{Ad}$ and hence, it can be viewed as an element of $\Omega^{2}(\mathcal{M},\mathrm{Ad}(P))$. The action of $BF$-theory is then defined by
\begin{align}\mathcal{S}_{\mathrm{BF}}[B,A]:=\int_{\mathcal{M}}\! \operatorname{tr}(B\wedge F[A]),\end{align}
where $\mathrm{tr}(\cdot\wedge\cdot):\Omega^{k}(\mathcal{M},\mathrm{Ad}(P))\times\Omega^{l}(\mathcal{M},\mathrm{Ad}(P))\to\Omega^{k+l}(\mathcal{M})$ is the wedge-product induced by the bundle metric $\langle\cdot,\cdot\rangle_{\mathrm{Ad}(P)}\in\Gamma^{\infty}(\mathrm{Ad}(P)^{\ast}\otimes\mathrm{Ad}(P)^{\ast})$, which in turn is induced by the bilinear form $\langle\cdot,\cdot\rangle_{\mathfrak{g}}$ on the Lie algebra $\mathfrak{g}$. If $\langle\cdot,\cdot\rangle_{\mathfrak{g}}$ is positive definite and $G$ simple and compact, then $\langle\cdot,\cdot\rangle_{\mathfrak{g}}$ is necessarily a negative multiple of the Killing form of $\mathfrak{g}$ \cite[p.118f.]{Hamilton}, which is also the reason why we use
 the notation ``$\mathrm{tr}$'' above. The reason why these type of theories are called ``$BF$-theories'' is obvious from the structure of the integrand in the action. Alternatively, one can also interpret the name as coming from ``\textit{background free}''. The equations of motion of $BF$-theory can be derived by varying the action, which yields
\begin{equation}\begin{aligned}0\stackrel{!}{=}\delta \mathcal{S}_{\mathrm{BF}}[B,A]&=\int_{\mathcal{M}}\! \operatorname{tr}(\delta B\wedge F[A]+B\wedge \mathrm{d}_{A}\delta A)=\\&=\int_{\mathcal{M}}\! \operatorname{tr}(\delta B\wedge F[A]+(-1)^{d-1}\mathrm{d}_{A}B\wedge\delta A),\end{aligned}\end{equation}
where we used that $\delta F[A]=\mathrm{d}_{A}\delta A$ as well as integration by parts in the last step. A more rigorous derivation can be found in Appendix \ref{GaugeSymmetriesBF}. Following this, the equations of motion are given by 
\begin{align}F[A]=\mathrm{d}A+\frac{1}{2}[A\wedge A]=0\hspace{1cm}\text{and}\hspace{1cm}\mathrm{d}_{A}B=0.\end{align}
The first one tells us that the connection $A$ is flat, or in physical terms, that the field strength corresponding to the gauge field $A$ vanishes. The interpretation of the second equation is more subtle and explained below. The action of $BF$-theory is clearly gauge-invariant: Let $f\in\mathcal{G}(P)$ be a gauge transformation, i.e. a principal bundle automorphism. Then the fields transform as
\begin{align}A\mapsto f^{\ast}A=\mathrm{Ad}_{\sigma_{f}^{-1}}\circ A+\sigma_{f}^{\ast}\mu_{G}\hspace{1cm}\text{and}\hspace{1cm}B\mapsto f^{\ast}B=\mathrm{Ad}_{\sigma_{f}^{-1}}\circ B,\end{align}
where $\sigma_{f}\in C^{\infty}(P,G)^{G}$ is the map defined by $f(p):=p\cdot\sigma_{f}(p)$ for all $p\in P$ and where $\mu_{G}\in\Omega^{1}(G,\mathfrak{g})$ denotes the Maurer-Cartan form on $G$. Here we have viewed $B$ as an element of $\Omega^{d-2}(P,\mathfrak{g})$ via the isomorphism $\Omega^{d-2}_{\mathrm{hor}}(P,\mathfrak{g})^{\mathrm{Ad}}\cong\Omega^{d-2}(\mathcal{M},\mathrm{Ad}(P))$. As a consequence of the first transformation rule above, the curvature transforms as
\begin{align}F[A]\mapsto \mathrm{Ad}_{\sigma_{f}^{-1}}\circ F[A].\end{align}
Using this, it is clear that the $BF$-action is invariant under gauge transformations, since $\langle\cdot,\cdot\rangle_{\mathfrak{g}}$ is $\mathrm{Ad}$-invariant. To describe the gauge transformations more explicitly, recall that we can define a local gauge field via $A_{s}:=s^{\ast}A\in\Omega^{1}(U,\mathfrak{g})$, where $s:U\to P$ denotes some local gauge defined on an open subset $U\subset\mathcal{M}$. Furthermore, we can think of $B$ as a $\mathfrak{g}$-valued $(d-2)$-form on $U$, by defining $B_{s}:=s^{\ast}B\in\Omega^{d-2}(U,\mathfrak{g})$. In particular, if $s_{i}:U_{i}\to P$ and $s_{j}:U_{j}\to P$ are two local gauges and if $G$ is a matrix Lie group, then the gauge transformations are given by
\begin{align}A_{s_{i}}=g_{ji}^{-1}\cdot A_{s_{j}}\cdot g_{ji}+g_{ji}^{-1}\cdot\mathrm{d}g_{ji}\hspace{1cm}\text{and}\hspace{1cm}B_{s_{i}}=g_{ji}^{-1}\cdot B_{s_{j}}\cdot g_{ji},\end{align}
where $g_{ji}:U_{i}\cap U_{j}\to G$ is such that $s_{i}(x)=s_{j}(x)\cdot g_{ji}(x)$ for all $x\in U_{i}\cap U_{j}$. Furthermore, it turns out that $BF$-theories also has yet another symmetry, which can be interpreted as some kind of translational invariance. To be precise, the $BF$-action is invariant under the following transformations:
\begin{align}A\mapsto A\hspace*{1cm}\text{and}\hspace*{1cm}B\mapsto B+\mathrm{d}_{A}\eta,\end{align}
where $\eta\in\Omega^{d-3}(\mathcal{M},\mathrm{Ad}(P))$. This is basically a consequence of the Bianchi identity $\mathrm{d}_{A}F[A]=0$, as the following short calculation shows:
\begin{equation}\begin{aligned}\mathcal{S}_{\mathrm{BF}}[B,A]\mapsto\int_{\mathcal{M}}\,\operatorname{tr}((B+\mathrm{d}_{A}\eta)\wedge F[A])=\int_{\mathcal{M}}\,\operatorname{tr}(B\wedge F[A]-\mathrm{d}_{A}F[A]\wedge \eta)=\mathcal{S}_{\mathrm{BF}}[B,A],\end{aligned}\end{equation}
where we used integration by parts in the last to last step. If $A$ is flat, then we can write any $B\in\Omega^{d-2}(\mathcal{M},\mathrm{Ad}(P))$ satisfying $\mathrm{d}_{A}B=0$ locally as $B=\mathrm{d}_{A}\eta$ for some $\eta\in\Omega^{d-3}(\mathcal{M},\mathrm{Ad}(P))$. As a consequence, all solutions of the equations of motion of $BF$-theory are locally equivalent up to gauge transformations. For a more detailed discussion of the gauge transformations together with rigorous proofs see Appendix \ref{GaugeSymmetriesBF}.\\
\\
It is immediate from our previous discussion that $3$-dimensional gravity in the triadic Palatini formalism (Equation (\ref{triad3d})) is a particular case of a $BF$-theory, where $G$ is given by $\mathrm{SU}(2)$ (or alternatively $\mathrm{SO}(3)$\footnote{The choice of taking $\mathrm{SO}(3)$ or its double cover $\mathrm{SU}(2)$ (and similarly $\mathrm{SO}(1,2)$ or $\mathrm{Sl}(2,\mathbb{R})$ in the Lorentzian case) does not affect the classical theory, but it does matter at the quantum level. If we want couple fermions to gravity, we have to use the double cover and hence, one often argues that this is a more natural choice \cite{BaezBFTheory}.}) in the Riemannian case and by $\mathrm{SO}(1,2)$ in the Lorentzian case and where the co-triad $e$ takes the role of the $B$-field above. Furthermore, one can show that the two gauge transformations of $BF$-theory are related to diffeomorphism invariance of gravity. More precisely, the action of diffeomorphisms is on-shell recovered by a combination of the gauge transformations discussed above \cite[p.28f.]{GFTDiff2}, \cite[p.2f.]{FreidelDiffeo}. In four dimensions, the situation is different, since not every $\mathrm{Ad}(P)$-valued $2$-form can be written as $e\wedge e$ for some $1$-form $e$. As a consequence, one has to add extra constraints forcing $B$ to be of the form $e\wedge e$. 

\subsection{Chern-Simons Theory and 3D Gravity}\label{CS}
In the previous section, we have discussed topological $BF$-theories and have seen that $3$-dimensional general relativity in the first order formalism can be understood as a particular case of such theories. The goal of this section is to show that $3$-dimensional gravity is also related to yet another very important type of topological field theories of Schwarz type, namely to Chern-Simons theories, as firstly observed in  \cite{ChernSimonsGravity}. This is a very interesting observation, because the quantization of Chern-Simons theory is well understood and therefore can be used to define non-perturbative $3$-dimensional quantum gravity. This approach was firstly discussed in \cite{WittenChernSimons}. A more recent review by E. Witten on the subject is \cite{Witten2}. General references for the mathematics of Chern-Simons theories are \cite{FreedChernSimons1,FreedChernSimons2}. For the relation to gravity, see for example \cite{CarlipQG,CompereBook,Donnay,KiranChernSimons}.\\
\\
We start by discussing the general setting and framework. Let $G$ be a Lie group with Lie algebra $\mathfrak{g}$ and let $\pi:P\to\mathcal{M}$ be a principal $G$-bundle over a smooth orientable manifold $\mathcal{M}$. Furthermore, let $\langle\cdot,\cdot\rangle_{\mathfrak{g}}$ be an $\mathrm{Ad}$-invariant, non-degenerate and symmetric bilinear form on $\mathfrak{g}$. Chern-Simons theory is a topological field theory defined using the ``Chern-Simons $3$-form'' 
\begin{align}\mathrm{CS}[A]:=\operatorname{tr}(A\wedge F[A])-\frac{1}{6}\mathrm{tr}(A\wedge [A\wedge A])=\operatorname{tr}(A\wedge\mathrm{d}A)+\frac{1}{3}\mathrm{tr}(A\wedge [A\wedge A])\in\Omega^{3}(P),\end{align}
introduced in \cite{ChernSimonsForm}, which has the defining property $\mathrm{d}\mathrm{CS}[A]=\mathrm{tr}(F[A]\wedge F[A])$, where $A\in\Omega^{1}(P,\mathfrak{g})$ is a connection $1$-form and where $\operatorname{tr}(\cdot\wedge\cdot)$ denotes the induced wedge-product defined using the inner product $\langle\cdot,\cdot\rangle_{\mathfrak{g}}$ on $\mathfrak{g}$, i.e.
\begin{align}\mathrm{tr}(\omega\wedge\eta):=\sum_{a,b=1}^{\mathrm{dim}(G)}(\omega^{a}\wedge\eta^{b})\langle T_{a},T_{b}\rangle\in\Omega^{k+l}(P)\end{align}
for all $\omega\in\Omega^{k}(P,\mathfrak{g})$ and for all $\eta\in\Omega^{l}(P,\mathfrak{g})$, where $\{T_{a}\}_{a}$ denotes a basis of $\mathfrak{g}$ and where $\omega^{a}\in\Omega^{k}(P)$ and $\eta^{b}\in\Omega^{l}(P)$ denote the coordinate forms with respect to this basis. One can easily verify that this definition is independent of the choice of basis. Note that if $G$ is compact and simple and $\langle\cdot,\cdot\rangle_{\mathfrak{g}}$ positive definite, then $\langle\cdot,\cdot\rangle_{\mathfrak{g}}$ is necessarily a negative multiple of the Killing form of $\mathfrak{g}$, which is also the reason why we use the notation ``$\mathrm{tr}$'' above. As a next step, recall that the bundle $P$ is trivial if and only if it admits a smooth global section. In particular, this is the case if $G$ is compact and simply-connected and if $\mathcal{M}$ has dimension $\leq 3$ \cite[p.14]{FreedChernSimons1}. The ``Chern-Simons action'' is then defined by 
\begin{align}\mathcal{S}_{\mathrm{CS}}[s,A]:=\frac{k}{4\pi}\int_{\mathcal{M}}\,s^{\ast}\mathrm{CS}[A],\end{align}
where $s:\mathcal{M}\to P$ is a global gauge and where $k\in\mathbb{C}$ denotes some constant, called the ``\textit{level}'' of the theory. At this point, the definition clearly depends on the choice of global gauge. Let $f\in\mathcal{G}(P)$ be a gauge transformation. Then, after a straightforward calculation, one finds that
\begin{align}\mathcal{S}_{\mathrm{CS}}[f\circ s,A]-\mathcal{S}_{\mathrm{CS}}[s,A]=\mathcal{S}_{\mathrm{CS}}[s,f^{\ast}A]-\mathcal{S}_{\mathrm{CS}}[s,A]=-\frac{k}{24\pi}\int_{\mathcal{M}}\,s^{\ast}\mathrm{tr}(\theta\wedge [\theta\wedge\theta])\end{align}
with $\theta:=\sigma_{f}^{\ast}\mu_{G}$, where $\mu_{G}\in\Omega^{1}(G,\mathfrak{g})$ denotes the Maurer-Cartan form on $G$ and where $\sigma_{f}\in C^{\infty}(P,G)^{G}$ is the map defined by $f(p)=p\cdot\sigma_{f}(p)$ for all $p\in P$. Let us now assume that $\langle\cdot,\cdot\rangle_{\mathfrak{g}}$ is normalized such that the closed $3$-form $-\frac{k}{24\pi}\mathrm{tr}(\mu_{G}\wedge [\mu_{G}\wedge\mu_{G}])\in\Omega^{3}(G)$ represents an integral cohomology class in $H^{3}(G;\mathbb{R})$. If this is the case, we see that the Chern-Simons action $S_{\mathrm{CS}}[s,A]$ is independent of the choice of global gauge modulo $\mathbb{Z}$. Hence, we have a well-defined gauge-invariant action $\mathcal{S}_{\mathrm{CS}}:\mathcal{C}(P)\to\mathbb{C}/\mathbb{Z}$, where $\mathcal{C}(P)\subset\Omega^{1}(P,\mathfrak{g})$ denotes the set of connection $1$-forms on $P$, which is an infinite-dimensional affine space over $\Omega^{1}_{\mathrm{hor}}(P,\mathfrak{g})^{\mathrm{Ad}}$, defined by
\begin{align}\mathcal{S}_{\mathrm{CS}}[A]:=\{S_{\mathrm{CS}}[s,A]\mid s\in\Gamma^{\infty}(\mathcal{M},P)\}\in\mathbb{C}/\mathbb{Z}\end{align}
for all $A\in\mathcal{C}(P)$. The equations of motion of Chern-Simons theory can easily be derived and are given by
\begin{align}F[A]=\mathrm{d}A+\frac{1}{2}[A\wedge A]=0.\end{align} 
As already mentioned, $3$-dimensional gravity formulated in the first-order formalism can be understood as a Chern-Simons theory defined with the corresponding isometry group of spacetime. We will present this relation in the following for a Lorentzian signature. The Riemannian case is completely analogues. To start with, let us consider the $(1+2)$-dimensional Poincaré group $\mathrm{Iso}(1,2):=\mathbb{M}^{3}\rtimes O(1,2)$, where $(\mathbb{M}^{3},\eta)$ denotes the Minkowski space with $\eta=\mathrm{diag}(1,-1,-1)$. Furthermore, let us choose generators $\{P_{a}\}$ of the translations and generators $\{J_{a}\}$ of the Lorentz group. Then the commutators of the Lie algebra $\mathfrak{iso}(1,2)$ are given by
\begin{align}
[J^{a},J^{b}]&=\varepsilon^{abc}J_{c} \\
[J^{a},P^{b}]&=\varepsilon^{abc}P_{c} \\
[P^{a},P^{b}]&=0.
\end{align}
A non-degenerate and $\mathrm{Ad}$-invariant symmetric bilinear form in this Lie algebra\footnote{There is also another bilinear form one can consider. See for example the original article \cite{WittenChernSimons} for a discussion.} is then given by $\operatorname{tr}(J^{a}P^{b})=\eta^{ab}$ and $\operatorname{tr}(J^{a}J^{b})=\operatorname{tr}(P^{a}P^{b})=0$. Now recall that $3$-dimensional gravity without cosmological constant in the first order formalism is given by
\begin{align}\label{3dAction}\mathcal{S}_{\mathrm{3d}}[e,\omega]=\frac{1}{2\kappa}\int_{\mathcal{M}}2e^{a}\wedge \bigg(\mathrm{d}\omega_{a}+\frac{1}{2}\varepsilon_{abc}\omega^{b}\wedge\omega^{c}\bigg),\end{align}
where $e^{a}=e^{a}_{\mu}\mathrm{d}x^{\mu}\in\Omega^{1}(U)$ denotes the co-triad and where $\omega^{a}=\frac{1}{2}\varepsilon^{abc}\omega_{bc}$ is defined using the spin-connection $\omega^{ab}=\omega^{ab}_{\mu}\mathrm{d}x^{\mu}\in\Omega^{1}(U)$ in some chart $(U,\varphi=\{x^{\mu}\})$ of $\mathcal{M}$. To compare this action with Chern-Simons theory, let us define a local connection $1$-form $A\in\Omega(U,\mathfrak{iso}(1,2))$ via 
\begin{align}A:=e^{a}P_{a}+\omega^{a}J_{a}.\end{align}
Using this connection $1$-form, a straightforward calculation shows that
\begin{align}\operatorname{tr}(A\wedge\mathrm{d}A)=e^{a}\wedge\mathrm{d}\omega_{a}+\omega^{a}\wedge\mathrm{d}e_{a}\hspace{0.4cm}\text{and}\hspace{0.4cm}\operatorname{tr}(A\wedge [A\wedge A])=6e^{a}\wedge \bigg (\frac{1}{2}\varepsilon_{abc}\omega^{b}\wedge\omega^{c}\bigg).\end{align}
Combining these two equations, we find the following expression for the Chern-Simons $3$-form corresponding to the connection $1$-form $A$:
\begin{align}\mathrm{CS}[A]=2e^{a}\wedge\bigg (\mathrm{d}\omega^{a}+\frac{1}{2}\varepsilon_{abc}\omega^{b}\wedge\omega^{c}\bigg )+\mathrm{d}(e^{a}\wedge\omega_{a}).\end{align}
As a consequence, $3$-dimensional gravity without a cosmological constant is up to a boundary term equivalent to a Chern-Simons theory of the Poincaré group $\mathrm{Iso}(1,2)$, using the $\mathfrak{iso}(1,2)$-valued connection form $A$ as defined above. To be precise, we have that 
\begin{align}\mathcal{S}_{\mathrm{3d}}[e,\omega]=\mathcal{S}_{\mathrm{CS}}[A(e,\omega)]+\int_{\partial\mathcal{M}}\,i^{\ast}(e^{a}\wedge\omega_{a}),\end{align}
where $i:\partial\mathcal{M}\to\mathcal{M}$ is the obvious embedding and where the level $k$ in the Chern-Simons action is identified with $k=c^{4}/4G$. In the case of $3$-dimensional gravity with a cosmological constant $\Lambda<0$, we have to add a corresponding volume term $\sqrt{-g}\mathrm{d}^{4}x=\frac{1}{3!}\varepsilon_{abc}e^{a}\wedge e^{b}\wedge e^{c}$ to the action (\ref{3dAction}), i.e.
\begin{align}\label{ed,coscons}\mathcal{S}_{\mathrm{3d},\Lambda}[e,\omega]=\frac{1}{2\kappa}\int_{\mathcal{M}}\bigg\{2e^{a}\wedge \bigg(\mathrm{d}\omega_{a}+\frac{1}{2}\varepsilon_{abc}\omega^{b}\wedge\omega^{c}\bigg)-\frac{\Lambda}{3}\varepsilon_{abc}e^{a}\wedge e^{b}\wedge e^{c}\bigg\}.\end{align}
In this case, we have to look at the corresponding isometry group of $3$-dimensional anti-de-Sitter space, which is given by $\mathrm{O}(2,2)$. Note that the Lie algebra of its connected component $\mathrm{SO}^{+}(2,2)$ is isomorphic to $\mathfrak{sl}(2,\mathbb{R})\times\mathfrak{sl}(2,\mathbb{R})$\footnote{On the level of Lie groups, there is an isomorphism $\mathrm{SO}^{+}(2,2)\cong(\mathrm{Sl}(2,\mathbb{R})\times\mathrm{Sl}(2,\mathbb{R}))/\mathbb{Z}_{2}$, since $\mathrm{Sl}(2,\mathbb{R})\times\mathrm{Sl}(2,\mathbb{R})$ is the double cover of $\mathrm{SO}^{+}(2,2)$, which is the identity component of the split orthogonal group $\mathrm{O}(2,2)$. In general, the double cover of $\mathrm{SO}^{+}(p,q)$ is by definition the spin group $\mathrm{Spin}(p,q)$ and in the case $p=q=2$, we have that $\mathrm{Spin}(2,2)\cong\mathrm{Sl}(2,\mathbb{R})\times\mathrm{Sl}(2,\mathbb{R})$. \cite[p.365ff.]{RudolphSchmidt2}}. The Lie algebra $\mathfrak{sl}(2,\mathbb{R})\cong\mathfrak{so}(1,2)$, can be described by
\begin{align}[T^{a},T^{b}]=\varepsilon^{abc}T_{c}.\end{align}
Let us define two local $\mathfrak{sl}(2,\mathbb{R})$-valued connection $1$-forms 
\begin{align}A_{\pm}:=A_{\pm}^{a}T_{a}:=\bigg (\omega^{a}\pm\frac{1}{\lambda}e^{a}\bigg )T_{a}\in\Omega^{1}(U,\mathfrak{sl}(2,\mathbb{R})),\end{align}
where we have parametrized $\Lambda=-1/\lambda^{2}$ for some $\lambda\in\mathbb{R}\textbackslash\{0\}$. As an $\mathrm{Ad}$-invariant non-degenerate symmetric bilinear form on $\mathfrak{sl}(2,\mathbb{R})$ we can use the Killing form, normalized as $\operatorname{tr}(T_{a}T_{b})=\frac{1}{2}\eta_{ab}$. Through a straightforward calculation as above, we get that 
\begin{equation}\begin{aligned}\mathrm{CS}[A_{+},A_{-}]:&=\mathrm{CS}[A_{+}]-\mathrm{CS}[A_{-}]=\\&=\frac{1}{\lambda}\bigg\{2e^{a}\wedge\bigg (\mathrm{d}\omega^{a}+\frac{1}{2}\varepsilon_{abc}\omega^{b}\wedge\omega^{c}\bigg )+\frac{1}{3\lambda^{2}}\varepsilon_{abc}e^{a}\wedge e^{b}\wedge e^{c}\bigg\}+\mathrm{d}(e^{a}\wedge\omega_{a}).\end{aligned}\end{equation}
Comparing this expression with Equation (\ref{ed,coscons}), we see that $3$-dimensional gravity with a cosmological constant $\Lambda<0$ can be written as the difference of two $\mathfrak{so}(1,2)$-Chern-Simons theories together with a boundary term, i.e.
\begin{align}\mathcal{S}_{\mathrm{3d},\Lambda}[e,\omega]=\mathcal{S}_{\mathrm{CS}}[A_{+}(e,\omega)]-\mathcal{S}_{\mathrm{CS}}[A_{-}(e,\omega)]+\int_{\partial\mathcal{M}}\,i^{\ast}(e^{a}\wedge\omega_{a}),\end{align}
where the constant $k$ in the definition of Chern-Simons theories is given by $k=\lambda c^{4}/4G$. In the case of a positive cosmological constant $\Lambda=-1/\lambda^{2}>0$ with $\lambda\in i\mathbb{R}\textbackslash\{0\}$, we have to consider $3$-dimensional de-Sitter space, whose isometry group is given by $\mathrm{O}(1,3)$. Here we can use that $\mathfrak{so}(1,3)\cong\mathfrak{sl}(2,\mathbb{C})$\footnote{On the level of Lie groups, there is an isomorphism $\mathrm{SO}^{+}(1,3)\cong\mathrm{Sl}(2,\mathbb{C})/\mathbb{Z}_{2}$, since $\mathrm{Spin}(1,3)\cong\mathrm{Sl}(2,\mathbb{C})$ is the double cover of the proper orthochronous Lorentz group $\mathrm{SO}^{+}(1,3)$.}. The Lie algebra of $\mathrm{sl}(2,\mathbb{C})$ is given by
\begin{align}[T^{a},T^{b}]=\varepsilon^{abc}T_{c}.\end{align}
Similarly to before, let us define two local $\mathfrak{sl}(2,\mathbb{C})$-valued connection $1$-forms 
\begin{align}A_{\pm}:=\bigg (\omega^{a}\pm\frac{i}{\lambda}e^{a}\bigg )T_{a}\in\Omega^{1}(U,\mathfrak{sl}(2,\mathbb{C})).\end{align}
Note that they two are not independent, because $A^{a}_{+}=\overline{A}^{a}_{-}$ in order to have a real action. This also shows that the gauge group in this case is not $\mathrm{Sl}(2,\mathbb{C})\times\mathrm{Sl}(2,\mathbb{C})$, but rather $\mathrm{Sl}(2,\mathbb{C})$, as claimed above. As an $\mathrm{Ad}$-invariant non-degenerate inner product on $\mathfrak{sl}(2,\mathbb{C})$ we again use the Killing form, normalized as $\operatorname{tr}(T_{a}T_{b})=\frac{1}{2}\eta_{ab}$. Then we get that
\begin{align}\mathcal{S}_{\mathrm{3d},\Lambda}[e,\omega]=\mathcal{S}_{\mathrm{CS}}[A_{+}(e,\omega)]-\mathcal{S}_{\mathrm{CS}}[A_{-}(e,\omega)]+\int_{\partial\mathcal{M}}\,i^{\ast}(e^{a}\wedge\omega_{a}),\end{align}
where the level is now given by $k=i\lambda c^{4}/4G$. As a last remark, note that for the Riemannian case, the Chern-Simons gauge groups are $\mathrm{Iso}(3):=\mathbb{R}^{3}\rtimes \mathrm{O}(3)$ ($\Lambda=0$), $\mathrm{SO}(1,3)$ ($\Lambda<0$) and $\mathrm{SO}(4)\cong(\mathrm{SU}(2)\times\mathrm{SU}(2))/\mathbb{Z}_{2}$ ($\Lambda>0$). In all six cases (Lorentzian/Riemannian and $\Lambda$\begin{footnotesize} $\gtreqqless$\end{footnotesize} $0$), only the Chern-Simons gauge group corresponding to Riemannian gravity with $\Lambda>0$ is a \textit{compact} Lie group. This situation is at the quantum level described by the Turaev-Viro spin foam model based on the quantum group $U(\mathfrak{su}(2))_{q}$, which we will discuss later.
\chapter{Spin Foam Models for 3D Quantum Gravity}\label{Chap2}
After the discussion of classical gravity, we now move on to quantum gravity. We will focus on manifestly background-independent approaches, in which we aim to explain the very origin of the geometry of spacetime. As a consequence, we are not using a background manifold, but rather purely pregeometric, discrete and algebraic data. We will start with the discussion of the ``Ponzano-Regge spin foam model'', which is a very rich and extensively studied spin foam model for $3$-dimensional quantum gravity. Afterwards, we briefly discuss the main structure of related spin foam models, which describe different situations, like the Lorentzian case and the ``Turaev-Viro model'', which includes a positive cosmological constant. We conclude this chapter by explaining how to characterize the boundary Hilbert space by means of so-called ``spin network states'', which allow us to suitably define transition amplitudes of these models. 

\section{The Ponzano-Regge Spin Foam Model}
The Ponzano-Regge model is a state-sum model based on the Lie group $\mathrm{SU}(2)$. It is the first spin foam model ever proposed and describes 3-dimensional Riemannian quantum gravity without a cosmological constant. Historically, it was proposed by G. Ponzano and T. Regge in 1968 \cite{PonzanoReggeModel} in their study of the asymptotic properties of Wigner's $6j$-symbol, which reproduces the classical Regge action of $3$-dimensional general relativity in the large spin limit. Since then, it has been related to many other approaches of quantum gravity, like quantum Chern-Simons theory and loop quantum gravity. Many details about this model can be found in the extensive series \cite{FreidelPonzanoRegge1,FreidelPonzanoRegge2,FreidelPonzanoRegge3} as well as in \cite{BarrettPonzanoRegge}, \cite[Ch.3]{GoellerThesis} and \cite[Ch.2]{OritiThesis}. For a general introduction to spin foam models see for example \cite{BaezBFTheory,PerezSM,RovelliVidotto} and references therein.

\subsection{Definition of the Model}
Let $\mathcal{M}$ be a compact and oriented $3$-dimensional manifold together with a triangulation $\Delta$, which always exists according to the Theorem of Moise. A short discussion of simplicial complexes and triangulations can be found in Appendix \ref{SimplicalComplexes}. If $\mathcal{M}$ has a boundary, then this triangulation naturally induces a triangulation on its boundary. To be precise, every $k$-cell in $\Delta$ touching the boundary yields a $(k-1)$-cell in the triangulation on the boundary. The induced triangulation of the boundary is a subcomplex of $\Delta$ and we will be denoted by $\partial\Delta$. The set of all vertices, edges, faces and tetrahedra in $\Delta$ will be denoted by $\mathcal{V}$, $\mathcal{E}$, $\mathcal{F},\mathcal{T}\subset\Delta$, respectively. The subsets of vertices, edges and faces living purely on the boundary will be denoted by $\partial\mathcal{V},\partial\mathcal{E},\partial\mathcal{F}\subset\partial\Delta$ and their complements by $\mathring{\mathcal{V}}:=\mathcal{V}\textbackslash\partial\mathcal{V}$, $\mathring{\mathcal{E}}:=\mathcal{E}\textbackslash\partial\mathcal{E}$ and $\mathring{\mathcal{F}}:=\mathcal{F}\textbackslash\partial\mathcal{F}$. The partition function of the Ponzano-Regge model is defined as the sum over all assignments of irreducible representations of $\mathrm{SU}(2)$, labelled by spins $j\in\mathbb{N}_{0}/2$, to edges of the bulk triangulation $\Delta\textbackslash\partial\Delta$ multiplied by an amplitude, which only depends on purely algebraic and combinatorial data:

\begin{Definition}\label{PolzanoRegge} (Ponzano-Regge Spin Foam Model)\newline
Let $\mathcal{M}$ be a compact and oriented manifold with boundary together with a triangulation $\Delta$. Then the partition function of the Ponzano-Regge model is formally defined by 
\begin{align*}\mathcal{Z}_{\mathrm{PR}}[\Delta,\{j_{e}\}_{e\in\partial\mathcal{E}}]:=\sum_{j\in(\mathbb{N}_{0}/2)^{\mathring{\mathcal{E}}}}\prod_{e\in\mathring{\mathcal{E}}}(-1)^{2j_{e}}(2j_{e}+1)\hspace{-5pt}\prod_{f\in\mathring{\mathcal{F}}}(-1)^{j_{f1}+j_{f2}+j_{f3}}\prod_{t\in\mathcal{T}}\begin{Bmatrix}j_{t1}&j_{t2}&j_{t3}\\j_{t4}&j_{t5}&j_{t6}\end{Bmatrix},\end{align*}
where the indices $fi$ denote the $3$ edges corresponding to the triangle $f\in\mathring{\mathcal{F}}$ and the indices $ti$ denote the 6 edges corresponding to the tetrahedron $t\in\mathcal{T}$.
\end{Definition}

Note that the partition function is a functional of boundary data. The weight for each tetrahedron $t\in\mathcal{T}$ is given by a Wigner's 6j-symbol\footnote{For the definition of $6j$-symbols and a discussion of its properties see for example \cite{FuchsSchweigert,QTAM}.}. In order to use a coherent convention for the $6j$-symbols, we label the edges of some given tetrahedron $t\in\mathcal{T}$ as in Figure \ref{TetrahedronLabelling} below.

\begin{figure}[H]
\centering
\includegraphics[scale=1.4]{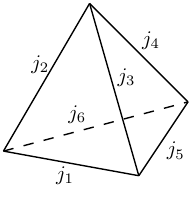}
\caption{Tetrahedron with edges labelled by spins.\label{TetrahedronLabelling}}
\end{figure}

However, note there is some freedom in labelling the edges, because the $6j$-symbol is symmetric under permutations of the vertices of the tetrahedron. In terms of the $6j$-symbol this means that it is symmetric under interchanging any two of the three columns and symmetric under interchanging the upper and the lower argument of any two columns. Not every weight in the Ponzano-Regge partition function is non-zero. For example, the $6j$-symbol vanishes whenever the ``\textit{triangle condition}'' $j_{t1}\in\{\vert j_{t2}-j_{t3}\vert,\vert j_{t2}-j_{t3}\vert+1,\dots, j_{t2}+j_{t3}\}$ is not fulfilled. Still, the Ponzano-Regge partition function is only formally defined and in general divergent. Therefore, a suitable regularization procedure is required, which we will discuss in Section \ref{GaugeFixPR}. Furthermore, as it stands, it is not clear if the Ponzano-Regge partition function is topologically invariant, i.e. if it does depend on the chosen triangulation $\Delta$ on $\mathcal{M}$. One way to approach this question is by recalling the following fact: Let $\mathcal{M}$ be a $d$-dimensional piecewise-linear manifold, as defined in Appendix \ref{TriangTheorem}. Then any two triangulations of $\mathcal{M}$ can be related by a finite sequence of so-called ``\textit{Pachner moves}'' \cite{Pachner,BarrettPachner}. In three dimensions, there are exactly two different types of Pachner moves. The $1-4$ Pachner move replaces a tetrahedron with four tetrahedra by adding a vertex in the barycentre. The $3-2$ Pachner move decomposes two tetrahedra, which have one common face, into three tetrahedra. Of course, also the inverse processes are possible. The two types of Pachner moves are sketched in Figure \ref{PachnerMoves} below.

\begin{figure}[H]
\centering
\includegraphics[scale=1]{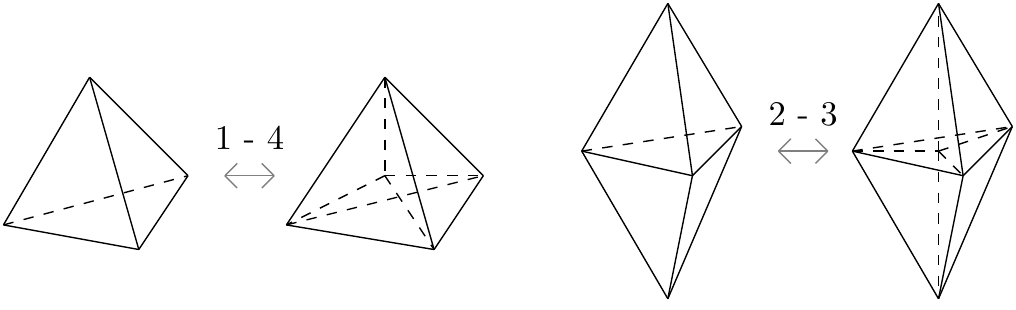}
\caption{A $1-4$ Pachner move and a $2-3$ Pachner move. \label{PachnerMoves}}
\end{figure}

By ``\textit{Pachner's Theorem}'' \cite{Pachner}, any two triangulations of a $d$-dimensional PL-manifold can be related by a finite sequence of Pachner moves. Therefore, in order to prove topological invariance, we could try to show that the Ponzano-Regge partition function is invariant under these two types of moves. It turns out that the Ponzano-Regge partition function is always invariant under the $2-3$ Pachner move, even without regularization. This is due to the ``\textit{Biedenharn-Elliot identity}\footnote{Mathematically, the Biedenharn-Elliot identity is a particular form of the pentagon identity for the tensor product of $\mathfrak{sl}(2,\mathbb{C})$-modules. See also \cite[p.291ff.]{FuchsSchweigert} for more details.}'' of the $6j$-symbol, which states that 
\begin{align}\begin{Bmatrix}j_{1}&j_{2}&j_{3}\\k_{1}&k_{2}&k_{3}\end{Bmatrix}\begin{Bmatrix}j_{1}&j_{2}&j_{3}\\h_{1}&h_{2}&h_{3}\end{Bmatrix}=\sum_{j}S_{j}\begin{Bmatrix}k_{2}&h_{2}&j\\h_{3}&k_{3}&j_{1}\end{Bmatrix}\begin{Bmatrix}k_{3}&h_{3}&j\\h_{1}&k_{1}&j_{2}\end{Bmatrix}\begin{Bmatrix}k_{1}&h_{1}&j\\h_{2}&k_{2}&j_{3}\end{Bmatrix},\end{align}
where $S_{j}=(-1)^{j+\sum_{i=1}^{3}(j_{i}+k_{i}+h_{i})}(2j+1)$ and where the sum goes over $j$ in steps of $1$ between $\operatorname{max}\{\vert k_{1}-h_{1}\vert,\vert k_{2}-h_{2}\vert,\vert k_{3}-h_{3}\vert\}$ and $\operatorname{min}\{k_{1}+h_{1}, k_{2}+h_{2}, k_{3}+h_{3}\}$. It can easily be seen that this identity just represents a $2-3$ Pachner move, where the tetrahedrons are labelled as explained in Figure \ref{TetrahedronLabelling} above. The invariance under the $1-4$ Pachner move is in general not true. As an example, consider a single tetrahedron. If we add a vertex in its barycentre, then we explicitly add a divergence to our partition function, since we add a ``bubble'' to the complex, as explained in Section \ref{GaugeFixPR} in more details. This was already observed in the original paper by G. Ponzano and T. Regge \cite{PonzanoReggeModel}. In this paper, the authors suggested a simple cut-off regularization, where the sum over assignments of spins to edges only sums over finitely many spins. The authors then proved that the partition function becomes invariant under a $1-4$ Pachner move. However, it turns out that with this naive regularization, the Biedenharn-Elliot identity cannot longer be applied and hence, we can not prove the invariance under both types of Pachner moves at the same time. In order to address this problem, we first have to find a suitable regularization procedure. In the end, it turns out that the divergences of the Ponzano-Regge partition function are related to the gauge-symmetries of the model, which are the residual symmetries of the continuous $BF$-action and hence are related to diffeomorphism invariance of gravity. As a consequence, we first of all have to find a suitable gauge-fixing procedure, which we will discuss in Section \ref{GaugeFixPR}. After this, it can be shown that the partition function exists whenever the manifold $\mathcal{M}$ fulfils a certain topological property. If this is the case, one can rewrite the partition function as an integral with measure given by the ``Reidemeister torsion'' \cite{BarrettReidemeister,BarrettPonzanoRegge}. This also proves topological invariance of the Ponzano-Regge partition function. More details about this can be found in Section \ref{Reidemeister}.

\subsection{Large j-limit of Wigner's 6j-Symbols and Relation to Gravity}
As already mentioned in the beginning of this chapter, the Ponzano-Regge model was historically motivated from the observation that Wigner's $6j$-symbols reproduce the classical Regge action for $3$-dimensional Riemannian general relativity in the large spin limit. The precise asymptotic formula conjectured by G. Ponzano and T. Regge in 1968 \cite{PonzanoReggeModel} for the $6j$-symbol corresponding to a tetrahedron $t\in\mathcal{T}$ is given by
\begin{align}\begin{Bmatrix}j_{t1}&j_{t2}&j_{t3}\\j_{t4}&j_{t5}&j_{t6}\end{Bmatrix}\overset{j\to\infty}{\approx}\frac{1}{\sqrt{12\pi V_{t}}}\cos\bigg (\mathcal{S}_{R,t}\bigg[j_{e}+\frac{1}{2}\bigg]+\frac{\pi}{4}\bigg ),\end{align}
where $V_{t}$ denotes the volume of the tetrahedron $t$, whose edge lengths are given by $j_{e}+1/2$. The functional $\mathcal{S}_{R,t}[l_{e}]$ denotes the classical Regge action \cite{Regge}, \cite[Sec.4.3]{RovelliVidotto} of the tetrahedron $t$, which is a functional of the edge lengths $l_{e}$ and defined via 
\begin{align}\mathcal{S}_{R,t}[l_{e}]:=\sum_{i=1}^{6}l_{ti}\cdot\theta(l_{ti}),\end{align}
where $\theta(l_{e})$ is the deficit angle at the edge $e$. Note that the limit $j\to \infty$ is a semi-classical one. This can directly be seen by the fact that the spins $j_{e}$ correspond to the edge lengths $l_{e}$ via $l_{e}=(j_{e}+1/2)l_{p}$, where $l_{p}$ denotes the Planck length. In other words, the limit $j_{e}\to\infty$ correspond to the limit $l_{p}\propto\sqrt{\hbar}\to 0$ and hence to $\hbar\to 0$. This asymptotic formula was rigorously proven by J. Roberts in 1999 \cite{RobertsPonzanoRegge}. For an alternative proof using integrals over group variables see \cite{FreidelAsymtotics6j}. \\
\\
In the limit $j\to\infty$, the sum over all spins can be regarded as an integral over the edges. The product over all $6j$-symbols then result into $\mathrm{exp}(i\mathcal{S}_{R}[l_{e}])$ and $\mathrm{exp}(-i\mathcal{S}_{R}[l_{e}])$, where $\mathcal{S}_{R}[l_{e}]=\sum_{t\in\mathcal{T}}\mathcal{S}_{R,t}[l_{e}]$ is the full Regge-action of $3$-dimensional Riemannian general relativity. As a result, we see that the Ponzano-Regge partition function describes the formal quantum partition function
\begin{align}\mathcal{Z}\approx\int\,\mathcal{D}g\,e^{\frac{i}{\hbar}\int_{\mathcal{M}}\,\mathrm{d}x\,\sqrt{-g}R_{g}}.\end{align}
Of course, this not totally precise, because we also get the contribution coming from $-i\mathcal{S}_{R}[l_{e}]$ as well as some interference terms. Furthermore, one needs to know if there exists a nice continuum limit of the theory. The reason why we get two exponentials, one with $-i\mathcal{S}_{R,t}$ and one with $+i\mathcal{S}_{R,t}$, is related to the fact that we are quantizing the triads and not the metric in the Ponzano-Regge spin foam model. Let us consider a triad $e$. Then it is always possible to act with the time-reversal operator $\mathcal{T}$ on $e$, which flips the sign of the $0$-component $e^{0}$. The Einstein-Hilbert action is clearly invariant under this flip of sign. However, the action written in terms of triads is not. In other words, the two triads $e$ and $\mathcal{T}e$ describe exactly the same Lorentzian metric, although the gravitational field defined by $e$ and $\mathcal{T}e$ has a different action. More generally, we can also act with an arbitrary Lorentz transformation on triads. Such transformations do not affect the action written in terms of the metric, however it does affect the first-order action. As a consequence, when quantizing gravity using the triads instead of the metric, we always get an additional contribution coming from the additional sign difference. The contribution coming from the time-reversal exactly corresponds to an additional term proportional to $\mathrm{exp}(-i\mathcal{S})$ in the Feynman path integral. Furthermore, the phase factor $\pi/4$ in the asymptotic formula can be understood geometrically: It can be interpreted as the ``\textit{Maslov index}'' of the saddle point approximation, which always appears when there is a flip of sign in the momentum between two saddle points \cite{Maslov}. For more details, see the treatment in \cite[p.115ff.]{RovelliVidotto} and references therein.

\subsection{Discretized Partition Function of Gravity and Ponzano-Regge Model}\label{DerPR}
The goal of this section is to explain the relation between the Ponzano-Regge partition function defined in the last section and $3$-dimensional quantum gravity in a more direct way. To be precise, we will see that the Ponzano-Regge model can be understood as the discretization of the quantum partition function of $3$-dimensional Riemannian general relativity without cosmological constant formulated in the triadic first-order formalism, as firstly discussed in \cite{FreidelPRDisc}, based on previous works cited therein. To start with, recall that the corresponding action is given by
\begin{align}\mathcal{S}_{\mathrm{3d}}[e,\omega]=\frac{1}{2\kappa}\int_{\mathcal{M}}\!\varepsilon_{abc}e^{a}\wedge F^{bc}[\omega]=\frac{1}{2\kappa}\int_{\mathcal{M}}\!\operatorname{tr}(e\wedge F[\omega]),\end{align}
where $e^{a}=e^{a}_{\mu}\mathrm{d}x^{\mu}$ denote the triad and where $F^{ab}[\omega]:=\mathrm{d}\omega^{ab}+{\omega^{a}}_{c}\wedge\omega^{cb}$ denotes the curvature corresponding to the spin connection $\omega^{ab}:=\omega^{ab}_{\mu}\mathrm{d}x^{\mu}$. In order to quantize this theory, we have to define the generating functional, i.e. the path integral\footnote{This is of course purely formal and we will not discuss any mathematical formulations of the path integral.} 
\begin{align}\label{partition} \mathcal{Z}_{\mathrm{3d}}[\mathcal{M}]=\int\mathcal{D}e\mathcal{D}\omega\,\mathrm{exp}\bigg (\frac{i}{2\kappa}\int_{\mathcal{M}}\!\operatorname{tr}(e\wedge F[\omega])\bigg ).\end{align}
In order to have a proper definition of the generating functional, one should restrict the path integral to gauge fields $e$ and $\omega$, which are not equivalent by gauge transformations, in order to not overcount the equivalent configurations of the system. We will ignore this for now and will come back to it later. If our manifold $\mathcal{M}$ has a boundary $\partial\mathcal{M}$, then we keep the spin-connection fixed on the boundary. Hence, the partition function becomes a functional on the boundary data given by a connection $\partial\omega$ on $\partial\mathcal{M}$, i.e. 
\begin{align}\mathcal{Z}_{\mathrm{3d}}[\mathcal{M},\partial\omega]=\int_{i^{\ast}\omega=\partial\omega}\mathcal{D}e\mathcal{D}\omega\,\mathrm{exp}\bigg (\frac{i}{2\kappa}\int_{\mathcal{M}}\!\operatorname{tr}(e\wedge F[\omega])\bigg ),\end{align}
where $i:\partial\mathcal{M}\to\mathcal{M}$ denotes the obvious inclusion. The reason why we keep the spin-connection fixed on the boundary is that the variation of the classical action on-shell reads
\begin{align}\delta \mathcal{S}_{\mathrm{3d}}[e,\omega]_{\text{on-shell}}=\frac{1}{2\kappa}\int_{\mathcal{\partial M}}\!\operatorname{tr}(e\wedge \delta\omega),\end{align}
and hence vanishes when we fix the spin-connection on $\partial\mathcal{M}$. In principal, one could also keep the triad fixed on the boundary. This would however lead to an additional term in the action analogously to the Gibbons-Hawking-York boundary term in the Einstein-Hilbert action, which would make the whole discussion from a gauge-theoretic point of view more complicated, since we would need to restore gauge-invariance at the boundary, for example by introducing new degrees of freedom. \cite[p.17]{GoellerThesis}\\
\\
In order to discretize this theory, let us choose a triangulation $\Delta$ on $\mathcal{M}$. An important point to note here is that we are not restricted to use a triangulation and we are also allowed to choose a more general type of cellular decomposition\footnote{See Appendix \ref{SimplicalComplexes} for an overview of cellular decompositions and simplicial complexes.}. However, in this section we would like to recover the historical Ponzano-Regge model containing $6j$-symbols assigned to tetrahedra and hence, we have to restrict to simplicial complexes. In the following, it is useful to consider the ``\textit{dual complex}'' corresponding to the triangulation of $\mathcal{M}$, which is a (in general not simplicial) cellular decomposition of $\mathcal{M}$, in which every $k$-simplex of $\Delta$ is replaced by an associated $(3-k)$-cell as follows: 
\begin{itemize} 
\item[(1)]There is vertex in the dual complex at the barycentre of each tetrahedron (=$3$-simplex) of the triangulation. 
\item[(2)]The $1$-cell dual to a triangle (=$2$-simplex) of the triangulation connects the two vertices, which are dual to the two tetrahedra, which have this triangle as a common face. Note that the dual $1$-cell meets the triangle exactly at one point. 
\item[(3)]Continuing this process up to dimension, we associate a dual $2$-cell to each edge (=$1$-simplex), which intersect the corresponding edge at exactly one point, as well as a $3$-cell, which is dual to a vertex (=$0$-simplex) of the triangulation.
\end{itemize}
If $\mathcal{M}$ is a manifold with boundary, then we can define a dual complex consisting of an interior part and a boundary part. The interior part is constructed as explained above, i.e. to every $k$-simplex in $\Delta$, we get a $(3-k)$-cell in the interior dual complex. The $1$-skeleton of this complex, i.e. the set of dual vertices and edges, is a graph with open ends, since all the cells dual to simplices in $\partial\Delta$ intersect the boundary. The boundary dual complex is defined to be the dual complex of the boundary triangulation $\partial\Delta$. The full dual complex, which we will denote by $\Delta^{\ast}$, is then obtained by gluing the boundary dual cells to the open ends of the interior dual cell. The result is a well-defined cellular complex with an induced boundary subcomplex $\partial\Delta^{\ast}=(\partial\Delta)^{\ast}$. Let us stress that while $(\partial\Delta)^{\ast}=\partial\Delta^{\ast}$, it is not true that every bulk dual cell in $\Delta^{\ast}$ is dual to a \textit{bulk} simplex in $\Delta$, since also the boundary simplices in $\partial\Delta$ are dual to dual cells in the interior. Figure \ref{DualTetrahedron} below shows two simple triangulations of the $3$-ball and their dual complexes.

\begin{figure}[H]
\centering
\subfloat{\includegraphics[trim=1cm 3.3cm 0 0,scale=0.28]{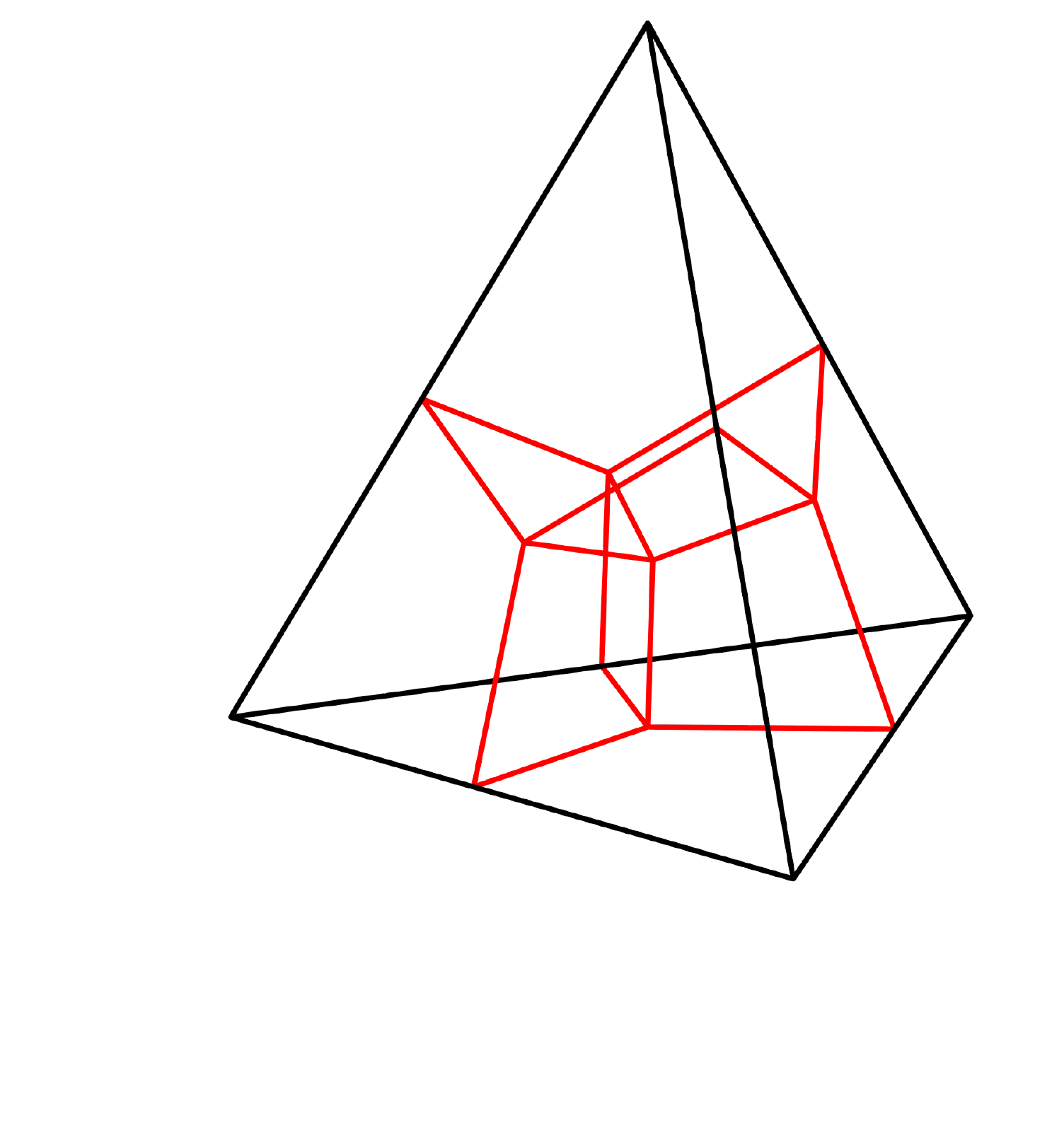}}\hspace{1cm}\subfloat{\includegraphics[trim=1cm 3.3cm 0 0,width=0.255\textwidth]{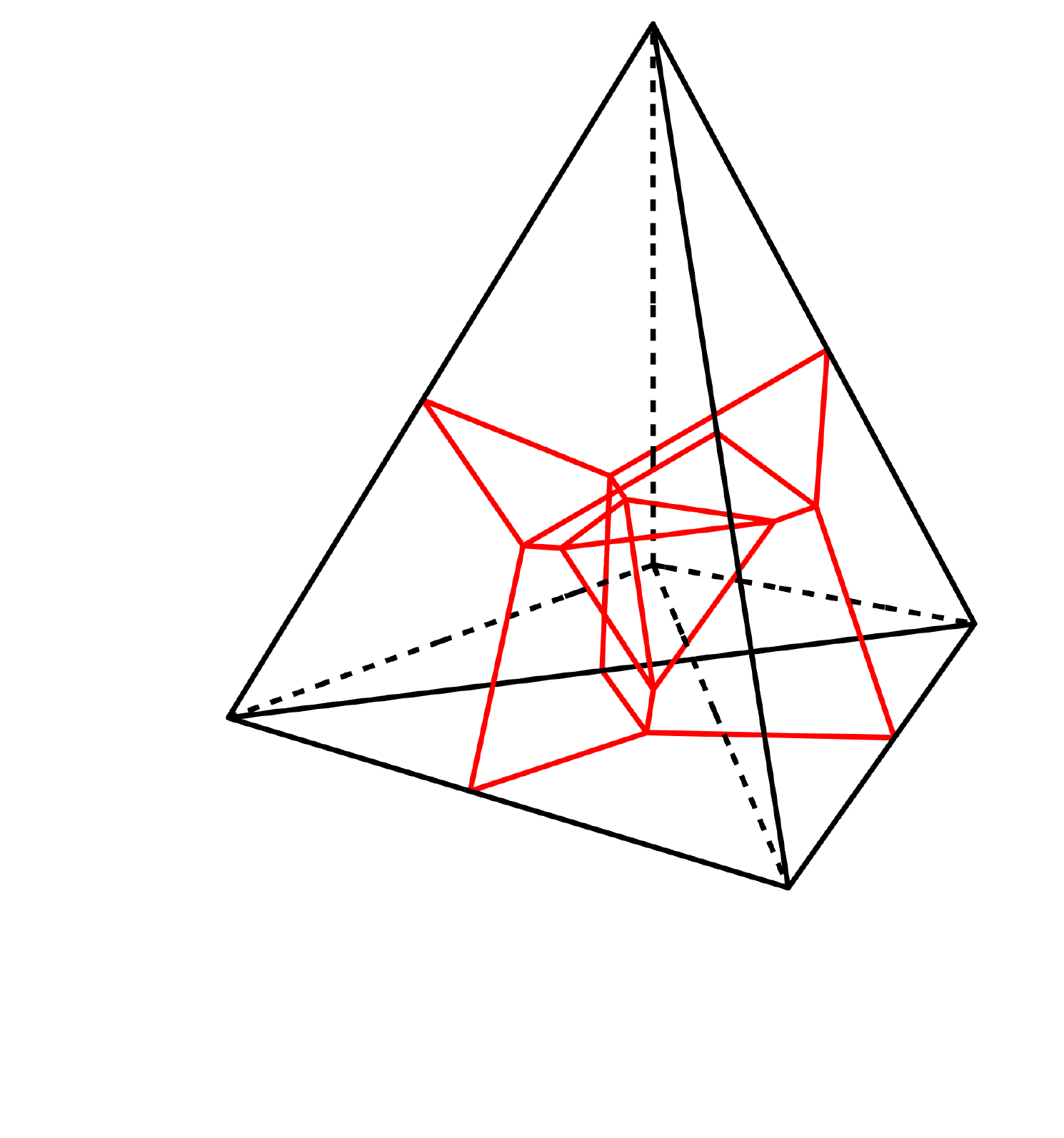}} 
\caption{A single tetrahedron (l.h.s.) and the triangulation obtained by performing a $(1-4)$-Pachner move (r.h.s.) together with the corresponding dual 1-skeletons drawn in red. The dual complex on the l.h.s. has $5,10,10$ and $4$ dual vertices, edges, faces and $3$-cells, from which $4$, $6$, $4$ and $0$ belong to the boundary and the dual complex on the r.h.s. has $8,16,14$ and $5$ dual vertices, edges, faces and $3$-cells, from which $4$, $6$, $4$ and $0$ belong to the boundary. \label{DualTetrahedron}}
\end{figure}

Similar to the simplices of the triangulation, let us denote the set of all vertices, edges and faces in $\Delta^{\ast}$ by $\mathcal{V}^{\ast}$, $\mathcal{E}^{\ast},\mathcal{F}^{\ast}\subset\Delta^{\ast}$, the dual vertices, edges and faces contained in $\partial\Delta^{\ast}$ by the same symbols with a ``$\partial$'' in front and their complements by the symbol ``$\circ$''. Now, let us assume that the triangulation is oriented. This means that we assign an orientation to each edge and face of the triangulation $\Delta$. This assignments also induces an orientation of the dual edges and dual faces, i.e. edges and faces in $\Delta^{\ast}$. Since every edge corresponds to some face, we also need to define a relative orientation between them. We do this by defining a function $\varepsilon:\mathcal{E}\times\mathcal{F}\to\{\pm 1\}$ by $\varepsilon(e,f)=1$ if the orientation of $e$ and $f$ agrees, by $\varepsilon(e,f)=-1$ if the orientations do not agree and by $\varepsilon(e,f)=0$ otherwise. Similar, we define $\varepsilon$ for the dual complex. Now, recall that the triads are related to the metric via $\eta_{ab}e_{\mu}^{a}e_{\nu}^{b}=g_{\mu\nu}$. Since the metric describes distances and length, it is natural to assign it on the edges of the triangulation $\Delta$. Furthermore, triads are Lie-algebra valued $1$-forms, which are naturally integrated along $1$-dimensional objects. Therefore, we assign an element $X_{e}$ of the Lie algebra to each edge $e\in\mathcal{E}$, which represents the integral of the triad over the edge $e$. Of course, equivalently we may also assign a Lie algebra element $X_{f^{\ast}}$ to all the bulk faces $f\in\mathring{\mathcal{F}}^{\ast}$ of the dual complex $\Delta^{\ast}$. The connection $\omega$ is also a 1-form and we assign a group element $g_{e^{\ast}}$ to each dual edge $e^{\ast}\in\mathcal{E}^{\ast}$, which represents the holonomy $\mathcal{P}\mathrm{exp}(\int_{e^{\ast}}\omega)$ along this edge. The curvature is then naturally defined to be the product of all the group elements $g_{e^{\ast}}$, which lie on the edges of the dual complex corresponding to some face $f^{\ast}\in\mathring{\mathcal{F}}^{\ast}$ in the bulk dual complex $\Delta^{\ast}\textbackslash\partial\Delta^{\ast}$. In other words, the curvature is represented by the holonomy around a face, which also makes sense in light of the Ambross-Singer theorem, which relates the holonomy of a connection of a principal bundle with the corresponding curvature form. To sum up, we have the following assignments:
\begin{itemize}
\item The triad $e$ is replaced by an element $X_{f^{\ast}}\in\mathfrak{su}(2)$ assigned to each face of the bulk dual complex $\Delta^{\ast}\textbackslash\partial\Delta^{\ast}$.
\item The spin connection $\omega$ is replaced by group element $g_{e^{\ast}}\in\mathrm{SU}(2)$ assigned to each edge $e^{\ast}$ of the dual complex $\Delta^{\ast}$.
\item The curvature $2$-form is now represented by $G_{f^{\ast}}\in\mathrm{SU}(2)$ on each face $f^{\ast}$ of the bulk dual complex $\Delta^{\ast}\textbackslash\partial\Delta^{\ast}$ and is a product of the group elements $g_{e^{\ast}}$, which are living on the dual edges contained in $f^{\ast}$. To be precise, we set
\begin{align}G_{f^{\ast}}:=\overrightarrow{\prod_{e^{\ast}\subset f^{\ast}}}g_{e^{\ast}}^{\varepsilon(e^{\ast},f^{\ast})},\end{align}
where the arrow over the product means that we have to take the product such that it respects the orientation of the face. Furthermore, the product has to be understood as starting from some given edge of $f^{\ast}$. The choice of starting point, however, does not matter, since in the end we will see that we get a $\delta$-function forcing flatness everywhere.
\end{itemize}

The above considerations lead to the following definition for the discretized partition function:

\begin{Definition}\label{disc} (Discretized Quantum Partition Function of 3D Gravity)\newline
Let $\mathcal{M}$ be a compact and oriented $3$-dimensional manifold with boundary $\partial\mathcal{M}$ together with a triangulation $\Delta$. Then we define the discretization of the formal quantum partition function for first-order gravity to be the functional
\begin{align*}\mathcal{Z}[\Delta,\{g_{e^{\ast}}\}_{e^{\ast}\in\partial\mathcal{E}^{\ast}}]\hspace{-2pt}:=\hspace{-3pt}\int_{\mathrm{SU}(2)^{\vert\mathring{\mathcal{E}}^{\ast}\vert}}\!\bigg (\prod_{e^{\ast}\in\mathring{\mathcal{E}}^{\ast}}\hspace{-4pt}\mathrm{d}g_{e^{\ast}}\bigg )\hspace{-4pt}\int_{\mathfrak{su}(2)^{\vert\mathring{\mathcal{F}}^{\ast}\vert}}\!\bigg (\prod_{f^{\ast}\in\mathring{\mathcal{F}}^{\ast}}\hspace{-4pt}\mathrm{d}X_{f^{\ast}}\bigg )\mathrm{exp}\bigg (i\cdot\mathrm{tr}\bigg\{\sum_{f^{\ast}\in\mathring{\mathcal{F}}^{\ast}}\hspace{-4pt}X_{f^{\ast}}G_{f^{\ast}}\bigg\}\hspace{-2pt}\bigg),\end{align*}
where $\mathrm{d}g$ denotes the normalized Haar measure on $\mathrm{SU}(2)$ (see Appendix \ref{HaarMeasureAppendix}) and where $\mathrm{d}X$ denotes the $3$-dimensional Lebesgue measure under the isomorphism $\mathfrak{su}(2)\cong\mathbb{R}^{3}$.
\end{Definition}

The goal for the rest of this section is to explicitly compute the integrals in order to reproduce the historical Ponzano-Regge partition function as defined in Definition \ref{PolzanoRegge}. As a first step, we compute the integrals corresponding to the discretized triads, i.e. the integrals over the Lie algebra elements. This can easily be done with the help of the following lemma:

\begin{Lemma}\label{deltatoshow}
Let $g\in\mathrm{SU}(2)$ be arbitrary. Then:
\begin{align*}\int_{\mathfrak{su}(2)}\,\mathrm{d}X\,\mathrm{exp}(i\cdot\mathrm{tr}(Xg))\propto (\delta_{\mathrm{SU}(2)}(g)+\delta_{\mathrm{SU}(2)}(-g)).\end{align*}\end{Lemma}

\begin{proof}
The proof of this Lemma is straightforward and the claim follows by expressing the $\mathrm{SU}(2)$-delta function in terms of the ordinary delta function using some appropriate parametrization of $\mathrm{SU}(2)$. More details can be found in \cite[Appendix B]{FreidelPonzanoRegge1}.
\end{proof}

Note that $(\delta_{\mathrm{SU}(2)}(g)+\delta_{\mathrm{SU}(2)}(-g))\propto\delta_{\mathrm{SO}(3)}(g)$. Therefore, we would get a $\mathrm{SO}(3)$-delta function instead of the $\mathrm{SU}(2)$-delta function. However, in order to reproduce to the historical $\mathrm{SU}(2)$-Ponzano-Regge amplitude, we need a single $\mathrm{SU}(2)$-delta function. Therefore, we have to get rid of the $\delta$-function with $-g$ as argument. This can be achieved by slightly changing the amplitude in the path integral, as explained in Appendix B of \cite{FreidelPonzanoRegge1}, i.e. using the formula
\begin{align}\int_{\mathfrak{su}(2)}\,\mathrm{d}X\,\mathrm{exp}(i\mathrm{tr}(Xg))(1\pm \varepsilon(g))\propto \delta_{\mathrm{SU}(2)}(\pm g),\end{align}
where $\varepsilon(g):=\operatorname{sign}(\cos(\theta))$ with $\theta\in [0,\pi]$ denotes the angle of $g$ using the parametrization $g=e^{i\theta n^{i}\sigma_{i}}$ with the Pauli-matrices $\sigma_{i}$, where $\vec{n}\in\mathbb{R}^{3}$ is some normal vector. The reason why we get the $\mathrm{SO}(3)$-$\delta$-function instead of the $\mathrm{SU}(2)$-$\delta$-function is that our choice of the discretized action for 3-dimensional Riemannian gravity was rather naive and completely loses track of the $\mathrm{SU}(2)$ nature of the model. To be precise, the identification of choosing the Lebesgue measure on $\mathfrak{su}(2)\cong\mathbb{R}^{3}$ was rather rough. A more natural candidate for the Fourier-transform on $\mathrm{SU}(2)$, which takes more care on the $\mathrm{SU}(2)$-nature of the model, is provided by the formula
\begin{align}\delta_{\mathrm{SU}(2)}(g)=\int_{\mathbb{C}^{2}}\,\mathrm{d}^{4}z\,\frac{\vert z\vert^{2}-1}{\pi^{2}}e^{-\vert z\vert^{2}+\langle z\vert g\vert z\rangle}\end{align}
introduced in \cite{DupuisSpinors}, where the integral is over spinors $z\in\mathbb{C}^{2}$ and where $\vert z\vert^{2}=z^{\dagger}z$ and $\langle z\vert g\vert z\rangle:=z^{\dagger}gz$ for all $z\in\mathbb{C}^{2}$ and $g\in\mathrm{SU}(2)$. Using this more refined version of the Fourier transform on $\mathrm{SU}(2)$, it can be shown that one recovers the full $\mathrm{SU}(2)$-Ponzano-Regge model. We therefore ignore the contribution of $\delta_{\mathrm{SU}(2)}(-g)$ at this point. More details about this issue may also be found in \cite[p.53]{GoellerThesis} and \cite{DupuisSpinors} as well as references therein. Using Lemma \ref{deltatoshow} for each integration over some Lie algebra element in the partition function of Definition \ref{disc} and ignoring the prefactors, we arrive at the following result:

\begin{Proposition}\label{step2} (Group Representation of the Partition Function)\newline
Let $\mathcal{M}$ be a compact and oriented $3$-dimensional manifold with boundary $\partial\mathcal{M}$ together with a triangulation $\Delta$. Then the discretized partition function $\mathcal{Z}$ can be written as
\begin{align*}\mathcal{Z}[\Delta,\{g_{e^{\ast}}\}_{e^{\ast}\in\partial\mathcal{E}^{\ast}}]=\int_{\mathrm{SU}(2)^{\vert\mathring{\mathcal{E}}^{\ast}\vert}}\bigg (\prod_{e^{\ast}\in\mathring{\mathcal{E}}^{\ast}}\!\mathrm{d}g_{e^{\ast}}\bigg )\prod_{f^{\ast}\in\mathring{\mathcal{F}}^{\ast}}\delta_{\mathrm{SU}(2)}(G_{f^{\ast}}).\end{align*}
\end{Proposition}

\begin{Remarks}\begin{itemize}\item[]
\item[(a)]By looking at the generating functional (\ref{partition}), note that the triad field $e$ has the form of a Lagrange multiplier, which enforces flatness of the model. As a consequence, we can also formally integrate over the triad $e$, which results into
\begin{align}\label{partition2}\mathcal{Z}_{\mathrm{3d}}[\mathcal{M},\partial\omega]=\int_{i^{\ast}\omega=\partial\omega}\,\mathcal{D}\omega\,\delta(F[\omega]).\end{align}
Using this expression for the quantum partition function, one may interpret the formula in Proposition \ref{step2} as the discretized version of (\ref{partition2}). 
\item[(b)]By looking at the group representation above, we directly see that the partition function does not depend on the choice of orientations of dual edges and face as well as the chosen starting points of faces, since the Haar measure of compact groups is ``\textit{unimodular}'', i.e. it satisfies $\mathrm{d}g=\mathrm{d}g^{-1}$, the delta function is cyclic, i.e. $\delta_{\mathrm{SU}(2)}(gh)=\delta_{\mathrm{SU}(2)}(hg)$ for all $g,h\in\mathrm{SU}(2)$, and the delta function satisfies $\delta_{\mathrm{SU}(2)}(g^{-1})=\delta_{\mathrm{SU}(2)}(g)$ for all $g\in\mathrm{SU}(2)$. The later statement follows from cyclicity and the fact that $g$ and $g^{-1}$ are contained in the same conjugacy class of $\mathrm{SU}(2)$.
\end{itemize}\end{Remarks}

As a next step, we have to integrate over the group variables. For this, it is useful to decompose the $\delta$-function above using the Theorem of Peter-Weyl, which we review in Appendix \ref{PeterWeylAppendix}. Of course, what we call the ``$\delta$-function'' is strictly speaking a distribution and the notation as a function makes only sense if we write it within an integration. Formally, the $\delta$-function of a compact Lie group $G$ can be decomposed as 
\begin{align}\delta_{G}(g)=\sum_{\rho\in\Lambda}\mathrm{dim}(\rho)\chi_{\rho}(g),\end{align}
where $\Lambda$ is the set of all (equivalence classes) of irreducible unitary representations of $G$. In the case of $\mathrm{SU}(2)$, all unitary and irreducible representations (up to equivalence) are labelled by a spin $j\in\mathbb{N}_{0}/2$. Plugging this formula into the previous expression for the discretized partition function, we find 
\begin{align}\label{step3}\mathcal{Z}[\Delta,\{g_{e^{\ast}}\}_{e^{\ast}\in\partial\mathcal{E}^{\ast}}]=\sum_{j\in(\mathbb{N}_{0}/2)^{\mathring{\mathcal{F}}^{\ast}}}\int_{\mathrm{SU}(2)^{\vert\mathring{\mathcal{E}}^{\ast}\vert}}\bigg (\prod_{e^{\ast}\in\mathring{\mathcal{E}}^{\ast}}\!\mathrm{d}g_{e^{\ast}}\bigg )\prod_{f^{\ast}\in\mathring{\mathcal{F}}^{\ast}} \,(2j_{f^{\ast}}+1)\cdot\chi_{j_{f^{\ast}}}(G_{f^{\ast}}),\end{align}
where the sum goes over all assignments of irreducible representation to the bulk dual faces $\mathring{\mathcal{F}}^{\ast}\subset\Delta^{\ast}\textbackslash\partial\Delta^{\ast}$. The characters of the $\mathrm{SU}(2)$-representations can be written explicitly in terms of Wigner's D-matrices, i.e.
\begin{align}\chi_{j}(g)=\sum_{m=-j}^{j}D^{j}_{mm}(g).\end{align}
As a consequence, the terms $\chi_{j_{f^{\ast}}}(G_{f^{\ast}})$ in the expression above are a product of $D$-matrices, which are contracted accordingly. In order to recover the historical Ponzano-Regge spin foam model, let us now restrict to manifolds with empty boundary, for simplicity, because in order to arrive at the Ponzano-Regge model with boundary, we have to assign spins to the boundary dual edges. This can be done by introducing a suitable boundary Hilbert state, which we will discuss in Section \ref{BoundarySpinFoam}. Since we are using a simplicial decomposition, there are three dual faces sharing each dual edge, because there are three edges bounding a triangle $f\in\mathcal{F}$. Therefore, we may use the following formula, relating the representation matrices with two Wigner's $3j$-symbols:
\begin{align}\int_{\mathrm{SU}(2)}\,D^{j_{1}}_{m_{1}n_{1}}(g)D^{j_{2}}_{m_{2}n_{2}}(g)D^{j_{3}}_{m_{3}n_{3}}(g)\,\mathrm{d}g=\begin{pmatrix}j_{1} & j_{2} & j_{3}\\m_{1} & m_{2} & m_{3}\end{pmatrix}\begin{pmatrix}j_{1} & j_{2} & j_{3}\\n_{1} & n_{2} & n_{3}\end{pmatrix}\end{align}
Note that this expression is only non-zero if $\vert j_{2}-j_{3}\vert\leq j_{1}\leq j_{2}+j_{3}$ and/or $m_{1}+m_{2}+m_{3}=0=n_{1}+n_{2}+n_{3}$. With this formula in mind, it is clear that the partition function (\ref{step3}) will consist of a bunch of $3j$-symbols, which are suitable contracted among themselves, after integrating over the group variables $g_{e^{\ast}}$. Each dual edge $e^{\ast}$ produces two $3j$-symbols, as already mentioned above. As a consequence, the contraction must happen on the dual vertices. At each dual vertex, there are precisely four dual edges meeting together and hence four $3j$-symbols. Therefore, we have to contract four $3j$-symbols at each dual vertex in such a way that the result in the end is an invariant under $\mathrm{SU}(2)$. The only such possibility is Wigner's $6j$-symbol. Plugging this into Equation (\ref{step3}) and carefully keeping track of signs yields exactly the historical Ponzano-Regge partition function, as defined in Definition \ref{PolzanoRegge}, written in terms of the dual complex $\Delta^{\ast}$:
\begin{equation}\label{PRSpin}\begin{aligned}\mathcal{Z}&[\Delta]=\hspace{-2pt}\sum_{j\in(\mathbb{N}_{0}/2)^{\mathcal{F}^{\ast}}}\prod_{f\in\mathcal{F}^{\ast}}(-1)^{2j_{f^{\ast}}}(2j_{f^{\ast}}+1)\hspace{-2pt}\prod_{e^{\ast}\in\mathcal{E}^{\ast}}(-1)^{\sum_{i=1}^{3}j_{e^{\ast}i}}\hspace{-2pt}\prod_{v^{\ast}\in\mathcal{V}^{\ast}}\begin{Bmatrix}j_{v^{\ast}1}&j_{v^{\ast}2}&j_{v^{\ast}3}\\j_{v^{\ast}4}&j_{v^{\ast}5}&j_{v^{\ast}6}\end{Bmatrix},\end{aligned}\end{equation}
where $e^{\ast}i$ denote the three dual faces corresponding to the dual edge $e^{\ast}$ and where $v^{\ast}i$ denote the six dual faces corresponding to some dual vertex $v^{\ast}$. The reason why we get an additional minus sign for half-integer spins, i.e. the factor $(-1)^{2j}$ in the above formula, is a little bit subtle. In the end, it is related to the choice of ``spherical category'', introduced in \cite{BarrettSphericalCategory}, for $\mathrm{SU}(2)$. It turns out that in the case of $\mathrm{SU}(2)$, there are exactly two different possibilities. The choice used for the Ponzano-Regge model is responsible for this additional sign factor. In other words, one gets the ``\textit{quantum dimension}'' $(-1)^{2j}(1+2j)$ instead of $(1+2j)$ for each representation, labelled by a spin $j$. More details about this issue and a detailed explanation of the appearance of this additional sign factor can be found in \cite[p.21ff.]{BarrettPonzanoRegge}.

\subsection{Divergences and Gauge-Fixing}\label{GaugeFixPR}
As a starting point, let us take the ``group representation'' of the Ponzano-Regge model, i.e. the partition function introduced in Proposition \ref{step2}:
\begin{align}\mathcal{Z}_{\mathrm{PR}}[\Delta,\{g_{e^{\ast}}\}_{e^{\ast}\in\partial\mathcal{E}^{\ast}}]=\int_{\mathrm{SU}(2)^{\vert\mathring{\mathcal{E}}^{\ast}\vert}}\bigg (\prod_{e^{\ast}\in\mathring{\mathcal{E}}^{\ast}}\!\mathrm{d}g_{e^{\ast}}\bigg )\prod_{f^{\ast}\in\mathring{\mathcal{F}}^{\ast}}\delta_{\mathrm{SU}(2)}(G_{f^{\ast}})\end{align}
As already mentioned, this expression is only formally defined and in general divergent. Mathematically, this is because of some redundant $\delta$-functions, which in turn is a consequence of the existence of ``\textit{bubbles}'', which are $3$-cell in the dual complex $\Delta^{\ast}$ not touching the boundary. In other words, the bubbles are precisely the 3-cells of $\Delta^{\ast}$, which are dual to bulk vertices of $\Delta$. In our case, we assign a $\delta$-function to each face of such a $3$-cell, however, it turns out that for each such polyhedron there is one redundant $\delta$-function, because its information is already contained in the $\delta$-function of the other faces. In the end, we will get a contribution of $\delta_{\mathrm{SU}(2)}(\mathds{1})$, where $\mathds{1}$ denotes the identity of $\mathrm{SU}(2)$, for each bubble after integrating. Therefore, the partition function of the Ponzano-Regge model is (partially) divergent due to the existence of so-called ``\textit{bubble divergences}''. As a consequence, it is natural to define the set of problematic edges. The collection of all such edges in $\Delta$, or equivalently dual faces in $\Delta^{\ast}\textbackslash\partial\Delta^{\ast}$, is usually called the ``\textit{tardis}''. Recall that the $\delta$-function can also be decomposed as a sum over all spin of $\mathrm{SU}(2)$. Therefore, an edge is contained in the tardis, if and only if this sum is not constrained to be finite by all the data on the boundary and the relations of the spin. By definition, if we have a non-tardis triangulation, then the Ponzano-Regge partition function is finite. Of course, the tardis itself is by no means unique. An important point to mention is that the Ponzano-Regge partition function might still diverge even when removing all redundant delta functions. An example is the so-called ``Bing's house with three rooms'' as discussed in \cite[p.13ff.]{BarrettPonzanoRegge}. This provides an example of a triangulation of a $3$-ball, which does not have any internal vertex and hence no bubbles in the dual complex, but which is still not a non-tardis triangulation, since its Ponzano-Regge partition function diverges. A closer analysis of the degree of divergence of general ``flat spin foam models'' was done in \cite{BonzomCellularCohomolgy,BonzomTwistedCohomology,Bonzom3}.\\
\\
It turns out that the existence of redundant delta functions is a consequence of the gauge-symmetries of the Ponzano-Regge model. Therefore, we have to find a suitable gauge-fixing procedure. For this, recall that $BF$-theory has two symmetries: The $\mathrm{SU}(2)$ gauge symmetry as well as the translational symmetry, as discussed in Section \ref{BF}. The discrete version of the $\mathrm{SU}(2)$ gauge symmetry is then parametrized by an element $k_{v^{\ast}}\in\mathrm{SU}(2)$ on each vertex of the dual complex $\Delta^{\ast}$ and given by \cite[Sec.3.1.1.]{FreidelPonzanoRegge1}
\begin{align}g_{e^{\ast}}&\mapsto k^{-1}_{\mathrm{t}(e^{\ast})}g_{e^{\ast}}k_{\mathrm{s}(e^{\ast})}\\
X_{f^{\ast}}&\mapsto k^{-1}_{\mathrm{st}(f^{\ast})}X_{f^{\ast}}k_{\mathrm{st}(f^{\ast})},\end{align}
where $s(e^{\ast}),t(e^{\ast})\in\mathcal{V}^{\ast}$ denotes the source and target dual vertex of the dual edge $e^{\ast}$ and where $\mathrm{st}(f^{\ast})\in\mathcal{V}^{\ast}$ denotes the dual vertex chosen to be the starting point of the product defining the curvature $G_{f^{\ast}}$ associated to the dual face $f^{\ast}$. As a consequence, the group element $G_{f^{\ast}}$ transforms as
\begin{align}G_{f^{\ast}}\mapsto k^{-1}_{\mathrm{st}(f^{\ast})}G_{f^{\ast}}k_{\mathrm{st}(f^{\ast})}.\end{align}
Looking at the discretized action for $3$-dimensional Riemannian gravity, which is given by
\begin{align}\label{DiscAction}\mathcal{S}_{\mathrm{3d,disc}}[X_{f^{\ast}},G_{f^{\ast}}]=\sum_{f^{\ast}\in\mathring{\mathcal{F}}^{\ast}}\operatorname{tr}(X_{f^{\ast}}G_{f^{\ast}}),\end{align}
and using cyclicity of the trace, it is clear that the action is invariant under these transformations, as it should be. The discrete residual of the translational symmetry of the continuous $BF$-action is more complicated. First of all, recall that the this symmetry is due to the Bianchi identity. At a discrete level, the Bianchi identity can be written as \cite{DiscBianchi}
\begin{align}\label{Bianchi}\prod_{e\text{ with }v\subset e}(k_{v}^{e})^{-1}G_{e}^{\varepsilon(e,v)}k_{v}^{e}=\mathds{1},\end{align}
where the product is over all edges starting/ending at the vertex $v\in\mathcal{V}$ and where $k_{v}^{e}\in\mathrm{SU}(2)$ is a group element parametrizing the parallel transport from $v$ to $\mathrm{st}(f^{\ast})$, where $f^{\ast}$ is the face dual to $e$. Note that these elements depends on the elements $g_{e^{\ast}}$. As shown in \cite[Sec.3.1.1.]{FreidelPonzanoRegge1}, the discrete translational symmetry can then be expressed as 
\begin{align}g_{e^{\ast}}&\mapsto g_{e^{\ast}}\\
X_{e} &\mapsto X_{e}+\varepsilon(e,v)(U_{e}^{v}(k_{v}^{e}\Phi_{v}(k_{v}^{e})^{-1}))-[\Omega_{e}^{v},(k_{v}^{e}\Phi_{v}(k_{e}^{v})^{-1})]\end{align}
for some Lie algebra element $\Omega_{e}^{v}\in\mathfrak{su}(2)$ and some scalar $U_{e}^{v}\in\mathbb{C}$. Using the discrete Bianchi identity (\ref{Bianchi}), one can show that the discrete action (\ref{DiscAction}) is invariant under these transformations. In order to fix the gauge-symmetries, we can us a standard procedure from lattice gauge theory, i.e. by choosing ``\textit{maximal trees}''. A maximal tree, or ``spanning tree'', $T$ is a subgraph of some given graph $\Gamma$ such that the following two conditions are satisfied:
\begin{itemize}
\item[(1)]$T$ is ``\textit{maximal}'', i.e. it touches every vertex in $\Gamma$.
\item[(2)]$T$ is a ``\textit{tree}'', i.e. it is connected and acyclic (=it contains no closed paths).
\end{itemize}
Using such a maximal tree, the idea is to use the gauge symmetries at the vertices to fix the elements living on the edges of the tree. In the specific case of the Ponzano-Regge model, we choose a internal maximal tree $T$ touching the boundary exactly once, i.e. it only includes one vertex of $\partial\Delta$, as well as a maximal tree $T^{\ast}$ in the dual complex $\Delta^{\ast}$. Using the gauge symmetries, we fix the elements living on the tree to be 
\begin{align}&\forall e\in T: X_{e}=0\\&\forall e^{\ast}\in T^{\ast}: g_{e^{\ast}}=\mathds{1}.\end{align}
Note that setting $X_{e}=0$ for every edge $e$ in the maximal tree $T$ exactly corresponds to removing one $\delta$-function associated to the dual faces corresponding to the edges in $T$. In other words, we are exactly removing the redundant $\delta$-functions of the bubbles explained in the previous part, which (partially) cause the divergences. In order to write down the gauge-fixed version of the Ponzano-Regge model, we have to compute the corresponding Fadeev-Popov determined. However, it can be shown that in the case of the Ponzano-Regge model this is actually equal to one \cite{FreidelDiffeo}. We finally arrive at the following definition for the gauge-fixed Ponzano-Regge partition function:

\begin{Definition} (Gauge-Fixed Ponzano-Regge Partition Function)\newline
Let $\mathcal{M}$ be a compact and oriented manifold with boundary together with a triangulation $\Delta$. Furthermore, let $T$ be an internal maximal tree in $\Delta$ touching the boundary once and $T^{\ast}$ be a maximal tree in $\Delta^{\ast}$. Then the gauge-fixed partition function of the Ponzano-Regge model is defined by
\begin{align*}\mathcal{Z}_{\mathrm{PR}}[\Delta,T,T^{\ast},\{g_{e^{\ast}}\}_{e^{\ast}\in\partial\mathcal{E}^{\ast}}]:=\int_{\mathrm{SU}(2)^{\vert\mathring{\mathcal{E}}^{\ast}\vert}}\bigg (\prod_{e^{\ast}\in\mathring{\mathcal{E}}^{\ast}}\!\mathrm{d}g_{e^{\ast}}\bigg )\,\bigg (\prod_{e^{\ast}\in T^{\ast}}\delta(g_{e^{\ast}})\bigg )\prod_{e\in\mathcal{E}\textbackslash T}\delta_{\mathrm{SU}(2)}(G_{e}).\end{align*}
\end{Definition}

A very important point to note is that the gauge-fixing procedure affects also the boundary. In principle, the Ponzano-Regge partition function is a functional, which depends on the group elements assigned to the dual edges on the boundary. However, the maximal tree $T^{\ast}$ contains in general also some of the dual edges of the boundary and hence, we see that the partition function above contains also $\delta$-functions, which contain all the elements $\{g_{e^{\ast}}\}_{e^{\ast}\in\partial\mathcal{E}^{\ast}}$ for which $g_{e^{\ast}}\in T^{\ast}$. In other words, after gauge-fixing, the Ponzano-Regge partition function becomes a functional of the \textit{gauge-fixed boundary data}. As a last remark, we have to ask the question if we have enough delta functions remaining to impose the flatness of the connection everywhere. It turns out that this is indeed the case:

\begin{Lemma}$\forall f^{\ast}\text{ dual to }e\in\mathcal{E}\textbackslash T: G_{f^{\ast}}=\mathds{1}$ $\Rightarrow$ $\forall f^{\ast}\in\mathring{\mathcal{F}}^{\ast}:G_{f^{\ast}}=\mathds{1}$\end{Lemma}

\begin{proof}A proof of this Lemma can be found in \cite[p.5f.]{FreidelPonzanoRegge2} as well as in \cite[p.26f.]{BarrettPonzanoRegge}.\end{proof}

\subsection{Existence and Topological Invariance}\label{Reidemeister}
As already mentioned at the beginning, even after the gauge-fixing procedure described above the Ponzano-Regge partition function might still be divergent. This is a general fact for so-called ``flat spin foam models'' \cite{BonzomCellularCohomolgy,BonzomTwistedCohomology,Bonzom3}. As stated and proven in \cite{BarrettReidemeister,BarrettPonzanoRegge}, one can show that the partition function of the Ponzano-Regge model exists as a distribution if and only if the ``second twisted cohomology group'' of a certain graph in the interior is trivial for every flat connection. Furthermore, the authors showed that in case this existence criterion is fulfilled, then the gauge-fixed Ponzano-Regge partition function can be written as an integral over the space of all flat connections in the interior modulo gauge transformations, where the measure is given by the so-called ``\textit{Reidemeister-torsion}''. See \cite{DupoisReidemeister} for details about this quantity. Note that this is not surprising, since it was already observed in \cite{BlauThompsonRaySinger} that topological $BF$-theories can be understood as an integration on the moduli space of flat connection with measure provided by the so-called ``Ray-Singer torsion'' \cite{TFTBook}. As a consequence, the partition function of quantum $BF$-theory is finite whenever the Ray-Singer torsion is well defined and since the Reidemeister torsion is the discrete counterpart of the Ray-Singer torsion, we have a similar situation in the discrete case, which is described by the Ponzano-Regge partition function. Since the Reidemeister torsion is a topological invariant, this also proves that the Ponzano-Regge partition function, if it exists, is independent of the choice of triangulation as well as the choice of maximal trees and hence, only depends on the choice of manifold and the assigned boundary data.

\section{Other Spin Foam Models for 3D Quantum Gravity}
As discussed before, the Ponzano-Regge spin foam model describes $3$-dimensional Riemannian quantum gravity without a cosmological constant. In this section, we briefly discuss two related spin foam models, namely the analogues spin foam model for $3$-dimensional Lorentzian quantum gravity, introduced in \cite{FreidelLorentzianPonzanoRegge}, as well as the Turaev-Viro model \cite{TuraevViroModel}, which described $3$-dimensional Riemannian quantum gravity in presence of a positive cosmological constant.

\subsection{The Lorentzian Ponzano-Regge Model}
The Ponzano-Regge model discussed so far describes $3$-dimensional Riemannian quantum gravity. The Lorentzian case can be described analogously. The derivation starting from the classical action in the triadic first-order formalism is technically more involved in the Lorentzian case. However, note that the complication is not conceptual, but rather a consequence of the fact that the gauge group for $3$-dimensional Lorentzian gravity, i.e. the Lie group $\mathrm{SO}(1,2)$, is non-compact as opposed to $\mathrm{SU}(2)$ in the Riemannian case. As a consequence, we have to find a suitable way to gauge-fix the model, in order to avoid additional divergences caused by the non-compactness of the group. In order to derive the partition function starting from the partition function of the classical action, we can principally start in the same way as in the case of the Ponzano-Regge model. After choosing some cellular decomposition and discretization of the corresponding variables, we have to start with the expression
\begin{align}\label{Lorentzian}\mathcal{Z}[\Delta,\{g_{e^{\ast}}\}_{e^{\ast}\in\partial\mathcal{E}^{\ast}}]=\int_{\mathrm{Sl}(2,\mathbb{R})^{\vert\mathring{\mathcal{E}}^{\ast}\vert}}\!\bigg (\prod_{e^{\ast}\in\mathring{\mathcal{E}}^{\ast}}\mathrm{d}g_{e^{\ast}}\bigg )\,\prod_{f^{\ast}\in\mathring{\mathcal{F}}^{\ast}}\delta_{\mathrm{Sl}(2,\mathbb{R})}(G_{f^{\ast}})\end{align}
as before, where we used the isomorphism $\mathrm{SO}^{+}(1,2)\cong\mathrm{Spin}(1,2)/\mathbb{Z}_{2}\cong\mathrm{Sl}(2,\mathbb{R})/\mathbb{Z}_{2}$. The main technical difference is the decomposition of the $\delta$-functions using the Theorem of Peter-Weyl. As explained in more details in \cite{FreidelLorentzianPonzanoRegge}, there are three different types of unitary irreducible representations appearing in this decomposition:
\begin{itemize}
\item[(1)]The principal series $T_{i\rho-\frac{1}{2},\varepsilon}$ labelled by a parameter $\rho>0$ with $\varepsilon\in\{0,\frac{1}{2}\}$, whose Casimir is given by $C_{\rho}=\rho^{2}+\frac{1}{4}$.
\item[(2)]The holomorphic discrete series $T_{l}^{+}$ with $l\in\mathbb{N}_{0}/2$ and Casimir $C_{l}=-l(l+1)\leq 0$.
\item[(3)]The anti-holomorphic discrete series $T_{l}^{-}$ with $l\in\mathbb{N}_{0}/2$ and Casimir $C_{l}=-l(l+1)\leq 0$.
\end{itemize}
Using this, the decomposition of the $\delta$-function reads
\begin{align}\delta_{\mathrm{Sl}(2,\mathbb{R})}(g)\hspace*{-2pt}=\hspace*{-2pt}\frac{1}{\pi^{2}}\bigg(\sum_{l\in\mathbb{N}_{0}/2}\hspace*{-4pt}(2l\hspace*{-1pt}+\hspace*{-1pt}1)[\chi_{l}^{+}(g)\hspace*{-1pt}+\hspace*{-1pt}\chi_{l}^{-}(g)]\hspace*{-1pt}+\hspace*{-4pt}\sum_{\varepsilon\in\{0,\frac{1}{2}\}}\int_{0}^{\infty}\hspace*{-2pt}\mathrm{d}\rho\,2\rho\,\mathrm{tanh}(\pi\rho+i\varepsilon\pi)\chi_{\rho,\varepsilon}(g)\bigg),\end{align}
where $\chi_{l}^{+}$, $\chi_{l}^{+}$ and $\chi_{\rho,\varepsilon}$ denote the corresponding characters. With this decomposition, we can proceed as in the case of the Ponzano-Regge model, by plugging this formula for the $\delta$-function into Formula (\ref{Lorentzian}). Note however that there is a very important difference: The integrand is a sum of characters in this case and as such, it is invariant under gauge transformation. As a consequence, there are some redundant integrals. In order to get rid of them, we have to apply some suitable gauge-fixing procedure, which can be done by choosing a maximal tree in the dual complex and setting every edge in this tree equal to some fixed group element, similarly as before. There was no need for this additional gauge-fixing procedure in the Riemannian Ponzano-Regge model precisely because $\mathrm{SU}(2)$ is a compact Lie group, which means that it has volume equal to one, when choosing the normalized Haar measure. Another subtlety in the Lorentzian case is that the $6j$-symbols of the Lie group $\mathrm{Sl}(2,\mathbb{R})$ cannot be defined by a contraction of four $3j$-symbols. This is also due to the non-compactness of the group, since such contractions would lead to an infinite result. Hence, one usually defines the $6j$-symbols of non-compact groups as the matrix elements of unitary operators defined on the space of invariant operators acting on the tensor product of four unitary irreducible representations. For the explicit expression of the resulting partition function for the Lorentzian Ponzano-Regge spin foam model and other details see \cite{FreidelLorentzianPonzanoRegge}, \cite[p.82ff.]{OritiThesis} and references therein.

\subsection{Cosmological Constant and the Turaev-Viro Spin Foam Model}
Another spin foam model, which is closely related to the Ponzano-Regge model, is the Turaev-Viro spin foam model proposed in 1992 by V. Turaev and O. J. Viro \cite{TuraevViroModel}. It is defined completely analogues to the Ponzano-Regge model, when we replace the group $\mathrm{SU}(2)$ by the quantum group $U(\mathfrak{su}(2))_{q}$, where $q\in\mathbb{C}$ is a root of unity, i.e. $q=\mathrm{exp}(\pi i/r)$ for $r\geq 3$\footnote{For an extensive discussion of quantum groups and their representation theory see for example \cite{KasselQuantumGroups}.}. As a consequence, the sum over representations is finite and hence one can directly prove topological invariance. It can be shown that the Turaev-Viro model corresponds to a state sum model describing 3-dimensional Riemannian quantum gravity with a positive cosmological constant given by $\Lambda\propto r^{-2}$. However, there is not yet a direct relationship between the Turaev-Viro model and a discretization of 3-dimensional Riemannian gravity with cosmological constant, as in the case of the Ponzano-Regge model.\\
\\
Let $\mathcal{M}$ be a compact and oriented $3$-dimensional manifold with boundary $\partial\mathcal{M}$. As before, we choose a triangulation $\Delta$ of $\mathcal{M}$ and denote the set of all vertices, edges, faces and tetrahedra in the triangulation $\Delta$ by $\mathcal{V}$, $\mathcal{E}$, $\mathcal{F}$ and $\mathcal{T}$, respectively. Furthermore, we define the $q$-deformation parameter by $q:=e^{\pi i/r}$ for some integer $r\geq 3$, i.e. $q^{2r}=1$. The Turaev-Viro spin foam model is defined by the partition function
\begin{align}\mathcal{Z}_{q}[\{j_{e}\}_{e\in\partial\mathcal{E}}]:=N_{q}^{-\vert\mathcal{V}\vert}\cdot\sum_{j\in (\mathbb{N}_{0}/2)^{\mathring{\mathcal{E}}}}\prod_{e\in\mathring{\mathcal{E}}}(-1)^{2j_{e}}d_{q}(j_{e})\prod_{t\in\mathcal{T}}\begin{Vmatrix}j_{t1}&j_{t2}&j_{t3}\\j_{t4}&j_{t5}&j_{t6}\end{Vmatrix}_{q},\end{align}
where $N_{q}$ denotes some rescaling constant for each vertex $v\in\mathcal{V}$, which depends on $q$. For any $n\in\mathbb{N}$, we define the corresponding $q$-deformed integer by
\begin{align}[n]_{q}:=\frac{q^{n}-q^{-n}}{q^{1}-q^{-1}}.\end{align}
The $q$-deformed dimension $d_{q}(j)$ used above is then defined by $d_{p}(j):=[d(j)]_{q}=[2j+1]_{q}$. The bracket in the above formula is given by
\begin{align}\begin{Vmatrix}j_{1}&j_{2}&j_{3}\\j_{4}&j_{5}&j_{6}\end{Vmatrix}_{q}:=i^{2\sum_{k=1}^{6}j_{k}}\begin{Bmatrix}j_{1}&j_{2}&j_{3}\\j_{4}&j_{5}&j_{6}\end{Bmatrix}_{q},\end{align}
where the bracket on the right-hand side denotes the $q$-deformed $6j$-symbols, whose definition can for example be found in the original article \cite{TuraevViroModel}. The sum in the definition of the Turaev-Viro invariant is over all assignments of irreducible representations of $U(\mathfrak{su}(2))_{q}$ to edges in the triangulation of the interior of the manifold. Note that for our choice of $q$, there are only finitely many irreducible representations. To be precise, the spins are limited to the range $0\leq j\leq \frac{r-2}{2}$. As a consequence, the above formula is finite. It can also be shown that it is invariant under changes of triangulations of the interior of the manifold. In other words, the Turaev-Viro amplitude is a topological invariant, which only depends on the boundary data, i.e. the assignments of spins to the edges on the boundary. \\
\\
In order to see how this invariant is related to the Ponzano-Regge model, note that we can also absorb the sign-factor corresponding to the faces in the historical Ponzano-Regge model (Definition \ref{PolzanoRegge}) by reformulating the partition function as
\begin{align}\mathcal{Z}_{\mathrm{PR}}^{\prime}[\Delta,\{j_{e}\}_{e\in\partial\mathcal{E}}]=\sum_{j\in (\mathbb{N}_{0}/2)^{\mathring{\mathcal{E}}}}\prod_{e\in\mathring{\mathcal{E}}}(-1)^{2j_{e}}(2j_{e}+1)\prod_{t\in\mathcal{T}}\begin{Vmatrix}j_{t1}&j_{t2}&j_{t3}\\j_{t4}&j_{t5}&j_{t6}\end{Vmatrix},\end{align}
where we use a similar notation as above, i.e.
\begin{align}\begin{Vmatrix}j_{1}&j_{2}&j_{3}\\j_{4}&j_{5}&j_{6}\end{Vmatrix}:=i^{2\sum_{k=1}^{6}j_{k}}\begin{Bmatrix}j_{1}&j_{2}&j_{3}\\j_{4}&j_{5}&j_{6}\end{Bmatrix}.\end{align}
This expression is equivalent to the historical Ponzano-Regge partition function, up to some powers of $i$, which only depends on boundary data \cite[p.6]{BarrettPonzanoRegge}. This expression looks exactly the same as the Turaev-Viro model, where in addition everything is labelled with the deformation parameter $q$. In order to see how they are related, note that one can show that
\begin{align}\begin{Bmatrix}j_{1}&j_{2}&j_{3}\\j_{4}&j_{5}&j_{6}\end{Bmatrix}_{q}=\begin{Bmatrix}j_{1}&j_{2}&j_{3}\\j_{4}&j_{5}&j_{6}\end{Bmatrix}+\mathcal{O}(r^{-2})\end{align}
as well as
\begin{align}[2j+1]_{q}=(2j+1)+\mathcal{O}(r^{-2}).\end{align}
As a consequence, we see that the weight factor of the Turaev-Viro invariant reduces to the weight factor of the Ponzano-Regge partition function $\mathcal{Z}^{\prime}_{\mathrm{PR}}$ in the limit $r\to\infty$, or $q\to 1$, i.e.
\begin{align}\prod_{e\in\mathring{\mathcal{E}}}(-1)^{2j_{e}}[2j_{e}+1]_{q}\prod_{t\in\mathcal{T}}\begin{Vmatrix}j_{t1}&j_{t2}&j_{t3}\\j_{t4}&j_{t5}&j_{t6}\end{Vmatrix}_{q}\overset{q\to 1}{\approx}\prod_{e\in\mathring{\mathcal{E}}}(-1)^{2j_{e}}(2j_{e}+1)\prod_{t\in\mathcal{T}}\begin{Vmatrix}j_{t1}&j_{t2}&j_{t3}\\j_{t4}&j_{t5}&j_{t6}\end{Vmatrix}.\end{align}
The upper limit $(r-2)/2$ of spins in the Turaev-Viro model plays the same role as a cutoff in the Ponzano-Regge partition function, i.e. by defining the cutoff-regularized Ponzano-Regge partition function as
\begin{align}\mathcal{Z}_{\mathrm{PR}}[\Delta,\{j_{e}\}_{e\in\partial\mathcal{E}}]=\lim_{\Lambda\to\infty}\mathcal{Z}_{\mathrm{PR},\Lambda}[\Delta,\{j_{e}\}_{e\in\partial\mathcal{E}}],\end{align}
where $\mathcal{Z}_{\mathrm{PR},\Lambda}[\Delta,\{j_{e}\}_{e\in\partial\mathcal{E}}]$ denotes
\begin{align}\mathcal{Z}_{\mathrm{PR},\Lambda}[\Delta,\{j_{e}\}_{e\in\partial\mathcal{E}}]:=N_{\Lambda}^{-\vert\mathcal{V}\vert}\sum_{j\in (\mathbb{N}_{0}/2)^{\mathring{\mathcal{E}}}}^{\Lambda}\prod_{e\in\mathring{\mathcal{E}}}(-1)^{2j_{e}}(2j_{e}+1)\prod_{t\in\mathcal{T}}\begin{Vmatrix}j_{t1}&j_{t2}&j_{t3}\\j_{t4}&j_{t5}&j_{t6}\end{Vmatrix}\end{align}
and where $N_{\Lambda}$ is some rescaling constant for each vertex $v\in\mathcal{V}$ depending on the cutoff $\Lambda$. Using this analogy, one would naively guess that the Turaev-Viro invariant reduces to the Ponzano-Regge partition function in the limit $q\to 1$. However, it is not always possible to take this limit. This is basically due to the fact that the number of terms in the sum increases without limit for many triangulations. As discussed in \cite[p.10ff.]{BarrettPonzanoRegge}, an example where the limit can be taken, are non-tardis triangulations, mentioned in Section \ref{GaugeFixPR}.\\
\\
As already mentioned above, it is not yet known how to relate the Turaev-Viro model directly with a discretization of the quantum partition function for Riemannian $3$-dimensional gravity with a positive cosmological constant, since it is a priori not clear how to integrate over a quantum group. The correspondence between the Turaev-Viro model and gravity can be seen in two different ways. First of all, we can look at the asymptotic behaviour of the quantum $6j$-symbols in the semi-classical limit $j\to\infty$, as we have also done in the case of the Ponzano-Regge model. It turns out that there is the following asymptotic formula for the $q$-$6j$-symbol corresponding to a tetrahedron $t\in\mathcal{T}$ with edges labelled by $ti$ for $i\in\{1,\dots,6\}$:
\begin{align}\begin{Bmatrix}j_{t1}&j_{t2}&j_{t3}\\j_{t4}&j_{t5}&j_{t6}\end{Bmatrix}_{q}\overset{j\to\infty}{\approx}\frac{C(r,j_{ti})}{\sqrt{V_{t}}}\cos\bigg (\mathcal{S}[r,l_{ti}]+\frac{\pi}{4}\bigg ),\end{align}
where $V_{t}$ denotes the volume of the tetrahedron $t$ with edge lengths given by $l_{ti}:=j_{ti}+1/2$ and where $C(r,j_{ti})$ denotes a constant which in principle could depend on the spins $j_{ti}$ at order $\mathcal{O}(r^{-2})$ and which depends on $r$ in such a way that it approaches $C\approx 1/\sqrt{12\pi}$ in the limits $q\to 1$ and $j\to\infty$. The functional $\mathcal{S}[r,j_{ti}]$ exactly corresponds to the classical Regge action $\mathcal{S}_{R,\Lambda,t}[l_{e}]$ of the tetrahedron $t$ with lengths equal to $l_{ti}=j_{ti}+1/2$ and with cosmological constant given by $\Lambda=(2\pi/r)^{2}>0$ in the large $r$-and large $j$-limit, i.e.
\begin{align}\mathcal{S}[r,j_{ti}]\overset{r\to\infty,j\to\infty}{\approx}\mathcal{S}_{R,\Lambda,t}\bigg [l_{ti}:= j_{ti}+\frac{1}{2}\bigg]=\sum_{i=1}^{6}(l_{ti}\theta_{ti}-\Lambda V_{t}),\end{align}
where $\theta_{ti}$ denote the deficit angles corresponding to the edges $ti$ of the tetrahedron $t\in\mathcal{T}$. Note also that this limit reduces to the case discussed before for the Ponzano-Regge model, when $q\to 1$, as it should. For a derivation of this formula see \cite{MizoguchiTada}. For a rigours proof with a precise definition of the numerator in the asymptotic formula above see \cite{TaylorWoodward}. The second way of relating the Turaev-Viro invariant with gravity is by comparing it with the Chern-Simons formulation of $3$-dimensional gravity. Recall that $3$-dimensional general relativity with a positive cosmological constant can be formulated as a Chern-Simons theory based on the Lie group $\mathrm{SO}(4)\cong(\mathrm{SU}(2)\times\mathrm{SU}(2))/\mathbb{Z}_{2}$. The classical action of $3$-dimensional gravity with cosmological constant $\Lambda$ in the triadic first-order formalism is given by
\begin{align}\mathcal{S}_{\mathrm{3d},\Lambda}[e,\omega]\propto\int_{\mathcal{M}}\bigg\{2e^{a}\wedge \bigg(\mathrm{d}\omega_{a}+\frac{1}{2}\varepsilon_{abc}\omega^{b}\wedge\omega^{c}\bigg)-\frac{\Lambda}{3}\varepsilon_{abc}\,e^{a}\wedge e^{b}\wedge e^{c}\bigg\}.\end{align}
Let us define two $\mathfrak{su}(2)$-valued connection $1$-forms 
\begin{align}A_{\pm}:=\bigg (\omega^{a}\pm\frac{i}{\lambda}e^{a}\bigg )T_{a},\end{align}
where we parametrized $\Lambda=-1/\lambda^{2}>0$ and where $T_{a}$ denote the generators of $\mathfrak{su}(2)$. Then, as discussed in Section \ref{CS}, we get (up to a boundary term) 
\begin{align}\mathcal{S}_{\mathrm{3d},\Lambda}[e,\omega]=\mathcal{S}_{\mathrm{CS}}[A_{+}]-\mathcal{S}_{\mathrm{CS}}[A_{-}],\end{align}
where $\mathcal{S}_{\mathrm{CS}}$ denotes the Chern-Simons action functional. At the quantum level, the correspondence between gravity and Chern-Simons theory is a little bit more subtle. From the relation between the two theories on the classical level, one would expect that
\begin{align}\label{CSvsGravity}\mathcal{Z}_{\mathrm{3d},\Lambda}(\mathcal{M})=\int\,\mathcal{D}\omega\mathcal{D}e\,e^{i\mathcal{S}_{\mathrm{3d},\Lambda}[e,\omega]}=\int\,\mathcal{D}A_{+}\mathcal{D}A_{-}\,e^{i(\mathcal{S}_{\mathrm{CS}}[A_{+}]-\mathcal{S}_{\mathrm{CS}}[A_{-}])}=\vert\mathcal{Z}_{CS}(\mathcal{M})\vert^{2}.\end{align}
However, it is difficult to make this precise, since the path integral is only formally defined. A first important step towards a more precise implementation of this equality was the seminal paper by E. Witten \cite{WittenJonesPolynomial}, in which he defines a topological invariant related to the Jones polynomial, which is given by the partition function of Chern-Simons theory. In 1991, V. Turaev and N. J. Reshetikhin constructed rigorously invariants of 3–manifolds using some techniques from the study of quantum groups, which they conjectured to be equivalent to Witten's invariant \cite{TuraevReshetikhin}. It was then shown by J. Roberts that the Turaev-Viro invariant is exactly the modulus squared of the Reshetikhin-Turaev invariant \cite{RobertsTuraevViro}. This precisely corresponds to the formal relation (\ref{CSvsGravity}) above. In other words, the Turaev-Viro model can be viewed as the modulus square of the $\mathrm{SU}(2)$-Chern-Simons partition function. \cite[Ch.3]{OritiThesis}

\section{Spin Networks and Transition Amplitudes}
In the previous two sections, we have encountered several spin foam models for $3$-dimensional gravity in different situations. The goal of this section is to introduce ``spin networks'', which are used to describe the boundary Hilbert space of our theory or the Hilbert space of lattice gauge theory in general. Furthermore, we will show how to define transition amplitudes of the Ponzano-Regge and related spin foam models.
 
\subsection{Abstract Spin Networks}
Spin network states were firstly introduced by R. Penrose as a purely combinatorial description of the geometry of spacetime \cite{PenroseSpinNetwork}. Later on it was realized by C. Rovelli and L. Smolin \cite{RovelliSmolin} that spin networks can be used to describe states in loop quantum gravity. However, although Penrose originally intended spin networks to describe \textit{spacetime}, they are more suitable to describe the geometry of \textit{space}. In fact, this is also the way how they are used in canonical loop quantum gravity: LQG is based on canonical quantization and hence the states of the theory only describe space at some fixed time. Dynamics enters the theory in form of certain constraints. As stated by J. C. Baez, spin networks provide a\\
\\
\textit{``mathematically rigorous and intuitively compelling picture of the kinematical aspects of quantum gravity.''} \cite[p.1]{BaezSpinFoamModels}\\
\\
Spin networks can either be defined abstractly or embedded in some manifold. Before stating the definition, recall that a ``\textit{directed graph}'' is a pair $\gamma=(\mathcal{V}_{\gamma},\mathcal{E}_{\gamma})$ consisting of a set $\mathcal{V}_{\gamma}$, whose elements are called ``\textit{vertices}'' and a set $\mathcal{E}_{\gamma}\subset\mathcal{V}_{\gamma}\times\mathcal{V}_{\gamma}$, whose elements are called ``\textit{edges}'', together with an ``\textit{orientation}'', i.e. a choice of ``\textit{source and target maps}'' $s,t:\mathcal{E}_{\gamma}\to\mathcal{V}_{\gamma}$, which assign to each edge a source and a target vertex. If $\gamma$ is an oriented graph, we call a given edge $e$ ``\textit{incoming}'' to a vertex $v$, if $t(e)=v$ and ``\textit{outgoing}'' to $v$, if $s(e)=v$. The set of all incoming edges to a vertex $v\in\mathcal{V}_{\gamma}$ will be denoted by $\mathcal{T}(v)$ and the set of all outgoing edges by $\mathcal{S}(v)$. Furthermore, we call a graph ``\textit{finite}'', if both the vertex and edge sets are finite sets. Abstract spin networks for some compact Lie group $G$ are then defined as follows:

\begin{Definition} (Abstract Spin Networks)\newline
A ``spin network'' is a triple $(\gamma,\rho,i)$ consisting of the following data:
\begin{itemize}
\item[(1)]A finite directed graph $\gamma=(\mathcal{V}_{\gamma},\mathcal{E}_{\gamma})$.
\item[(2)]A map $\rho:\mathcal{E}_{\gamma}\ni e\mapsto (\mathcal{H}_{e},\rho_{e})$, which assigns a unitary irreducible representation $\rho_{e}:G\to\mathrm{Aut}(\mathcal{H}_{e})$ of $G$ to each edge $e$ of $\gamma$.
\item[(3)]A map $i:\mathcal{V}_{\gamma}\ni v\mapsto i_{v}\in\mathrm{Int}(\bigotimes_{e \in\mathcal{T}(v)}\mathcal{H}_{e},\bigotimes_{e^{\prime}\in\mathcal{S}(v)}\mathcal{H}_{e^{\prime}})$, which assigns to each vertex $v$ of $\gamma$ an intertwiner of the form 
\begin{align*}i_{v}:\bigotimes_{e \in\mathcal{T}(v)}\mathcal{H}_{e}\to\bigotimes_{e^{\prime}\in\mathcal{S}(v)}\mathcal{H}_{e^{\prime}}.\end{align*}
\end{itemize}
\end{Definition}

Note that a spin network does not have to be connected: A finite directed graph $\gamma$ may have several disconnected components. If this is the case, we can decompose the spin network on $\gamma$ into a disjoint union of spin networks defined on the connected components of the graph $\gamma$.

\subsection{Spin Network States in Lattice Gauge Theory}
Spin networks (and in fact also spin foams \cite{BaezSpinFoamModels}) appear naturally in the discussion of gauge theory on a graph \cite{BaezGaugeTheory}. For this, consider a compact and connected Lie group $G$ as well as a principal $G$-bundle $P$ over a compact oriented $d$-dimensional manifold $\mathcal{M}\cong\mathbb{R}\times \Sigma$, where $\Sigma$ is a spacelike hypersurface representing space. Furthermore, let $\gamma=(\mathcal{V}_{\gamma},\mathcal{E}_{\gamma})$ be a finite directed graph embedded in $\Sigma$. If we assume that the manifold $\Sigma$ is triangulated with some triangulation $\Delta$, then we can take $\gamma$ to be the (dual) $1$-skeleton, i.e. the set of all (dual) vertices and (dual) edges of $\Delta$. As usual in lattice gauge theory, we discretize a connection $1$-form of our principal bundle by assigning a group element $g\in G$ to each edge of the graph $\gamma$. In other words, the space of discrete connections is given by 
\begin{align}\mathcal{A}_{\gamma}:=\{\{g_{e}\}_{e\in\mathcal{E}_{\gamma}}\in G^{\vert\mathcal{E}_{\gamma}\vert}\}= G^{\vert\mathcal{E}_{\gamma}\vert}.\end{align}
As before, these group elements represent the holonomies along the corresponding edges. A gauge transformation can be described via a group element $k_{v}$ living on the vertices of $\gamma$. As discussed before, such a gauge transformation acts as 
\begin{align}g_{e}\mapsto k^{-1}_{t(e)}g_{e}k_{s(e)}\end{align}
on some group element $g_{e}$, where $t,s:\mathcal{E}_{\gamma}\to\mathcal{V}_{\gamma}$ denote the target and source map of $\gamma$. Hence, the set of gauge transformations is given by
\begin{align}\mathcal{G}_{\gamma}:=\{\{k_{v}\}_{v\in\mathcal{V}_{\gamma}}\in G^{\vert\mathcal{V}_{\gamma}\vert}\}= G^{\vert\mathcal{V}_{\gamma}\vert}.\end{align}
As a consequence, the space in which we are interested is the quotient space $\mathcal{A}_{\gamma}/\mathcal{G}_{\gamma}$. The quantization of this space is then given by the Hilbert space $L^{2}(\mathcal{A}_{\gamma}/\mathcal{G}_{\gamma})$ for some suitable measure. In this case, since $\mathcal{A}_{\gamma}=G^{\vert\mathcal{E}_{\gamma}\vert}$, we can take the measure to be the product of $\vert\mathcal{E}_{\gamma}\vert$ copies of the normalized Haar measure of $G$. This measures induces a measure on the quotient $\mathcal{A}_{\gamma}/\mathcal{G}_{\gamma}$ by using the quotient map and hence we have a well-defined Hilbert space $L^{2}(\mathcal{A}_{\gamma}/\mathcal{G}_{\gamma})$. Note that this Hilbert space is natural isomorphic to the $\mathcal{G}$-invariant subspace of $L^{2}(\mathcal{A}_{\gamma})$. In other words, an element in $L^{2}(\mathcal{A}_{\gamma}/\mathcal{G}_{\gamma})$ can be identified with a function $\psi:\mathcal{A}_{\gamma}\to\mathbb{C}$ satisfying 
\begin{align}\Vert\psi\Vert^{2}_{L^{2}}:=\int_{G^{\vert\mathcal{E}_{\gamma}\vert}}\,\bigg(\prod_{e\in\mathcal{E}_{\gamma}}\mathrm{d}g_{e}\bigg)\,\vert\psi(\{g_{e}\}_{e\in\mathcal{E}_{\gamma}})\vert^{2}<\infty,\end{align} 
where $\mathrm{d}g$ denotes the normalized Haar measure of the Lie group $G$, as well as 
\begin{align}\psi(\{g_{e}\}_{e\in\mathcal{E}_{\gamma}})=\psi(\{k_{t(e)}^{-1}g_{e}k_{s(e)}\}_{e\in\mathcal{E}_{\gamma}})\end{align}
for all $\{k_{v}\}_{v\in\mathcal{V}_{\gamma}}\in\mathcal{G}_{\gamma}$. The scalar product corresponding to the Hilbert space $L^{2}(\mathcal{A}_{\gamma}/\mathcal{G}_{\gamma})$ is given by the usual $L^{2}$-inner product, i.e. 
\begin{align}\langle\psi,\varphi\rangle_{L^{2}}:=\int_{G^{\vert\mathcal{E}_{\gamma}\vert}}\,\bigg(\prod_{e\in\mathcal{E}_{\gamma}}\mathrm{d}g_{e}\bigg)\, \overline{\psi(\{g_{e}\}_{e\in\mathcal{E}_{\gamma}})}\varphi(\{g_{e}\}_{e\in\mathcal{E}_{\gamma}})\end{align}
for all $\psi,\varphi\in L^{2}(\mathcal{A}_{\gamma}/\mathcal{G}_{\gamma})$. Let us now discuss the relation of the Hilbert space $L^{2}(\mathcal{A}_{\gamma}/\mathcal{G}_{\gamma})$ and spin networks of the embedded graph $\gamma$. In the end, what we will see is that spin networks define an orthonormal basis of this Hilbert space. We will follow the derivation in \cite{BaezGaugeTheory}. Let us denote the set of all unitary irreducible representations of $G$ by $\Lambda$ (including the $1$-dimensional trivial one). To be precise, we should denote by $\Lambda$ the set of all equivalence classes of unitary irreducible representation of $G$. Nevertheless, we will usually write $\rho\in\Lambda$, meaning that we pick a representation $\rho$ representing such an equivalence class. At each edge of $\gamma$, we have the Hilbert space $L^{2}(G)$. Now, take the ``\textit{bi-regular representation}'', which is a unitary representation of $G\times G$ on $L^{2}(G)$, i.e. a map $\tau:G\times G\to \mathcal{U}(L^{2}(G))$ where $\mathcal{U}(\mathcal{H})$ denotes the set of all unitary operator of some Hilbert space $\mathcal{H}$. It is defined as
\begin{align}(\tau(g_{1},g_{2})(f))(g):=f(g_{2}^{-1}gg_{1})\end{align}
for all $g_{1},g_{2},g\in G$ and $f\in L^{2}(G)$. Note that this representation is unitary precisely because the normalized Haar measure is bi-invariant on compact Lie groups. As a consequence, we can decompose this representation using the Theorem of Peter-Weyl:
\begin{align}\label{PW}L^{2}(G)\cong\bigoplus_{\rho\in\Lambda}\mathcal{H}_{\rho}\otimes \mathcal{H}^{\ast}_{\rho},\end{align}
where $\mathcal{H}_{\rho}$ denote the Hilbert space on which a representation $\rho\in\Lambda$ is defined and where $\mathcal{H}_{\rho}^{\ast}$ denotes the dual vector space, which is the underlying vector space of the dual representation $\rho^{\ast}$, which is also contained in $\Lambda$. More details about this decomposition can be found in Appendix \ref{PeterWeylAppendix}. As an immediate consequence of the above isomorphism, we get that
\begin{align}L^{2}(\mathcal{A}_{\gamma})\cong\bigotimes_{e\in\mathcal{E}_{\gamma}}\bigoplus_{\rho\in\Lambda}\mathcal{H}_{\rho}\otimes \mathcal{H}^{\ast}_{\rho},\end{align}
where a gauge transformation $\{k_{v}\}_{v\in\mathcal{V}}\in\mathcal{G}_{\gamma}$ acts on the space on the right-hand side via
\begin{align}\bigotimes_{e\in\mathcal{E}_{\gamma}}\bigoplus_{\rho\in\Lambda}\rho(k_{s(e)})\otimes \rho^{\ast}(k_{t(e)}).\end{align}
As discussed in \cite{BaezGaugeTheory}, we can use this expression to write the Hilbert space $L^{2}(\mathcal{A}_{\gamma}/\mathcal{G}_{\gamma})$ as 
\begin{align}\label{inv}L^{2}(\mathcal{A}_{\gamma}/\mathcal{G}_{\gamma})\cong\bigoplus_{\rho\in\Lambda^{\mathcal{E}_{\gamma}}}\bigotimes_{v\in\mathcal{V}_{\gamma}}\mathrm{Inv}(v,\rho),\end{align}
where $\mathrm{Inv}(v,\rho)$ denotes for all $v\in\mathcal{V}_{\gamma}$ the set of all invariant elements of the representation
\begin{align}\bigotimes_{e\in\mathcal{S}(v)}\rho_{e}\otimes\bigotimes_{e\in\mathcal{T}(v)}\rho_{e}^{\ast}.\end{align}
In order to relate this to intertwiners, recall that every unitary irreducible representation of some compact Lie group is finite-dimensional. Using the discussion in Appendix \ref{InvEleInter}, there is an isomorphism
\begin{align}\mathrm{Inv}(v,\rho)\cong\mathrm{Int}\bigg(\bigotimes_{e\in\mathcal{S}(v)}\mathcal{H}_{\rho_{e}},\bigotimes_{e\in\mathcal{T}(v)}\mathcal{H}_{\rho_{e}}\bigg ).\end{align}
Let us summarize the obtained results in the following proposition:

\begin{Proposition} There is the following isomorphism of Hilbert spaces:
\begin{align*}L^{2}(\mathcal{A}_{\gamma}/\mathcal{G}_{\gamma})\cong\bigoplus_{\rho\in\Lambda^{\mathcal{E}_{\gamma}}}\bigotimes_{v\in\mathcal{V}_{\gamma}}\mathrm{Int}\bigg(\bigotimes_{e\in\mathcal{S}(v)}\mathcal{H}_{\rho_{e}},\bigotimes_{e\in\mathcal{T}(v)}\mathcal{H}_{\rho_{e}}\bigg ).\end{align*}
\end{Proposition}

As a consequence, we see that the Hilbert space $L^{2}(\mathcal{A}_{\gamma}/\mathcal{G}_{\gamma})$ is actually spanned by spin network states. More explicitly, if $(\gamma,\rho,i)$ is a spin network as defined in the previous section, we obtain an element in $L^{2}(\mathcal{A}_{\gamma}/\mathcal{G}_{\gamma})$ via
\begin{align}\psi(\{g_{e}\}_{e\in\mathcal{E}}):=\bigg (\bigotimes_{v\in\mathcal{V}_{\gamma}}i_{v}\bigg )\bullet_{\gamma}\bigg(\bigotimes_{e\in\mathcal{E}_{\gamma}}\rho_{e}(g_{e})\bigg ),\end{align}
where $\bullet_{\gamma}$ means contracting at each vertex $v\in\mathcal{V}_{\gamma}$ the upper indices of the matrices corresponding to the incoming edges in $v$, the lower indexes of the matrices assigned to the outgoing edges in $v$ and the corresponding upper and lower indices of the intertwiners $i_{v}$. \cite{BaezBFTheory,ProvenziThesis}

\subsection{Boundary Hilbert Space and Transition Amplitudes}\label{BoundarySpinFoam}
Let $\mathcal{M}$ be a compact and oriented $3$-dimensional manifold with boundary $\partial\mathcal{M}$ together with a triangulation $\Delta$. Again, we also consider the corresponding dual complex, which we denote by $\Delta^{\ast}$. Now, consider as our graph the dual $1$-skeleton of the boundary, i.e. the set of all dual edges and dual vertices in $\partial\Delta^{\ast}$. As discussed in the previous part, the boundary Hilbert space of this system is given by
\begin{align}\mathcal{H}_{\partial}:=L^{2}(\mathrm{SU}(2)^{\vert\partial\mathcal{E}^{\ast}\vert}/\mathrm{SU}(2)^{\vert\partial\mathcal{V}^{\ast}\vert}),\end{align}
where $\partial\mathcal{V}^{\ast}$ and $\partial\mathcal{E}^{\ast}$ are the sets of all dual boundary vertices and dual boundary edges, respectively, and where the measure is given by $\vert\partial\mathcal{E}^{\ast}\vert$ copies of the normalized Haar measure of $\mathrm{SU}(2)$. As before, we may identify this space with the $\mathrm{SU}(2)^{\vert\partial\mathcal{V}^{\ast}\vert}$-invariant subset of $L^{2}(\mathrm{SU}(2)^{\vert\partial\mathcal{E}^{\ast}\vert})$, i.e. the set of all $L^{2}$-integrable functions of the type $\psi:\mathrm{SU}(2)^{\vert\partial\mathcal{E}^{\ast}\vert}\to\mathbb{C}$ satisfying
\begin{align}\psi(\{g_{e^{\ast}}\}_{e^{\ast}\in\partial\mathcal{E}^{\ast}})=\psi(\{k_{t(e^{\ast})}^{-1}g_{e^{\ast}}k_{s(e^{\ast})}\}_{e^{\ast}\in\partial\mathcal{E}^{\ast}})\end{align}
for all $\{k_{v^{\ast}}\}_{v^{\ast}\in\partial\mathcal{V}^{\ast}}\in\mathrm{SU}(2)^{\vert\partial\mathcal{V}^{\ast}\vert}$. Using this, the transition amplitudes of the Ponzano-Regge model for some given boundary spin network state is defined as follows:

\begin{Definition}\label{TransAmplPR} (Transition Amplitudes of the Ponzano-Regge Model)\newline
Let $\Psi$ be an abstract spin network living on the boundary dual $1$-skeleton of the triangulation $\Delta$ and $\psi\in\mathcal{H}_{\partial}$ be the corresponding spin network function. Then we define the transition amplitude of the Ponzano-Regge spin foam model with respect to this boundary state by
\begin{align*}\langle\mathcal{Z}_{\mathrm{PR}}\vert \Psi\rangle:=\langle\mathcal{Z}_{\mathrm{PR}}[\Delta]\vert \psi\rangle_{L^{2}}=\int_{\mathrm{SU}(2)^{\vert\partial\mathcal{E}^{\ast}\vert}}\,\bigg (\prod_{e^{\ast}\in\partial\mathcal{E}^{\ast}}\mathrm{d}g_{e^{\ast}}\bigg )\,\mathcal{Z}_{\mathrm{PR}}[\Delta,\{g_{e^{\ast}}\}_{e^{\ast}\in\partial\mathcal{E}^{\ast}}]\cdot \psi(\{g_{e^{\ast}}\}_{e^{\ast}\in\partial\mathcal{E}^{\ast}}),\end{align*}
where $\mathcal{Z}_{\mathrm{PR}}[\Delta,\{g_{e^{\ast}}\}_{e^{\ast}\in\partial\mathcal{E}^{\ast}}]$ denotes the group representation of the Ponzano-Regge partition function as introduced in Proposition \ref{step2}.\end{Definition}

Note the absence of the complex-conjugate in this expression, which usually comes from the $L^{2}$-inner product. This is due to the fact that $\mathcal{Z}_{\mathrm{PR}}$ is real-valued, which follows from the fact that $\overline{\delta_{\mathrm{SU}(2)}}=\delta_{\mathrm{SU}(2)}$. The interpretation of this quantity is as follows: If $\gamma$ has two disconnected components, then $\langle\mathcal{Z}_{\mathrm{PR}}\vert \Psi\rangle$ computes the probability amplitude between the two states described by the components, where the transition is weighted by the Ponzano-Regge partition function. If $\gamma$ has a single boundary component, then $\langle\mathcal{Z}_{\mathrm{PR}}\vert\Psi\rangle$ can be interpreted as the probability for transition of the state described by $\Psi$ from the vacuum. \cite{FreidelPonzanoRegge1}, \cite[p.70f.]{GoellerThesis}
\chapter{Coloured Group Field Theory in 3D and Transition Amplitudes}\label{Chap3}
Previously, we have discussed various spin foam models for $3$-dimensional quantum gravity, like the Ponzano-Regge and Turaev-Viro model. In this Chapter, we discuss yet another approach to quantum gravity, namely ``Group Field Theories'' (short ``GFTs''), which are closely related to spin foam models and to other approaches of quantum gravity, like LQG as well as matrix and tensor models. GFTs are combinatorially non-local field theories defined on (copies of) a Lie group (or quantum group, homogeneous space). They have a similar combinatorial structure as tensor models, but are enriched with group theoretic data allowing for proposing additional symmetry properties of their fields and for the application of methods from group theory. On the other hand, we can also apply methods developed in field theory. We will be mostly interested in the ``Boulatov model'', which is a simplicial GFT for $3$-dimensional Riemannian quantum gravity without a cosmological constant. As we will see in the following, the Feynman amplitudes of this model turn out to be the transition amplitudes of the Ponzano-Regge spin foam model and hence, the transition amplitudes of the Boulatov model generically involve a sum over all bulk topologies for some given boundary manifold, as opposed to the Ponzano-Regge model, where we have fixed both the boundary and bulk topology. The main goal of this chapter is to develop a formalism in order to deal with this sum in a more systematic way. For this, it turns out that the ``coloured'' version of the Boulatov model is most convenient, since its Feynman graphs are ``coloured graphs'', i.e. graphs admitting a proper edge-colouring. This colouring encodes a full simplicial complex, which is generically dual to a (normal) pseudomanifold. In the first part of this chapter, we define the coloured Boulatov model and discuss some of its properties. We then move to a detailed discussion of the topology of coloured graphs, where we will take great advantage from tools developed in ``crystallization theory'' \cite{GagliardiBoundaryGraph}, a branch of geometric topology. Last but not least, we will discuss transition amplitudes of this model and show how to write them in a more systematic way.
 
\section{The Coloured Simplicial Boulatov Model}
Let us start by discussing the Boulatov model for $3$-dimensional Riemannian quantum gravity. We first of all review the definition and combinatorial structure of the original (uncoloured) model by D. V. Boulatov from 1992 and show how it is related to the Ponzano-Regge spin foam model. We then discuss its coloured version and explain several advantages of using colouring when discussing tensor models and GFTs. For a general introduction into group field theories see for example  \cite{FreidelGFTOverview,Oriti06,Oriti07,OritiGFTApproach,Oriti10,KarjewskiGFT,OritiMicroscopicDynamics,CarrozzaGFT} and references therein. For the relationship with spin foam models and LQG see \cite{OritiLQG1,OritiLQG2}.

\subsection{Group Field Theory in 3D: The Boulatov Model}
The Boulatov model, proposed by D. V. Boulatov in 1992 \cite{BoulatovModel}, is a group field theory based on the Lie group $\mathrm{SU}(2)$ (or alternatively $\mathrm{SO}(3)$), which describes $3$-dimensional Riemannian quantum gravity without a cosmological constant. The model is defined to be a quantum field theory based on three copies of the Lie group $\mathrm{SU}(2)$. The main objects of the theory are fields of the type $\varphi\in L^{2}(\mathrm{SU}(2)^{3})$ with values either in $\mathbb{R}$ or $\mathbb{C}$ satisfying the $\mathrm{SU}(2)$ gauge invariance property
\begin{align}\label{InvCond}\forall h\in\mathrm{SU}(2):\varphi(hg_{1},hg_{2},hg_{3})=\varphi(g_{1},g_{2},g_{3})\end{align}
for all $g_{1},g_{2},g_{3}\in\mathrm{SU}(2)$. Note that the choice of imposing either right or left translation is just a convention. The Boulatov model in its complex version\footnote{Note that in the original article \cite{BoulatovModel}, the author considered only real-valued fields. The difference between real- and complex-valued fields will become clear later.} is then defined by the classical action
\begin{equation}\label{Boulact}\begin{aligned}\mathcal{S}_{\lambda}[\varphi,\overline{\varphi}]:=&\int_{\mathrm{SU}(2)^{3}}\,\bigg (\prod_{i=1}^{3}\mathrm{d}g_{i}\bigg )\,\varphi(g_{1},g_{2},g_{3})\varphi(g_{1},g_{2},g_{3})\\&-\frac{\lambda}{4!}\int_{\mathrm{SU}(2)^{6}}\,\bigg (\prod_{i=1}^{6}\mathrm{d}g_{i}\bigg )\,\varphi(g_{1},g_{2},g_{3})\varphi(g_{3},g_{4},g_{5})\varphi(g_{5},g_{2},g_{6})\varphi(g_{6},g_{4},g_{1})+c.c.,\end{aligned}\end{equation}
where $\mathrm{d}g$ denotes the normalized Haar measure on $\mathrm{SU}(2)$ and where $\lambda$ is some coupling. Geometrically, a field $\varphi(g_{1},g_{2},g_{3})$ encodes the kinematics of a quantum triangle described by three dual edges labelled with $g_{1},g_{2},g_{3}$. In other words, the GFT field $\varphi$ lives on the space of possible geometries of the triangle. With this interpretation, the kinetic term of the action can be interpreted as the gluing of two triangles and the interaction term can be understood as the gluing of $4$ triangles such that they form a tetrahedron, as the following figure shows:

\begin{figure}[H]
\captionsetup[subfigure]{labelformat=empty}
\centering
\subfloat[$\varphi(g_{1},g_{2},g_{3})$]{\includegraphics[width=0.2\textwidth]{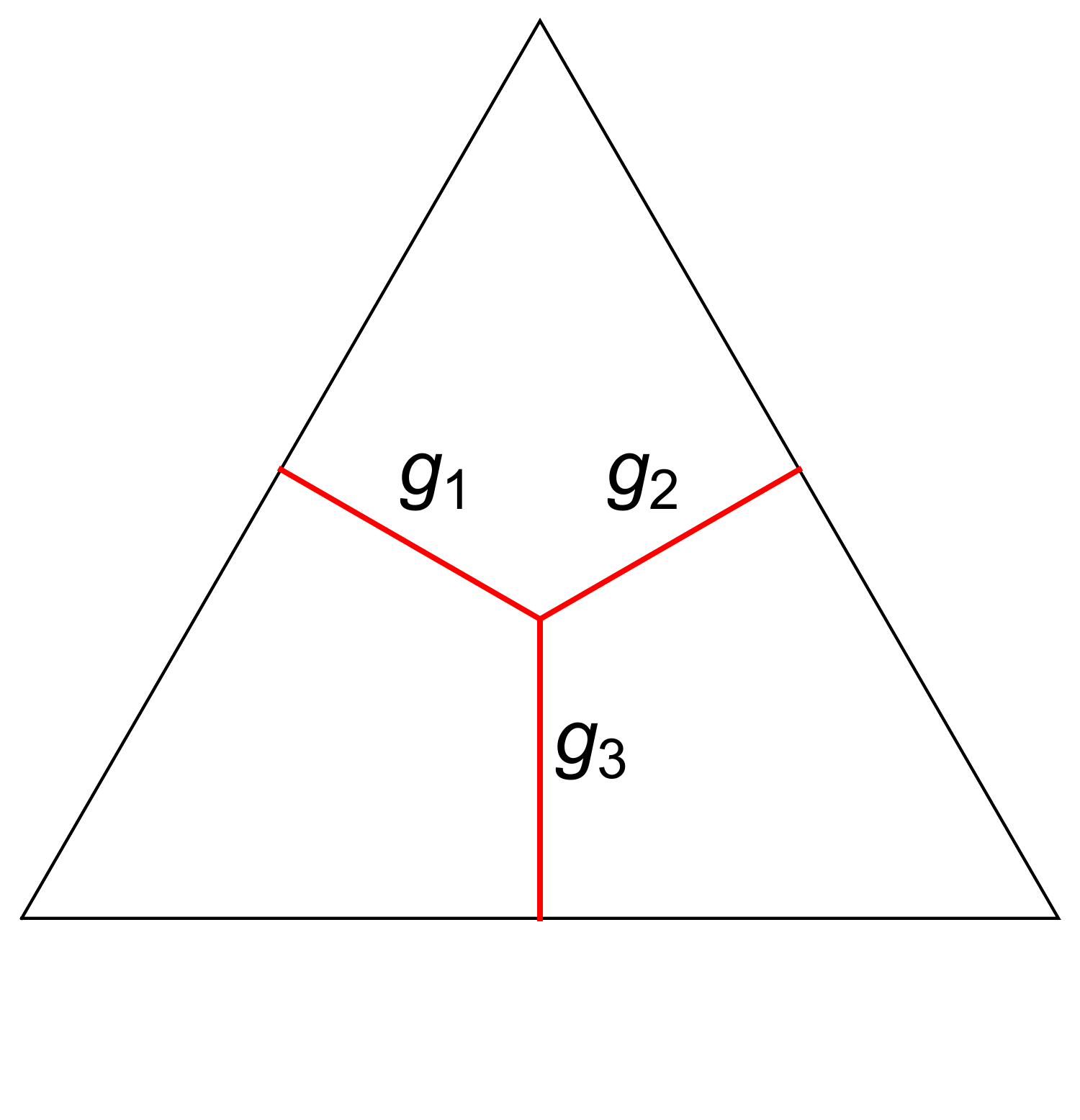}}\hspace{1cm}
\subfloat[$\mathcal{L}_{\lambda,\mathrm{int}}\text{[}\varphi\text{]}\propto\varphi^{4}$]{\includegraphics[trim=1cm 2cm 0 0,width=0.25\textwidth]{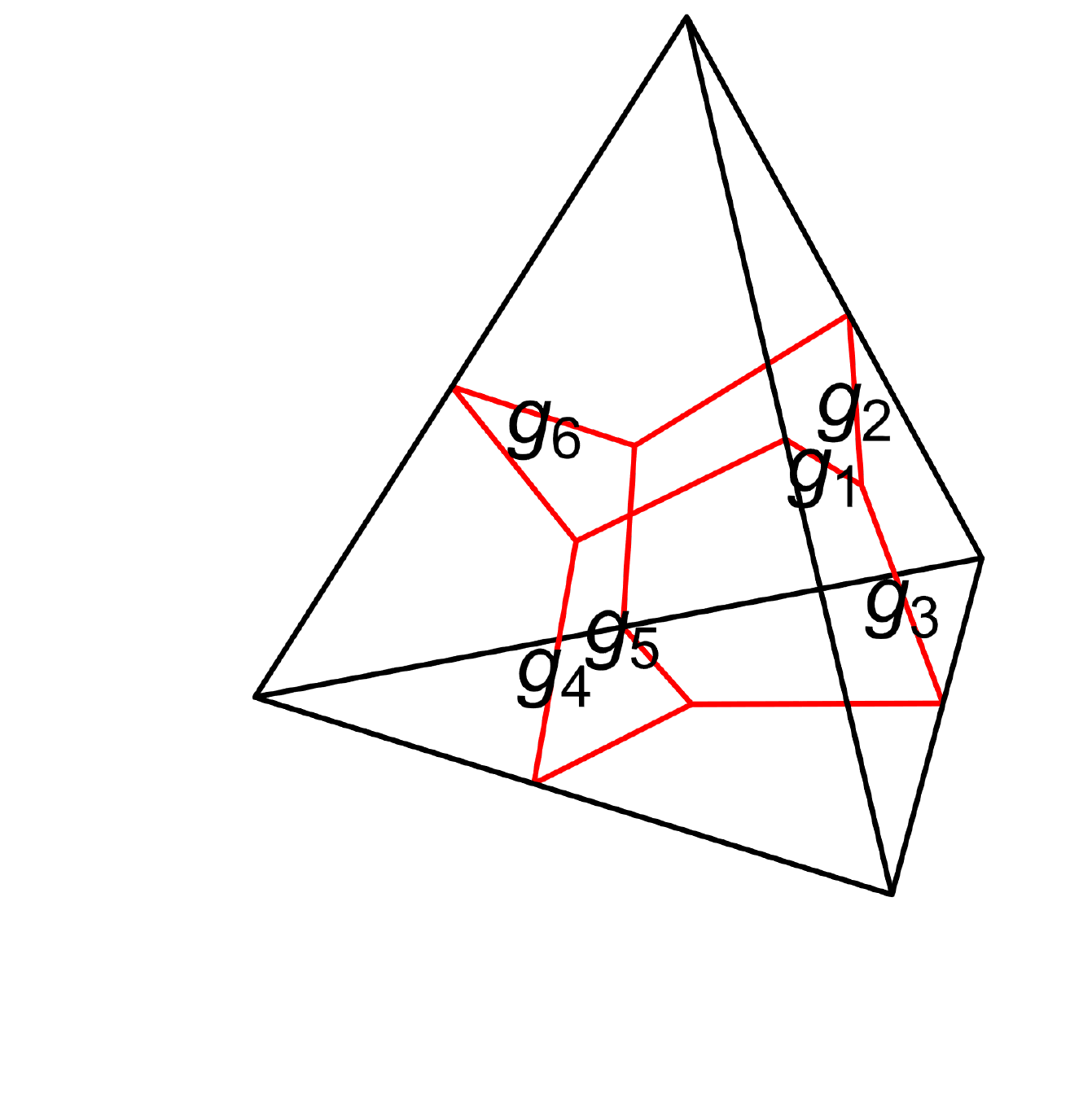}}   
\caption{A triangle described by a field $\varphi$ and a tetrahedron with boundary dual edges drawn in red corresponding to the interaction term of the Boulatov model.}
\end{figure}

With this geometric picture in mind, it also becomes clear how to interpret the symmetry property of Equation (\ref{InvCond}): Recall that at the discrete level, a gauge transformation lives on the vertices of the dual complex. If we view the triangle described by $\varphi(g_{1},g_{2},g_{3})$ in the dual picture, then it is just a single vertex with three outgoing edges to which we have assigned the group elements $g_{1},g_{2},g_{3}$. A gauge transformation hence acts by multiplication with a group element $h$ assigned to this vertex. Therefore, we conclude that the above symmetry property is nothing else than $\mathrm{SU}(2)$ gauge invariance proposed for each triangle.\\
\\
There are several different version of the Boulatov model using a slightly different combinatorics of their field arguments and with different symmetry properties of the Boulatov fields, which in turn leads to a slightly different combinatorics of the corresponding Feynman diagrams. Often one assumes that the field $\varphi$ is invariant under even permutations of arguments, i.e.
\begin{align}\varphi(g_{1},g_{2},g_{3})=\varphi(g_{2},g_{3},g_{1})=\varphi(g_{3},g_{1},g_{2})\end{align}
for all $g_{1},g_{2},g_{3}\in\mathrm{SU}(2)$. This is for example used in the original article \cite{BoulatovModel} by D. V. Boulatov. In the complex case, one often assumes in addition that the fields are orientation-reversing under complex-conjugation, i.e.
\begin{align}\label{prop}\overline{\varphi(g_{1},g_{2},g_{3})}=\varphi(g_{3},g_{2},g_{1}).\end{align}
The reason for this is that one can show that the Feynman graphs in this case are always dual to \textit{orientable} simplicial complexes \cite{DePietriPetronio}.\\
\\
Using methods from group theory and harmonic analysis, we can also take different representations of the fields, which essentially corresponds to different choices of describing the Hilbert space of a quantum triangle. Taking fields of the type $\varphi\in L^{2}(\mathrm{SU}(2)^{3})$ satisfying the gauge invariance condition (\ref{InvCond}) is usually called the ``\textit{group representation}''. Instead of working with fields based on the Lie group $\mathrm{SU}(2)$, we may equivalently use fields defined on the corresponding Lie algebra $\mathfrak{su}(2)$, by using the ``non-commutative group Fourier transform'', which we review in Appendix \ref{NCFourier}: If we take some field $\varphi\in L^{2}(\mathrm{SU}(2)^{3})$, then we get a corresponding field in the ``\textit{Lie algebra representation}'' by taking the Fourier transform, i.e.
\begin{align}\hat{\varphi}(x_{1},x_{2},x_{3}):=\int_{\mathrm{SU}(2)^{3}}\bigg(\prod_{i=1}^{3}\mathrm{d}g_{i}\bigg)\,\varphi(g_{1},g_{2},g_{3})e_{g_{1}^{-1}}(x_{1})e_{g_{2}^{-1}}(x_{2})e_{g_{3}^{-1}}(x_{3}),\end{align}
where $e:T^{\ast}\mathrm{SU}(2)\to\mathbb{C}$ are the ``plane-waves'' of $\mathrm{SU}(2)$, as discussed in the appendix. The fields $\hat{\varphi}$ are elements of $L^{2}_{\star}(\mathfrak{su}(2)^{3})$, which is the image of the Fourier transform defined on $L^{2}(\mathrm{SU}(2)^{3})$. Using the isomorphism $\mathfrak{su}(2)\cong\mathbb{R}^{3}$, we can also think of $x_{i}$ being the three vectors describing the edges of the triangle, as sketched in the following figure:

\begin{figure}[H]
\captionsetup[subfigure]{labelformat=empty}
\centering
\subfloat[$\hat{\varphi}(x_{1},x_{2},x_{3})$]{\includegraphics[width=0.2\textwidth]{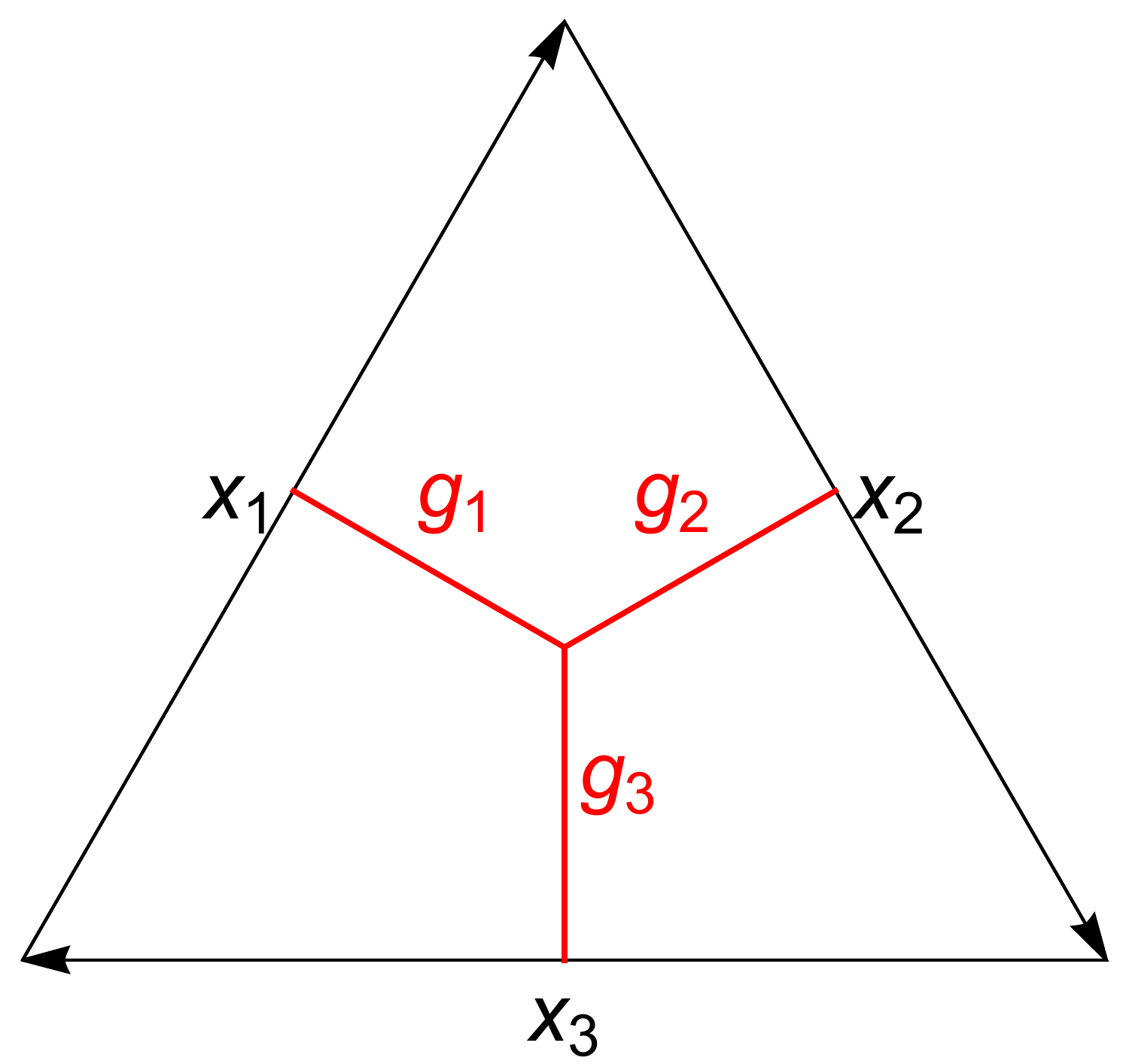}} 
\caption{A triangle described by a field $\hat{\varphi}$ depending on the edge vectors $x_{i}$. }
\end{figure}

In order for three vectors to define a triangle, they should of course not be linearly independent, but rather should satisfy the closure constraint
\begin{align}x_{1}+x_{2}+x_{3}=0.\end{align}
Now, it turns out that this closure constraint exactly corresponds to the $\mathrm{SU}(2)$ gauge invariance (\ref{InvCond}) in the group representation. More precisely, it holds that 
\begin{align}\widehat{P\varphi}=C\star\hat{\varphi}\end{align}
with $C(x_{1},x_{2},x_{3}):=\delta(x_{1}+x_{2}+x_{3})$, where $\delta$ is the delta function of the space $L^{2}_{\star}(\mathfrak{su}(2))$, as defined in Appendix \ref{NCFourier}, and where $P:L^{2}(\mathrm{SU}(2)^{3})\to L^{2}(\mathrm{SU}(2)^{3})$ is the projection operator defined by 
\begin{align}(P\varphi)(g_{1},g_{2},g_{3}):=\int_{\mathrm{SU}(2)}\,\mathrm{d}h\,\varphi(hg_{1},hg_{2},hg_{3}).\end{align}
After a straightforward calculation, one finds that the action in this representation can be written as
\begin{equation}\begin{aligned}&\mathcal{S}_{\lambda}[\hat{\varphi},\hat{\overline{\varphi}}]:=\int_{\mathfrak{su}(2)^{3}}\,\bigg (\prod_{i=1}^{3}\frac{\mathrm{d}^{3}x_{i}}{(2\pi)^{3}}\bigg )\,\hat{\overline{\varphi}}(x_{1},x_{2},x_{3})\star\hat{\varphi}(x_{1},x_{2},x_{3})\\&-\frac{\lambda}{4!}\int_{\mathfrak{su}(2)^{6}}\,\bigg (\prod_{i=1}^{6}\frac{\mathrm{d}^{3}x_{i}}{(2\pi)^{3}}\bigg )\,\hat{\varphi}(x_{1},x_{2},x_{3})\star\hat{\varphi}(x_{3},x_{4},x_{5})\star\hat{\varphi}(x_{5},x_{2},x_{6})\star\hat{\varphi}(x_{6},x_{4},x_{1})+c.c.,\end{aligned}\end{equation}
where $\mathrm{d}^{3}x$ denotes the Lebesgue measure on $\mathfrak{su}(2)\cong\mathbb{R}^{3}$ and where the star product has to be understood as $\hat{\varphi}(x_{i})\star\hat{\varphi}(x_{i}):=(\hat{\varphi}\star\hat{\varphi}_{-})(x_{i})$ for repeated arguments with $\hat{\varphi}_{-}(x_{i}):=\hat{\varphi}(-x_{i})$. More details can be found in the original article \cite{OritiNonCom} as well as in \cite[p.21ff.]{OritiMicroscopicDynamics}.\\
\\
Yet another way to represent the Boulatov fields is the ``\textit{spin representation}'', which is obtained by decomposing the fields using the Theorem of Peter-Weyl. This essentially amounts to decomposing the Hilbert space of a quantum triangle in the group representations into its irreducible subspaces. The fields are decomposed as
\begin{equation}\begin{aligned}\varphi(g_{1},g_{2},g_{3})=\sum_{j_{1},j_{2},j_{3}}\sum_{m_{i},k_{i}}\phi_{j_{1}j_{2}j_{3}}^{m_{1}m_{2}m_{3}}D^{j_{1}}_{m_{1}k_{1}}(g_{1})D^{j_{2}}_{m_{2}k_{2}}(g_{2})D^{j_{3}}_{m_{3}k_{3}}(g_{3})\begin{pmatrix}j_{1} & j_{2} & j_{3}\\k_{1} & k_{2} & k_{3}\end{pmatrix},\end{aligned}\end{equation}
where the complex coefficients $\phi_{j_{1}j_{2}j_{3}}^{m_{1}m_{2}m_{3}}\in\mathbb{C}$ correspond to the spin-representation of the field. The action in this representation can be written as
\begin{equation}\begin{aligned}\mathcal{S}_{\lambda}[\phi,\overline{\phi}]=&\sum_{j_{1},j_{2},j_{3}}\sum_{m_{i}}\vert\phi_{j_{1}j_{2}j_{3}}^{m_{1}m_{2}m_{3}}\vert^{2}-\frac{\lambda}{4!}\sum_{j_{1},\dots,j_{6}}\sum_{m_{i}}(-1)^{\sum_{i=1}^{6}(j_{i}-m_{i})}\times\\&\times\phi_{j_{1}j_{2}j_{3}}^{-m_{1}-m_{2}-m_{3}}\phi_{j_{3}j_{4}j_{5}}^{m_{3}-m_{4}m_{5}}\phi_{j_{5}j_{2}j_{6}}^{-m_{5}m_{2}m_{6}}\phi_{j_{6}j_{4}j_{1}}^{-m_{6}m_{4}m_{1}}\begin{Bmatrix}j_{1}&j_{2}&j_{3}\\j_{4}&j_{5}&j_{6}\end{Bmatrix}+c.c.\end{aligned}.\end{equation}
For a derivation of this formula see the original article \cite[p.3f.]{BoulatovModel}. Also other representations have been proposed, like the ``\textit{vertex representation}'', discussed in \cite{GFTDiff2,GFTVertex}.

\subsection{Quantum Dynamics and Relation to the Ponzano-Regge Model}\label{DynamicsGFT}
The dynamics of a GFT is defined by the perturbative expansion of the partition function in the coupling $\lambda$ in terms of Feynman diagrams. Formally, we have that
\begin{align}\label{PertExp}\mathcal{Z}=\int\,\mathcal{D}\varphi\mathcal{D}\overline{\varphi}\,e^{-\mathcal{S}_{\lambda}[\varphi,\overline{\varphi}]}=\sum_{\Gamma}\frac{\lambda^{N_{\Gamma}}}{\mathrm{sym}(\Gamma)}\mathcal{A}_{\Gamma},\end{align}
where the sum is over all $2$-complexes $\Gamma$ which are dual to $3$-dimensional simplicial complexes $\Delta$, $N_{\Gamma}$ denotes the number of interaction vertices of $\Gamma$, $\mathrm{sym}(\Gamma)$ denotes the symmetry factor and $\mathcal{A}_{\Gamma}$ denotes the Feynman amplitude of $\Gamma$. The type of simplicial complexes appearing in the sum depends crucially on the chosen symmetry assumption of the Boulatov fields. The calculation of the amplitude $\mathcal{A}_{\Gamma}$ corresponding to some Feynman diagram $\Gamma$ can be done in all the different representations discussed in the previous part. In the end, we will see that in all three representations, we reproduce the discretized quantum partition function of $3$-dimensional Riemannian quantum gravity in different formulations.\\
\\
In the group representation, we have to start with the action as defined in Equation (\ref{Boulact}). As in standard QFT, the propagator of the theory is given by the inverse of the kinetic kernel, which in the case of the Boulatov model is given by
\begin{align}\mathcal{K}^{-1}(\{g_{i}\},\{\widetilde{g}_{i}\})=\int_{\mathrm{SU}(2)}\,\mathrm{d}h\,\prod_{i=1}^{3}\delta(g^{-1}_{i}h\widetilde{g}_{i})=\mathcal{K}(\{g_{i}\},\{\widetilde{g}_{i}\}).\end{align}
Also the interaction vertex can directly be read of the action and is given by 
\begin{align}\mathcal{V}(\{g_{ij}\})=\int_{\mathrm{SU}(2)^{4}}\,\bigg(\prod_{i=1}^{4}\mathrm{d}h_{i}\bigg )\,\prod_{i\neq j}\delta(g_{ij}^{-1}h_{i}^{-1}h_{j}g_{ji}),\end{align}
where $g_{ij}$ is the group element living on the triangle labelled by $i$ assigned to the dual edge connecting the triangles $i$ and $j$. The elements $h_{i}$ can be viewed as group elements assigned to the interior dual edges connecting the barycentre of the tetrahedron with the barycentre of the triangle labelled by $i$. We now can compute the amplitudes as in standard QFT. Basically the only mathematical formula which we need in order to do so is the following identity:
\begin{align}\label{formula}\int_{\mathrm{SU}(2)}\,\mathrm{d}k\,\delta(g^{-1}h_{1}k)\delta(k^{-1}h_{2}\widetilde{g})=\delta(g^{-1}h_{1}h_{2}\widetilde{g})\end{align} 
Let $\Gamma$ now be a fixed $2$-complex, which is dual to some $3$-dimensional simplicial complex $\Delta$ without boundary. As usual, we denote the set of all edges in $\Delta$ by $\mathcal{E}$, the set of all faces in $\Delta$ by $\mathcal{F}$, as well as the set of all dual edges in $\Gamma$ by $\mathcal{E}^{\ast}$ and the set of all dual faces in $\Gamma$ by $\mathcal{F}^{\ast}$. Next, using the position-space Feynman rules, we get an integration for each dual edge in $\mathcal{E}^{\ast}$ as well as the corresponding vertex kernel $\mathcal{V}(\{g_{ij}\})$. Furthermore, we get for each interaction vertex a factor of $\lambda$ as well as an overall symmetry factor for the diagram $\Gamma$. In our convention above (Formula (\ref{PertExp})), we have already factored out this two contributions and hence we ignore it in the following. Now, two triangles are glued together by a propagator and hence we get a factor of $\mathcal{K}(\{g_{i}\},\{\widetilde{g}_{i}\})$ for each gluing. After integrating over all the group elements denoted by $g_{ij}$ assigned to the boundary dual edges, we are left with a delta function for each dual face $f^{\ast}$ containing a product over all the group elements $h_{i}$ assigned to the dual edges contained in $f^{\ast}$. In other words, the Boulatov model imposes flatness of the
(discrete) curvature assigned to each dual 2-face. As a consequence, we will end up with the following expression for the amplitude $\mathcal{A}_{\Gamma}$ corresponding to the $2$-complex $\Gamma$: 
\begin{align}\mathcal{A}_{\Gamma}=\int_{\mathrm{SU}(2)^{\vert\mathcal{E}^{\ast}\vert}}\!\bigg(\prod_{e^{\ast}\in\mathcal{E}^{\ast}}\mathrm{d}g_{e^{\ast}}\bigg)\,\prod_{f^{\ast}\in\mathcal{F}^{\ast}}\delta\bigg(\overrightarrow{\prod_{e^{\ast}\subset f^{\ast}}}h_{e^{\ast}}^{\varepsilon(e^{\ast},f^{\ast)}}\bigg),\end{align}
where $\varepsilon(e^{\ast},f^{\ast})$ is equal to $1$ if the orientation of $e^{\ast}$ and $f^{\ast}$ agrees and $-1$ otherwise. Now, this is exactly the Ponzano-Regge partition function in its group representation (see Proposition (\ref{step2})). Similarly, if we work with the Lie algebra representation, one can show that the resulting amplitude becomes (\cite{OritiNonCom} and \cite[p.29]{OritiMicroscopicDynamics})
\begin{align}\mathcal{A}_{\Gamma}=\int_{\mathrm{SU}(2)^{\vert\mathcal{E}^{\ast}\vert}}\!\bigg(\prod_{e^{\ast}\in\mathcal{E}^{\ast}}\mathrm{d}g_{e^{\ast}}\bigg)\int_{\mathfrak{su}(2)^{\vert\mathcal{F}^{\ast}\vert}}\!\bigg(\prod_{f^{\ast}\in\mathcal{F}^{\ast}}\mathrm{d}X_{f^{\ast}}\bigg)\,\mathrm{exp}\bigg (i\cdot\mathrm{tr}\bigg\{\sum_{f^{\ast}\in\mathcal{F}^{\ast}}X_{f^{\ast}}G_{f^{\ast}}\bigg\}\bigg ).\end{align}
This is exactly the discretized path integral of $3$-dimensional Riemannian gravity in its $\mathrm{BF}$-formulation (see Definition \ref{disc}). Doing the same calculation in the spin representation yields the partition function of the Ponzano-Regge spin foam model (see Definition \ref{PolzanoRegge} and Equation (\ref{PRSpin})) corresponding to the triangulation $\Delta$ \cite[p.4]{BoulatovModel}:
\begin{align}\mathcal{A}_{\Gamma}=\sum_{j\in (\mathbb{N}_{0}/2)^{\vert\mathcal{F}^{\ast}\vert}}\prod_{f\in\mathcal{F}^{\ast}}(-1)^{2j_{f^{\ast}}}(2j_{f^{\ast}}+1)\prod_{e^{\ast}\in\mathcal{E}^{\ast}}(-1)^{\sum_{i=1}^{3}j_{e^{\ast}i}}\prod_{v^{\ast}\in\mathcal{V}^{\ast}}\begin{Bmatrix}j_{v^{\ast}1}&j_{v^{\ast}2}&j_{v^{\ast}3}\\j_{v^{\ast}4}&j_{v^{\ast}5}&j_{v^{\ast}6}\end{Bmatrix},\end{align}
where the sum is over all assignments of spins to dual faces and where $\mathcal{V}^{\ast}$ is the set of all dual vertices in $\Gamma$, i.e. the set of all tetrahedra of the simplicial complex $\Delta$ dual to $\Gamma$. All in all, we see that the partition function for the Boulatov model generates the sum over all $2$-complexes, dual to $3$-dimensional simplicial complexes, weighted by Feynman amplitudes, which exactly correspond to the Ponzano-Regge partition function. Of course, the expression for the partition function is formal and the Feynman amplitudes corresponding to the Ponzano-Regge partition function are in general divergent, as discussed in previous parts. \\
\\
Last but not least, let us mention briefly a general way to represent Feynman diagrams of the Boulatov model by using so-called ``\textit{stranded diagrams}''. The interaction vertices in this representation are drawn as follows:

\begin{figure}[H]
\captionsetup[subfigure]{labelformat=empty}
\centering
\subfloat{\includegraphics[width=0.18\textwidth]{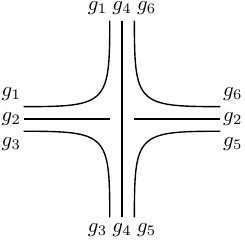}}\hspace{1cm}
\subfloat{\includegraphics[trim=1cm 2cm 0 0,width=0.2\textwidth]{Boulatov1b.pdf}}   
\caption{Interaction vertices in the stranded diagram picture.}
\end{figure}

Each strand represents a face, which bounds the corresponding tetrahedron, and each free line in the diagram represents a boundary dual edge. The propagator, which glues to faces together, allows to connect the distinct strands, as shown in the following figure:

\begin{figure}[H]
\captionsetup[subfigure]{labelformat=empty}
\centering
\subfloat{\includegraphics[width=0.2\textwidth]{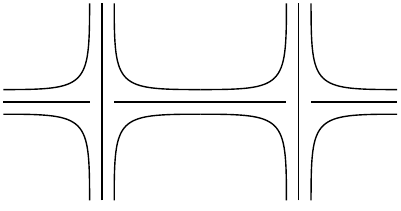}}\hspace{0.5cm}
\subfloat{\includegraphics[trim=0 1cm 0 0,width=0.2\textwidth]{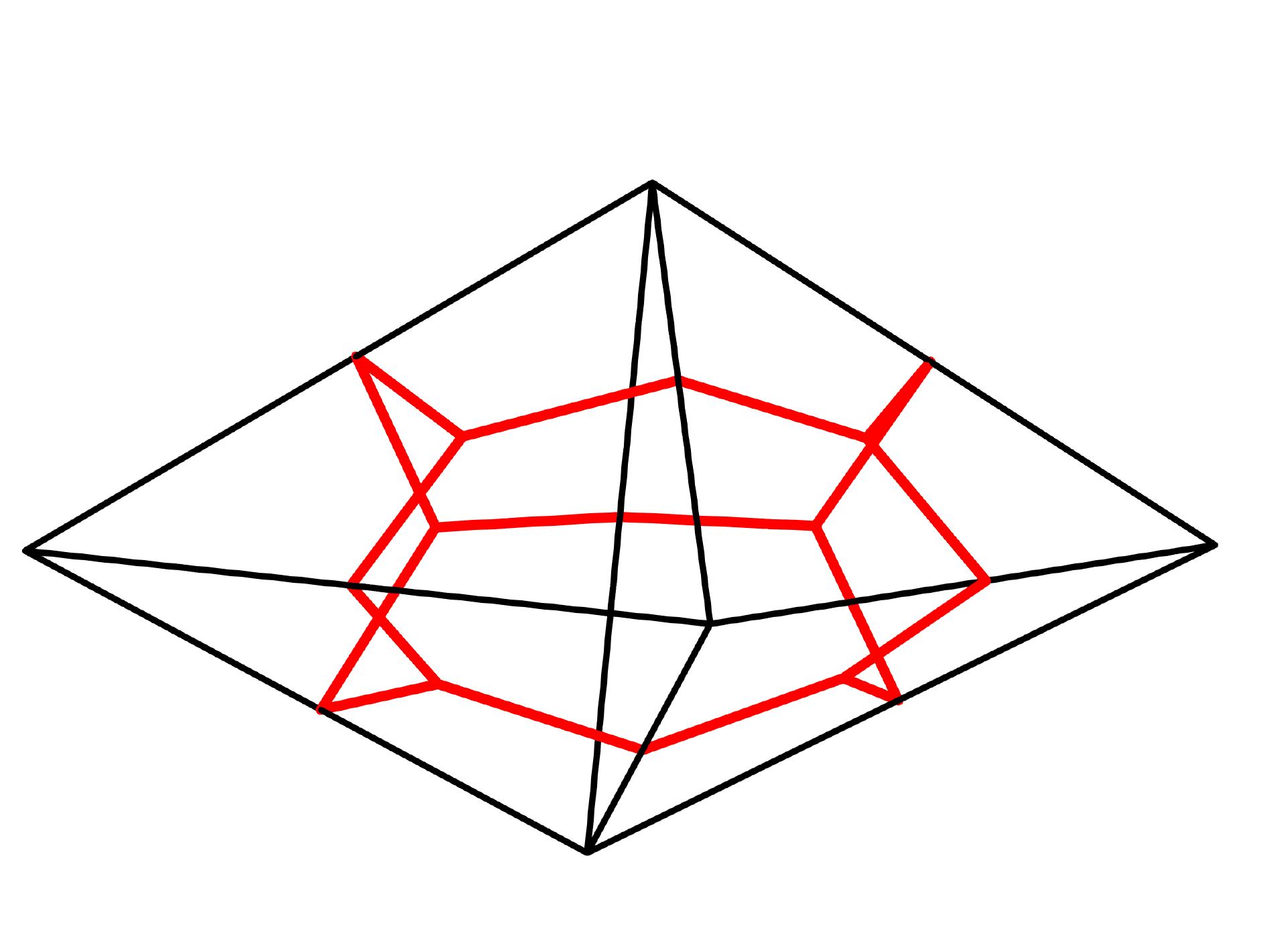}} \hspace{0.5cm}
\subfloat{\includegraphics[width=0.2\textwidth]{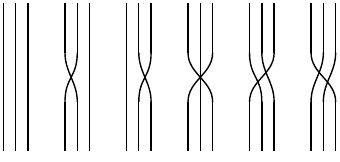}}   
\caption{Gluing of Simplices in the stranded diagram picture.}
\end{figure}

Depending on the chosen additional symmetry properties, the glued strands are also allowed to have twists, as shown on the right-hand side above. An example of a stranded diagram for the triangulation of the $3$-ball with four tetrahedra, obtained by performing a $(1-4)$-Pachner move to a single tetrahedron, is drawn below. 
\begin{figure}[H]
\captionsetup[subfigure]{labelformat=empty}
\centering
\subfloat{\includegraphics[width=0.23\textwidth]{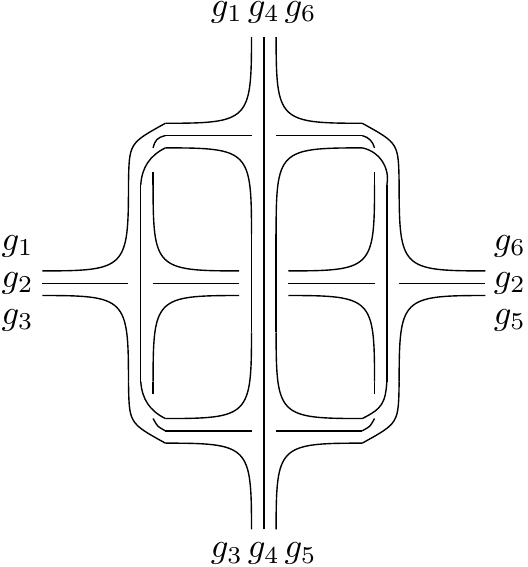}}\hspace{1cm}
\subfloat{\includegraphics[trim=1cm 1.8cm 0 0,width=0.23\textwidth]{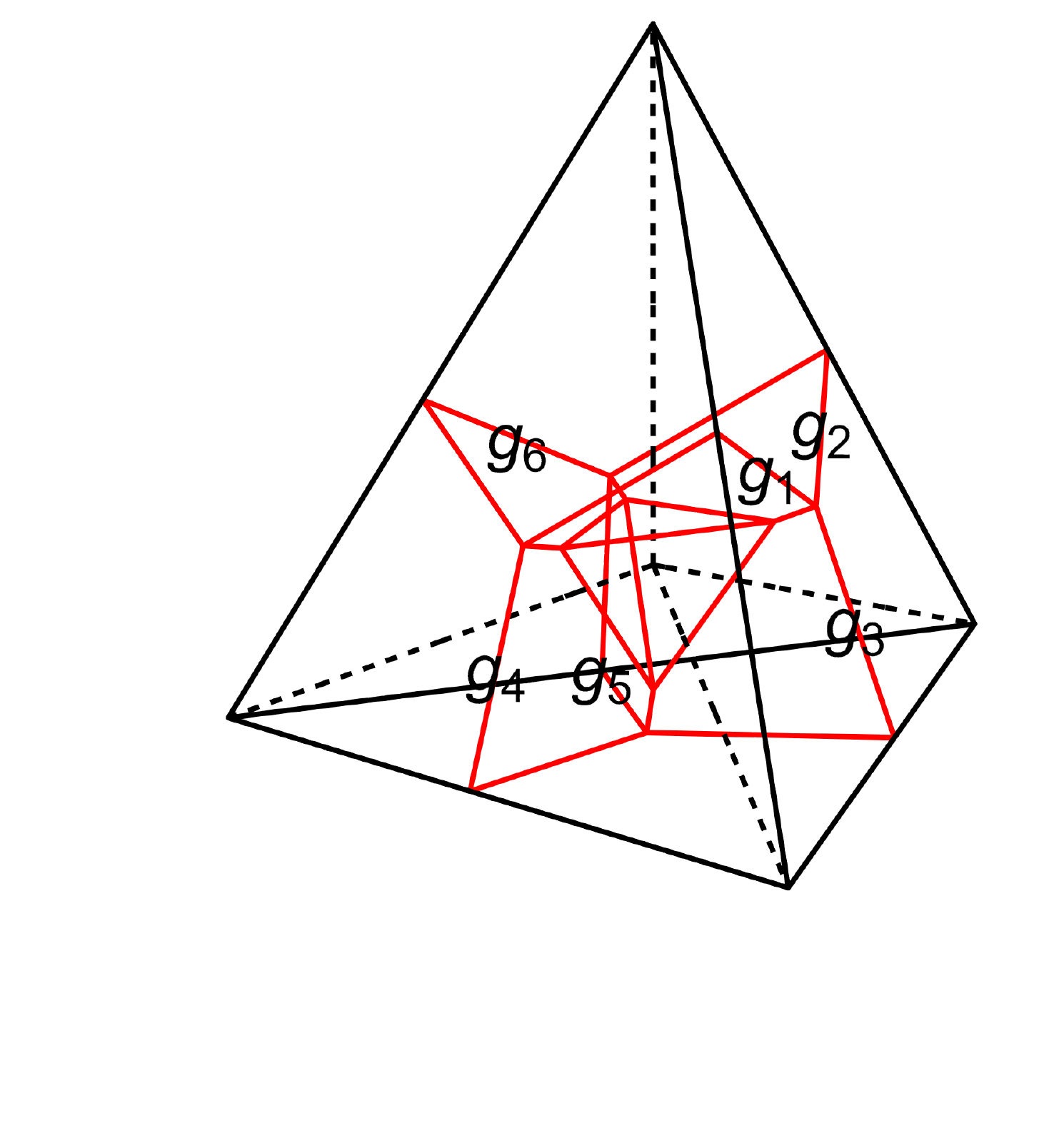}}   
\caption{Stranded diagram for the $(1-4)$-Pachner move and corresponding triangulation with dual $1$-skeleton drawn in red.}
\end{figure}

\subsection{The coloured Boulatov Model and its Feynman Graphs}
Let us now turn to the coloured version of the Boulatov model. More precisely, we will consider the ``\textit{complex, bosonic and coloured Boulatov model of simplicial type}'', which is a QFT based on four $\mathbb{C}$-valued bosonic scalar fields $\{\varphi_{l}\}_{l=0,\dots,3}\subset L^{2}(\mathrm{SU}(2)^{3},\mathbb{C})$, which are $\mathrm{SU}(2)$ gauge invariant, i.e.
\begin{align}\label{GaugeCol}\forall h\in\mathrm{SU}(2):\varphi_{l}(hg_{1},hg_{2},hg_{3})=\varphi_{l}(g_{1},g_{2},g_{3})\end{align} 
for all $g_{1},g_{2},g_{3}\in\mathrm{SU}(2)$ and for all $l\in\{0,1,2,3\}$. The action of this model is given by 
\begin{equation}\begin{aligned}\mathcal{S}_{\lambda}&[\varphi_{l},\overline{\varphi}_{l}]:=\sum_{l=0}^{3}\int_{\mathrm{SU}(2)^{3}}\bigg (\prod_{i=1}^{3}\mathrm{d}g_{i}\bigg )\,\vert\varphi_{l}(g_{1},g_{2},g_{3})\vert^{2}\\&-\lambda\int_{\mathrm{SU}(2)^{6}}\bigg (\prod_{i=1}^{6}\mathrm{d}g_{i}\bigg )\,\varphi_{0}(g_{1},g_{2},g_{3})\varphi_{1}(g_{3},g_{4},g_{5})\varphi_{2}(g_{5},g_{2},g_{6})\varphi_{3}(g_{6},g_{4},g_{1})+c.c.\end{aligned}\end{equation}
The action includes in total four kinetic terms, one for each colour, and there is no mixing of colours in the propagation. Note that we \textit{do not} assume any other symmetry properties or a relation between the fields and their complex conjugate, since the fact that we only produce \textit{orientable} simplicial complexes as our Feynman graphs already follows from the complex nature of the model and from the colouring, as we will see later. Hence, the action includes two distinct interaction terms, one for the $\varphi$-fields and one for the $\overline{\varphi}$-fields. Historically, the coloured Boulatov model was introduced by R. Gurau in 2011 \cite{GurauColouredGFT}. However, note that the author in this publication defined the fields to be Grassmannian. The reason for this is that the model in this case has an additional $\mathrm{SU}(4)$ colour symmetry. However, this symmetry does not change the combinatorics and the structure of the Feynman graphs of the model, and hence we will work with the bosonic theory without loss of generality\footnote{The bosonic coloured Boulatov model has also been discussed earlier, e.g. in \cite{BosonicGFT,GFTDiff2,GFTVertex}.}. Since we have not assumed any additional relation between the $\varphi$- and $\overline{\varphi}$-fields, we have two distinct interaction vertices, which correspond to the two different choices of orientation of a tetrahedron. The two interaction vertices are drawn in their stranded diagram and simplicial picture below.

\begin{figure}[H]
\captionsetup[subfigure]{labelformat=empty}
\centering
\subfloat[$\mathcal{L}_{\lambda,\mathrm{int}}\text{[}\varphi_{l}\text{]}\propto\varphi_{l}^{4}$]{\includegraphics[width=0.2\textwidth]{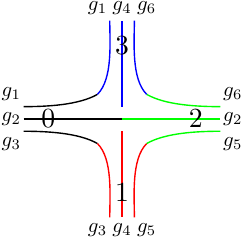}}
\subfloat{\includegraphics[trim=1cm 2cm 0 0,width=0.25\textwidth]{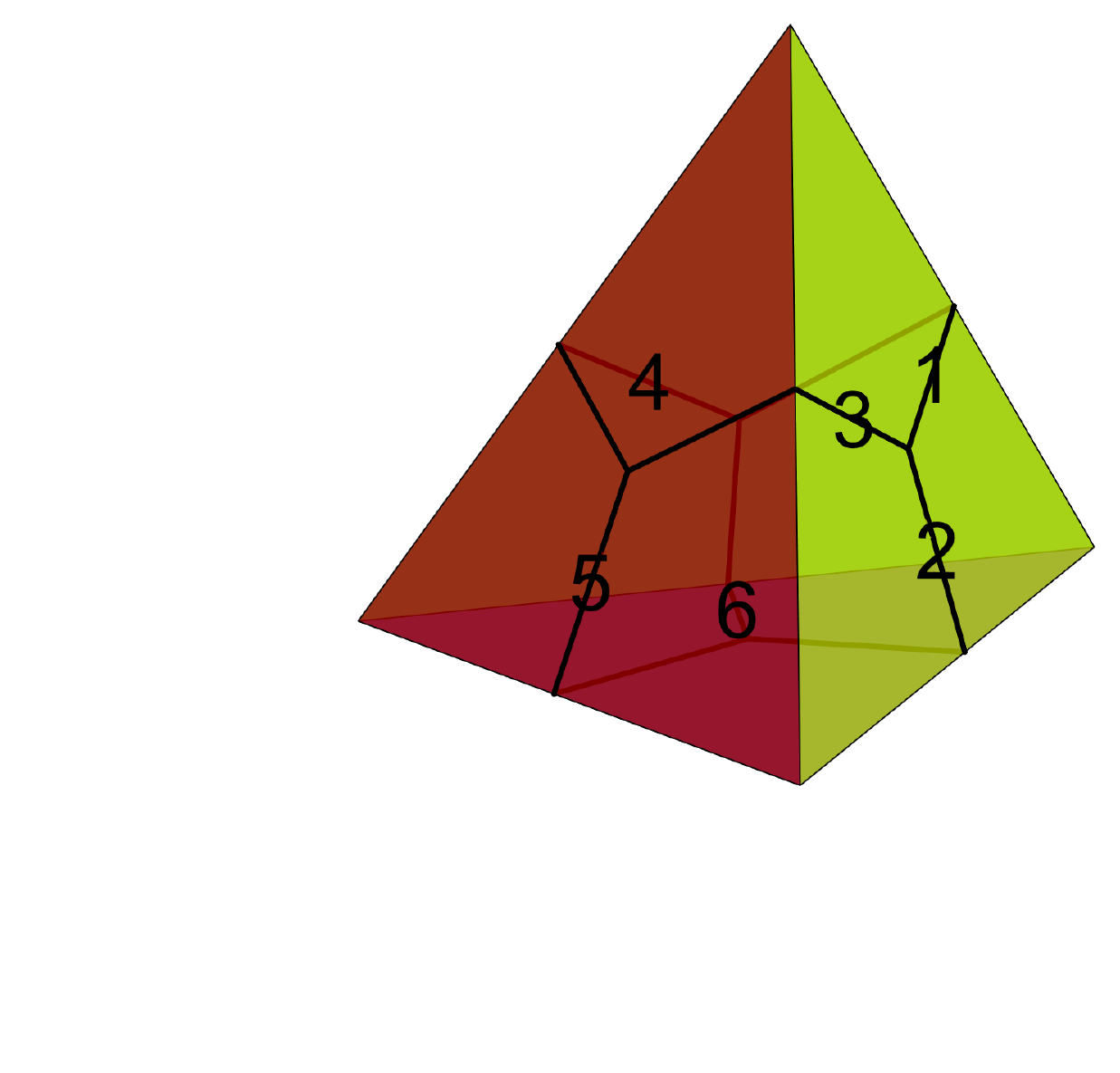}}\hspace{0.5cm}
\subfloat[$\mathcal{L}_{\lambda,\mathrm{int}}\text{[}\overline{\varphi}_{l}\text{]}\propto\overline{\varphi}_{l}^{4}$]{\includegraphics[width=0.2\textwidth]{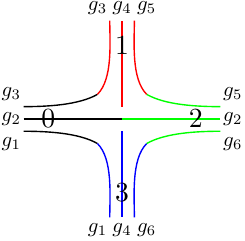}}
\subfloat{\includegraphics[trim=1cm 2cm 0 0,width=0.25\textwidth]{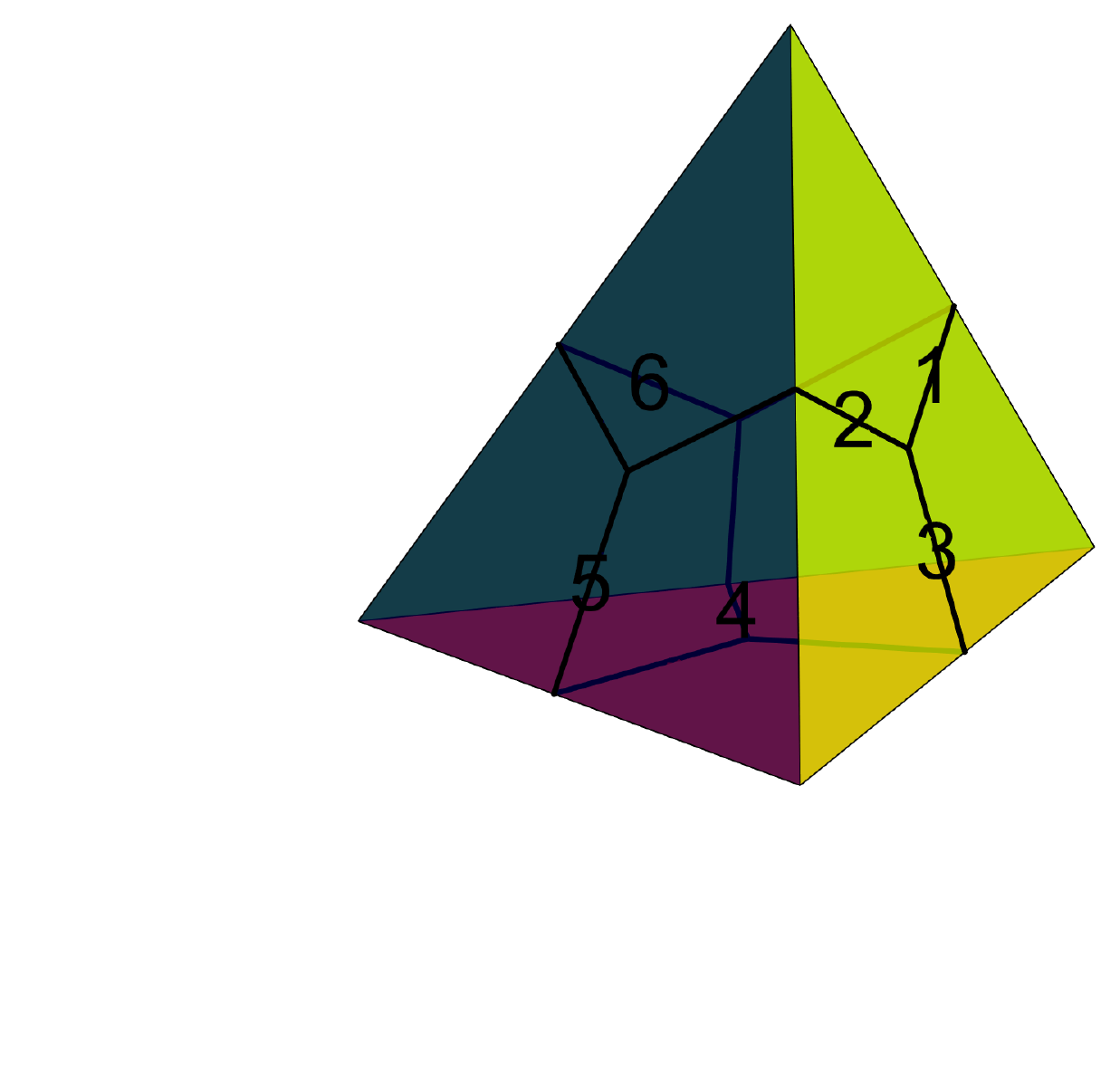}}\hspace{0.5cm}  
\caption{Interaction vertices of the coloured Boulatov model.}
\end{figure}

The kinetic term only allows us to glue two tetrahedra of different types to each other, since it is of the form $\sum_{l}\varphi_{l}\overline{\varphi}_{l}$. Hence, the Feynman graphs are \textit{bipartite} in this case, which means that they have two types of vertices and every edge connects vertices of different types. Furthermore, since we have not assumed any symmetry properties of the field arguments, we can glue two faces belonging two different tetrahedra only in a \textit{unique} way: The colouring of faces induces a vertex colouring, which is obtained by labelling each vertex with the colour of the triangle on the opposite. The gluing of two faces is then such that all the colours of vertices agree. Hence, we see that in the stranded diagram picture, a free line with two colours $i,j\in\{0,1,2,3\}$ is always glued to a free line with the same two colours. In other words, the strand structure of the Feynman diagrams is rigid in this case and their are no twists within the strands. As a consequence, the strand structure is redundant and we can collapse each strand to a single edge of the same colour, as drawn in the figure below.

\begin{figure}[H]
\captionsetup[subfigure]{labelformat=empty}
\centering{\includegraphics[width=0.9\textwidth]{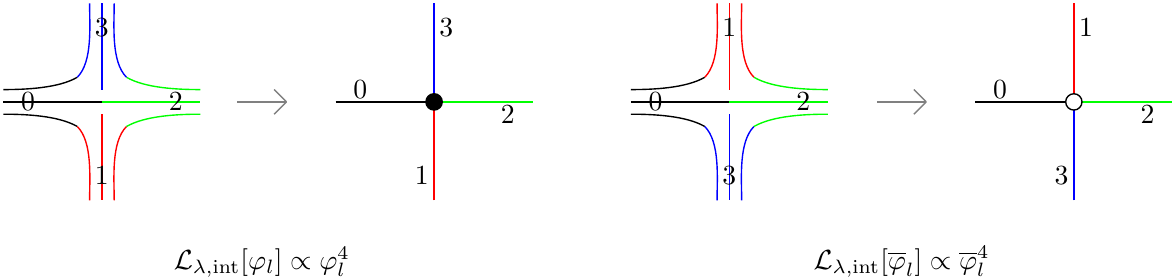}} 
\caption{Feynman graphs of the coloured Boulatov model can equivalently be viewed as coloured graphs.}
\end{figure}

To sum up, the Feynman graphs of the coloured Boulatov model can also be viewed as \textit{proper edge-coloured} and \textit{bipartite} graphs. As we will see, these graphs have a very clear geometrical interpretation and allow us to easily reconstruct the full simplicial complex. Furthermore, they are generically dual to oriented and normal pseudomanifolds.

\subsection{Why do we use a Coloured Model?}
Before studying the topology of proper edge-coloured graphs (possibly with boundary) in detail, let us discuss some reasons why we use the coloured version of the Boulatov model. First of all, in general uncoloured tensor models and GFTs, the perturbative expansion of observables is usually quite hard to control, since the theory produces in general very singular types of complexes. The Feynman graphs of such theories are results of an arbitrary gluing of simplices. However, not every gluing of simplices is an abstract simplicial complex in the mathematical sense (see Definition \ref{SimplComp} in Appendix \ref{SimplicalComplexes}). Of course, an arbitrary gluing of simplices $\Delta$ always satisfies the condition that $\sigma\in\Delta$ and $\tau\subset\sigma$ implies $\tau\in\Delta$. However, by definition, a simplicial complex $\Delta$ is a subset of the powerset of some finite vertex set $\mathcal{V}$. As a consequence, a $d$-simplex is by definition a \textit{set} with cardinality $d+1$. Now, if we glue simplices arbitrarily together, then we could end up with identifying two a priori distinct vertices. As a consequence, such a simplex would become a multiset rather than a set. This might not be a problem, since there might still exists some homeomorphism to a well-behaved simplicial complex, however, the point is that this might not always be possible. As proven in \cite{GurauColouredGFTPseudo}, it turns out that such identifications lead to highly non-trivial topologies, which are not even dual to pseudomanifolds, and which tend to dominate the perturbative expansion in power-counting. As we will discuss in more details later, it turns out that the simplicial complexes dual to coloured graphs are always dual to normal pseudomanifolds and hence, this subtelty does not occur in the coloured case\footnote{It should be mentioned that their are some objections to the claim that this is a specific result for \textit{coloured} GFTs. \cite{SmerlakGurauObjection}}. Another purely technical reason why working with coloured graphs turns out to be an advantage is that there are already many results concerning the topology of coloured graphs known in the mathematical literature. For the uncoloured Boulatov model, the Feynman graphs are in general stranded diagrams with a possibly highly non-trivial strand structures. The topology of such general graphs is not so well known and much harder to study in general. Since we are mainly interested in transition amplitudes of these models, which generically involve a sum over topologies, working with coloured models is hence much more convenient.\\
\\
Besides many known mathematical properties of coloured graphs, there are also many results concerning the physics of coloured models. Examples include the 1/N expansion of coloured simplicial tensor models and GFTs \cite{GurauLargeN1,GurauLargeN2,GurauLargeN3} showing that the dominant contribution to the free energy in the large $N$-limit are ``melonic diagrams'', which are special graphs with spherical topology, as well as the critical behaviour and continuum limit of coloured tensor models \cite{CritTM}. Meanwhile, there was a second class of coloured models developed, which only use a single uncoloured tensor with the most general invariant interactions, labelled by bubbles \cite{CritTM3,GurauBook}. The Feynman graphs of these models are also proper edge-coloured graphs, where the colouring in this setting appears as a canonical book-keeping device and not as a fundamental feature as for simplicial coloured models. It was shown that these models have a similar large N-expansion as simplicial coloured models. Furthermore, one was able to show a universality result in the sense that the tensors are distributed by a Gaussian in the large N-limit \cite{CritTM2}. The two different classes of coloured models have also been shown to be closely related. More precisely, integrating out all coloured tensors except for one in a simplicial coloured tensor model gives precisely a model containing a single coloured tensor with bubble interactions, each weighted by the same coupling \cite{CritTM3}. Finally, let us mention that many result in GFT renormalization theory have been obtained by studying coloured models, i.e. see the review \cite{CarrozzaGFT}.\\
\\
Last but not least, let us mention a possibly physical reason why to use coloured models, namely diffeomorphism invariance. One of the main questions in manifestly background-independent approaches, such as LQG, spin foam models and GFT, is to show how the dynamics reduces to general relativity in a semi-classical and continuum limit. For this, it is essential to understand how diffeomorphism invariance is incorporated in the pregeometric language used. This invariance does not need to be exact at the quantum level, however, it should become exact at least in the continuum and classical limit. In \cite{GFTDiff1,GFTDiff2}, the authors constructed an explicit set of transformations for fields in the coloured Boulatov model, which leaves the action invariant, using the Drinfeld (quantum) double of $\mathrm{SO}(3)$. It was then shown that these transformations are related to the discrete residual action of diffeomorphisms in simplicial gravity path integrals. In their study, it turns out that the colouring is \textit{crucial} in the definition of the proposed GFT diffeomorphism symmetry: Removing the colouring results into a breaking of the symmetry and one can only define a restricted class of transformations for uncoloured fields. Furthermore, this residual symmetry does not admit a clear simplicial gravity interpretations. The authors in \cite{GFTDiff2} end their discussion with the following conclusion:\\
\\
\textit{``In its light, we recognize the colored Boulatov GFT model as the correct GFT description of 3d quantum gravity.''} \cite[p.26]{GFTDiff2}\\
\\
To sum up, this section provided us with many reasons why working with coloured tensor models and GFTs is more convenient from a purely technical point of view, but also gives us some motivations why colouring might be a necessary feature of GFT models of quantum gravity. Hence, we will now turn to a detailed study of the topology of coloured graphs.

\section{Topology of Coloured Graphs with Boundary}
As seen before, the Feynman diagrams of coloured tensor models and GFTs are edge-coloured graphs. It turns out that the topology of these graphs is not only studied in quantum gravity, but also in crystallization theory, which is a branch of geometric and combinatorial topology. Many results have been obtained in this field since it was pioneered by M. Pezzana and his group in the late 1960s and 1970s. In this section, we discuss some general concepts and important results from the topology of coloured graphs, combining notions which are used both in quantum gravity and crystallization theory. We will mainly focus on the general notion of coloured graphs representing pseudomanifolds with boundaries. For a general review of the topology of coloured graphs in the context of coloured tensor models and GFTs see for example the reviews \cite{GurauColouredTensorModelsReview,GurauColouredTensorModelsReview2} as well as the textbook \cite{GurauBook}. For surveys on crystallization theory see for example \cite{GagliardiBoundaryGraph,CTReview,Review2018} and references therein.

\subsection{Coloured Graphs and Pseudomanifolds}
In the following, a ``\textit{graph}'' is a pair $\gamma=(\mathcal{V},\mathcal{E})$, where $\mathcal{V}$ is a set called the ``\textit{vertex set}'' and where $\mathcal{E}$ is a multiset\footnote{Graphs in which $\mathcal{E}$ is allowed to be a multiset are also often called ``\textit{multigraphs}'' in order to distinguish them from ``\textit{simple graphs}'', where the edge set is a proper set. Multigraphs with loops are then usually called ``\textit{pseudographs}'', although note that some authors use the terms ``multigraphs'' and ``pseudographs'' as synonyms. In the following, ``graph'' will always mean ``multigraph without loops''.} containing sets of the form $\{v,w\}\in\mathcal{V}\times\mathcal{V}$, called the ``\textit{edge set}''. Allowing $\mathcal{E}$ to be a multiset means that two vertices can be connected by several edges. However, note that an edge $e=\{v,w\}\in\mathcal{V}\times\mathcal{V}$ is by definition a proper set, which means that we do not allow for ``\textit{loops}'', i.e. edges starting and ending at the same vertex. Let us firstly define closed coloured graphs, which are the vacuum diagrams of coloured tensor models and GFTs:

\begin{Definition}\label{ClosedColouredGraph} (Closed Coloured Graph)\newline
In the following, a ``closed $(d+1)$-coloured graph'' $\mathcal{G}=(\mathcal{V}_{\mathcal{G}},\mathcal{E}_{\mathcal{G}})$ with $d\in\mathbb{N}$ is meant to be a finite, balanced bipartite, $(d+1)$-valent and proper $(d+1)$-edge-coloured graph.
\end{Definition}

We should explain some notions of graph theory, which we have used in this definition: The graph $\mathcal{G}$ is called ``\textit{bipartite}'', if there is a partition $\mathcal{V}_{\mathcal{G}}=V_{\mathcal{G}}\cup\overline{V}_{\mathcal{G}}$, such that every edge connects a vertex in $V_{\mathcal{G}}$ with a vertex in $\overline{V}_{\mathcal{G}}$. If in addition $\vert V_{\mathcal{G}}\vert=\vert \overline{V}_{\mathcal{G}}\vert$, then the graph is called ``\textit{balanced}''. Furthermore, a ``\textit{$(d+1)$-edge-colouring}'' is a map of the form $\gamma:\mathcal{E}_{\mathcal{G}}\to\mathcal{C}_{d}$, where $\mathcal{C}_{d}$ is some set with cardinality $\vert\mathcal{C}_{d}\vert=d+1$, which we call ``\textit{colour set}''. In the following, we will usually label the colours by choosing $\mathcal{C}_{d}:=\{0,\dots,d\}$. Such an edge-colouring is called ``\textit{proper}'', if $\gamma(e_{1})\neq\gamma(e_{2})$ for all edges $e_{1},e_{2}\in\mathcal{E}_{\mathcal{G}}$, which are adjacent to the same vertex $v\in\mathcal{V}_{\mathcal{G}}$, i.e. $v\in e_{1}\cap e_{2}$. In other words, the colouring is injective on each vertex of the graph.

\begin{Remarks}\label{RemarksClosedColouredGraphs}\begin{itemize}\item[]
\item[(a)]Note that we actually do not need to assume closed coloured graphs to be \textit{balanced}, since a bipartite, $(d+1)$-valent and proper $(d+1)$-edge-coloured graph $\mathcal{G}$ is always balanced. To see this, choose a colour $i\in\mathcal{C}_{d}$ and consider the graph $\mathcal{G}^{\prime}$ obtained from $\mathcal{G}$ by deleting all edges of colours $\mathcal{C}_{d}\textbackslash\{i\}$. This graph is disconnected and each connected component consists of two vertices of different type, which are connected by an edge of colour $i$. Hence, we see that the two types of vertices always come in pairs.
\item[(b)]From a physical point of view, the fact that our Feynman diagrams are bipartite graphs comes from the complex nature of the model. For the real version of the coloured Boulatov model, we would also allow for non-bipartite coloured graphs.
\end{itemize}\end{Remarks} 

As a next step, let us generalize the previous notion of coloured graphs to ``open coloured graphs'', which are the non-vacuum diagrams of coloured tensor models and GFTs:

\begin{Definition}\label{OpenColouredGraph} (Open Coloured Graph)\newline
An open $(d+1)$-coloured graph is a finite, bipartite and proper $(d+1)$-edge-coloured graph $\mathcal{G}=(\mathcal{V}_{\mathcal{G}},\mathcal{E}_{\mathcal{G}})$ with the following extra property: The vertex set admits a decomposition $\mathcal{V}=\mathcal{V}_{\mathcal{G},\mathrm{int}}\cup\mathcal{V}_{\mathcal{G},\partial}$, where $\mathcal{V}_{\mathcal{G},\mathrm{int}}$ consists of $(d+1)$-valent vertices, called ``internal vertices'', and where $\mathcal{V}_{\mathcal{G},\partial}$ consists of $1$-valent vertices, which we call ``boundary vertices''.
\end{Definition}

As a consequence, the edge set of an open $(d+1)$-coloured graph $\mathcal{G}$ can be decomposed as $\mathcal{E}=\mathcal{E}_{\mathcal{G},\mathrm{int}}\cup\mathcal{E}_{\mathcal{G},\partial}$, where edges in $\mathcal{E}_{\mathcal{G},\mathrm{int}}$, which we call ``\textit{internal edges}'', connect two internal vertices and an edge in $\mathcal{E}_{\mathcal{G},\partial}$ connects an internal vertex with a boundary vertex. We call an edge of the latter type ``\textit{external leg}''. 

\begin{Remarks}\begin{itemize}\item[]
\item[(a)]An open coloured graph is in general not \textit{balanced} bipartite, i.e. $\vert V_{\mathcal{G}}\vert=\vert\overline{V}_{\mathcal{G}}\vert$ is in general not fulfilled. As an example, take the open $(3+1)$-coloured graph representing a single tetrahedron, i.e. the graph with a single internal vertex and four adjacent external legs.
\item[(b)]There are also other conventions for open graphs in the literature: Some authors define open graphs to be ``pregraphs'', in which external legs are defined to be half-edges, i.e. they do not end at a $1$-valent vertex (e.g. in \cite{SenseTM}). Furthermore, in crystallization theory open graphs are usually defined without external legs at all, i.e. they define graphs with two types of vertices: ``Internal'' $(d+1)$-valent vertices and vertices with valency $\leq d$, which they then call ``boundary vertices'', i.e. see \cite[p.124]{GagliardiBoundaryGraph} and \cite{Gagliardi87}.
\end{itemize}\end{Remarks}

Having defined the notion of coloured graphs, we define a corresponding simplicial complex\footnote{Strictly speaking, we get something which is slightly more general than an abstract simplicial complex. For example, consider the graph with two vertices, which are connected by $(d+1)$-edges. This graph describes two $d$-simplices, which have the exact same vertex set and hence, the set of simplices is no longer a set, but rather a multiset. In the crystallization theory literature, this slightly more general objects are called ``\textit{pseudo(simplicial)-complexes}''. We won't make such a distinction and just speak about ``simplicial complexes''} in the following way, as already mentioned in our discussion of the coloured Boulatov model:

\begin{Definition} Let $\mathcal{G}=(\mathcal{V}_{\mathcal{G}},\mathcal{E}_{\mathcal{G}})$ be some open $(d+1)$-coloured graph. Then we define a simplicial complex $\Delta_{\mathcal{G}}$ in the following way:
\begin{itemize}
\item[(1)]Assign a $d$-simplex $\sigma_{v}$ to each vertex $v\in\mathcal{V}_{\mathcal{G}}$ and colour the $(d-1)$-faces of $\sigma_{v}$ by $d+1$ colours. This induces a vertex colouring, where each   vertex is labelled by the colour of the $(d-1)$-face on the opposite side.
\item[(2)]If two vertices $v,w$ in $\mathcal{G}$ are connected by an edge of colour $i\in\mathcal{C}_{d}$, we glue the two tetrahedra together along their $(d-1)$-face of colour $i$ in the unique ways such that all the colours of vertices agree.
\end{itemize}
We shall denote the corresponding geometric realization by $\vert\Delta_{\mathcal{G}}\vert$.
\end{Definition}

The interpretation of coloured graphs is as follows: If $\mathcal{G}$ is a closed coloured graph, then $\mathcal{G}$ is nothing else than the dual $1$-skeleton of the simplicial complex $\Delta_{\mathcal{G}}$. If $\mathcal{G}$ is an open coloured graph, then $\mathcal{G}$ is the \textit{interior} dual $1$-skeleton, i.e. the dual 1-skeleton without all the dual vertices and dual edges living purely on the boundary. For an example see Figure \ref{OpenGraphFig} below. In general, the 1-skeleton of some complex is of course not enough to specify the full simplicial complex and hence also the topology, since there is no information about how two simplices are glued together. However, the colouring of the graph carries all the informations about higher-dimensional dual cells and about their nested structure and hence, it encodes a \textit{full} simplicial complex. We will discuss this point in more details in Section \ref{SecBubbles}. If $\mathcal{G}$ is an open coloured graph, then the boundary dual $1$-skeleton can be read of as follows:

\begin{Definition} (Boundary Graph)\newline
Let $\mathcal{G}$ be an open $(d+1)$-coloured graph. Then we define the ``boundary graph'' $\partial\mathcal{G}$ as follows: There is a vertex in $\partial\mathcal{G}$ for each external leg in $\mathcal{G}$ and each such vertex has a colour coming from the colour of the corresponding external leg. Two vertices of colours $i$ and $j$ are connected by a bicoloured edge of colour $ij$ whenever there is a bicoloured path in $\mathcal{G}$ with colours $i,j$ starting and ending at the corresponding external legs.
\end{Definition}

\begin{Remarks}\label{RemarkBoundaryGraph}Note that the structure of a boundary graph is in general very different from closed coloured graphs defined previously. The boundary graph of some open $(d+1)$-coloured graph is of course always $d$-valent, however, in general neither proper edge-coloured nor bipartite. What is true instead is that every vertex has a colour $i\in\mathcal{C}_{d}$ and all $d$ edges adjacent to a vertex of colour $i$ are bicoloured with colours $ij$ for $j\in\mathcal{C}_{d}\textbackslash\{i\}$. Note also that this implies that the vertex colouring is such that two vertices of colour $i$ and $j$ can only be connected by an edge of colour $ij$.
\end{Remarks}

The following figure shows an example of an open $(3+1)$-coloured graph together with its boundary graph $\partial\mathcal{G}$ and its simplicial complex $\Delta_{\mathcal{G}}$. In this example one can also directly see the interpretation of $\mathcal{G}$ as the interior dual $1$-skeleton and of $\partial\mathcal{G}$ as the boundary dual $1$-skeleton\footnote{We usually will omit drawing the $1$-valent boundary vertices of open graphs explicitly in order to make the concept of external legs more visible.}.

\begin{figure}[H]
\captionsetup[subfigure]{labelformat=empty}
\centering
\subfloat[$\mathcal{G}$]{\includegraphics[width=0.25\textwidth]{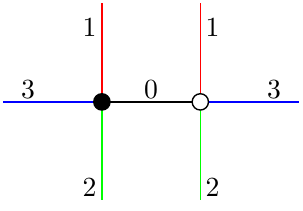}}\hfill
\subfloat[$\partial\mathcal{G}$]{\includegraphics[width=0.3\textwidth]{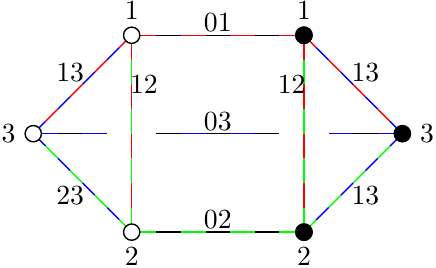}}\hfill
\subfloat[$\Delta_{\mathcal{G}}$]{\includegraphics[width=0.3\textwidth]{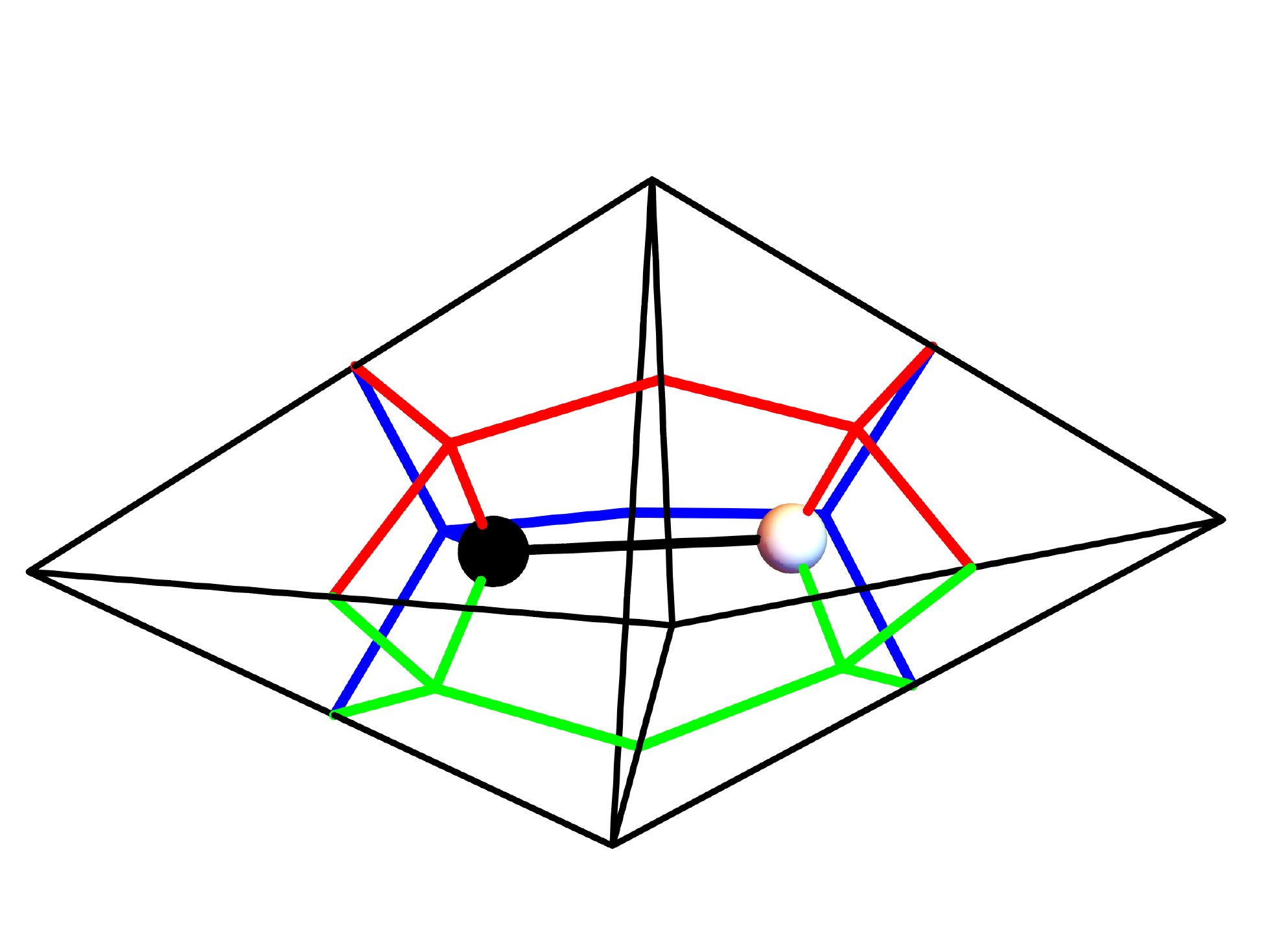}}    
\caption{An open $(3+1)$-coloured graph $\mathcal{G}$ with its boundary graph $\partial\mathcal{G}$ and its corresponding simplicial complex $\Delta_{\mathcal{G}}$ (drawn with its dual $1$-skeleton).\label{OpenGraphFig}}
\end{figure}

As already mentioned a couple of times, one of the great advantages of working with coloured models is the fact that we only produce pseudomanifolds and no other types of topological singularities. This is summarized in the following theorem:

\begin{Theorem} Let $\mathcal{G}$ be an open $(d+1)$-coloured graph. Then $\vert\Delta_{\mathcal{G}}\vert$ is an orientable and normal pseudomanifold with boundary.\end{Theorem}

\begin{proof}The proof that a graph represents a normal pseudomanifold  for the closed case can be found in \cite[p.20ff.]{GurauColouredGFTPseudo}. A generalization for the boundary case is straightforward. For orientability, see for example \cite{GagliardiBoundaryGraph} and \cite{Caravelli}. For the definition of normal pseudomanifolds see Appendix \ref{Pseudomanifolds}.\end{proof}

\begin{Remark}If we consider \textit{real} coloured GFTs, then we are also producing non-orientable manifolds, since for coloured graphs, orientability is equivalent to bipartiteness. In that sense, working with complex models seems to be more natural from a physical point of view.\end{Remark}

As already mentioned in Remark \ref{RemarkBoundaryGraph}, the structure of a boundary graph is quite different from a coloured graph. In the following, it will be much more convenient to restrict only to those open coloured graphs, for which the boundary graph becomes again a closed coloured graph in the sense of Definition \ref{ClosedColouredGraph}. This can be achieved by working only with coloured graphs, for which all the external legs have the same colour. In fact, our definition of open graphs and their boundary graphs, which originates from \cite{GurauBoundaryGraph}, is slightly more general then the notion of coloured graphs used in the crystallization theory literature, in which one usually defines coloured graphs directly in such a way that all external legs have the same colour, e.g. see \cite{GagliardiCorbodant,GagliardiBoundaryGraph,Gagliardi87,GagliardiMultRes}. Let us record the following properties for this special class of open coloured graphs:

\begin{Proposition} Let $\mathcal{G}$ be an open $(d+1)$-coloured graph with the property that all external legs have the same colour. Then the boundary graph $\partial\mathcal{G}$ is a closed $d$-coloured graph as defined in Definition \ref{ClosedColouredGraph} and $\mathcal{G}$ is \textit{balanced} bipartite.
\end{Proposition}

\begin{proof}If all external legs of $\mathcal{G}$ have the same colour, lets say $0$, then there is no information encoded in the vertex colouring of $\partial\mathcal{G}$ and we can ignore it. Furthermore, all the edges of $\partial\mathcal{G}$ are coloured by $0i$ for some $i\in\mathcal{C}_{d}\textbackslash\{0\}$ and hence, we can just colour them by $i$. This shows that $\partial\mathcal{G}$ admits an obvious proper $d$-edge colouring $\gamma_{\partial}:\mathcal{E}_{\partial\mathcal{G}}\to\mathcal{C}_{d-1}^{\ast}$ induced by the colouring $\gamma$ of $\mathcal{G}$, where the colour set in this case is $\mathcal{C}_{d-1}^{\ast}:=\{1,\dots,d\}$. To see that $\partial\mathcal{G}$ is bipartite, just observe that every edge in $\partial\mathcal{G}$ comes from a bicoloured path of $\mathcal{G}$, which starts and ends at an external leg of the same colour. Hence, the number of edges contained in this path is odd, which means that the number of vertices contained in this path is even. It follows that the source and target vertex of an edge of $\partial\mathcal{G}$ are of different kind. For the second claim, note that in this case the graph $\mathcal{G}^{\prime}$ obtained from $\mathcal{G}$ by deleting all the edges of colour $0$ is a (possibly disconnected) $d$-valent and proper $d$-edge coloured graph and such a graph is always balanced (see Remark \ref{RemarksClosedColouredGraphs}(a)). Hence, also the internal vertices come in pairs, which proves the claim.
\end{proof}

\begin{Remark}It is important to note that we restrict only to proper edge-coloured boundary graphs and hence only to open graphs with the property that all external legs have the same colour for purely technical reasons. However, physically speaking, this is a restriction in the class of boundary states we consider.\end{Remark}

From now we will mainly work with this restricted class of open coloured graphs. Therefore, it is convenient to establish the following notation:

\begin{Definition} (Notation for a Special Class of Coloured Graphs)\newline
We will denote by $\mathfrak{G}_{d}$ the set of all open $(d+1)$-coloured graphs in which all external legs have the same colour $0$. For the subset of closed $(d+1)$-coloured graphs we write $\overline{\mathfrak{G}}_{d}\subset\mathfrak{G}_{d}$.\end{Definition}

\begin{Remark}By definition, if $\mathcal{G}\in\mathfrak{G}_{d}$, then $\partial\mathcal{G}\in\overline{\mathfrak{G}}_{d-1}$ and $\partial(\partial\mathcal{G})$ is the empty graph.\end{Remark}

Last but not least, let us discuss a notion of equivalence for coloured graphs: For this recall that an isomorphism of two (multi)graphs $\Gamma_{1}=(\mathcal{V}_{1},\mathcal{E}_{1})$ and $\Gamma_{2}=(\mathcal{V}_{2},\mathcal{E}_{2})$ is a bijective map $\varphi:\mathcal{V}_{1}\to\mathcal{V}_{2}$ such that two vertices $v,w\in\mathcal{V}_{1}$ are adjacent in $\Gamma_{1}$ by $k$ edges if and only if $\varphi(v)$ and $\varphi(w)$ are adjacent by the same number of edges in $\Gamma_{2}$. If such a map exists, we say that $\Gamma_{1}$ and $\Gamma_{2}$ are isomorphic and write $\Gamma_{1}\cong\Gamma_{2}$. In other words, two isomorphic graphs can be brought into the same form by shifting the vertices and edges accordingly, since they have the same number of vertices and the same connectivity of edges. In general, two isomorphic coloured graphs might not describe the same simplicial complex, since they might have a different colouring, which results into a different gluing of simplices. This is again a consequence of the fact that the (dual) 1-skeleton is not enough to specify the topology. Therefore, it is natural to define the following notion of isomorphism for coloured graphs:

\begin{Definition} (Isomorphism of Coloured Graphs)\newline
Let $\mathcal{G}_{i}=(\mathcal{V}_{i},\mathcal{E}_{i})\in\mathfrak{G}_{d}$ with $i\in\{1,2\}$ be two open $(d+1)$-coloured graphs with proper edge-colourings $\gamma_{i}:\mathcal{E}_{i}\to\mathcal{C}_{i}$, where $\mathcal{C}_{i}$ with $\vert\mathcal{C}_{i}\vert=d+1$ are the corresponding colour sets. Then, we call an isomorphism of graphs $\varphi:\mathcal{V}_{1}\to\mathcal{V}_{2}$ ``isomorphism of coloured graphs'', or simply ``colour-isomorphism'', if there exists a bijection $\psi:\mathcal{C}_{1}\to\mathcal{C}_{2}$ such that $\gamma_{2}\circ\varphi=\psi\circ\gamma_{1}$.\end{Definition}

In other words, two coloured graphs are isomorphic if they are isomorphic as (multi)graphs and if their colourings are related by a recolouring of their edges in a bijective way. Of course, two coloured graphs describe the same simplicial complex if and only if they are colour-isomorphic. The following figure shows two closed $(2+1)$-coloured graphs, which are isomorphic as graphs, but \textit{not} as coloured graphs:

\begin{figure}[H]
\centering
\includegraphics[scale=1.4]{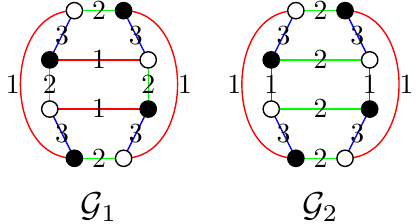}
\caption{Two closed $(2+1)$-coloured graphs $\mathcal{G}_{1}$ and $\mathcal{G}_{2}$, which are isomorphic as graphs, but not as coloured graphs.}
\end{figure}

The two graphs $\mathcal{G}_{1},\mathcal{G}_{2}\in\overline{\mathfrak{G}}_{2}$ are clearly isomorphic as graphs, however, they are obviously not colour-isomorphic, since we just exchanged the colours of four edges of colour $1$ and $2$. As a consequence, the two graphs clearly describe different simplicial complexes. In fact, the corresponding geometric realizations are not even homeomorphic: The graph $\mathcal{G}_{1}$ represents the $2$-torus $T^{2}=S^{1}\times S^{1}$, whereas the graph $\mathcal{G}_{2}$ represents the $2$-sphere $S^{2}$. This can either be seen by calculating their Euler characteristics using the formula explained in Remark \ref{EulerColour}(b) below, or by explicitly drawing the corresponding complexes, as shown in the figure below.

\begin{figure}[H]
\centering
\includegraphics[trim=1.5cm 0.5cm 0 1cm,scale=1.4]{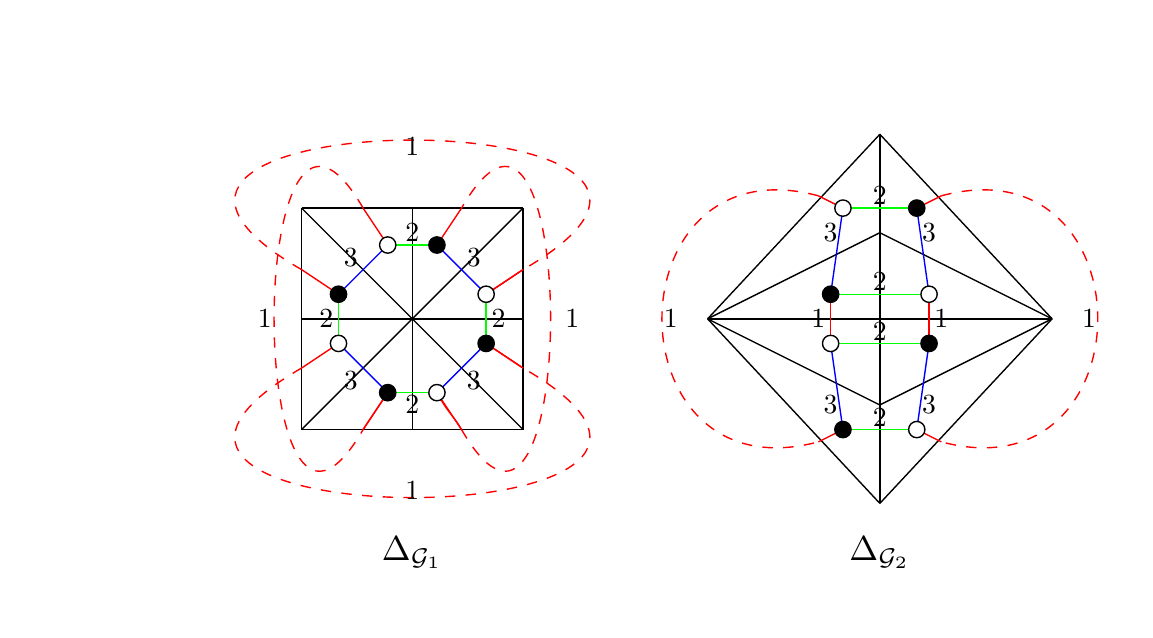}
\caption{The simplicial complexes $\Delta_{\mathcal{G}_{1}}$ and $\Delta_{\mathcal{G}_{2}}$ together with their dual $1$-skeletons corresponding to the graphs $\mathcal{G}_{1}$ and $\mathcal{G}_{2}$. The remaining edges of the triangulations are glued together as indicated by the dotted dual edges.}
\end{figure}

This is also an explicit example of the fact mentioned above that the 1-skeleton alone is not enough to determine the topology of the corresponding simplicial complex.

\subsection{Bubbles and their Multiplicities}\label{SecBubbles}
In the last section, we have seen that a coloured graph does not only encode the dual 1-skeleton of some simplicial complex, but in fact the \textit{full} simplicial complex, since it carries all the informations about the gluing of higher-dimensional dual cells. In this section, we discuss more precisely how these higher-dimensional cells and there nested structure are encoded in the colouring of our Feynman diagrams. Let us start with the following definition:

\begin{Definition} (Bubbles)\newline
Let $\mathcal{G}\in\mathfrak{G}_{d}$ be an open $(d+1)$-coloured graph and $i_{1},\dots,i_{k}\in\mathcal{C}_{d}$ with $i_{1}<\dots<i_{k}$ and $k\in\{0,\dots,d\}$. We call a connected component of the graph obtained by deleting all the edges of colours $\mathcal{C}_{d}\textbackslash\{i_{1},\dots,i_{k}\}$ a ``$k$-bubble of colours $i_{1},\dots,i_{k}$''. We denote the $k$-bubbles of colours $i_{1},\dots,i_{k}$ by $\mathcal{B}^{i_{1}\dots i_{k}}_{(\rho)}$, where $\rho$ labels the various bubbles of the same colours. The set of all $k$-bubbles will be denoted by $\mathcal{B}^{[k]}$.\end{Definition}

\begin{Remark}Note that the set of $0$-bubbles is precisely the vertex set $\mathcal{V}_{\mathcal{G}}$ of $\mathcal{G}$. This would in principle also include the $1$-valent boundary vertices. However, as a convention, we will in the following only take the $(d+1)$-valent internal vertices as $0$-bubbles.\end{Remark}

By the previous remark, the $0$-bubbles of some coloured graph are exactly the internal vertices of the graph and hence, we see that the $0$-bubbles exactly correspond to the $d$-simplices of the simplicial complex. Furthermore, $1$-bubbles are exactly the edges and hence the $(d-1)$-simplices of the complex. $2$-bubbles are usually called ``\textit{faces of the graph}'' and one can easily convince oneself that they correspond to the $(d-2)$-simplices of the complex. More generally, it turns out that one can extend this discussion to all dimensions:

\begin{Proposition} (Interpretation of Bubbles)\newline
There is a one-to-one correspondence between the $k$-bubbles of some open $(d+1)$-coloured graph $\mathcal{G}\in\mathfrak{G}_{d}$ and the $(d-k)$-simplices of the corresponding simplicial complex $\Delta_{\mathcal{G}}$. \end{Proposition}

\begin{proof}It is not too hard to see that a $k$-bubble $\mathcal{B}$ is exactly the graph, which is dual to the link of a $(d-k)$-simplex $\sigma$ of $\Delta_{\mathcal{G}}$, i.e.
\begin{align}\Delta_{\mathcal{B}}=\mathrm{Lk}_{\Delta_{\mathcal{G}}}(\sigma).\end{align}
More precisely, recall that the colouring of the $d+1$ faces of each $d$-simplex in the complex induces a colouring of vertices. Now, a $k$-simplex $\sigma$ has $(k+1)$ vertices, which have some colours, lets say $\{i_{1},\dots,i_{k},i_{k+1}\}\subset\mathcal{C}_{d}$. The link of $\sigma$ is by definition a $(d-1-k)$-dimensional complex, which is dual to a $(d-k)$-coloured graph. This $(d-k)$-coloured graph is exactly a $(d-k)$-bubble in $\mathcal{G}$ with colours $\mathcal{C}_{d}\textbackslash\{i_{1},\dots,i_{k+1}\}$.
\end{proof}

\begin{Remarks}\label{EulerColour}\begin{itemize}\item[]
\item[(a)]From a mathematical point of view, the simplicial complex dual to a coloured graph $\mathcal{G}\in\mathfrak{G}_{d}$ is defined via a bijective map of the form $\varphi:\Delta_{\mathcal{G},d}\to\mathcal{V}_{\mathcal{G},\mathrm{int}}=\mathcal{B}^{[0]}$. The previous proposition tells us that this map induces actually a family of bijective maps of the form $\varphi_{k}:\Delta_{\mathcal{G},k}\to\mathcal{B}^{[d-k]}$ such that $\varphi_{k=d}=\varphi$. Note also that these maps are inclusion reversing: Consider a $k$-simplex $\sigma$ and let $\tau$ be some face of $\sigma$. Then $\varphi_{k}(\sigma)$ is a $(d-k)$-bubble within the $(d-l)$-bubble $\varphi_{l}(\tau)$. Hence, the colouring does not only include informations about higher-dimensional dual cells but also about their nested structure. In other words, a coloured graphs encodes all the $k$-skeletons for every $0\leq k\leq d$ and hence encodes a full simplicial complex.
\item[(b)]The concepts of bubbles allows us to directly read of the Euler characteristic of the simplicial complex dual to some graph, namely by counting all its bubbles. More precisely, we have that 
\begin{align}\chi(\Delta_{\mathcal{G}})=\sum_{j=0}^{d}(-1)^{j}\mathcal{B}^{[d-j]}\end{align}
for some open $(d+1)$-coloured graph $\mathcal{G}\in\mathfrak{G}_{d}$.
\end{itemize}\end{Remarks}

If $\mathcal{G}$ is an open $(d+1)$-coloured graph in $\mathfrak{G}_{d}$, then its $k$-bubbles for some $k>1$ are clearly elements of $\mathfrak{G}_{k-1}$. In general, such bubbles can either be open graphs or closed graphs. If some $k$-bubble $\mathcal{B}$ is open, then the corresponding $(d-k)$-simplex lives purely on the boundary of the simplicial complex $\Delta_{\mathcal{G}}$ and if $\mathcal{B}\in\overline{\mathfrak{G}}_{k-1}$, then the corresponding simplex lives in the interior of $\Delta_{\mathcal{G}}$ (possibly touching the boundary). The following figure shows an open $(3+1)$-coloured graph $\mathcal{G}$, called the ``\textit{elementary melonic $3$-ball}'', together with all its $3$-bubbles corresponding to the four vertices of the simplicial complex dual to $\mathcal{G}$:

\begin{figure}[H]
\centering
\includegraphics[scale=1]{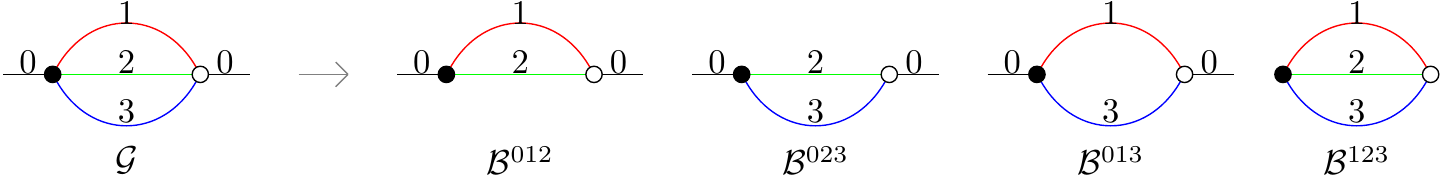}
\caption{The elementary melonic 3-ball $\mathcal{G}$ (l.h.s.) and its four $3$-bubbles.\label{BubblesFig}}
\end{figure}

In this example, three $3$-bubbles are open graphs and correspond to the three vertices of $\Delta_{\mathcal{G}}$ living on the boundary, whereas one $3$-bubbles is closed and corresponds to an internal vertex in $\Delta_{\mathcal{G}}$. Furthermore, the graph has in total six $2$-bubbles (=faces), from which three are open. Hence, the simplicial complex has six edges, from which three are purely on the boundary. All in all, we see that the Euler characteristic of this graph is 
\begin{align}\chi(\Delta_{\mathcal{G}})=\mathcal{B}^{[3]}-\mathcal{B}^{[2]}+\mathcal{B}^{[1]}-\mathcal{B}^{[0]}=4-6+5-2=1.\end{align} 
In fact, it is not too hard to see that the simplicial complex $\Delta_{\mathcal{G}}$ actually represents a $3$-ball. In this case, we also see that all the $3$-bubbles are either $2$-spheres or $2$-balls (=disks). In general, the topology of bubbles can also be non-trivial. As it turns out, the topology of these bubbles can be used to decide whether a coloured graph describes a manifold or a pseudomanifold:

\begin{Proposition}\label{ManifoldsGraphs} Let $\mathcal{G}\in\mathfrak{G}_{d}$ be an open $(d+1)$-coloured graph. Then $\vert\Delta_{\mathcal{G}}\vert$ is a manifold if and only if all the $d$-bubbles of $\mathcal{G}$ represent either $(d-1)$-spheres or $(d-1)$-balls.\end{Proposition}

\begin{proof}It is clear that every triangulation with the property that all the links of its vertices (=the $d$-bubbles of the corresponding graph) represent spheres or balls is a manifold, since in this case, the triangulation is a piecewise-linear structure, as discussed in Appendix \ref{TriangTheorem}. For the reverse, see the review \cite{GagliardiBoundaryGraph} and references therein.\end{proof}

\begin{Remark}As discussed in Appendix \ref{TriangTheorem}, not every triangulation of a manifold is a piecewise-linear one for dimension $d>4$. Hence, the previous Proposition also tells us that only piecewise-linear manifolds are representable by coloured graphs. \end{Remark}

Previously, we have defined the boundary graph $\partial\mathcal{G}$ corresponding to some open $(d+1)$-coloured graph $\mathcal{G}\in\mathfrak{G}_{d}$ and said that the underlying graph is exactly the boundary dual $1$-skeleton of the complex $\Delta_{\mathcal{G}}$. Since $\partial\mathcal{G}$ is a closed $d$-coloured graph, we can construct the corresponding simplicial complex $\Delta_{\partial\mathcal{G}}$. Naively, we would guess that this simplicial complex is exactly the boundary of the simplicial complex dual to $\mathcal{G}$, i.e. $\Delta_{\partial\mathcal{G}}=\partial\Delta_{\mathcal{G}}$. However, it turns out that $\partial\Delta_{\mathcal{G}}$ is in general just a quotient of the simplicial complex $\Delta_{\partial\mathcal{G}}$ obtained by identifying some of its simplices. This is actually well known in the crystallization theory and goes under the name ``multiple residues\footnote{The term ``residues'' is used in crystallization theory for what is called ``bubbles'' in the quantum gravity literature.}'' \cite{GagliardiExistence,Gagliardi87,GagliardiMultRes}. Let us discuss this point in more details. To start with, consider the following closed $(2+1)$-coloured graph $\gamma\in\overline{\mathfrak{G}}_{2}$, called the ``\textit{pillow graph}'', which we fix to be our boundary graph:

\begin{figure}[H]
\centering
\includegraphics[scale=1.4]{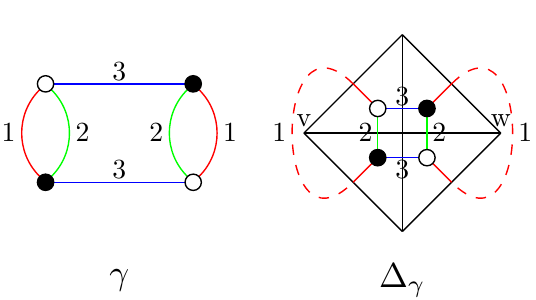}
\caption{A closed $(2+1)$-coloured graph $\gamma$ (l.h.s.) together with its simplicial complex $\Delta_{\gamma}$ (r.h.s.). The dotted red lines indicate the gluing of the remaining edges.\label{MultRes}}
\end{figure}

The graph clearly describes a $2$-sphere, as can be seen by looking at the simplicial complex $\Delta_{\gamma}$ dual to $\gamma$ drawn on the right-hand side above. Now, consider the following two open $(3+1)$-coloured graphs $\mathcal{G}_{1},\mathcal{G}_{2}\in\mathfrak{G}_{3}$:

\begin{figure}[H]
\centering
\includegraphics[scale=1.4]{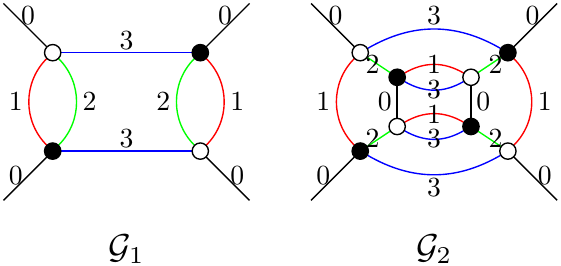}
\caption{Two open $(3+1)$-coloured graphs $\mathcal{G}_{1}$ and $\mathcal{G}_{2}$ with boundary graph given by the graph $\gamma$ drawn above.\label{G1G2MultRes}}
\end{figure}

Both of these graphs satisfy $\partial\mathcal{G}_{1}=\partial\mathcal{G}_{2}=\gamma$. One can easily see that the boundary of the simplicial complex $\Delta_{\mathcal{G}_{1}}$, which describes a $3$-ball, is given by the complex $\Delta_{\gamma}$, i.e. \begin{align}\partial\Delta_{\mathcal{G}_{1}}=\Delta_{\partial\mathcal{G}=\gamma}.\end{align} 
However, as we will argue now, this is not the case for the simplicial complex dual to $\mathcal{G}_{2}$. For this, note that the graph $\mathcal{G}_{2}$ has in total four $3$-bubbles, from which three are open graphs. One of them, the $3$-bubble of colour $012$, has two disconnected boundary components, as drawn in the figure below.

\begin{figure}[H]
\centering
\includegraphics[scale=1.4]{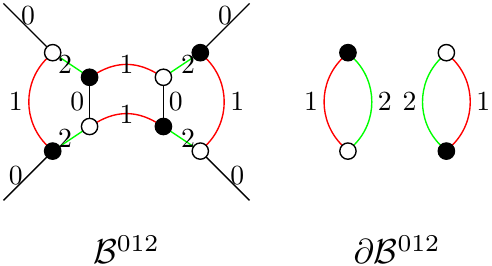}
\caption{The unique $3$-bubble $\mathcal{B}^{123}$ of colour $012$ of the graph $\mathcal{G}_{2}$ together with its boundary graph $\partial\mathcal{B}^{123}$, which is a subgraph of $\gamma=\partial\mathcal{G}_{2}$.}
\end{figure}

As explained above, the $3$-bubbles of some open $(3+1)$-coloured graph $\mathcal{G}$ correspond to the vertices of the simplicial complex $\Delta_{\mathcal{G}}$, whereas the $2$-bubbles of its closed $(2+1)$-coloured boundary graph $\partial\mathcal{G}$ correspond to the vertices of the complex $\Delta_{\partial\mathcal{G}}$. Hence, we see that in the above example, the two vertices dual to the two $2$-bubbles of colour $12$ of $\gamma$ are identified in the simplicial complex $\Delta_{\mathcal{G}_{2}}$, since they both correspond to the same $3$-bubble in $\mathcal{G}_{2}$. In other words, the boundary of the simplicial complex $\Delta_{\mathcal{G}_{2}}$ is the complex obtained by identifying the two vertices $v$ and $w$ of the complex $\Delta_{\gamma}$ drawn on the right-hand side in Figure \ref{MultRes}, i.e. it we can write
\begin{align}\partial\Delta_{\mathcal{G}_{2}}=\Delta_{\gamma}/_{v\sim w}.\end{align}
The geometric realization of this complex is the ``\textit{pinched torus}'', i.e. the pseudomanifold obtained by identifying two distinct points on a $2$-sphere. Hence, let us make the following Definition:

\begin{Definition} (Multiplicity of Bubbles)\newline
Let $\mathcal{G}$ be an open $(d+1)$-coloured graph. We call the number of boundary components of some bubble $\mathcal{B}$ the ``multiplicity of $\mathcal{B}$'' and denote it by $\mathrm{mult}(\mathcal{B})$. If $\mathrm{mult}(\mathcal{B})\in\{0,1\}$, then we call the bubble ``simple''.\end{Definition}

\begin{Remarks}\label{RemarksMultRes}\begin{itemize}\item[]
\item[(a)]If an open graph $\mathcal{G}$ is such that all its bubbles are simple, then clearly $\Delta_{\partial\mathcal{G}}=\partial\Delta_{\mathcal{G}}$.
\item[(b)]If $\mathcal{G}$ is dual to a manifold, then the bubbles are always simple, since they are dual to spheres and balls. In other words, these ``pinching effects'' on the boundary complex can only occur for pseudomanifolds. Of course, this is not an if and only if statement and there are also pseudomanifolds with the property $\Delta_{\partial\mathcal{G}}=\partial\Delta_{\mathcal{G}}$. 
\item[(c)]Note that $k$-bubbles with $k\leq 2$ have at most one boundary component and hence, $(2+1)$-coloured graphs always have the property $\Delta_{\partial\mathcal{G}}=\partial\Delta_{\mathcal{G}}$. In fact, one can easily see that all $(2+1)$-coloured graph represent manifolds, since all its $2$-bubbles are either circles or lines and hence, this can also be seen as a special case of Remark (b) above.
\end{itemize}\end{Remarks}

Using the discussion of the example above, one can easily see that there is the following general relationship between the complex of the boundary graph and the boundary of the simplicial complex of the corresponding open graph:

\begin{Proposition} (Boundary Complex of a General Open Graph)\newline
Let $\mathcal{G}$ be an open $(d+1)$-coloured graph with boundary graph $\mathcal{G}$. Then
\begin{align*}\partial\Delta_{\mathcal{G}}=\Delta_{\partial\mathcal{G}}/\sim,\end{align*}
where $\sim$ identifies for each non-simple $(k+1)$-bubble $\mathcal{B}$ of $\mathcal{G}$ with $k\in\{2,\dots d\}$ the corresponding $k$-simplices belonging to the various boundary components of $\mathcal{B}$.
\end{Proposition} 

Let us end this section with a few remarks: First of all, note that the appearance of this additional pinching effects on the boundary had to be expected, since by its definition, the boundary graph only takes the $1$-skeleton of the complex $\partial\Delta_{\mathcal{G}}$ into account. Of course, due to its colouring, the boundary graph encodes a full simplicial complex, however, it does not include these additional informations of possible identifications of $k$-simplices with $k\leq d-3$. In other words, the boundary graph describes the ``\textit{desingularized}'' boundary of the complex $\Delta_{\mathcal{G}}$. However, note that this additional information \textit{is} included in the graph $\mathcal{G}$. Hence, every open coloured graph still describes a unique simplicial complex and vice versa. Another way to see this phenomena is to realize that not every pseudomanifold is representable by a coloured graph. This is what happens in the example discussed above, since the boundary of the complex $\Delta_{\mathcal{G}_{2}}$ is the ``pinched torus''. Now, the pinched torus is a pseudomanifold, which is \textit{not} normal, because the link of its singular point is not connected. Hence, it is not representable as a coloured graph. However, the simplicial complex $\Delta_{\mathcal{G}_{2}}$, whose boundary is the pinched torus, is representable as a coloured graph, namely by the graph $\mathcal{G}_{2}$. In general, note that the boundary of a normal pseudomanifold is in general again a pseudomanifold, which does not need to be normal, even if it is connected.

\subsection{A Glance of Crystallization Theory: Existence of Coloured Graphs}
In general, every open $(d+1)$-coloured graph represents a pseudomanifold with boundary. However, it is a priori not clear for which type of topologies their exists a coloured graph describing them. In this section, the goal is to review some central objects and results from crystallization theory, which show that every manifold admits a special type of coloured graph representing it. First of all, let us introduce the notion of ``contracted graphs'', which are dual to triangulations with the smallest possible number of vertices:

\begin{Definition} (Contracted Graphs)
\begin{itemize}
\item[(1)]A closed $(d+1)$-coloured graph $\mathcal{G}\in\overline{\mathfrak{G}}_{d}$ is called ``contracted'', if it admits exactly one $d$-bubble without colour $i$ for all $i\in\mathcal{C}_{d}$, i.e. $\mathcal{B}^{[d]}=d+1$. 
\item[(2)]Let $\mathcal{G}\in\mathfrak{G}_{d}$ be an open $(d+1)$-coloured graph. Furthermore, let us denote by $C(\partial\mathcal{G})$ the number of connected components of $\partial\mathcal{G}$. Then $\mathcal{G}$ is called ``$\partial$-contracted'', if there is exactly one $d$-bubble without colour $0$ and exactly $C(\partial\mathcal{G})$ $d$-bubbles without colour $i$ for all $i\in\mathcal{C}_{d}\textbackslash\{0\}$, i.e. $\mathcal{B}^{[d]}=1+d\cdot C(\partial\mathcal{G})$.
\end{itemize}
\end{Definition}

As an example, the two graphs in Figure \ref{G1G2MultRes} are $\partial$-contracted and the corresponding boundary graph, drawn in Figure \ref{MultRes}, is contracted. Using this definition, a ``crystallization'' of a piecewise-linear, compact and connected manifold is defined in the following way \cite{GagliardiBoundaryGraph}:

\begin{Definition} (Crystallization)\newline
Let $\mathcal{M}$ be a compact PL-manifold with or without boundary. Then, we call a coloured graph $\mathcal{G}\in\mathfrak{G}_{d}$ a ``crystallization of $\mathcal{M}$'', if the following two conditions are fulfilled:
\begin{itemize}
\item[(1)]$\mathcal{G}$ represents $\mathcal{M}$, i.e. $\vert\Delta_{\mathcal{G}}\vert$ is homeomorphic to $\mathcal{M}$.
\item[(2)]$\mathcal{G}$ is contracted if $\partial\mathcal{M}=\emptyset$ and $\partial$-contracted otherwise.\end{itemize}\end{Definition}

Of course, if $\mathcal{G}$ is a crystallization of a manifold with boundary, then $\partial\mathcal{G}$ is a crystallization of its boundary components. We can now state one of the central theorems of crystallization theory, as well as its generalization for manifolds with boundary, which shows that for manifolds there always exists a coloured graph and in fact even a crystallization representing it:

\begin{Theorem}\label{Pezzana} (Existence of Crystallizations)
\begin{itemize}
\item[(1)]Every closed and connected $d$-dimensional PL-manifold admits a crystallization.
\item[(2)]Let $\mathcal{M}$ be a compact and connected $d$-dimensional PL-manifold with (possibly disconnected) boundary. Then for every crystallization $\gamma$ of the boundary $\partial\mathcal{M}$, there exists a crystallization $\mathcal{G}$ such that $\partial\mathcal{G}$ is colour-isomorphic to $\gamma$.
\end{itemize}
\end{Theorem}

\begin{proof}The first statement, is usually called ``Theorem of Pezzana'' and originates from \cite{Pezzana}. A sketch of the proof in English can be found in \cite{GagliardiFerri}. The second claim can be found in \cite{GagliardiExistence} (for manifolds with connected boundary) and in \cite[p.66]{GagliardiCorbodant} (general case). The idea of the proof is basically to explicitly construct a contracted triangulation out of a given piecewise-linear (pseudo)triangulation.\end{proof}

\subsection{Connected Sums of Manifolds and Coloured Graphs}\label{ConSum}
One way to build new manifolds out of some given manifolds is provided by performing their ``\textit{connected sum}''. The connected sum of two closed and connected $d$-dimensional manifolds $\mathcal{M}$ and $\mathcal{N}$ is the manifold, usually denoted by $\mathcal{M}\#\mathcal{N}$, which is defined by cutting out an open $d$-ball $B_{1}$ inside $\mathcal{M}$ and an open $d$-ball $B_{2}$ inside $\mathcal{N}$ and by gluing the so-created boundary $(d-1)$-spheres together. If both $\mathcal{M}$ and $\mathcal{N}$ are oriented, then we should assume in addition that the ``\textit{gluing map}'' $\varphi:\partial B_{1}\to\partial B_{2}$ is orientation-reversing. The resulted manifold $M\# N$ then comes equipped with an obvious orientation, which agrees with the orientations chosen on $\mathcal{M}$ and $\mathcal{N}$, and is usually called the ``\textit{oriented} connected sum''. One should note that this construction depends in general on the chosen orientations of $\mathcal{M}$ and $\mathcal{N}$. Famous examples are the non-homeomorphic manifolds $\mathbb{C}P^{2}\#\mathbb{C}P^{2}$ and $\mathbb{C}P^{2}\#\overline{\mathbb{C}P^{2}}$, where $\overline{\mathbb{C}P^{2}}$ is the manifold $\mathbb{C}P^{2}$ with the opposite orientation. In fact, these two manifolds are not even homotopy equivalent. If both $\mathcal{M}$ and $\mathcal{N}$ admit orientation-reversing self-homeomorphisms, then one can easily show that the connected sum does not depend on the chosen orientations. In particular, this is the case for compact and orientable surfaces, however, already fails to be true for general $3$-manifolds. On the other hand, the connected sum does in general not depend on all the other choices, i.e. the choice of balls and the choice of (orientation-reversing) homeomorphism\footnote{For topological manifolds, this was not clear for a long time, as it is consequence of the highly non-trivial ``\textit{Annulus Theorem}'', which was proven by R. C. Kirby in 1969 \cite{KirbyAT} for $d\geq 5$ and by F. Quinn in 1982 \cite{Quinn} for the case $d=4$. For $d\leq 3$ it is a consequence of the triangulation theorems of Radó (1924) and Moise (1952), reviewed in the appendix.}. \\
\\
For manifolds with boundary, there are actually two different types of connected sums one can define. Let $\mathcal{M}$ and $\mathcal{N}$ be two connected and compact manifolds with connected boundary.
\begin{itemize}
\item[(1)]\textbf{(Internal) connected sum}: Let us choose two closed $d$-balls $B_{1}$ and $B_{2}$ inside $\mathcal{M}$ and $\mathcal{N}$, such that they do not intersect the boundaries of $\mathcal{M}$ and $\mathcal{N}$. Then the (internal) connected sum is the manifold denoted by $\mathcal{M}\#\mathcal{N}$, which is obtained by cutting out the interior of the balls from $\mathcal{M}$ and $\mathcal{N}$ and gluing the two created boundary spheres together. As a consequence, it holds that $\partial (\mathcal{M}\#\mathcal{N})=(\partial\mathcal{M})\coprod (\partial\mathcal{N})$. Furthermore, note that the $d$-sphere is the neutral element of this operation, i.e. $\mathcal{M}\# S^{d}\cong\mathcal{M}$ for all $\mathcal{M}$ as above. The internal connected sum contains the connected sum of manifolds without boundaries as a special case. Furthermore, it also contains the mixed case, i.e. the case where one manifold has empty and the other manifold non-empty boundary.
\item[(2)]\textbf{Boundary connected sum}: Let us choose two closed $(d-1)$-dimensional balls $B_{1}$ and $B_{2}$ inside $\partial\mathcal{M}$ and $\partial\mathcal{N}$. Then the boundary connected sum is the manifold denoted by $\mathcal{M}\#_{\partial}\mathcal{N}$, which is obtained by identifying the two balls to each other. Note that it holds that $\partial (\mathcal{M}\#_{\partial} \mathcal{N})=(\partial\mathcal{M})\# (\partial\mathcal{N})$. Furthermore, note that the closed $d$-ball $B^{d}$ is the neutral element of this operation, i.e. $\mathcal{M}\# B^{d}\cong\mathcal{M}$ for all $d$-manifolds $\mathcal{M}$ as above.\end{itemize}

We can define in both cases also the oriented version by choosing orientations on $\mathcal{M}$ and $\mathcal{N}$ and by assuming that the gluing maps are orientation-reversing. Again, this depends in general on the choice of orientations of $\mathcal{M}$ and $\mathcal{N}$. Let us define the following operation on coloured graphs:

\begin{Definition} (Connected Sums and Coloured Graphs)\newline
Let $\mathcal{G}_{1},\mathcal{G}_{2}\in\mathfrak{G}_{d}$ be two open $(d+1)$-coloured graphs representing manifolds. Then, let us define the following graph: Lets take a vertex $v$ of $\mathcal{G}_{1}$ and a vertex $w$ of $\mathcal{G}_{2}$ of different types (i.e. one black and one white). Then, we denote by $\mathcal{G}_{1}\#_{\{v,w\}}\mathcal{G}_{2}$ the graph obtained by deleting the two vertices and gluing the ``hanging'' pairs of edges together respecting their colouring. We call this graph the ``graph connected sum of $\mathcal{G}_{1}$ and $\mathcal{G}_{2}$ at $v$ and $w$''.
\end{Definition}

\begin{Remark}Note that if both vertices $v$ and $w$ do admit an adjacent external leg, then the procedure would produce a disconnected part containing a single edge of colour $0$ connecting two boundary vertices. In this case, we do not include this additional disconnected piece in the definition of $\mathcal{G}_{1}\#_{\{v,w\}}\mathcal{G}_{2}$. \end{Remark}

Two examples are drawn in Figure \ref{FigConSum} below. It turns out that these graphs describe in many cases the connected sums of the corresponding manifolds. Some situations where this is the case are summarized in the Theorem below. To state the Theorem more properly, let us introduce the following terminology: We call an internal vertex of some open $(d+1)$-coloured graph ``\textit{strictly internal}'', if all the $d$-bubbles to which the vertex belongs, are closed. In other words, a vertex in some open coloured graph is strictly internal if and only if all faces of all dimensions of the corresponding $d$-simplex are not on the boundary. Of course, any strictly internal vertex does not admit adjacent external legs, but the reverse is in general not true. Using this terminology, we can state the following theorem:

\begin{Theorem}\label{ThmConSum} (Connected Sums and Coloured Graphs)\newline
Let $\mathcal{G}_{1},\mathcal{G}_{2}\in\mathfrak{G}_{d}$ be two open $(d+1)$-coloured graphs representing manifolds $\mathcal{M}_{1}$ and $\mathcal{M}_{2}$. Furthermore, let $v$ be a vertex of $\mathcal{G}_{1}$ and $w$ be a vertex of $\mathcal{G}_{2}$. Then:
\begin{itemize}
\item[(1)]If both $v$ and $w$ admit an adjacent external leg, then $\mathcal{G}_{1}\#_{\{v,w\}}\mathcal{G}_{2}$ represents the oriented boundary connected sum $M_{1}\#_{\partial} M_{2}$.
\item[(2)]If both vertices $v$ and $w$ do not admit an adjacent external leg and if at least one of them is strictly internal, then $\mathcal{G}_{1}\#_{\{v,w\}}\mathcal{G}_{2}$ represents the oriented internal connected sum $M_{1}\# M_{2}$.
\item[(3)]If $v$ is an strictly internal vertex and $w$ admits an adjacent external leg, then $\mathcal{G}_{1}\#_{\{v,w\}}\mathcal{G}_{2}$ represents the manifold $(\mathcal{M}_{1}\# B^{d})\#_{\partial}\mathcal{M}_{2}$, where $B^{d}$ denotes the closed $d$-ball.
\end{itemize}
\end{Theorem}

\begin{proof}The detailed proof can be found in \cite[p.12f. and p.15]{GagliardiConnectedSum}. Geometrically, the proof is actually quite obvious: As an example, in case (2), we delete an internal $d$-simplex in one of the triangulations and another $d$-simplex (possibly touching the boundary with some of its faces of dimension $<d-1$) in another complex. Now, since a $d$-simplex represents a $d$-ball, removing these simplices results into removing balls inside the corresponding manifolds. Furthermore, connecting the hanging pair of edges of the coloured graph obtained by deleting these two vertices precisely corresponds to gluing the created boundary $d$-spheres together. Taking the two vertices of different types ensures that the gluing map is orientation-reversing.\end{proof}

\begin{Remarks}\begin{itemize}\item[]
\item[(a)]Note again the subtlety of the choice of orientation: We already started with \textit{oriented} manifolds, since we took bipartite graphs at the beginning. Of course, we might choose the same manifold with the opposite orientation, by interchanging the types of vertices in the corresponding graphs. The resulting connected sums might not be homeomorphic, as the example with $\mathbb{C}P^{2}$ mentioned above shows.
\item[(b)]One has to be careful, because its seems that this theorem is sometimes stated imprecisely in the literature. As an example, in \cite[p.445]{CasaliConSum}, it is stated that the graph-connected sum $\mathcal{G}_{1}\#_{\{v,w\}}\mathcal{G}_{2}$ always represents the internal connected sum of the corresponding manifolds, whenever both $v$ and $w$ do not admit adjacent external legs. However, if one drops the assumption that at least one of these two vertices is \textit{strictly} internal, one can easily construct counter-examples to this claim.
\end{itemize}\end{Remarks}

The figure below shows two examples of the previous theorem. Figure (a) shows the boundary connected sum of two disks, which is again a disk, and figure (b) shows an example of the (internal) connected sum of two disks, which is homeomorphic to the cylinder $S^{1}\times [0,1]$, i.e. the unique (up to homeomorphism) surface with genus zero and two boundary components.

\begin{figure}[H]
\centering
\includegraphics[scale=1.4]{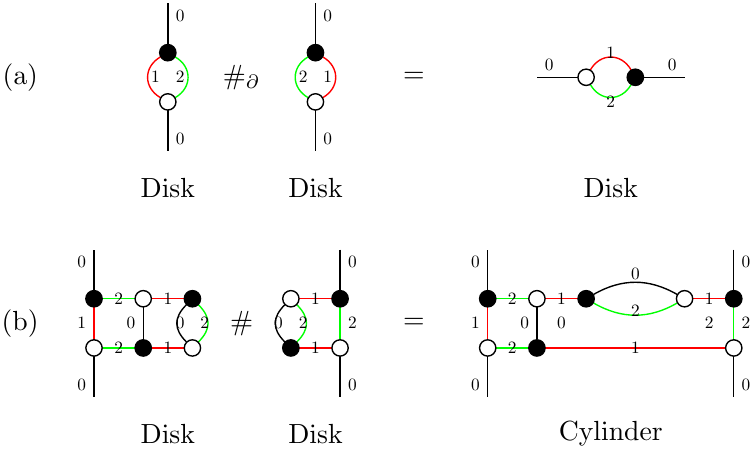}
\caption{Two examples of graph-connected sums of open $(2+1)$-coloured graphs representing the disk and their topological interpretation. \label{FigConSum}}
\end{figure}

The fact that these graph indeed describe disks and cylinders can be seen by explicitly calculating their Euler characteristic as well as the number of boundary components.\\
\\
Let us collect two immediate consequences of the theorem above, concerning the mixed case of the graph-connected sum of a closed graph with an open graph:

\begin{Corollary}\label{CorConSum} Let $\mathcal{G}_{1}\in\overline{\mathfrak{G}}_{d}$ be a closed $(d+1)$-coloured graph representing a manifold $\mathcal{M}_{1}$ and $\mathcal{G}_{2}\in\mathfrak{G}_{d}$ be an open $(d+1)$-coloured graph representing a manifold $\mathcal{M}_{2}$. Furthermore, let $v$ be a vertex of $\mathcal{G}_{1}$ and $w$ be a vertex of $\mathcal{G}_{2}$. Then:
\begin{itemize}
\item[(1)]If $w$ is an internal vertex, which does not admit an adjacent external leg, then $\mathcal{G}_{1}\#_{\{v,w\}}\mathcal{G}_{2}$ represents the oriented internal connected sum $M_{1}\# M_{2}$.
\item[(2)]If $\mathcal{M}_{1}\cong S^{d}$ and if $w$ is an internal vertex, which admits an adjacent external leg, then $\mathcal{G}_{1}\#_{\{v,w\}}\mathcal{G}_{2}$ represents the manifold $\mathcal{M}_{2}$.
\end{itemize}
\end{Corollary}

\begin{proof}Claim (1) follows directly from Theorem \ref{ThmConSum}(2), since in a closed graph every vertex is strictly internal. For claim (2), recall that the sphere is the neutral element of $\#$ whereas the ball is the neutral element of $\#_{\partial}$ and hence, by \ref{ThmConSum}(3), $\mathcal{G}_{1}\#_{\{v,w\}}\mathcal{G}_{2}$ represents the manifold $(S^{d}\# B^{d})\#_{\partial}\mathcal{M}_{2}\cong B^{d}\#_{\partial}\mathcal{M}_{2}\cong\mathcal{M}_{2}$.\end{proof}

\subsection{Combinatorial and Topological Moves}
In general, many different triangulations will lead to the same manifold and hence, we would like to have a technique in order to compare various triangulations to each other. There are several possibilities to do so. For PL-manifolds, there is for example ``Pachner's theorem'', as discussed previously, which states that two PL-manifolds are PL-homeomorphic if and only if they are related by a finite sequence of ``Pachner moves''. In $3$-dimensions, every triangulation gives rise to a PL-structure and there are only two different types of Pachner moves, i.e. the $(1-4)$- and the $(2-3)$-move (and their inverses). For our purpose, this is however not the right choice of moves: First of all, in GFT and tensor models, we produce much more singular topologies, even in the case of coloured models and hence, Pachner's theorem might be not applicable. Furthermore, in complex coloured models, we have to find a procedure which respects the structure coming from the colouring and orientation, i.e. such moves should still lead to proper colourable and bipartite triangulations. For example, applying a $(1-4)$-Pachner to some tetrahedron in a Feynman graph of our complex model would not be possible, since the resulting complex is not bipartite anymore. It turns out that a suitable set of moves is given by so-called ``dipole moves'', which were introduced in \cite{GagliardiFerri}. Let us start by defining the central objects of this section:

\begin{Definition} (Dipoles)\newline
Let $\mathcal{G}\in\mathfrak{G}_{d}$ be an open $(d+1)$-coloured graph, such that $\vert\mathcal{V}_{\mathcal{G},\mathrm{int}}\vert>2$. We call a subgraph $d_{k}$ consisting of two internal vertices $v,w\in\mathcal{V}_{\mathcal{G},\mathrm{int}}$, which are connected by $k$ edges of colours $i_{1},\dots,i_{k}\in\mathcal{C}_{d}$, ``$k$-dipole of colours $i_{1},\dots,i_{k}$'', if the two $(d+1-k)$-bubbles of colour $\mathcal{C}_{d}\textbackslash\{i_{1},\dots,i_{k}\}$ containing $v$ and $w$, respectively, are distinct.\end{Definition}

In the language of tensor models, a $k$-dipole is hence nothing else than an elementary melonic $(k-1)$-sphere within a graph, separating bubbles with the complementary colours. If some coloured graphs admits a dipole, then we define another graph by ``contracting the dipole'':

\begin{Definition} (Dipole Contraction)\newline
Let $\mathcal{G}\in\mathfrak{G}_{d}$ be an open $(d+1)$-coloured graph and $d_{k}$ a $k$-dipole within $\mathcal{G}$ with vertices $v,w$. Then we define a graph $\mathcal{G}/d_{k}\in\mathfrak{G}_{d}$ to be the graph obtained by deleting the two vertices $v$ and $w$ and by connecting the ``hanging pairs'' of edges respecting their colouring. We say that ``$\mathcal{G}/d_{k}$ is obtained by contracting the $k$-dipole $d_{k}$ in $\mathcal{G}$''. The inverse process is called ``creating a dipole''.
\end{Definition} 

Three examples of $1$-dipole moves in open $(3+1)$-coloured graphs are drawn in the figure below.

\begin{figure}[H]
\centering
\includegraphics[scale=1.1]{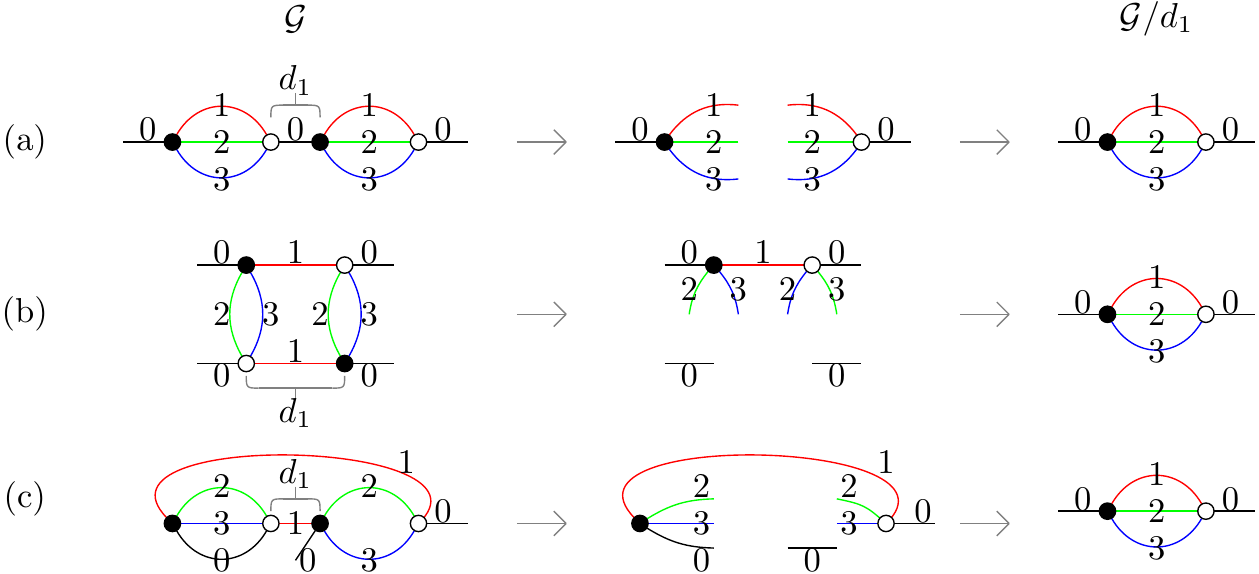}
\caption{Three examples of $1$-dipole contractions in open $(3+1)$-coloured graphs.\label{FigDipoleMove}}
\end{figure}

These examples represent three different cases: Both vertices involved in the dipole do not admit adjacent external legs (a), both vertices do admit adjacent external legs (b) and the mixed case, i.e. one vertex has an adjacent external leg and the other one does not (c). Note that we change the boundary graph only in the second case.

\begin{Remarks}\label{DipoleConSum}\begin{itemize}\item[]
\item[(a)]If both vertices $v$ and $w$ admit an adjacent external leg, then the procedure would produce a disconnected part containing a single edge of colour $0$ connecting two boundary edges, which, as a definition,  we drop from the contracted graph, e.g. see Figure \ref{FigDipoleMove}(b). 
\item[(b)]Note that performing a $k$-dipole move in some open $(d+1)$-coloured graph $\mathcal{G}$ is geometrically one and the same as performing the graph-connected sum of two distinct $(d+1-k)$-bubbles within the graph $\mathcal{G}$.\end{itemize}\end{Remarks}

Next, let us proof the following result, which shows that the boundary complex is left untouched in a dipole move whenever at least one of the bubbles separated is closed:

\begin{Proposition}\label{PropDipoleBound} (Boundary Complex an Dipole Moves)\newline
Let $\mathcal{G}\in\mathfrak{G}_{d}$ be an open $(d+1)$-coloured graph and $d_{k}$ a $k$-dipole within $\mathcal{G}$. If at least one of the two $(d+1-k)$-bubbles separated by the dipole is closed, then $\partial\mathcal{G}=\partial(\mathcal{G}/d_{k})$ and also $\partial\Delta_{\mathcal{G}}=\partial\Delta_{\mathcal{G}/d_{k}}$.\end{Proposition}

\begin{proof}If colour $0$ is involved in the dipole, then the claim is trivial, since the two $(d+1-k)$-bubbles separated by the dipole are closed in this case and hence, all the bicoloured paths starting and ending at boundary vertices are still there after the contraction. Hence, let us assume without loss of generality that the colours involved in the $k$-dipole are $1,\dots,k$. Furthermore, let us assume that the $(d+1-k)$-bubble $\mathcal{B}_{v}^{d+1-k}$, which is separated by the dipole and contains $v$, is closed. The general situation is sketched in the following figure:

\begin{figure}[H]
\centering
\includegraphics[scale=1.3]{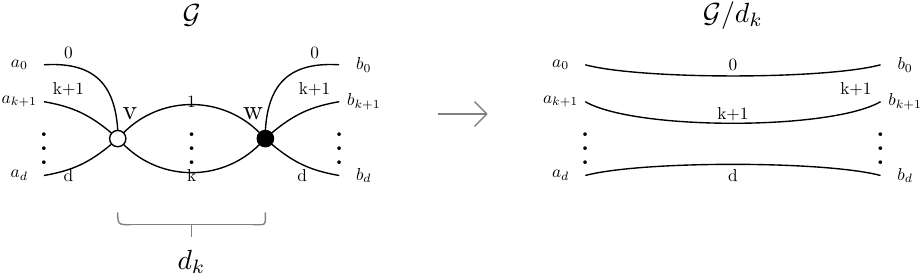}
\caption{A general $k$-dipole contraction.}
\end{figure}

Note that the vertices $a_{i}$ do not necessarily have to be distinct and similarly for the $b_{i}$'s. Furthermore, $b_{0}$ could in principle be a $1$-valent boundary vertex. Now, clearly all the bicoloured paths starting and ending at an external leg of colours $0i$ with $i\in\{1,\dots,k\}$, which are going through the dipole, necessarily contain the vertices $a_{0}$ and $b_{0}$ and still exist after contracting $d_{k}$. Now, let us consider a bicoloured path containing the vertex $w$ of colours $0j$ with $j\in\{k+1,\dots,d\}$. Such a path connects the vertex $b_{0}$ with $w$ and the vertex $w$ with $b_{j}$. Now, if we contract the dipole $d_{k}$, then this bicoloured path still exists precisely because we have assumed that the $(d+1-k)$-bubble $\mathcal{B}_{v}^{d+1-k}$ is closed: The path in $\mathcal{G}/d_{k}$ connects the vertex $b_{0}$ with $a_{0}$, then the vertex $a_{0}$ is connected with $a_{j}$ by a bicoloured path of colours $0j$ and the vertex $a_{j}$ is connected with $b_{j}$. This shows that all the all the non-cyclic faces of $\mathcal{G}$ are still contained in $\mathcal{G}/d_{k}$. Furthermore, it also clear that we do not produce new non-cyclic faces, since the number of external legs is left untouched in case one of the two vertices does not admit an adjacent external leg. Hence, we conclude that $\partial\mathcal{G}=\partial(\mathcal{G}/d_{k})$. Now, in general, equivalent boundary boundary graphs does in general not imply that the two complexes $\Delta_{\mathcal{G}}$ and $\Delta_{\mathcal{G}/d_{k}}$ have the same boundaries, as we have seen in our discussion in Section \ref{SecBubbles}. However, observe that a $k$-dipole move induces a $k$-dipole move in all the bubbles containing some of the edges of colours $1,\dots,k$ involved in the dipole. Hence, by a similar argument, all these bubbles will have the same boundary graphs and hence the same multiplicity after contracting the dipole. In fact, when performing a $k$-dipole move of colour $1,\dots,k$, we just reduce the number of $(d+1-k)$-bubbles with colours $1,\dots,k$ by a closed one and leave the number and multiplicity of all the other $(d+1-k)$-bubbles untouched. This proves that $\partial\Delta_{\mathcal{G}}=\partial\Delta_{\mathcal{G}/d_{k}}$.
\end{proof}

Dipole moves allow us to define the notion of ``\textit{combinatorial equivalent}'' graphs, i.e. graphs which are related by a finite sequence of dipole moves. However, in general, two combinatorial equivalent graph are not homeomorphic. This can easily be seen by the fact that performing a dipole move is the same as performing the connected sum of two submanifolds, as mentioned in Remark \ref{DipoleConSum}(b). Hence, whenever one of these two submanifolds is not a sphere or ball, a dipole move changes the topology. Let us introduce the following terminology:

\begin{Definition} (Proper Dipole Moves)\newline
Let $\mathcal{G}\in\mathfrak{G}_{d}$ be an open $(d+1)$-coloured graph and $d_{k}$ a $k$-dipole within $\mathcal{G}$. We say that $d_{k}$ is ``proper'', if $\vert\Delta_{\mathcal{G}}\vert$ and $\vert\Delta_{\mathcal{G}/d_{k}}\vert$ represent the same manifold (up to PL-homeomorphism).\end{Definition}

As an example, all the dipole moves drawn in Figure \ref{FigDipoleMove} turn out to be proper, since all the graphs represent $3$-balls. More generally, as proven in \cite{GagliardiFerri} for the case of graphs representing closed manifolds and in \cite{Gagliardi87} for the case of general graphs with boundary, one can define two classes of dipole moves, which are always proper:

\begin{Theorem}\label{DipoleProper} (Conditions for Proper Dipole Moves)\newline
Let $\mathcal{G}\in\mathfrak{G}_{d}$ be an open $(d+1)$-coloured graph and $d_{k}$ a $k$-dipole within $\mathcal{G}$.
\begin{itemize}
\item[(1)]If at least one of the $(d+1-k)$-bubbles separated by the dipole is an element in $\overline{\mathfrak{G}}_{d+1-k}$ and represents a $(d-k)$-sphere, then $d_{k}$ is proper. We will call such dipoles ``internal proper dipoles''.
\item[(2)]If both vertices contained in the dipole $d_{k}$ admit adjacent external legs and at least one of the $(d+1-k)$-bubbles separated by the dipole represents a $(d-k)$-ball, then $d_{k}$ is proper. We will call such dipoles ``non-internal proper dipoles''.
\end{itemize}
\end{Theorem}

\begin{proof}The Theorem appeared first in its general formulation for graphs with boundaries in \cite[p.58]{Gagliardi87}. In this paper, one can also find an elegant proof, in which one uses the explicit geometric description of dipole moves. For a proof of the special case of closed graphs, see the earlier paper \cite[p.92]{GagliardiFerri}. Using the discussion of the graph-connected sum of the last section, one can actually give an alternative and very simple proof of the statement, as already observed in \cite[p.5]{CasaliPL}: In case (1), we basically just perform the (internal) connected sum of a spherical $(d+1-k)$-bubble with some other topology (possibly with boundary), which is trivial and hence leaves the topology invariant. This follows from Corollary \ref{CorConSum} (1) and (2). In case (2), we perform the boundary connected sum of some $(d+1-k)$-bubble representing a $(d-k)$-ball with some other topology and hence we do not change the topology either. This is a consequence of Theorem \ref{ThmConSum} (1).\end{proof}

Again, as an example, the three examples of Figure \ref{FigDipoleMove} do have these properties. More precisely, the dipoles in the graphs (a) and (c) are internal proper $1$-dipoles and the the dipole in (b) is a non-internal proper $1$-dipole.

\begin{Remarks}\begin{itemize}\item[]
\item[(a)]Note that by Proposition \ref{PropDipoleBound}, an internal proper dipole leaves the boundary complex invariant whereas a non-internal proper dipole changes the boundary complex explicitly, since it does reduce the number of boundary $(d-1)$-simplices by two.
\item[(b)]Every non-internal proper $k$-dipole move induces an internal proper $k$-dipole move on its boundary graph. The reverse of this statement is in general not true: A general dipole move on some boundary graph does not have to correspond to a dipole move in the corresponding open graph. However, it turns out that every proper dipole on the boundary graph corresponds to a ``wound move'' in its corresponding open graph. These moves are another set of combinatorial moves, which are discussed in Appendix \ref{DipolesAppendix}.\end{itemize}\end{Remarks}

Let us record the following immediate consequences concerning special cases of the Theorem of Gagliardi-Ferri presented above:

\begin{Corollary} 
Let $\mathcal{G}\in\mathfrak{G}_{d}$ be some $(d+1)$-coloured graph. 
\begin{itemize}
\item[(1)]Every $d$-dipole is proper. If $\mathcal{G}$ is closed, then also every $(d-1)$-dipole is proper.
\item[(2)]If $\mathcal{G}$ is closed and represents a manifold, then every dipole is proper. 
\item[(3)]If $\mathcal{G}$ represents a manifold, then every $k$-dipole involving the colour $0$ is proper.
\item[(4)]If $\mathcal{G}$ is open and represents a manifolds, then every $k$-dipole in which both vertices admit adjacent external legs is a non-internal proper one. 
\end{itemize}
\end{Corollary}

\begin{proof}This follows from the previous theorem as well as Proposition \ref{ManifoldsGraphs}, i.e. the fact that for manifolds all $d$-bubbles represent spheres or balls.\end{proof}

Up to now, we have introduced a set of moves for general coloured graphs, which leave the topology invariant. However, it is not yet clear if this set of moves are enough to relate any two coloured graphs describing the same topology to each other. However, it turns out that this is the case, by the following theorem due to M. R. Casali, which can be viewed as an analogous statement as Pachner's theorem:

\begin{Theorem}\label{Casali} (Equivalence Criterion of Casali)\newline
Let $\mathcal{G}_{1},\mathcal{G}_{2}\in\mathfrak{G}_{d}$ be two open $(d+1)$-coloured graphs representing manifolds. Then the manifolds $\mathcal{M}:=\vert\Delta_{\mathcal{G}_{1}}\vert$ and $\mathcal{M}_{2}:=\vert\Delta_{\mathcal{G}_{2}}\vert$ are PL-homeomorphic if and only if $\mathcal{G}_{1}$ and $\mathcal{G}_{2}$ are related by a finite sequence of proper dipole moves of the two types defined in Theorem \ref{DipoleProper}. Moreover, if $\partial\mathcal{G}_{1}\cong\partial\mathcal{G}_{2}$, then $\mathcal{M}_{1}$ and $\mathcal{M}_{2}$ are PL-homeomorphic if and only if $\mathcal{G}_{1}$ and $\mathcal{G}_{2}$ are related by finite sequence of internal proper dipole moves (up to colour-isomorphism).
\end{Theorem}

\begin{proof}The special case of graphs representing closed manifolds is already discussed in \cite{GagliardiFerri}. For the proof of the case of manifolds with boundary, see the ``Main Theorem'' in the original paper by M. R. Casali \cite[p.9]{CasaliPL}, which is based on the ``Main Theorem'' in \cite[p.138]{Casali}, which discusses a similar statement for the specific case of $d=3$ and without the additional part of graphs with the same boundary.\end{proof}

Note that Theorem \ref{DipoleProper} is not an ``if-and-only-if statement'' and there might also be dipoles, which are proper, but which do not satisfy any of the two assumption of Theorem \ref{DipoleProper}. However, by Theorem \ref{Casali}, these ``accidental'' proper dipole moves do not really play a role, since they can always be replaced by a finite sequence of proper internal and non-internal moves of the type introduced in Theorem \ref{DipoleProper}, since these two moves turned out to be enough to characterise topological equivalence of open coloured graphs. In Appendix \ref{DipolesAppendix}, we give some additional insight into this topic by discussing an explicit example of such an ``accidental'' proper dipole move and by discussing the effects of non-proper moves in general. Furthermore, we present another class of moves, so-called ``wound-moves'', which only exist for graphs with non-empty boundary and which might be interesting to look at in more details in some further work.

\section{Core Graphs and Transition Amplitudes}
After this quite extensive discussion of the topology of coloured graphs with boundary, we now move to the discussion of transition amplitudes of the coloured Boulatov model. We will start by discussing the amplitudes of the coloured Boulatov model, which are proportional to the Ponzano-Regge partition functions, in the language of coloured graphs developed previously. Afterwards, we define suitable boundary observables and show how to use them in order to define transition amplitudes. The main goal of the remaining chapter is then to rewrite this expression in a more systematic way by introducing an algorithm, which relates graphs with the same topology and amplitude. After some general remarks about the most general boundary state and about the sum over topologies, we end this chapter with the simplest possible example. 

\subsection{Rescaling of the Action and Regularization}
In the following, let us introduce a simple cut-off regularization of the coloured Boulatov model by introducing the regularized delta functions
\begin{align}\delta^{N}(g):=\sum_{j\in\{0,\dots,2N\}/2}(2j+1)\chi^{j}(g),\end{align}
where $\chi^{j}$ denotes the character of the unique (up to unitary equivalence) unitary irreducible representation of $\mathrm{SU}(2)$ with dimension $(2j+1)$ and where $N\in\mathbb{N}$ denotes some cut-off parameter. Furthermore, let us rescale the interaction term by rescaling the coupling as $\lambda\to\lambda/[\delta^{N}(\mathds{1})]^{1/2}$,
where $\mathds{1}\in\mathrm{SU}(2)$ denotes the identity of $\mathrm{SU}(2)$. Using this rescaling, the action becomes
\begin{equation}\begin{aligned}&\mathcal{S}_{\lambda}[\varphi_{l},\overline{\varphi}_{l}]:=\sum_{l=0}^{3}\int_{\mathrm{SU}(2)^{3}}\bigg (\prod_{i=1}^{3}\mathrm{d}g_{i}\bigg )\,\vert\varphi_{l}(g_{1},g_{2},g_{3})\vert^{2}\\&-\frac{\lambda}{\sqrt{\delta^{N}(\mathds{1})}}\int_{\mathrm{SU}(2)^{6}}\bigg (\prod_{i=1}^{6}\mathrm{d}g_{i}\bigg )\,\varphi_{0}(g_{1},g_{2},g_{3})\varphi_{1}(g_{3},g_{4},g_{5})\varphi_{2}(g_{5},g_{2},g_{6})\varphi_{3}(g_{6},g_{4},g_{1})+c.c.\end{aligned}\end{equation}
The same rescaling of the action has been done in \cite{GurauLargeN1} (and in an analogues way in \cite{GurauLargeN2,GurauLargeN3} for the higher-dimensional Ooguri-Boulatov type models\footnote{The Ooguri model is a GFT based on $\mathrm{SO}(4)$ \cite{OoguriModel}. It generalizes the combinatorial structure of the Boulatov model to the $4$d case, i.e. fields describe tetrahedra and the interaction term describes the gluing of five tetrahedra such that they form a $4$-simplex. However, since gravity in $4d$ is \textit{not} a pure $BF$-theory, this model does not describe gravity, but rather $4$d topological $BF$-theory. In order to construct GFT models for quantum gravity in $4$d, one has to add the corresponding simplicity constraint, i.e. see \cite{Oriti4D}. More generally, these type of models can be extended to arbitrary dimensions.}), because\\
\\
``\textit{in order to obtain a sensible large $N$-limit [of the free energy], the scaling of $\lambda$ and $\overline{\lambda}$ must be chosen such that the maximally divergent graphs have uniform degree of divergence at all orders}'' \cite[p.2]{GurauLargeN1}\\
\\
Another reason is that this scaling allows us to reduce coloured graphs to simpler graphs via proper internal $1$-dipole moves without changing the degree of divergence, as we will see below. 

\subsection{Partition Function and Amplitudes of Vacuum Diagrams}
As already mentioned, the Feynman amplitudes of the Boulatov model are exactly the Ponzano-Regge amplitudes weighted by factors coming from the interaction term. Let us write down the general expression for the amplitudes of vacuum diagrams as well as the partition function of the coloured Boulatov model explicitly, using the language of coloured graphs developed in previous sections. Let us start with the closed case: Let $\mathcal{G}\in\overline{\mathfrak{G}}_{3}$ be a closed $(3+1)$-coloured graph, where we denote by $\mathcal{V}_{\mathcal{G}}$, $\mathcal{E}_{\mathcal{G}}$ and $\mathcal{F}_{\mathcal{G}}$ the set of vertices, edges and faces (=$2$-bubbles) of $\mathcal{G}$. The amplitude corresponding to the graph $\mathcal{G}$ is given by the vertex factor multiplied by the Ponzano-Regge partition function, i.e.
\begin{align}\mathcal{A}^{\lambda}_{\mathcal{G}}=\bigg (\frac{\lambda\overline{\lambda}}{\delta^{N}(\mathds{1})}\bigg )^{\frac{\vert\mathcal{V}_{\mathcal{G}}\vert}{2}}\int_{\mathrm{SU}(2)^{\vert\mathcal{E}_{\mathcal{G}}\vert}}\,\bigg(\prod_{e\in\mathcal{E}_{\mathcal{G}}}\mathrm{d}h_{e}\bigg )\,\prod_{f\in\mathcal{F}_{\mathcal{G}}}\delta^{N}\bigg (\overrightarrow{\prod_{e\in f}}h_{e}^{\varepsilon(e,f)}\bigg ),\end{align}
where we write $e\in f$ for an edge belonging to the face $f$ and where $\varepsilon(e,f)$ is equal to $1$, if the orientation of $e$ and $f$ agrees and $-1$ otherwise. Recall that a bipartite graph can always be directed in a canonical way, e.g. by choosing the orientation such that each edge starts at a black vertex and ends at a white vertex. The generating functional of the coloured Boulatov model is then, in analogy to the uncoloured case, given by
\begin{align}\mathcal{Z}_{\mathrm{cBM}}=\int\,\bigg(\prod_{l=0}^{3}\,\mathcal{D}\varphi_{l}\mathcal{D}\overline{\varphi}_{l}\bigg)\,e^{-\mathcal{S}_{\lambda}[\varphi_{l},\overline{\varphi}_{l}]}=\sum_{\mathcal{G}\in\overline{\mathfrak{G}}_{3}}\frac{1}{\mathrm{sym}(\mathcal{G})}\mathcal{A}^{\lambda}_{\mathcal{G}},\end{align}
where $\mathrm{sym}(\mathcal{G})$ denotes the symmetry factor of the diagram $\mathcal{G}$, which is related to the colour-automorphism group of the graph $\mathcal{G}$, and where the sum is over all closed $(3+1)$-coloured graphs. The equality can be obtained by expressing the interaction part in the formal path integral as a Taylor series in the coupling and by explicitly computing Gaussian integrals, which results into the sum over all pairwise contractions of fields appearing in the interaction vertices, as explained in Section \ref{DynamicsGFT} for the uncoloured Boulatov model. Using this, we can also write down the ``\textit{free energy}'' of the model, which, as usual, is defined via $\mathcal{Z}_{\mathrm{cBM}}=e^{-F_{\mathrm{cBM}}}$ and hence given by
\begin{align}\mathcal{F}_{\mathrm{cBM}}[\lambda]=\sum_{\mathcal{G}\in\overline{\mathfrak{G}}_{3}\text{ connected }}\frac{1}{\mathrm{sym}(\mathcal{G})}\mathcal{A}^{\lambda}_{\mathcal{G}},\end{align}
where the sum in this case only involves \textit{connected} closed $(3+1)$-coloured graphs. As shown in \cite{GurauLargeN1}, the leading order graphs of this expansion in the large $N$-limit are so-called ``\textit{melonic diagrams}'', which are certain coloured graphs with spherical topology. A similar result has been obtained for higher-dimensional Ooguri-Boulatov-type models \cite{GurauLargeN2,GurauLargeN3}. See also \cite{CritTM,GurauColouredTensorModelsReview,GurauColouredTensorModelsReview2} for an extended discussion in the setting of simplicial coloured tensor models and \cite{CritTM2,GurauBook} for a discussion in the setting of coloured tensor models with bubble interactions.

\subsection{Boundary Observables and Transition Amplitudes}
Boundary states in GFT are described by \textit{spin networks}, similar to boundary states in spin foam models. Hence, we have to introduce suitable observables, which are endowed with the corresponding quantum geometric data. Since we are working with a proper field theory, such observables should be functionals of the fundamental GFT-fields and compatible with all the symmetries of our theory. Since we restricted our discussion only to open graphs with the property that all external legs have the same colour $0$, we define our boundary observables such that they only depend on the fields $\varphi_{0}$ and $\overline{\varphi}_{0}$. The general idea for the definition of these observables originates from \cite[p.6]{FreidelGFTOverview}, however, in the setting of coloured GFTs, the definition can be made much more precisely:

\begin{Definition} (GFT Boundary Observable)\newline
Let $\gamma=(\mathcal{V}_{\gamma},\mathcal{E}_{\gamma})\in\overline{\mathfrak{G}}_{2}$ be a closed $(2+1)$-coloured graph and $\Psi=(\gamma,\rho,i)$ be a spin network living on $\gamma$ with corresponding spin network function $\psi\in L^{2}(\mathrm{SU}(2)^{\vert\mathcal{E}_{\gamma}\vert}/\mathrm{SU}(2)^{\vert\mathcal{V}_{\gamma}\vert})$. Then, we define the ``GFT-boundary observable corresponding to $\Psi$'' to be the functional
\begin{align*}\mathcal{O}_{\Psi}[\varphi_{0},\overline{\varphi}_{0}]:=\int_{\mathrm{SU}(2)^{3\vert\mathcal{V}_{\gamma}\vert}}\bigg (\prod_{v\in\mathcal{V}_{\gamma}}\prod_{i=1}^{3}&\mathrm{d}g_{vi}\bigg )\psi(\{g_{s(e)\varphi(e)}g_{t(e)\varphi(e)}^{-1}\}_{e\in\mathcal{E}_{\gamma}})\times\\&\times\bigg (\prod_{v\in V_{\gamma}} \varphi_{0}(g_{v3},g_{v2},g_{v1})\bigg )\bigg (\prod_{v\in \overline{V}_{\gamma}}\overline{\varphi}_{0}(g_{v3},g_{v2},g_{v1})\bigg ),\end{align*}
where $g_{vi}$ are the group elements assigned to the half-edge of $\gamma$ of colour $i\in\{1,2,3\}$ connected to the vertex $v\in\mathcal{V}_{\gamma}$, where $\varphi:\mathcal{E}_{\gamma}\to\mathcal{C}_{2}^{\ast}=\{1,2,3\}$ is the proper edge-colouring of the graph $\gamma$ and where $s,t:\mathcal{E}_{\gamma}\to\mathcal{V}_{\gamma}$ denote the source and target maps, as usual.
\end{Definition}

\begin{Remarks}\begin{itemize}\item[]
\item[(a)]Note that we do not only fix a boundary graph, but already a boundary graph with a fixed orientation and colouring and hence with a fixed topology. This is an important difference to the general definition in \cite{FreidelGFTOverview}, since in the uncoloured version, we only use the boundary graph and as already mentioned, the (dual) $1$-skeleton alone is not enough to determine the topology. Hence, the GFT-boundary observables in the coloured Boulatov model encodes the full simplicial complex $\Delta_{\gamma}$ and hence determine a fixed boundary topology (up to the additional pinching effects for pseudomanifolds with non-simple bubbles discussed in Section \ref{SecBubbles}). 
\item[(b)]The colouring of the graph is encoded in the ordering of the group elements in the fields: Recall that we haven't assumed any kind of permutation symmetry of the fields and hence, the ordering of group elements in each field tell us precisely to which other colour they are glued, where the gluing of half-edges is encoded in the arguments of the spin network.
\item[(c)]Again, note that we restrict only to proper coloured boundary graphs, since we only use open graphs with the property that all external legs have the same colour. More generally, we could allow for general vertex-coloured and edge-bicoloured boundary graphs with the properties as explained in Remark \ref{RemarkBoundaryGraph}. In this case, the observable would depend on all the fields $\varphi_{0,1,2,3}$ and $\overline{\varphi}_{0,1,2,3}$. Hence, we would produce more general open graphs, i.e. open $(3+1)$-coloured graphs, which are not contained in $\mathfrak{G}_{3}$.
\end{itemize}\end{Remarks}

As a next step, let us write down the transition amplitude of the Ponzano-Regge model in the language of coloured graphs. For this, consider an open $(3+1)$-coloured graph $\mathcal{G}\in\mathfrak{G}_{3}$ with $\mathcal{G}=(\mathcal{V}_{\mathcal{G}},\mathcal{E}_{\mathcal{G}})$. Let us fix the following notation:

\begin{itemize}
\item[(1)]Recall that the vertex set can be decomposed as $\mathcal{V}_{\mathcal{G}}=V_{\mathcal{G}}\cup \overline{V}_{\mathcal{G}}$, by bipartiteness. Furthermore, we can decompose these two sets as $V_{\mathcal{G}}=V_{\mathcal{G},\mathrm{int}}\cup V_{\mathcal{G},\partial}$ and similarly $\overline{V}_{\mathcal{G}}=\overline{V}_{\mathcal{G},,\mathrm{int}}\cup \overline{V}_{\mathcal{G},\partial}$, where $ V_{\mathcal{G},\partial}, \overline{V}_{\mathcal{G},\partial}$ are the $1$-valent boundary vertices and where $V_{\mathcal{G},\mathrm{int}},\overline{V}_{\mathcal{G},\mathrm{int}}$ are the $4$-valent bulk vertices.
\item[(2)] The set of faces (=2-bubbles) of $\mathcal{G}$ is denoted by $\mathcal{F}_{\mathcal{G}}$. This set can also be decomposed as $\mathcal{F}_{\mathcal{G}}=\mathcal{F}_{\mathcal{G},\mathrm{int}}\cup\mathcal{F}_{\mathcal{G},\partial}$, where $\mathcal{F}_{\mathcal{G},\partial}$ is the set of all bicoloured paths starting and ending at a boundary vertex and where $\mathcal{F}_{\mathcal{G},\mathrm{int}}$ is the set of all bicoloured cycles, i.e. internal (closed) faces of $\mathcal{G}$.
\item[(3)] Let us denote the set of edges of the boundary graph $\partial\mathcal{G}$ by $\mathcal{E}_{\partial\mathcal{G}}$. Recall that this set is one-to-one correspondence with the set $\mathcal{F}_{\mathcal{G},\partial}$. We will denote this bijection by
\begin{equation}\begin{aligned}e:\partial\mathcal{F}_{\mathcal{G}}&\to\mathcal{E}_{\partial\mathcal{G}}\\f&\mapsto e(f).\end{aligned}\end{equation}
\end{itemize}

Now, let us choose a spin network $\Psi=(\partial\mathcal{G},\rho,i)$ living on the boundary graph $\partial\mathcal{G}$. The Ponzano-Regge partition function corresponding to the pseudomanifold $\vert\Delta_{\mathcal{G}}\vert$ is then the functional given by
\begin{equation}\begin{aligned}&\mathcal{Z}_{\mathrm{PR}}^{\mathcal{G}}[\{g_{e}\}_{e\in\mathcal{E}_{\partial\mathcal{G}}}]=\\&=\int_{\mathrm{SU}(2)^{\vert\mathcal{E}_{\mathcal{G}}\vert}}\bigg (\prod_{e\in\mathcal{E}_{\mathcal{G}}}\mathrm{d}h_{e}\bigg )\,\bigg \{\prod_{f\in\mathcal{F}_{\mathcal{G},\mathrm{int}}}\delta^{N}\bigg (\overrightarrow{\prod_{e\in f}}h_{e}^{\varepsilon(e,f)}\bigg )\bigg \}\bigg \{\prod_{f\in\mathcal{F}_{\mathcal{G},\partial}}\delta^{N}\bigg (g_{e(f)}^{\varepsilon(e(f),f)}\cdot\overrightarrow{\prod_{e\in f}}h_{e}^{\varepsilon(e,f)}\bigg )\bigg \},\end{aligned}\end{equation}
where the starting points of the delta functions in the product over non-cyclic faces are fixed to be one the corresponding boundary vertices. If we denote the corresponding spin network function by $\psi\in L^{2}(\mathrm{SU}(2)^{\vert\mathcal{E}_{\partial\mathcal{G}}\vert}/\mathrm{SU}(2)^{\vert\partial\mathcal{V}\vert})$, then the transition amplitude of the Ponzano-Regge model is given by
\begin{align}\langle\mathcal{Z}_{\mathrm{PR}}^{\mathcal{G}}\vert\Psi\rangle=\int_{\mathrm{SU}(2)^{\vert\mathcal{E}_{\partial\mathcal{G}}\vert}}\,\bigg (\prod_{e\in\mathcal{E}_{\partial\mathcal{G}}}\mathrm{d}g_{e}\bigg )\,\mathcal{Z}_{\mathrm{PR}}^{\mathcal{G}}[\{g_{e}\}_{e\in\mathcal{E}_{\partial\mathcal{G}}}]\cdot\psi(\{g_{e}\}_{e\in\mathcal{E}_{\partial\mathcal{G}}}),\end{align}
as discussed in Section \ref{BoundarySpinFoam}. Let us now discuss how to define transition amplitudes for the coloured Boulatov model. First of all, we make the following definition:

\begin{Definition} (Transition Amplitudes for the Boulatov Model)\newline
Let $\gamma\in\overline{\mathfrak{G}}_{2}$ be a closed $(2+1)$-coloured graph and $\Psi=(\gamma,\rho,i)$ be a spin network living on $\gamma$. Then the transition amplitude is defined by \begin{align*}\langle\mathcal{Z}_{\mathrm{cBM}}\vert \Psi\rangle:=\int\,\bigg(\prod_{l=0}^{3}\,\mathcal{D}\varphi_{l}\mathcal{D}\overline{\varphi}_{l}\bigg)\,e^{-\mathcal{S}_{\lambda}[\varphi_{l},\overline{\varphi}_{l}]}\mathcal{O}_{\Psi}[\varphi_{0},\overline{\varphi}_{0}].\end{align*}
\end{Definition}

Again, expanding the interaction term of the action in the coupling, we can write the formal path integral as a sum over Gaussian integral, which will lead to a sum over all pair-wise contractions of fields in the product of interaction Lagrangians and the given fields within the boundary observable. The computation is the same as for the partition function with the difference that now some of the faces contain the group elements assigned to the boundary, which are contained in the observable $\mathcal{O}_{\Psi}$. Renaming $g_{e}:=g_{s_{e}i}g_{t_{e}i}^{-1}$ for each edge $e\in\mathcal{E}_{\partial\mathcal{G}}$ of colour $i\in\{1,2,3\}$, we are hence left with the integration over all boundary edges, where the integrand is given by the spin network $\psi$ weighted by the Ponzano-Regge partition function for each Feynman diagram, together with prefactors coming from the interaction term. In other words, we can formally write
\begin{align}\label{transampl}\langle\mathcal{Z}_{\mathrm{cBM}}\vert \Psi\rangle=\int\,\bigg(\prod_{l=0}^{3}\,\mathcal{D}\varphi_{l}\mathcal{D}\overline{\varphi}_{l}\bigg)\,e^{-\mathcal{S}_{\lambda}[\varphi_{l},\overline{\varphi}_{l}]}\mathcal{O}_{\Psi}[\varphi_{0},\overline{\varphi}_{0}]=\sum_{\mathcal{G}\in\mathfrak{G}_{3}\text{ with }\partial\mathcal{G}=\gamma}\frac{1}{\mathrm{sym}(\mathcal{G})}\langle\mathcal{A}_{\mathcal{G}}^{\lambda}\vert\Psi\rangle,\end{align}
where the sum is over all open $(3+1)$-coloured graphs in $\mathfrak{G}_{3}$ with $\partial\mathcal{G}=\gamma$ and where the amplitude functionals $\mathcal{A}_{\mathcal{G}}^{\lambda}[\{g_{e}\}_{e\in\mathcal{E}_{\partial\mathcal{G}}}]$ are defined to be the Ponzano-Regge partition function $\mathcal{Z}_{\mathrm{PR}}^{\mathcal{G}}[\{g_{e}\}_{e\in\mathcal{E}_{\partial\mathcal{G}}}]$ together with the prefactors coming from the interaction term, i.e.
\begin{align}\mathcal{A}_{\mathcal{G}}^{\lambda}[\{g_{e}\}_{e\in\mathcal{E}_{\partial\mathcal{G}}}]:=\bigg (\frac{\lambda\overline{\lambda}}{\delta^{N}(\mathds{1})}\bigg )^{\frac{\vert\mathcal{V}_{\mathcal{G},\mathrm{int}}\vert}{2}}\mathcal{Z}_{\mathrm{PR}}^{\mathcal{G}}[\{g_{e}\}_{e\in\mathcal{E}_{\partial\mathcal{G}}}].\end{align}
In other words, the amplitudes $\langle\mathcal{A}_{\mathcal{G}}^{\lambda}\vert\Psi\rangle$ appearing in the expansion (\ref{transampl}) above are given by
\begin{align}\langle\mathcal{A}_{\mathcal{G}}^{\lambda}\vert\Psi\rangle=\bigg (\frac{\lambda\overline{\lambda}}{\delta^{N}(\mathds{1})}\bigg )^{\frac{\vert\mathcal{V}_{\mathcal{G},\mathrm{int}}\vert}{2}}\int_{\mathrm{SU}(2)^{\vert\mathcal{E}_{\partial\mathcal{G}}\vert}}\,\bigg (\prod_{e\in\mathcal{E}_{\partial\mathcal{G}}}\mathrm{d}g_{e}\bigg )\,\mathcal{Z}_{\mathrm{PR}}^{\mathcal{G}}[\{g_{e}\}_{e\in\mathcal{E}_{\partial\mathcal{G}}}]\cdot\psi(\{g_{e}\}_{e\in\mathcal{E}_{\partial\mathcal{G}}}).\end{align}

\begin{Remarks}\begin{itemize}\item[]
\item[(a)]The interpretation of the quantity $\langle\mathcal{Z}_{\mathrm{cBM}}\vert \Psi\rangle$ is the same as for the Ponzano-Regge model: If $\gamma$ has two boundary components, then it computes the probability amplitude between these two states, where in this case we sum over all topologies matching the given boundary topologies, each weighted by the Ponzano-Regge partition function. If $\gamma$ has a single boundary component, then $\langle\mathcal{Z}_{\mathrm{cBM}}\vert \Psi\rangle$ can be interpreted as the probability for transition of the state from the vacuum.
\item[(b)]More precisely, we should take the logarithm of the right-hand side in the definition of the transition amplitude $\langle\mathcal{Z}_{\mathrm{cBM}}\vert \Psi\rangle$, since then we only produce \textit{connected} open $(d+1)$-coloured graphs. However, in the following we will mainly work with a single connected boundary component and hence, all the disconnected parts we produce are closed graph. These additional vacuum diagrams are anyway cancelled by the normalization one usually puts in front of the path integral.
\end{itemize}\end{Remarks}

Last but not least, let us explain how to gauge-fix the Feynman amplitudes of the Boulatov model: Recall that the Boulatov model has a $\mathrm{SU}(2)$ gauge symmetry, as defined in Equation (\ref{GaugeCol}). It is not too hard to see that, on the level of its Feynman amplitudes, this symmetry exactly corresponds to the $\mathrm{SU}(2)$ gauge symmetry of the Ponzano-Regge model, as discussed in Section \ref{GaugeFixPR}, which is the discrete residual of the $\mathrm{SU}(2)$ gauge symmetry of the continuous $BF$-action. Hence, we can gauge-fix the amplitudes in the same way. If $\mathcal{G}$ is some open $(3+1)$-coloured graph matching our given boundary graph $\gamma$, then we choose a maximal tree $\mathcal{T}_{\mathcal{G}}$ in the dual $1$-skeleton of the complex $\Delta_{\mathcal{G}}$. In general, this tree is also allowed to include edges on the boundary. Hence, a tree in the open graph $\mathcal{G}$ is allowed to ``jump'' between two external legs, whenever they are connected by a bicoloured path. In this case, the corresponding boundary group element is included in the tree. Let us decompose $\mathcal{T}_{\mathcal{G}}=\mathcal{T}_{\mathcal{G},\mathrm{int}}\cup\mathcal{T}_{\mathcal{G},\partial}$. The gauge-fixed amplitude functional is then given by
\begin{equation}\begin{aligned}\mathcal{A}_{\mathcal{G}}^{\lambda,\mathrm{gauge-fixed}}&[\{g_{e}\}_{e\in\mathcal{E}_{\partial\mathcal{G}}}]=\int_{\mathrm{SU}(2)^{\vert\mathcal{E}_{\mathcal{G}}\vert}}\bigg (\prod_{e\in\mathcal{E}_{\mathcal{G}}}\mathrm{d}h_{e}\bigg )\,\bigg(\prod_{e\in\mathcal{T}_{G,\mathrm{int}}}\delta^{N}(h_{e})\bigg )\bigg(\prod_{e\in\mathcal{T}_{G,\partial}}\delta^{N}(g_{e})\bigg )\times\\&\times\bigg \{\prod_{f\in\mathcal{F}_{\mathcal{G},\mathrm{int}}}\delta^{N}\bigg (\overrightarrow{\prod_{e\in f}}h_{e}^{\varepsilon(e,f)}\bigg )\bigg \}\bigg \{\prod_{f\in\mathcal{F}_{\mathcal{G},\partial}}\delta^{N}\bigg (g_{e(f)}^{\varepsilon(e(f),f)}\cdot\overrightarrow{\prod_{e\in f}}h_{e}^{\varepsilon(e,f)}\bigg )\bigg \}.\end{aligned}\end{equation}
Recall that the Ponzano-Regge partition function has also another gauge-symmetry, namely the residual of the translational symmetry of the continuous $BF$-action. Since the corresponding gauge-group is not compact, this gauge-symmetry also leads to divergences, and hence is (partially) responsible for the divergences of the Ponzano-Regge model. Up to now, it is not known how to deal with this symmetry in the Boulatov model. One was able to construct a set of transformations of fields (in their Lie-algebra representation) \cite[p.11]{GFTDiff2}, which was shown to be related to the translational symmetry. However, this set of transformations is not a proper symmetry in the field-theoretic sense and hence, it is a priori not clear how to gauge-fix it. Furthermore, it has been observed that this symmetry is broken at the level of pseudomanifolds, since the translational symmetry in the Ponzano-Regge model is due to the discrete Bianchi identity, which only exists on spherical bubbles. Therefore, when talking about gauge-fixed amplitudes in the following, we will always mean $\mathrm{SU}(2)$ but not translational gauge fixed.

\subsection{Change of Amplitudes under Internal Dipole Moves}
In order to obtain a topological expansion of the transition amplitude, the idea is to group different coloured graphs with the same amplitude and the same boundary graph, which are describing the same topology, together. We have already seen in our discussion about the topology of coloured graphs that a suitable way to relate graphs in a topology-preserving way is given by proper dipole moves. Furthermore, if we restrict to internal dipoles, then also the boundary graphs are left invariant. Hence, let us analyse how the Feynman amplitudes of open coloured graphs change when performing an internal dipole move:

\begin{Proposition}\label{AmplChangeInt} (Change of Amplitudes under Proper Internal Dipole Moves)\newline
Let $\mathcal{G}\in\mathfrak{G}_{3}$ and $d_{k}$ with $k\in\{1,2,3\}$ be an internal proper $k$-dipole within $\mathcal{G}$. Then the amplitudes of $\mathcal{G}$ and $\mathcal{G}/d_{k}$ are related as follows:
\begin{itemize}
\item For $k=1$: $\mathcal{A}_{\mathcal{G}}^{\lambda}[\{g_{e}\}_{e\in\mathcal{E}_{\partial\mathcal{G}}}]=(\lambda\overline{\lambda})\mathcal{A}_{\mathcal{G}/d_{1}}^{\lambda}[\{g_{e}\}_{e\in\mathcal{E}_{\partial\mathcal{G}}}]$.
\item For $k=2$: $\mathcal{A}_{\mathcal{G}}^{\lambda}[\{g_{e}\}_{e\in\mathcal{E}_{\partial\mathcal{G}}}]=\frac{(\lambda\overline{\lambda})}{\delta^{N}(\mathds{1})}\mathcal{A}_{\mathcal{G}/d_{2}}^{\lambda}[\{g_{e}\}_{e\in\mathcal{E}_{\partial\mathcal{G}}}]$.
\item For $k=3$: $\mathcal{A}_{\mathcal{G}}^{\lambda}[\{g_{e}\}_{e\in\mathcal{E}_{\partial\mathcal{G}}}]=(\lambda\overline{\lambda})\mathcal{A}_{\mathcal{G}/d_{3}}^{\lambda}[\{g_{e}\}_{e\in\mathcal{E}_{\partial\mathcal{G}}}]$.
\end{itemize}
\end{Proposition}

\begin{proof}Let us start by proving the case of $k=1$. The strategy is the same as for the case of closed $(3+1)$-coloured graphs in \cite[p.6]{GurauLargeN1} and for general closed $(d+1)$-coloured graphs in \cite[p.11f.]{GurauLargeN3}: We try to relate the amplitude of $\mathcal{G}$ to the amplitude of $\mathcal{G}/d_{1}$ by applying suitable substitutions within the integrals. The difference in our case is that some of the delta functions will contain boundary group elements. In general, we should distinguish between the following two cases:
\begin{itemize}
\item[(1)]The dipole edge is not of colour $0$.
\item[(2)]The dipole edge is of colour $0$.
\end{itemize}
In the first case, the general situation looks as follows:
\begin{figure}[H]
\centering
\includegraphics[scale=1.6]{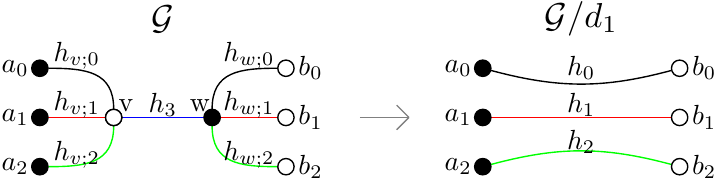}
\caption{A $1$-dipole contraction involving an edge of colour $3$ and group elements assigned to all the relevant edges.}
\end{figure}
We consider a $1$-dipole containing an edge of colour $3$, between two vertices $v$ and $w$. Furthermore, we assume that the $3$-bubble $\mathcal{B}_{va_{0}a_{1}a_{2}}^{012}$ of colour $012$ containing the vertex $v$ represents a $2$-sphere. The $3$-bubble of colour $012$ containing $w$ is allowed to be open and to have arbitrary topology. Note that the vertices $a_{i}$ do not have to be distinct and similar for the $b_{i}$'s. Furthermore, the vertex $b_{0}$ could in principle also be a $1$-valent boundary vertex. Next, let us denote the group elements living on the edges $va_{i}$ by $h_{v;i}$ and analogously the group elements living on the edges $b_{i}w$ by $h_{w;i}$. The group element living on the dipole edge $vw$ is denoted by $h_{3}$. The contribution of the dipole to the Ponzano-Regge functional is given by the following integrals:\newpage
\begin{equation}\begin{aligned}\int_{\mathrm{SU}(2)^{7}}&\,\mathrm{d}h_{3}\bigg (\prod_{i=0}^{2}\mathrm{d}h_{v;i}\mathrm{d}h_{w;i}\bigg )\\&\delta^{N}(h_{v;0}h_{3}^{-1}h_{w;0}H^{03}[g])\delta^{N}(h_{v;1}h_{3}^{-1}h_{w;1}H^{13})\delta^{N}(h_{v;2}h_{3}^{-1}h_{w;2}H^{23})\\ & \delta^{N}(h_{v;0}h_{v;1}^{-1}H_{v}^{01})\delta^{N}(h_{v;2}h_{v;0}^{-1}H_{v}^{02})\delta^{N}(h_{v;1}h_{v;2}^{-1}H_{v}^{12})\\&\delta^{N}(h_{w;1}^{-1}h_{w;0}H_{w}^{01}[g])\delta^{N}(h_{w;0}^{-1}h_{w;2}H_{w}^{02}[g])\delta^{N}(h_{w;2}^{-1}h_{w;1}H_{w}^{12}).\end{aligned}\end{equation}
The group elements $H^{i3}$ for $i\in\{0,1,2\}$ denote the products of the group elements assigned to the bicoloured path of colour $i3$ starting at $b_{i}$ and ending at $a_{i}$. The product $H^{03}$ could in principle contain a boundary group element, which is indicated by the notation $[g]$, since the corresponding face could be non-cyclic, i.e. contain two external legs. The group elements $H^{ij}_{v}$ with $i,j\in\{0,1,2\}$ and $i<j$ are the product of the remaining group elements of the edges belonging to the faces of colour $ij$ containing the vertex $v$. Since the $3$-bubble $\mathcal{B}_{v}^{012}$ is closed, all these faces are cyclic and hence these products do not contain boundary group elements. Last but not least, $H^{ij}_{w}$ with $i,j\in\{0,1,2\}$ and $i<j$ are the product of the remaining group elements of the edges belonging to the faces of colour $ij$ containing the vertex $w$. The faces of colour $01$ and $02$ could in principle be non-cyclic and hence $H^{01}_{w}$ and $H^{02}_{w}$ could again contain one of the boundary group elements, which we again indicate by $[g]$. \\
\\
As a first step, let us use the change the variables $h_{w;i}$ to $h_{i}:=h_{v;i}h_{3}^{-1}h_{w;i}$ for all $i\in\{0,1,2\}$ and hence $\mathrm{d}h^{\prime}_{w;i}=\mathrm{d}h_{w;i}$, by bi-invariance of the Haar measure. Using this transformation, we see that the integrand is no longer dependent on $h_{3}$ and hence, we can integrate trivially over it (recall that we use the \textit{normalized} Haar measure). The contribution from the dipole becomes
\begin{equation}\begin{aligned}\int_{\mathrm{SU}(2)^{6}}&\,\bigg (\prod_{i=0}^{2}\mathrm{d}h_{v;i}\mathrm{d}h_{i}\bigg )\\&\delta^{N}(h_{0}H^{03}[g])\delta^{N}(h_{1}H^{13})\delta^{N}(h_{2}H^{23})\\ & \delta^{N}(h_{v;0}h_{v;1}^{-1}H_{v}^{01})\delta^{N}(h_{v;2}h_{v;0}^{-1}H_{v}^{02})\delta^{N}(h_{v;1}h_{v;2}^{-1}H_{v}^{12})\\&\delta^{N}(h_{1}^{-1}h_{v;1}h^{-1}_{v;0}h_{0}H_{w}^{01}[g])\delta^{N}(h_{0}^{-1}h_{v;0}h^{-1}_{v;2}h_{2}H_{w}^{02}[g])\delta^{N}(h_{2}^{-1}h_{v;2}h^{-1}_{v;1}h_{1}H_{w}^{12}).\end{aligned}\end{equation}
As a next step, we can use the two delta functions $\delta^{N}(h_{v;0}h_{v;1}^{-1}H_{v}^{01})$ and $\delta^{N}(h_{v;2}h_{v;0}^{-1}H_{v}^{02})$ to integrate over the group elements $h_{v;1}$ and $h_{v;2}$, which results into the replacements $h_{v;1}:=H^{01}_{v}h_{v;0}$ and $h^{-1}_{v;2}:=h_{v;0}^{-1}H^{02}_{v}$ and hence, we are left with the contribution
\begin{equation}\begin{aligned}\int_{\mathrm{SU}(2)^{4}}&\,\mathrm{d}h_{0;v}\mathrm{d}h_{0}\mathrm{d}h_{1}\mathrm{d}h_{2}\\&\delta^{N}(h_{0}H^{03}[g])\delta^{N}(h_{1}H^{13})\delta^{N}(h_{2}H^{23})\\ &\delta^{N}(H^{01}_{v}H^{02}_{v}H^{12}_{v})\\
&\delta^{N}(h_{1}^{-1}H^{01}_{v}h_{0}H^{01}_{w}[g])\delta^{N}(h_{0}^{-1}H^{02}_{v}h_{2}H^{02}_{w}[g])\delta^{N}(h_{2}^{-1}H^{12}_{v}h_{1}H^{12}_{w}).\end{aligned}\end{equation}
We see that the integration over $h_{v;0}$ is now trivial and gives $1$ (since we use the normalized Haar measure for compact Lie groups) and hence we end up with the following expression:
\begin{equation}\begin{aligned}\int_{\mathrm{SU}(2)^{3}}&\,\mathrm{d}h_{0}\mathrm{d}h_{1}\mathrm{d}h_{2}\\&\delta^{N}(h_{0}H^{03}[g])\delta^{N}(h_{1}H^{13})\delta^{N}(h_{2}H^{23})\\ &\delta^{N}(H^{01}_{v}H^{02}_{v}H^{12}_{v})\\
&\delta^{N}(h_{1}^{-1}H^{01}_{v}h_{0}H^{01}_{w}[g])\delta^{N}(h_{0}^{-1}H^{02}_{v}h_{2}H^{02}_{w}[g])\delta^{N}(h_{2}^{-1}H^{12}_{v}h_{1}H^{12}_{w}).\end{aligned}\end{equation}
The interpretation of this result is the following: We integrate over three group elements $h_{0}$, $h_{1}$, $h_{2}$ which are the group elements living on the three edges $a_{i}b_{i}$, which we get after dipole contracting. This can also be seen by the replacements which we did in the integration, i.e. $h_{i}=h_{v;i}h^{-1}_{3}h_{w;i}$ for all $i\in\{0,1,2\}$. In other words, all the changes of variables in the calculation exactly corresponds to the cancellation of the $1$-dipole. As a consequence, we see that the first row of delta functions corresponds to the bicoloured paths $i3$ for $i\in\{0,1,2\}$ containing one of the three edges $a_{i}b_{i}$. Let us now turn to the third line: Before contracting the dipole, we had for each pair $ij$ with $i,j\in\{0,1,2\}$ and $i<j$ precisely two bicoloured faces in our integration, one containing $v$ and one containing $w$. After contracting the dipole, we get rid of the colour $3$ edge and connect all the lines with colours $i\in\{0,1,2\}$ to each other. As a consequence, we combine for each $i,j$ the two bicoloured paths, which before contracting the dipole were disconnected by the colour $3$ edge. To sum up, the third line of delta functions corresponds to all the faces with colour $i,j\in\{0,1,2\}$ of the graph containing two of the edges $a_{i}b_{i}$. At the end of the day, we see that the first and third line of our result above precisely corresponds to the contribution of the three edges $a_{i}b_{i}$ of the contracted graph $\mathcal{G}/d_{1}$. Hence, we have related the amplitude of $\mathcal{G}$ with the amplitude of $\mathcal{G}/d_{1}$ up to the additional factor $\delta^{N}(H^{01}_{v}H^{02}_{v}H^{12}_{v})$. To get rid of this term, we make use of the assumption that the bubble $\mathcal{B}^{012}_{va_{0}a_{1}a_{2}}$ is spherical. Now, the product $H^{01}_{v}H^{02}_{v}H^{12}_{v}$ corresponds to the exterior face of the planar ribbon graph obtained by cutting the vertex $v$ in the spherical $3$-bubble $\mathcal{B}_{va_{0}a_{1}a_{2}}^{012}$ and all the delta function corresponding to closed faces of this planar graph are also contained in the amplitude of $\mathcal{G}/d_{1}$. Hence, as explained in \cite[p.6]{GurauLargeN1}, \cite[p.12]{GurauLargeN3} in more details, $\delta^{N}(H^{01}_{v}H^{02}_{v}H^{12}_{v})$ can be replaced by $\delta^{N}(\mathds{1})$. As a consequence, taking into account that we reduce the number of internal vertices by two, we have that
\begin{align}\mathcal{A}_{\mathcal{G}}^{\lambda}[\{g_{e}\}_{e\in\mathcal{E}_{\partial\mathcal{G}}}]=\frac{\lambda\overline{\lambda}}{\delta^{N}(\mathds{1})}\delta^{N}(\mathds{1})\mathcal{A}_{\mathcal{G}/d_{1}}^{\lambda}[\{g_{e}\}_{e\in\mathcal{E}_{\partial\mathcal{G}}}]=(\lambda\overline{\lambda})\cdot\mathcal{A}_{\mathcal{G}/d_{1}}^{\lambda}[\{g_{e}\}_{e\in\mathcal{E}_{\partial\mathcal{G}}}],\end{align}
which concludes the proof. In the second case, i.e. the case where the dipole edge is of colour $0$, the proof is exactly the same with the difference that now all the faces containing the dipole edge could contain a boundary group element and none of the faces containing the vertex $w$, but not the vertex $v$. \\
\\
The proof for $2$- and $3$-dipole can be done completely analogues. In the case of $2$-dipoles, we do not get an extra term and the amplitudes can directly be related. Similarly, we can relate the amplitudes for the case of $3$-dipoles directly, however, we get an additional factor $\delta^{N}(\mathds{1})$ in this case from the fact that there is already one redundant face of the three faces containing the three edges involved in the dipole. This factor cancels again with the factor $1/\delta^{N}(\mathds{1})$ coming from the fact that we reduce the number of internal vertices by two.
\end{proof}

\begin{Remarks}\begin{itemize}\item[]
\item[(a)]Note that the result could have been guessed: Since proper internal dipole moves do not change the topology and the boundary complex, we know, by topological invariance of the Ponzano-Regge model, that the partition functions should be the same. Now, since we haven't fixed the translational symmetry, we get for each internal dual $3$-cell (=closed $3$-bubble) a divergence. Furthermore, observe that that the number of closed $3$-bubbles is reduced by one for $1$- and $3$-dipoles, whereas it stays the same for $2$-dipoles. Hence, we get rid of a factor of $\delta^{N}(\mathds{1})$ for $1$- and $3$-dipole contractions, which cancels the delta function coming from the fact that we change the number of internal vertices by two.
\item[(b)]Obviously, we get as a simple corollary that the same relations are true for the $\mathrm{SU}(2)$ gauge fixed amplitudes. More rigorously, let us choose a maximal tree in the uncontracted graph $\mathcal{G}$ such that the edges with group elements $h_{v;0}$, $h_{0;w}$ and $h_{3}$ are on the tree, but not the other edges involved in the dipole, i.e. the edges equipped with $h_{v;1,2}$ and $h_{w;1,2}$. This is of course always possible, since all the other vertices $a_{1,2}$ and $b_{1,2}$ are connected to the vertices $a_{0}$ and $b_{0}$ via some face, possibly involving boundary edges. Furthermore, choosing this tree is no loss of generality, since the amplitude is independent of the choice of tree, as discussed in some previous parts. Now, using this tree means to add $\delta^{N}(h_{v;0})$, $\delta^{N}(h_{w;0})$ and $\delta^{N}(h_{3})$ to the integral above. Doing the same steps, we get something like $\delta^{N}(h_{v;0})\delta^{N}(h_{3})\delta^{N}(h_{3}h_{v;0}^{-1}h_{0})$, which is the same as $\delta^{N}(h_{0,v})\delta^{N}(h_{3})\delta^{N}(h_{0})$. Now, recall that the elements $h_{0,v}$ and $h_{3}$ do not appear anywhere else in the integration after all the substitutions and hence we can integrate them trivially, which means that we are left with $\delta^{N}(h_{0})$. So, the result is the gauge-fixed amplitude of the contracted graph $\mathcal{G}/d_{1}$ where the tree is such that the edge equipped with $h_{0}$ is part of the tree, but not the other two edges with $h_{1,2}$
\end{itemize}\end{Remarks}

Let us briefly comment on the claim in the end of the proof above that the additional delta function $\delta^{N}(H^{01}_{v}H^{02}_{v}H^{12}_{v})$ appearing in the amplitude can always be replaced by $\delta^{N}(\mathds{1})$. As mentioned above, a technical proof can be found in the cited literature, however, let us show why this is intuitively clear. The general situation is sketched in the following figure:

\begin{figure}[H]
\centering
\includegraphics[scale=1.3]{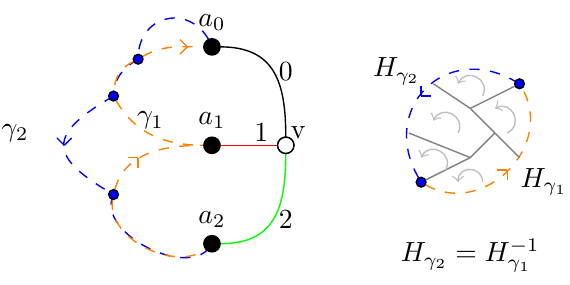}
\caption{The two paths corresponding to the faces equipped with $H^{01}_{v}$, $H^{02}_{v}$ and $H^{12}_{v}$.}
\end{figure}

The important point is that a spherical $3$-bubble is clearly a \textit{planar} ribbon graph, which means that it can be drawn in such a way that all its edges are only intersecting on their source and target points, i.e. there is no ``crossing'' of edges, and such that all regions bounded by some path of edges are faces, i.e. the paths are bicoloured. The left-hand side of the figure above shows two paths $\gamma_{1}$ and $\gamma_{2}$, where $\gamma_{1}$ connects the vertex $a_{2}$ with $a_{0}$ and is equipped with the product of group elements $H^{12}_{v}H^{01}_{v}$ and where $\gamma_{2}$ connects $a_{0}$ with $a_{2}$ and is equipped with the products of group elements $H^{02}_{v}$. Now, the point is that the path $\gamma_{2}$ can always be identified with the inverse of the path $\gamma_{1}$, which means that we can replace the group element $H^{02}_{v}$ by $(H^{12}_{v}H^{01}_{v})^{-1}$ and hence $\delta^{N}(H^{01}_{v}H^{02}_{v}H^{12}_{v})\to\delta^{N}(\mathds{1})$. This is follows from the fact that the graph is planar and the fact that all the delta functions corresponding to the closed faces of it are contained in the amplitude. More precisely, note that whenever the paths $\gamma_{1}$ and $\gamma_{2}$ deviate from each other, then they bound some finite region, as drawn on the right-hand side in the figure above. The group element $H_{\gamma_{1}}$ denotes the product of group elements corresponding to this part of the graph $\gamma_{1}$ and similarly, $H_{\gamma_{2}}$ denotes the product of group elements corresponding to this part of the graph $\gamma_{2}$. Within this region, all the delta functions corresponding to the closed faces will project to the boundary, which means that the group elements assigned to this parts of the paths will be identified, i.e. $H_{\gamma_{2}}=H_{\gamma_{1}}^{-1}$. Applying this to every region bounded by some parts of the paths, we end up with the claimed equality $\gamma_{1}=-\gamma_{2}$ and hence $H^{02}_{v}=(H^{12}_{v}H^{01}_{v})^{-1}$.

\subsection{Bubble Rooting and Core Graphs}
The main idea in order to obtain a topological expansion is to reduce some given open coloured graph to a simpler one with the same boundary graph by applying as much internal proper dipole moves as possible. This idea generalizes essentially the ``bubble rooting procedure'' of closed coloured graphs introduced in \cite{GurauLargeN1,GurauLargeN2,GurauLargeN3} in order to expand the free energy of coloured Boulatov-Ooguri type models to open coloures graphs. In order to obtain a well-defined rooting procedure, we should restrict only to proper $1$-dipoles, since two neighbouring $3$-bubbles of the same colour in some connected graph are always connected by a $1$-dipole, whereas for example two neighbouring $2$-bubbles are not necessarily connected by a $2$-dipole.\\
\\
 In the following, we work with the general $d$-dimensional case: Let $\gamma$ be a closed $d$-coloured graph, which we fix to be our boundary graph. Furthermore, let $\mathcal{G}\in\mathfrak{G}_{d}$ be an connected open $(d+1)$-coloured graph with $\partial\mathcal{G}=\gamma$. Now, lets take some colour $i\in\Delta_{d}=\{0,\dots,d\}$. Then there are two possibilities:

\begin{itemize}
\item[(1)]All $d$-bubbles without colour $i$ are closed and represent $d$-spheres. 
\item[(2)]There exists at least one $d$-bubbles without colour $i$, which is not spherical. Note that this includes both the case of $d$-bubbles, which are open, and $d$-bubbles which are closed, but not spherical. 
\end{itemize}

Note that for graphs with non-empty boundary contained in $\mathfrak{G}_{d}$ property (1) can only be the case for $i=0$. In case (1), we choose one of the spherical $d$-bubbles without colour $i$ as ``\textit{principal root}'' and denote it by $\mathcal{R}^{i}_{(1)}$. In case (2), we choose one of the non-spherical $d$-bubbles without colour $i$ as ``\textit{principal root}'' $\mathcal{R}^{i}_{(1)}$ and all other non-spherical $d$-bubbles without colour $i$ as ``\textit{branching roots}'', which we denote by $\mathcal{R}^{i}_{(\mu)}$ with some labelling parameter $\mu$. As a next step, we need the ``connectivity graph of colour $i$'', which is defined as follows:

\begin{Definition} (Connectivity Graphs)\newline
Let $\mathcal{G}$ be some open $(d+1)$-coloured graph. Then the ``connectivity graph of colour $i\in\Delta_{d}$'' is the graph $\mathcal{C}^{i}[\mathcal{G}]$ defined as follows:
\begin{itemize}
\item[(1)]There is a vertex in $\mathcal{C}^{i}[\mathcal{G}]$ for each $d$-bubbles without colour $i$ in $\mathcal{G}$.
\item[(2)]Two vertices in $\mathcal{C}^{i}[\mathcal{G}]$ are connected by an edge, if there is an edge of colour $i$ in $\mathcal{G}$ connecting the two $d$-bubbles corresponding to the two vertices.
\end{itemize}
\end{Definition}

Note that the connectivity graphs are in general ``pseudographs'', i.e. multigraphs in which an edge is also allowed to be a multiset, which means that there are tadpole lines. Furthermore, the connectivity graphs have in general no fixed valency. Now, let us the following algorithm:

\begin{itemize}
\item[(1)]Draw the $0$-connectivity graph and choose a maximal tree $\mathcal{T}^{0}$ in it. In general, there are two different types of vertices in this graph: The roots and all the other vertices, which correspond to the spherical $d$-bubbles. Now, every vertex is connected to the principal root by a unique path contained in the maximal tree. For each branch root, let us draw the incident edge belonging to the tree, which is contained in this path, as a dashed line. All the other edges we draw as solid lines.
\item[(2)]All the solid lines belonging to the tree are proper internal $1$-dipoles and we contract them in the graph $\mathcal{G}$. Repeating this procedure for all solid lines, we are left with either a unique $\hat{0}$-bubble, which is spherical, or with a bunch of non-spherical $\hat{0}$-bubbles.
\item[(3)]Next, we choose a maximal tree $\mathcal{T}^{1}$ in the $1$-connectivity graph in the graph obtained \textit{after} step (2). Note that this tree in general depends on the tree $\mathcal{T}^{0}$. Then, we repeat step (2), i.e. we contract the proper internal $1$-dipole corresponding to the solid lines.
\item[(4)]We repeat this procedure for all other colours by choosing trees $\mathcal{T}^{j}$ for all $j\in\{0,\dots,d\}$, which depend on the chosen trees $\mathcal{T}^{j-1},\dots,\mathcal{T}^{0}$. Again, after each choice of tree, we contract all the dipoles as in step (2).
\end{itemize}

The general procedure at some iteration $j\in\{0,\dots,d\}$ is sketched in the following figure:

\begin{figure}[H]
\centering
\includegraphics[scale=1.1]{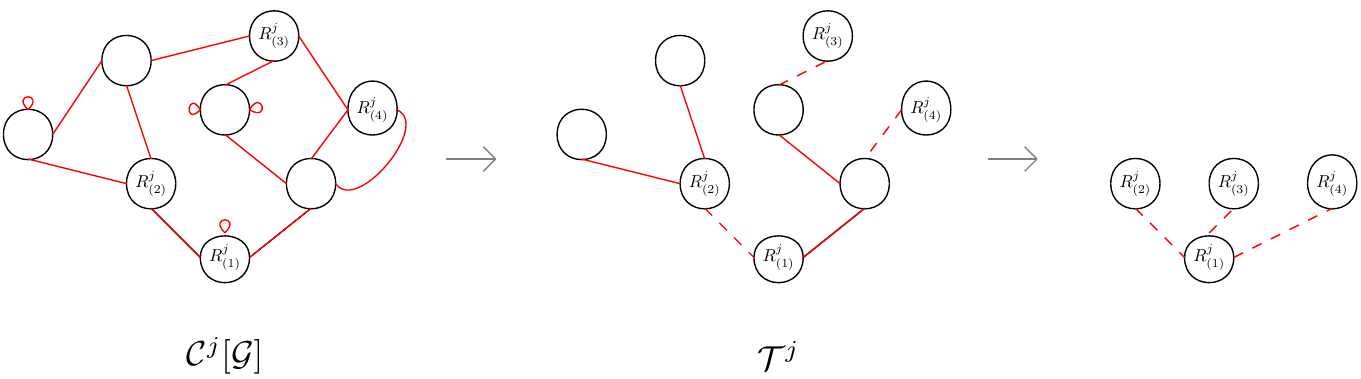}
\caption{The procedure per iteration $j$.}
\end{figure}

Note that the procedure at each iteration does not depend on the order of dipoles we contract in some fixed tree, since a proper internal $1$-dipole of colour $i$ does only reduce the number of $\hat{i}$-bubbles by a spherical one and hence, all the other $1$-dipoles of colour $i$ on this tree are still $1$-dipoles. Furthermore, note that the procedure is well defined, since when we contract the tree of colour $j$, we do not change the number of proper internal $1$-dipoles of colours $\neq j$. This is a consequence of the following lemma:

\begin{Lemma} Let $\mathcal{G}$ be an open $(d+1)$-coloured graph. A proper internal $1$-dipole move of colour $i\in\Delta_{d}$ does not change the number as well as the topologies of all $d$-bubbles of $\mathcal{G}$ involving colour $i$.\end{Lemma}

\begin{proof}It is clear that we do not touch the number of all these bubbles. For the second claim, observe that every proper internal $1$-dipole move in $\mathcal{G}$ consisting of an edge $e\in\mathcal{E}_{\mathcal{G}}$ of colour $i\in\Delta_{d}$ also corresponds to a proper internal $1$-dipole move within all the $d$-bubbles of $\mathcal{G}$ involving the edge $e$, because every $(d-1)$-bubble contained in the spherical bubble separated by the dipole is itself a closed bubble representing a $(d-2)$-sphere.\end{proof}

In other words, contracting the connectivity graph of some colour $j$ might change the connectivity graphs of colour $i\neq j$, but there is still either a unique spherical $\hat{i}$-bubble for $i<j$, or all the $\hat{i}$-bubbles for $i<j$ are non-spherical and hence, we do not produce any new proper internal $1$-dipole of colour $i<j$. Furthermore, the number of dipoles we can contract in the connectivity graphs with $i>j$ stays the same. After this procedure, we are left with a graph in which we cannot perform any more proper internal $1$-dipoles. We call such graph topological core graphs:

\begin{Definition} (Topological Core Graphs for Open Graphs)\newline
Let $\gamma$ be some closed $(2+1)$-coloured graph. A ``topological core graph with boundary $\gamma$ of order $p$'', where $2p=\vert\mathcal{V}_{\mathcal{G},\mathrm{int}}\vert$, is an open $(d+1)$-coloured graph $\mathcal{G}_{p}$ such that $\partial\mathcal{G}_{p}=\gamma$ and such that for every colour $i\in\Delta_{d}$ one of the following applies:
\begin{itemize}
\item[(1)]There is unique closed and spherical $d$-bubble without colour $i$.
\item[(2)]All $d$-bubbles without colour $i$ are either closed and non-spherical or open graphs.
\end{itemize}
\end{Definition}

\begin{Remark}Note that for closed graphs representing manifolds, the notion of topological core graphs is actually one and the same as the notion of crystallizations introduced earlier. To see this, recall that a crystallization of a closed $d$-manifold is a closed $(d+1)$-coloured graph, which has a unique $d$-bubble for each combination of $d$ colours. Hence, every crystallization of a closed manifold is clearly a core graph. However, also the reverse is clearly true: If some closed graph represents a manifold, then all its $d$-bubbles are $d$-sphere. As a consequence, every topological core graph dual to a closed manifold is a crystallization. For open graphs, however, this is not always the case. Recall that a crystallization of a $d$-manifold with boundary is a $\partial$-contracted open $(d+1)$-coloured graph representing it, which means that there is only one closed $d$-bubble of colours $\hat{0}$ and for each boundary component $(d-1)$ $d$-bubbles touching it. Obviously, every crystallization of an open manifold is a core graph. However, the reverse is in general not true, since in a core graph there might be several $d$-bubbles of the same colour, which all include external legs and hence cannot be spherical, since they represent $d$-balls. If the boundary graph $\gamma$ is itself a crystallization, then this direction becomes true, since for every $2$-bubble on the boundary there is precisely one open $3$-bubble in the open graph.
\end{Remark}

Let us now focus on the $3$-dimensional case. As already mentioned, the core graph, which we obtain from from some given open coloured graph by this bubble rooting procedure, does in general depend on the the chosen trees. However, there is the following result:

\begin{Proposition} (Amplitudes of Core Graphs obtain in the Procedure)\newline
Let $\mathcal{G}\in\mathfrak{G}_{3}$ be an open $(3+1)$-coloured graph and $\mathcal{G}_{c}$ some topological core graph obtained by rooting $\mathcal{G}$. Then
\begin{align*}\vert\mathcal{V}_{\mathcal{G}_{c},\mathrm{int}}\vert=\vert\mathcal{V}_{\mathcal{G},\mathrm{int}}\vert-2(\mathcal{B}^{[d]}-\mathcal{R}^{[d]}),\end{align*}
where $\mathcal{R}^{[d]}$ denotes the total number of roots, as well as
\begin{align*}\mathcal{A}^{\lambda}_{\mathcal{G}}[\{g_{e}\}_{e\in\partial\mathcal{G}}]=(\lambda\overline{\lambda})^{\mathcal{B}^{[d]}-\mathcal{R}^{[d]}}\mathcal{A}^{\lambda}_{\mathcal{G}_{c}}[\{g_{e}\}_{e\in\partial\mathcal{G}}].\end{align*}
\end{Proposition}

\begin{proof}As already mentioned above, the number of $1$-dipoles we contract is independent of the choice of trees and one can easily see that it is given by $\mathcal{B}^{[d]}-\mathcal{R}^{[d]}$. Hence, the first claim follows, since any $1$-dipole reduces the number of internal vertices by two. The second claim follows from repeatedly applying Proposition \ref{AmplChangeInt}. \end{proof}

Therefore, it makes sense to introduce the following notion of equivalence:

\begin{Definition} (Core Equivalence Classes)\newline
We call two open topological core graphs $\mathcal{G}_{p}$ and $\mathcal{G}_{q}$ with the the same boundary ``equivalent'', if the have the same order, i.e. $p=q$, the same topology as well as the same amplitude. This defines an equivalence class, which we call ``topological core equivalence class'' and denote by $[\mathcal{G}_{p}]$. We denote the set of all topological core equivalence classes with respect to some given boundary graph $\gamma$ by $\mathfrak{G}^{\mathrm{core}}_{\gamma}$.\end{Definition}

\subsection{Topological Expansion of the Transition Amplitude}
Using the bubble rooting procedure defined in the previous paragraph, we are now finally ready to write done a topological expansion for the transition amplitude of the coloured Boulatov model. Let $\gamma\in\overline{\mathfrak{G}}_{2}$ be some closed $(2+1)$-coloured graph of arbitrary topology and $\Psi=(\gamma,\rho,i)$ be some spin network state living on it. Then we can expand the transition amplitude (\ref{transampl}) in terms of the topological core equivalence classes as follows:

\begin{align}\langle\mathcal{Z}_{\mathrm{cBM}}\vert\Psi\rangle=\sum_{p=\vert\mathcal{V}_{\gamma}\vert/2}^{\infty}\,\sum_{[\mathcal{G}_{p}]\in \mathfrak{G}^{\mathrm{core}}_{p,\gamma}}C^{[\mathcal{G}_{p}]}(\lambda,\overline{\lambda})\langle\mathcal{A}^{\lambda}_{\mathcal{G}_{p}}\vert\Psi\rangle,\end{align}

where $\mathfrak{G}^{\mathrm{core}}_{p,\gamma}\subset\mathfrak{G}^{\mathrm{core}}_{\gamma}$ is the set of all core equivalence classes of order $p$ and where $C^{[\mathcal{G}_{p}]}(\lambda,\overline{\lambda})$ is a combinatorial factor, which counts all the factors of $\lambda\overline{\lambda}$ coming from graphs, which root back to some graph in the equivalence class $[\mathcal{G}_{p}]$ as well as their symmetry factor. More precisely, the factor it is given by

\begin{align}C^{[\mathcal{G}_{p}]}(\lambda,\overline{\lambda}):=\sum_{\mathcal{G}\in\mathfrak{G}_{3}\text{ with }\partial\mathcal{G}=\gamma\text{ and }\mathcal{G}\to [\mathcal{G}_{p}]}\frac{(\lambda\overline{\lambda})^{\frac{\vert\mathcal{V}_{\mathcal{G},\mathrm{int}}\vert}{2}-p}}{\mathrm{sym}(\mathcal{G})},\end{align}

where the sum is over all open $(3+1)$-coloured graphs with boundary graph $\gamma$, which can be rooted to one of the members of the core equivalence class $[\mathcal{G}_{p}]$.\\
\\
The expansion written above is a \textit{topological expansion}, which means that each term in the sum corresponds to some fixed topology. However, one should note that there are some redundancies within this expansion:

\begin{itemize}
\item[(1)]Two core graphs at the same order $p$ might have the same amplitude, but are not topological equivalent.
\item[(2)]Two core graphs at the same order $p$ might be topological equivalent, but still have different amplitudes. 
\item[(3)]Last but not least, it might happen that there are two core graphs with different orders $p\neq q$, which are topologically equivalent and also have the same amplitude. However, they are still in different core equivalence classes, since the have different orders.
\end{itemize}

To sum up, every topological core equivalence class represents a fixed topology, however, there are in general an infinite number of different distinct equivalence classes, which represent the same (pseudo)manifold.\\
\\
Before moving to the next topic, let us briefly say a few words about the lowest order diagram, i.e. the core graph with smallest possible number of vertices: A core graph at this order has to have at least $\vert\mathcal{V}_{\gamma}\vert$ internal vertices, since we work in the convention that all external legs have the same colour $0$. One can easily see that there is actually a unique open graph with boundary $\gamma$ and with $\vert\mathcal{V}_{\gamma}\vert$ internal vertices, which is obtained by adding an external leg of colour $0$ to each vertex in $\gamma$. Let us call this graph the ``smallest matching graph of $\gamma$'' and denote it by $\mathcal{G}_{\mathrm{SMG}}$. This graph is clearly a topological core graph, which is the single member of its equivalence class. In other words, we have that $\mathfrak{G}_{\vert\mathcal{V}_{\gamma}\vert/2,\gamma}^{\mathrm{core}}=\{[\mathcal{G}_{\mathrm{SMG}}]\}$ and $[\mathcal{G}_{\mathrm{SMG}}]=\{\mathcal{G}_{\mathrm{SMG}}\}$. If $\gamma$ does not represent a $2$-sphere, then $\mathcal{G}_{\mathrm{SMG}}$ cannot be a manifold, since its $3$-bubble of colour $123$ is equivalent to $\gamma$ and hence by assumption non-spherical. If $\gamma$ is a $2$-sphere, then one can easily check that $\mathcal{G}_{\mathrm{SMG}}$ always represents a manifold, since all its $3$-bubbles are either spheres or disks. Furthermore, a closer look reveals that $\mathcal{G}_{\mathrm{SMG}}$ generically represents a $3$-ball in this case:

\begin{Proposition}The smallest matching graph $\mathcal{G}_{\mathrm{SMG}}$ of an arbitrary closed $(2+1)$-coloured graph $\gamma$ with spherical topology represents a $3$-ball.\end{Proposition}

\begin{proof}First of all, let us focus on the boundary graph: Let us contract all the $1$-dipoles (all of them are proper in this case) within $\gamma$ to obtain a core graph $\gamma_{c}$. Now, the claim is that independently of how the spherical graph $\gamma$ looks like, $\gamma_{c}$ is always the elementary melonic $2$-sphere. To see this, recall that a core graph has for every colour $i$ either a unique spherical $\hat{i}$-bubble or all $\hat{i}$-bubbles are non-spherical. However, the latter option is not possible for closed $2$-dimensional graphs. This means that the core graph obtained by rooting $\gamma$ has exactly three $2$-bubbles, one for each pair of colours. Furthermore, since $\gamma$ represents a $2$-sphere, also the corresponding topological core graph represents a $2$-sphere. Hence, the topological core graph $\gamma_{c}$ corresponding to $\gamma$ has the following property:
\begin{align}2=\chi(\Delta_{\gamma_{c}})=\vert\mathcal{V}_{\gamma_{c}}\vert-\vert\mathcal{E}_{\gamma_{c}}\vert+\vert\mathcal{F}_{\gamma_{c}}\vert=-\frac{1}{2}\vert\mathcal{V}_{\gamma_{c}}\vert+3\end{align}
and hence $\vert\mathcal{V}_{\gamma_{c}}\vert=2$, where we used the fact that $3\vert\mathcal{V}_{\gamma_{c}}\vert=2\vert\mathcal{E}_{\gamma_{c}}\vert$. This shows that $\gamma_{c}$ has exactly two vertices and hence is the elementary melonic $2$-sphere. Now, the main point is that every $1$-dipole move in $\gamma$ corresponds to a non-internal proper $1$-dipole in $\mathcal{G}_{\mathrm{SMG}}$, by its definition. Hence, performing the same dipole contractions as in $\gamma$ to $\mathcal{G}_{\mathrm{SMG}}$, we will end up with the elementary melonic $3$-ball. In other words, we see that $\mathcal{G}_{\mathrm{SMG}}$ is related to a $3$-ball by a finite sequence of proper dipole moves. Hence, it represents a $3$-ball by itself.\end{proof}

\subsection{Some General Considerations}\label{GeneralConsiderations}
Let us briefly make some general remarks on the transition amplitude of the coloured Boulatov model. First of all, let us discuss what kind of boundary states are allowed. As already mentioned, every closed $(2+1)$-coloured graph represents a manifolds, i.e. $2$-dimensional pseudomanifolds, which are not manifolds, are not representable by coloured graphs. On the other hand, we know by the Theorem of Pezzana (Theorem \ref{Pezzana}(2)) that all closed surfaces admit a crystallization. It is actually quite easy to construct examples explicitly:

\begin{figure}[H]
\centering
\includegraphics[scale=1.6]{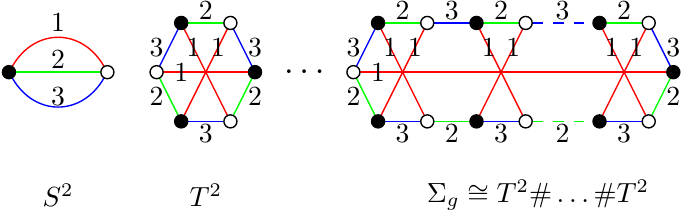}
\caption{Crystallizations of all orientable, closed surfaces.}
\end{figure}

The graph $\mathcal{G}_{g}$ corresponding to the genus $g$-surface for $g\geq 2$ on the r.h.s. are constructed via the graph connected sum as defined in Section \ref{ConSum}. That they indeed describe the surfaces $\Sigma_{g}$ can also be seen by calculating their Euler characteristic, which gives
\begin{align}\chi(\Delta_{\mathcal{G}_{g}})=\vert\mathcal{F}_{\mathcal{G}_{g}}\vert-\vert\mathcal{E}_{\mathcal{G}_{g}}\vert+\vert\mathcal{V}_{\mathcal{G}_{g}}\vert=3-3(1+2g)+(2+4g)=2-2g=\chi(\Sigma_{g}).\end{align}
By Casali's Theorem (Theorem \ref{Casali}), it follows that the most general boundary state of the coloured Boulatov model can be obtained by performing a finite sequence of proper dipole moves to a disjoint union the of the graphs drawn in the figure above.\\
\\
As a next remark, let us discuss which type of topologies do appear in the transition amplitude. First of all, recall that not every $3$-dimensional normal pseudomanifold with boundary is representable by a coloured graph. On the other hand, by Theorem \ref{Pezzana}(2), we know that for every compact and connected PL-manifold with boundary and every crystallization of its boundary there exists a crystallization matching the given boundary graph. Hence, we know that every compact, connected and orientable manifold whose boundary is homeomorphic to our given boundary topology appears in the transition amplitude, since every $3$-manifold has a unique PL-structure (up to PL-homeomorphism) according to the Theorem of Moise.\\
\\
Last but not least, let us mention a straightforward way to construct all open coloured graphs matching a given boundary graph $\gamma\in\overline{\mathfrak{G}}_{2}$. For this, take some open $(3+1)$-coloured graph $\mathcal{G}\in\mathfrak{G}_{3}$ with $\partial\mathcal{G}=\gamma$. Then, let us construct a closed graph $\mathcal{G}^{\prime}\in\overline{\mathfrak{G}}_{3}$ as follows: Assign a vertex to each external leg of $\mathcal{G}$ respecting bipartiteness and connect them with edges similarly as described by the boundary graph, i.e. connect two external legs by an edge of colour $i$ whenever there is a $0i$-path connecting them. The resulting graph $\mathcal{G}^{\prime}$ is a closed $(3+1)$-coloured graph, which admits a $3$-bubble of colour $123$, which coincides with our given boundary graph $\gamma$. An example of this procedure is drawn below.

\begin{figure}[H]
\centering
\includegraphics[scale=1.1]{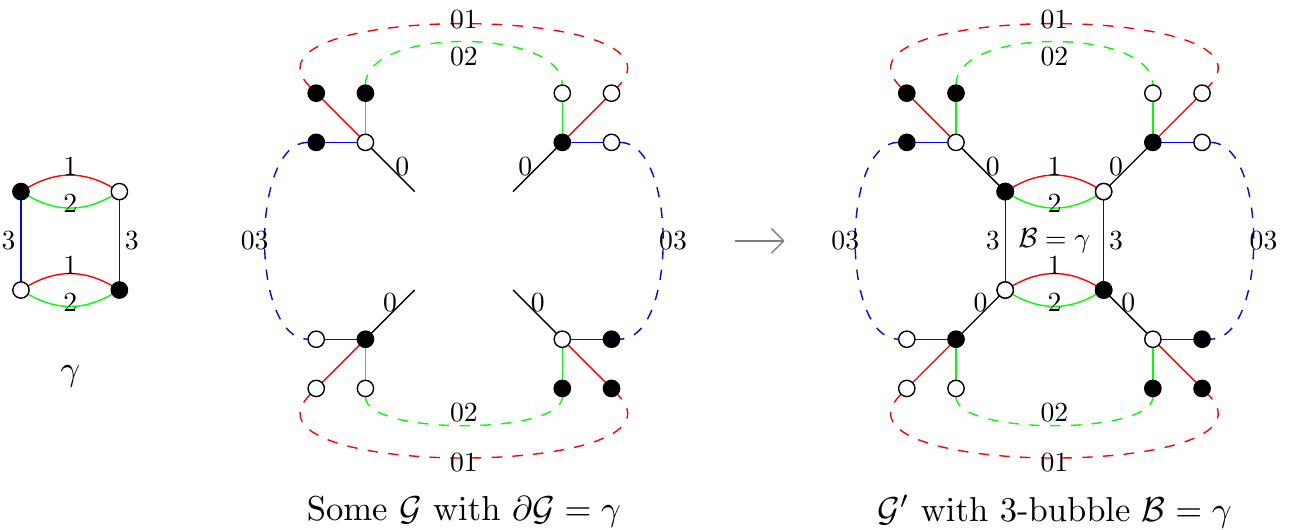}
\caption{The ``pillow graph'' $\gamma$ and the construction of the graph $\mathcal{G}^{\prime}$ described above. Dotted lines represent the corresponding bicoloured paths (=2-bubbles).}
\end{figure}

Reversing this process shows that every open $(3+1)$-coloured graph matching some given boundary graph $\gamma$ can be obtained from a closed $(3+1)$-coloured graph, which admits a $3$-bubble $\mathcal{B}=\gamma$, by deleting this bubble. Furthermore, if $\gamma$ represents the $2$-sphere, then every open graph $\mathcal{G}$ with $\partial\mathcal{G}=\gamma$ representing a manifold (resp. pseudomanifold) is obtained by cutting out a bubble $\mathcal{B}=\gamma$ from a closed $(3+1)$-coloured graph representing a manifold (resp. pseudomanifold).

\subsection{A Simple Example}\label{SimpleExample}
Let us discuss the simplest possible example, i.e. the simplest possible boundary topology represented by the simplest possible closed coloured graph, i.e. we choose the elementary melonic $2$-sphere\footnote{These graphs are also known as ``\textit{dipoles}'' in the literature. However, we have already used this name for the concept of dipole moves.} as our boundary graph. The boundary graph $\gamma$ together with its triangulation is drawn in the following figure:

\begin{figure}[H]
\centering
\includegraphics[scale=1.2]{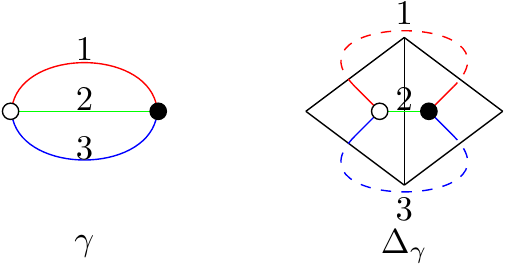}
\caption{The elementary melonic 2-sphere $\gamma$ (l.h.s.) and its corresponding simplicial complex (r.h.s.), where the gluing of edges is as indicated by the dashed lines.}
\end{figure}

The following figure shows five open $(3+1)$-coloured graphs with boundary given by $\gamma$. Each of them is a topological core graph and defines a topological core equivalence class, which is distinct to the classes represented by the other graphs. Furthermore, one can easily convince one-self that these are in fact all inequivalent core graphs up to core equivalence and up to order $p=3$:

\begin{figure}[H]
\centering
\includegraphics[scale=1]{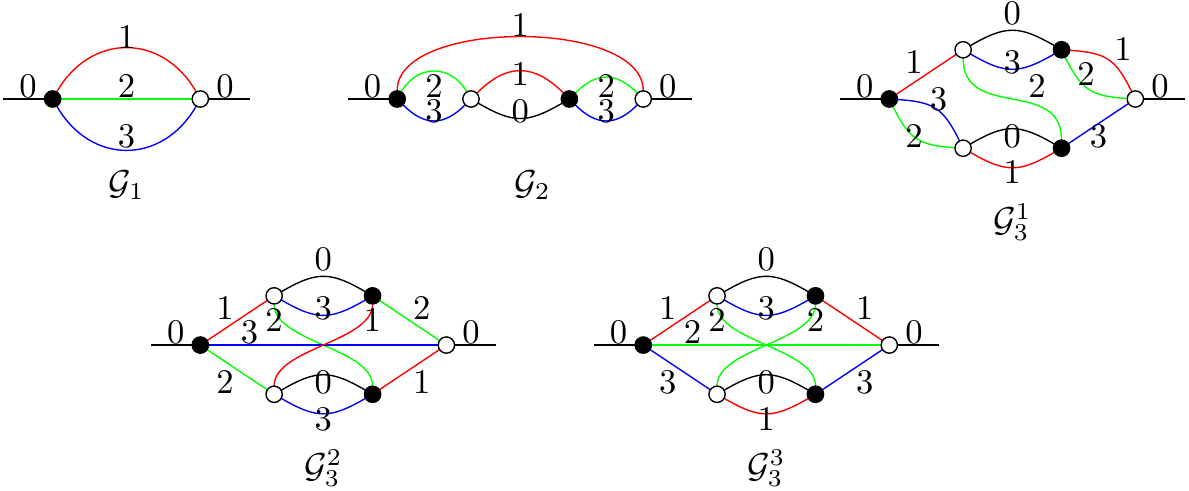}
\caption{All inequivalent core graphs with boundary graph $\gamma$ up to order $p=3$ and up to core equivalence.\label{SimpleExampleFig}}
\end{figure}

The core graphs $\mathcal{G}_{1}$, $\mathcal{G}_{2}$ and $\mathcal{G}_{3}^{1}$ represent $3$-balls: The graph $\mathcal{G}_{1}$ is just the elementary melonic $3$-ball and the graphs $\mathcal{G}_{2}$ and $\mathcal{G}_{3}$ can be reduced to $\mathcal{G}_{1}$ by performing proper internal $2$-dipole moves. The core graphs $\mathcal{G}_{3}^{2}$ and $\mathcal{G}^{3}_{3}$ represent both pseudomanifolds with spherical boundary, which are not manifolds, since both of them admit an internal $3$-bubble of colour $123$, which represents the $2$-torus. However, note also that the two pseudomanifolds represented by $\mathcal{G}_{3}^{2}$ and $\mathcal{G}^{3}_{3}$ are not the same, since the Euler characteristics of the  psudomanifold represented by $\mathcal{G}_{3}^{2}$ is $\chi(\Delta_{\mathcal{G}_{3}^{2}})=3$, whereas the Euler characteristic of the complex dual to $\mathcal{G}_{3}^{3}$ is $\chi(\Delta_{\mathcal{G}_{3}^{2}})=2$. \\
\\
A straightforward calculation gives the (un-gauge-fixed) amplitudes corresponding to the five core equivalence classes represented above. The results are as follows:
\begin{align}\mathcal{A}_{[\mathcal{G}_{1}]}^{\lambda}[\{g_{1},g_{2},g_{3}\}]&=(\lambda\overline{\lambda})\delta^{N}(g_{1}g_{3}^{-1})\delta^{N}(g_{2}g_{3}^{-1})\\\mathcal{A}_{[\mathcal{G}_{2}]}^{\lambda}[\{g_{1},g_{2},g_{3}\}]&=(\lambda\overline{\lambda})^{2}[\delta^{N}(\mathds{1})]^{-1}\delta^{N}(g_{1}g_{3}^{-1})\delta^{N}(g_{2}g_{3}^{-1})\\\mathcal{A}_{[\mathcal{G}_{3}^{1}]}^{\lambda}[\{g_{1},g_{2},g_{3}\}]&=(\lambda\overline{\lambda})^{3}[\delta^{N}(\mathds{1})]^{-2}\delta^{N}(g_{1}g_{3}^{-1})\delta^{N}(g_{2}g_{3}^{-1})\\\mathcal{A}_{[\mathcal{G}_{3}^{2}]}^{\lambda}[\{g_{1},g_{2},g_{3}\}]&=(\lambda\overline{\lambda})^{3}[\delta^{N}(\mathds{1})]^{-3}\delta^{N}(g_{1}g_{3}^{-1})\delta^{N}(g_{2}g_{3}^{-1})\int_{\mathrm{SU}(2)^{3}}\mathrm{d}x\mathrm{d}y\mathrm{d}z\delta^{N}(xyz(zyx)^{-1})\\\mathcal{A}_{[\mathcal{G}_{3}^{3}]}^{\lambda}[\{g_{1},g_{2},g_{3}\}]&=(\lambda\overline{\lambda})^{3}[\delta^{N}(\mathds{1})]^{-2}\delta^{N}(g_{1}g_{3}^{-1})\delta^{N}(g_{2}g_{3}^{-1})
\end{align}
The group element $g_{i}$ here denotes the group element assigned to the boundary edge of colour $i\in\{1,2,3\}$. If we fix the $\mathrm{SU}(2)$ gauge symmetry in the amplitudes of the graphs $\mathcal{G}_{1}$, $\mathcal{G}_{2}$ and $\mathcal{G}_{3}^{1}$, which represent $3$-balls, then we just get the spin network evaluation when applied to some boundary spin network state, since the amplitudes are then proportional to $\delta^{N}(g_{1})\delta^{N}(g_{2})\delta^{N}(g_{3})$, as expected. The remaining integral in the amplitude of graph $\mathcal{G}_{3}^{2}$ comes from the non-trivial bulk topology. Note also that the amplitudes of $\mathcal{G}_{2}$ and $\mathcal{G}_{3}^{1}$ can be obtained from the amplitude of $\mathcal{G}_{1}$, by applying Proposition \ref{AmplChangeInt}.\\
\\
By the previous discussion, we know that all manifolds with spherical boundary will appear in the transition amplitude. Hence, we now would like to analyse, how the amplitudes of manifolds look like in this example. First of all, observe that every compact, orientable and connected $3$-dimensional topological manifold $\mathcal{M}$ with boundary $\partial\mathcal{M}$ homeomorphic to the $2$-sphere can be obtained by cutting out the interior of an embedded ball inside some closed, orientable and connected $3$-manifold $\mathcal{N}$. Hence, every open $(3+1)$-coloured graph representing a manifold, whose boundary graph is given by the elementary melonic $2$-sphere $\gamma$, can be obtained by cutting an edge of colour $0$ in some closed $(3+1)$-coloured graph representing a closed $3$-manifold\footnote{In consistency with the general considerations above, we should more rigorously first insert a $3$-dipole into an edge of colour $0$ and cut the so-created $3$-bubble $\mathcal{B}=\gamma$ out. This if of course the same as to say that we simply cut an edge of colour $0$.}. Let us look at some examples. The following figure shows three closed topological core graphs representing the closed and orientable manifolds $S^{2}\times S^{1}$, $\mathbb{R}P^{3}\cong L(2,1)$ as well as the Lens space\footnote{The Lens space $L(p,q)$, where $p,q\in\mathbb{N}$ are coprime numbers (i.e. $\mathrm{gcd}(p,q)=1$), is a compact $3$-manifold defined to be the quotient of $S^{3}$ under the free group action $\mathbb{Z}_{p}\times S^{3}\supset (m,(z,\omega))\mapsto(e^{2\pi im/p}z,e^{2\pi imp/q}\omega)\in S^{3}$ of $\mathbb{Z}_{p}$ on $S^{3}$ viewed as a subset of $\mathbb{C}^{2}$. Note that $L(p,q_{1})$ and $L(p,q_{2})$ are homeomorphic if and only if $q_{1}=\pm q_{2}$ (mod $p$) or $q_{1}q_{2}=\pm 1$ (mod $p$). \cite{Lens}} $L(3,1)$:

\begin{figure}[H]
\captionsetup[subfigure]{labelformat=empty}
\centering
\subfloat[$S^{2}\times S^{1}$]{\includegraphics[width=0.15\textwidth]{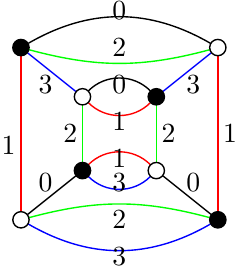}}\hspace{1cm}
\subfloat[$\mathbb{R}P^{3}$]{\includegraphics[width=0.18\textwidth]{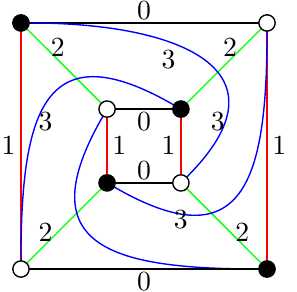}}\hspace{1cm}
\subfloat[$L(3,1)$]{\includegraphics[width=0.18\textwidth]{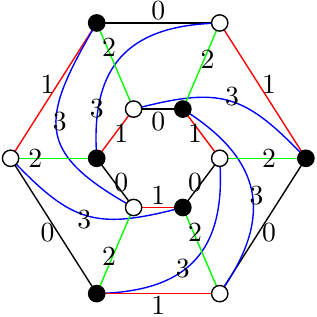}}
\caption{Three closed $(3+1)$-coloured graphs representing manifolds.}
\end{figure}

The graphs and the description of their topology are taken from the crystallization theory literature, e.g. see \cite[p.125/127]{GagliardiBoundaryGraph} and references therein. The graphs obtained by cutting an edge of colour $0$ are drawn in the following figure:

\begin{figure}[H]
\captionsetup[subfigure]{labelformat=empty}
\centering
\subfloat[$(S^{2}\times S^{1})\textbackslash\mathring{B}^{3}$]{\includegraphics[width=0.18\textwidth]{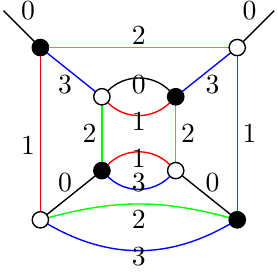}}\hspace{1cm}
\subfloat[$\mathbb{R}P^{3}\textbackslash\mathring{B}^{3}$]{\includegraphics[width=0.18\textwidth]{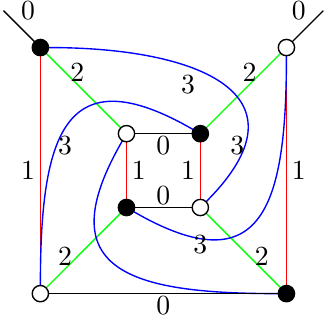}}\hspace{1cm}
\subfloat[$L(3,1)\textbackslash\mathring{B}^{3}$]{\includegraphics[width=0.18\textwidth]{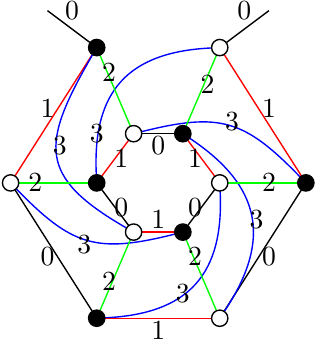}}
\caption{Three open $(3+1)$-coloured graphs representing manifolds with spherical boundary.}
\end{figure}

Choosing some maximal trees within these graphs, a straightforward calculation gives the following $\mathrm{SU}(2)$ gauge fixed amplitude functionals corresponding to the three manifolds with spherical boundary drawn above:
\begin{align}\mathcal{A}_{S^{2}\times S^{1}\textbackslash\mathring{B}^{3}}^{\lambda,\mathrm{gauge-fixed}}[\{g_{1},g_{2},g_{3}\}]&=(\lambda\overline{\lambda})^{4}[\delta^{N}(\mathds{1})]^{-2}\bigg(\prod_{i=1}^{3}\delta^{N}(g_{i})\bigg)\\\mathcal{A}_{\mathbb{R}P^{3}\textbackslash\mathring{B}^{3}}^{\lambda,\mathrm{gauge-fixed}}[\{g_{1},g_{2},g_{3}\}]&=(\lambda\overline{\lambda})^{4}[\delta^{N}(\mathds{1})]^{-3}\bigg(\prod_{i=1}^{3}\delta^{N}(g_{i})\bigg)\,\int_{\mathrm{SU}(2)}\,\mathrm{d}h\,\delta^{N}(h^{2})\\\mathcal{A}_{L(3,1)\textbackslash\mathring{B}^{3}}^{\lambda,\mathrm{gauge-fixed}}[\{g_{1},g_{2},g_{3}\}]&=(\lambda\overline{\lambda})^{6}[\delta^{N}(\mathds{1})]^{-5}\bigg(\prod_{i=1}^{3}\delta^{N}(g_{i})\bigg)\,\int_{\mathrm{SU}(2)}\,\mathrm{d}h\,\delta^{N}(h^{3})
\end{align}
We see that all of them are proportional to the spin network evaluation when applied to some boundary spin network state. In fact, let us now prove that this is true more generally:

\begin{Proposition} The amplitude of any open $(3+1)$-coloured open graph $\mathcal{G}$ representing a manifold with boundary graph $\partial\mathcal{G}=\gamma$ is proportional to 
\begin{align*}\delta^{N}(g_{1}g_{3}^{-1})\delta^{N}(g_{2}g_{3}^{-1}).\end{align*} 
Furthermore, the $\mathrm{SU}(2)$ gauge fixed amplitude of any such manifold is proportional to
\begin{align*}\delta^{N}(g_{1})\delta^{N}(g_{2})\delta^{N}(g_{3}).\end{align*}  
Hence, the transition amplitude with respect to some spin network $\psi\in L^{2}(\mathrm{SU}(2)^{3}/\mathrm{SU}(2)^{2})$ of any manifold appearing in the perturbative expansion of the Boulatov transition amplitude is proportional to the spin network evaluation $\psi(g_{1}=g_{2}=g_{3}=\mathds{1})$.\end{Proposition}

\begin{proof}
As explained above, every open $(3+1)$-coloured graph $\mathcal{G}\in\mathfrak{G}_{3}$ with $\partial\mathcal{G}=\gamma$ can be obtained by cutting an edge of colour $0$ in some closed graph $\mathcal{G}^{\prime}\in\overline{\mathfrak{G}}_{3}$ representing a manifold. Now, if we cut some edge of colour $0$ of a graph representing a manifold, then we can apply the following sequence of dipole moves:
\begin{figure}[H]
\centering
\includegraphics[scale=0.85]{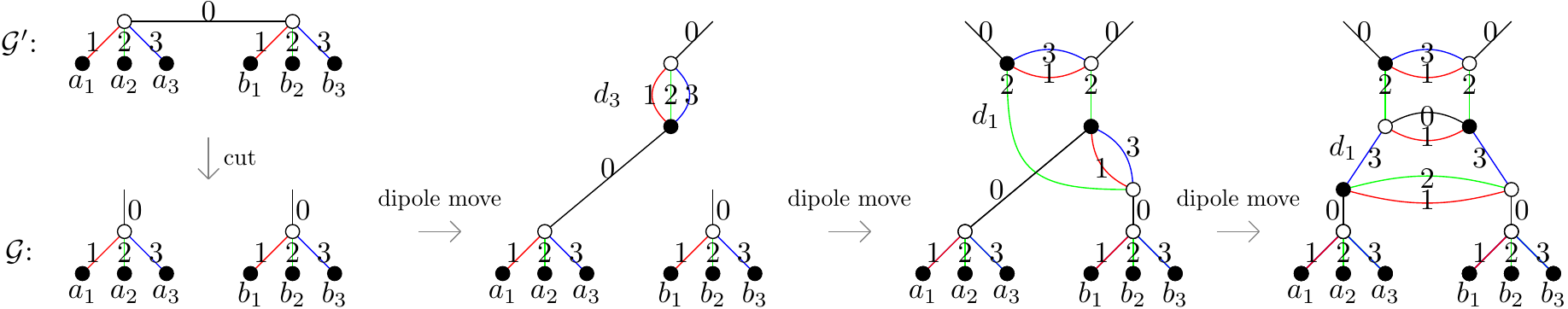}
\caption{An open graph obtained by cutting an edge and a series of dipole moves.}
\end{figure}
All the dipole moves performed above are proper internal ones: The first move is just a $3$-dipole creation. The dipole yielding the third graph from the second graph is proper precisely because we assumed $\mathcal{G}^{\prime}$ to represent a manifold: The closed $3$-bubble separated by the created dipole $d_{1}$ of colour $013$ is the graph obtained by adding a $2$-dipole of colour $13$ to the $3$-bubble of colour $013$ of the graph $\mathcal{G}^{\prime}$ containing $a_{1,3}$ and $b_{1,3}$ and hence represents a $2$-sphere. Similarly, we see that the $1$-dipole move yielding the fourth graph from the third one is proper. Using Proposition \ref{AmplChangeInt}, it follows that that the amplitudes of the graph on the r.h.s. is the same as the amplitude of $\mathcal{G}$ up to a factor of $(\lambda\overline{\lambda})^{3}$. Now, lets consider the relevant part of the graph containing all the non-cyclic faces leading to the boundary edges:
\begin{figure}[H]
\centering
\includegraphics[scale=1.2]{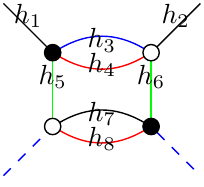}
\caption{Part of the graph on the r.h.s. above containing all the non-cyclic faces.}
\end{figure}
With this notation, the relevant parts of the amplitude are given by
\begin{equation}\begin{aligned}\int_{\mathrm{SU}(2)^{8}}\,\bigg(\prod_{i=1}^{8}\mathrm{d}h_{i}\bigg)\,&\delta^{N}(g_{1}h_{1}^{-1}h_{4}h_{2}^{-1})\delta^{N}(g_{2}h_{1}^{-1}h_{3}h_{2}^{-1})\delta^{N}(g_{3}h_{1}^{-1}h_{5}h_{7}^{-1}h_{6}h_{2}^{-1})\times\\&\times\delta^{N}(h_{3}h_{4}^{-1})\delta^{N}(h_{7}h_{8}^{-1})\delta^{N}(h_{4}h_{6}^{-1}h_{8}h_{5}^{-1})\Delta(h_{3},\dots,h_{8}),\end{aligned}\end{equation}
where $\Delta(h_{3},\dots,h_{8})$ denotes the product of other delta functions involving colour $3$ and the group elements $h_{3},h_{4},h_{6},h_{7},h_{8}$. Using some of the delta functions, this can be rewritten as
\begin{equation}\begin{aligned}\int_{\mathrm{SU}(2)^{8}}\,\bigg(\prod_{i=1}^{8}\mathrm{d}h_{i}\bigg)\,&\delta^{N}(g_{1}h_{1}^{-1}h_{4}h_{2}^{-1})\delta^{N}(g_{2}h_{1}^{-1}h_{4}h_{2}^{-1})\delta^{N}(g_{3}h_{1}^{-1}h_{4}h_{2}^{-1})\times\\&\times\delta^{N}(h_{3}h_{4}^{-1})\delta^{N}(h_{7}h_{8}^{-1})\delta^{N}(h_{4}h_{6}^{-1}h_{8}h_{5}^{-1})\Delta(h_{3},\dots,h_{8}),\end{aligned}\end{equation}
Next, let us make the substitution $h_{1}^{\prime}=h_{2}h_{4}^{-1}h_{1}$, which, using $\mathrm{d}h_{1}^{\prime}=\mathrm{d}h_{1}$ by bi-invariance of the Haar-measure, leads to 
\begin{equation}\begin{aligned}\int_{\mathrm{SU}(2)^{7}}\,\bigg(\prod_{i=1,i\neq 2}^{8}\mathrm{d}h_{i}\bigg)\,&\delta^{N}(g_{1}h_{1}^{-1})\delta^{N}(g_{2}h_{1}^{-1})\delta^{N}(g_{3}h_{1}^{-1})\times\\&\times\delta^{N}(h_{3}h_{4}^{-1})\delta^{N}(h_{7}h_{8}^{-1})\delta^{N}(h_{4}h_{6}^{-1}h_{8}h_{5}^{-1})\Delta(h_{3},\dots,h_{8}),\end{aligned}\end{equation}
where we renamed $h_{1}^{\prime}\to h_{1}$ and where the integral over $h_{2}$ dropped out. Integrating over $h_{1}$ yields the desired result, i.e. the proportionality to 
\begin{align}\delta^{N}(g_{1}g_{3}^{-1})\delta^{N}(g_{2}g_{3}^{-1}).\end{align} 
For the second claim, let us choose the maximal tree such that it includes the following edges:
\begin{figure}[H]
\centering
\includegraphics[scale=1.4]{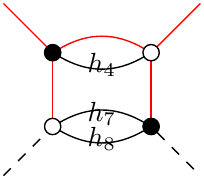}
\caption{Choice of maximal tree (in red) in the part of the graph on the r.h.s. above containing all the non-cyclic faces.}
\end{figure}
Using some of the delta functions corresponding to the internal faces, it is immediate that we can also set the group elements $h_{4}$, $h_{7}$ and $h_{8}$ equal to the identity. Hence, we trivially get that the $\mathrm{SU}(2)$ gauge fixed amplitude is proportional to 
\begin{align}\delta^{N}(g_{1})\delta^{N}(g_{2})\delta^{N}(g_{3})\end{align}  
which concludes the proof. 
\end{proof}

To sum up, for every manifold appearing in the transition amplitude with respect to the elementary melonic $2$-sphere as our boundary state we get the spin network evaluation together with some prefactor. As a consequence, the transition amplitude of the Boulatov model restricted to manifolds will factorize into the boundary part, which is just the spin-network evaluation, and some prefactor, which includes all the factors of $\lambda\overline{\lambda}$, $\delta^{N}(\mathds{1})$ and remaining integrals over bulk edges. Hence, this provides a perfect example for ``\textit{holography}'', which we will discuss in the next chapter, since in the end the result only depends on the assigned boundary data. By topological invariance of the Ponzano-Regge spin foam model, one can argue that a similar result will hold for other boundary graphs describing the $2$-sphere. However, there are still some open question: First of all, the previous result is not true for pseudomanifolds in general and their might be pseudomanifolds with different amplitudes. Furthermore, the previous result does not tell us what kind of topology is the most dominant contribution to the transition amplitude. Another interesting point for further works would be to explicitly compute the prefactor of the factorized transition amplitude, either for the whole manifold sector, or at least for some specific types of topologies appearing in the transition amplitude. In general, this factor might be divergent, but it would be interesting to try proving some Borel summability results, for example by adding a ``\textit{pillow}''-term to the action of the coloured Boulatov model, as done in \cite{Pillow}.  
\chapter{Transition Amplitudes with a Toroidal Boundary and Holography}\label{Chap4}
In the end of the last chapter, we have analysed the transition amplitude of the coloured Boulatov model for a particular choice of boundary graph, namely the elementary melonic $2$-sphere. In this chapter, we move on to the next to trivial boundary topology, which is given by the $2$-torus $T^{2}:=S^{1}\times S^{1}$, i.e. the unique (up to homeomorphism) closed and orientable surface with genus $g=1$. The transition amplitude of the Ponzano-Regge model on the solid torus $\overline{T}^{2}=D^{2}\times S^{1}$, where $D^{2}$ denotes the disk (=closed $2$-ball), has been studied in many details in \cite{GoellerThesis}. In particular, for a certain choice of boundary states with a clear geometrical interpretation, it was shown that one recovers the structure of the BMS characters in the asymptotic limit, which are the characters of the asymptotic symmetry group of general relativity, together with some quantum corrections. This is a very interesting observation with respect to the so-called ``\textit{holographic principle}'', which conjectures that the geometry and dynamics of some region of spacetime can fully be characterized by a ``dual theory'' living purely on the boundary. Although not rigorously proven, many evidences and implementations of this principle have been found in various different approaches to quantum gravity and hence, it is often used as a guiding principle. The first part of this chapter is hence devoted to a short overview of the history of holography in the context of quantum gravity and of some recent results obtained in the particular context of the Ponzano-Regge spin foam model. Afterwards, we construct some open $(3+1)$-coloured graphs representing the solid torus and calculate their transition amplitudes. The remaining part of this chapter is devoted to a discussion of graphs representing other manifolds with boundary given by the $2$-torus, which appear in the transition amplitude of the Boulatov model. Although we will not perform a full computation for a given boundary spin network state, this work will set the foundation for a closer analysis in some future work aiming at the exact computation of the Boulatov transition function with a toroidal boundary and the study of (quasi-local) holographic dualities in this context.

\section{Ponzano-Regge Spin Foam Model and Holography}
In the first part of this chapter, we discuss some general features of holography in the context of quantum gravity, focusing on its historical origin. Afterwards, we discuss some results regarding holographic dualities in the case of finite and bounded regions of $3$-dimensional spacetime described by the Ponzano-Regge spin foam model. More precisely, we review some results from the literature regarding (quasi-local) holographic dualities in the Ponzano-Regge model on the closed $3$-ball and on the solid torus. For a general overview see for example the first chapter of the PhD thesis \cite{Mele} as well as the review \cite{Bousso}. 

\subsection{Quantum Gravity and Holography}
The central idea of what is called the ``\textit{holographic principle''} nowadays is that the dynamics of quantum gravity in some spacetime region can purely be described by a ``\textit{dual}'' theory living on the boundary of this region. More precisely, the (strong) holographic principle states that for some region of spacetime $\mathcal{R}=\mathbb{R}\times\Sigma$, where $\Sigma$ is some spacelike hypersurface, with bulk and boundary Hilbert spaces $\mathcal{H}_{\Sigma}$ and $\mathcal{H}_{\partial\Sigma}$ and corresponding Hamiltonians $H_{\Sigma}$ and $H_{\partial\Sigma}$, there is an isomorphism 
\begin{align}\mathcal{I}_{\mathrm{hol}}:\mathcal{H}_{\Sigma}\to\mathcal{H}_{\partial\Sigma},\end{align}
such that $\mathcal{I}_{\mathrm{hol}}\circ H_{\Sigma}=H_{\partial\Sigma}$ \cite{SmolinHP}. When talking about boundary and spacetime regions, it is important to note that we do not restrict to asymptotic boundaries, but we also include the case of boundaries enclosing some finite and bounded region of spacetime. In order to explicitly distinguish them, let us in the following call the latter case ``\textit{quasi-local holography}''. Early motivations for holography in the context of gravity came from the physics of black holes. In particular, important motivations are the discovery of the entropy of a black hole by J. D. Bekenstein \cite{BekensteinBH}, which states that the entropy of a black hole $S_{\mathrm{BH}}$ is proportional to its surface area $A$ and \textit{not} to its volume, i.e.
\begin{align}S_{\mathrm{BH}}=\frac{A}{4}\frac{k_{\mathrm{B}}}{l_{p}^{2}},\end{align}
where $k_{\mathrm{B}}$ denotes Boltzmann's constant and $l_{p}=(G\hbar/c^{3})^{1/2}$ the Planck length, as well as the long-standing debate regarding the loss of information due to radiation initiated by S. W. Hawking in the early 1970s \cite{HawkingsBH}. The holographic principle itself goes beyond the physics of black holes and was originally formulated by G. 't Hooft \cite{Hooft} and L. Susskind \cite{Susskind}. The prime example of the holographic principle in practice is the famous ``\textit{AdS/CFT correspondence}'' firstly proposed by J. M. Maldacena in the late 1990s \cite{AdSCFT}. This correspondence conjectures that quantum gravity on $d$-dimensional anti-de Sitter space (AdS) is dual to a conformal field theory (CFT) living on the asymptotic boundary of the anti-de Sitter space at spatial infinity. Although it still an open problem whether the full conjecture is true, there are many results supporting the statement of the AdS/CFT correspondence. In particular, it was shown that the dual theory for $(2+1)$-dimensional gravity with a negative cosmological constant described by a $(\mathrm{SL}(2,\mathbb{R})\times\mathrm{Sl}(2,\mathbb{R}))$-Chern-Simons theory (see Section \ref{CS}) is given by a ``Liouville field theory'', which is a certain $2$-dimensional conformal field theory \cite{Liouville}.\\
\\
When talking about the holographic principle, it is also interesting to look at the corresponding symmetry groups of $d$-dimensional spacetime and its $(d-1)$-dimensional boundary. A first remarkable result was obtained in 1962 by H. Bondi, M. G. J. van der Burg, A. W. K. Metzner \cite{BMS1} as well as by R. K. Sachs \cite{BMS2}. The authors looked at the asymptotic symmetry group of asymptotically flat $4$-dimensional general relativity. Naively, one would expect that general relativity reduces to special relativity in this limit and hence that one recovers the Poincaré group. However, it turns out that one obtains an infinite-dimensional group, which contains the Poincaré group only as a finite-dimensional subgroup. This group is nowadays called the ``\textit{BMS-group}'' and is given by a semidirect product of the Lorentz group with the infinite-dimensional abelian group of ``supertranslations''. In the $3$-dimensional case, another seminal work was published by J. D. Brown and M. Henneaux in 1986 \cite{Brown}, in which they looked at the asymptotic symmetry group for $3$-dimensional gravity with a negative cosmological constant. Instead of $\mathrm{SO}(2,2)$, which is the symmetry group of $3$-dimensional anti-de Sitter space, they also recovered an infinite-dimensional group, namely the $2$-dimensional conformal group. This can also be seen as a particular instance of the conjectured AdS/CFT correspondence mentioned above. For more details about asymptotic symmetries in the $3$-dimensional case see for example \cite{Oblak}.

\subsection{Ponzano-Regge Model on the Ball and Holography}
Let us now discuss some explicit results regarding quasi-local holography in the Ponzano-Regge spin foam model. The most trivial case is the Ponzano-Regge model on the closed $3$-ball $B^{3}$ with boundary given by the $2$-sphere $S^{2}$. This situation has been studied in \cite{PRBall1,PRBall2}. We have already seen in our simple example in Section \ref{SimpleExample} that the Ponzano-Regge transition function of the $3$-ball is given by the ``\textit{spin network evaluation}'', i.e. for some spin network $\Psi$ living on the dual $1$-skeleton $\gamma=(\partial\mathcal{V}^{\ast},\partial\mathcal{E}^{\ast})$ of some triangulation $\Delta$ of the $3$-ball it holds that
\begin{align}\label{BallPR}\langle\mathcal{Z}^{\mathrm{gauge-fixed}}_{\mathrm{PR}}\vert\Psi\rangle=\psi(\{g_{e^{\ast}}=\mathds{1}\}_{e^{\ast}\in\partial\mathcal{E}^{\ast}}),\end{align}
where $\psi\in L^{2}(\mathrm{SU}(2)^{\vert\partial\mathcal{E}^{\ast}\vert}/\mathrm{SU}(2)^{\vert\partial\mathcal{V}^{\ast}\vert})$ denotes the corresponding spin network function. Gauge-fixed in this context means that we have fixed both the $\mathrm{SU}(2)$ and translational gauge symmetry, as explained in Section \ref{GaugeFixPR}. We have only shown this for the simplest possible coloured boundary graph representing a $2$-sphere, however, it is clear that this holds in general for every triangulations of the $3$-ball, by topological invariance. A detailed proof can be found in \cite[p.9ff.]{G4}. Note, however, that the result could also have been guessed, since every cycle in the $3$-ball is contractible. Now, let $\gamma=(\mathcal{V}_{\gamma},\mathcal{E}_{\gamma})$ be a $3$-valent, finite, directed, connected and planar graph. ``\textit{Planar}'' means that it can be embedded into the plane, or equivalently, into the $2$-sphere without any intersections of edges. As an example, $\gamma$ could be the (dual) $1$-skeleton of some triangulation of the $2$-sphere. It was then shown in \cite{PRBall2}, using some supersymmetric dualities, that there is the following relation:
\begin{align}\label{Ising}\mathcal{Z}_{\mathrm{Spin}}[\gamma,\{Y_{e}\}_{e\in\mathcal{E}_{\gamma}}]=\bigg(2^{2\vert\mathcal{V}_{\gamma}\vert}\prod_{e\in\mathcal{E}_{\gamma}}\mathrm{cosh}(y_{e})^{2}\bigg) (\mathcal{Z}_{\mathrm{Ising}}[\gamma,\{y_{e}\}_{e\in\mathcal{E}_{\gamma}}])^{-2},\end{align}
where $\{y_{e}\}_{e\in\mathcal{E}_{\gamma}}$ are coupling constants and where $Y_{e}:=\mathrm{tanh}(y_{e})$. The functional $\mathcal{Z}_{\mathrm{Ising}}$ denotes the partition function of the ``\textit{(generalized) Ising model}'' on the graph $\gamma$. More precisely, it is defined via
\begin{align}\mathcal{Z}_{\mathrm{Ising}}[\gamma,\{y_{e}\}_{e\in\mathcal{E}_{\gamma}}]:=\sum_{\sigma\in\{\pm 1\}^{\mathcal{V}_{\gamma}}}\mathrm{exp}\bigg(\sum_{e\in\mathcal{E}_{\gamma}}y_{e}\sigma(s(e))\sigma(t(e))\bigg),\end{align}
where the sum goes over all assignments of ``spins'' to vertices, i.e. $\sigma:\mathcal{V}_{\gamma}\to\{-1,1\}$, and where $s,t:\mathcal{E}_{\gamma}\to\mathcal{V}_{\gamma}$ denote the source and target maps of the graph $\gamma$, as usual. The functional $\mathcal{Z}_{\mathrm{Spin}}$ denotes the ``\textit{generating functional of spin network evaluation}'' introduced in \cite{FreidelHynbida}. To define it, let us choose a spin network $\Psi=(\gamma,\rho,i)$ on the given graph $\gamma$ with corresponding spin network function $\psi_{\{j_{e}\}_{e\in\mathcal{E}_{\gamma}}}\in L^{2}(\mathrm{SU}(2)^{\vert\mathcal{E}_{\gamma}\vert}/\mathrm{SU}(2)^{\vert\mathcal{V}_{\gamma}\vert})$. Since we are working on a $3$-valent graph with a fixed orientation, every intertwiner is necessarily a multiple of the Clebsch-Gordan coefficients and hence, the spin network only depends on the choice of representations, labelled by spins, assigned to edges. Then, we define 
\begin{align}\mathcal{Z}_{\mathrm{Spin}}[\gamma,\{Y_{e}\}_{e\in\mathcal{E}_{\gamma}}]:=\sum_{j\in(\mathbb{N}_{0}/2)^{\mathcal{E}_{\gamma}}}\omega(\{j_{e}\}_{e\in\mathcal{E}_{\gamma}})\bigg(\prod_{e\in\mathcal{E}_{\gamma}}Y_{e}^{2j_{e}}\bigg)\psi_{\{j_{e}\}_{e\in\mathcal{E}_{\gamma}}}(\{g_{e^{\ast}}=\mathds{1}\}_{e^{\ast}\in\partial\mathcal{E}^{\ast}}),\end{align}
where $\omega(\{j_{e}\}_{e\in\mathcal{E}_{\gamma}})$ are some coefficients, whose precise definition can be found in the original article as well as in \cite{FreidelHynbida}, which are chosen in such a way that the series converges. Comparing Equation (\ref{Ising}) with Equation (\ref{BallPR}), we hence get that 
\begin{align}\langle\mathcal{Z}_{\mathrm{PR}}^{\mathrm{gauge-fixed}}\vert\Psi^{\{y_{e}\}_{e\in\mathcal{E}_{\gamma}}}_{\mathrm{sup}}\rangle\propto 1/\mathcal{Z}_{\mathrm{Ising}}[\gamma,\{y_{e}\}_{e\in\mathcal{E}_{\gamma}}]^{2},\end{align}
where $\Psi_{\mathrm{sup}}$ is the superposition of spin networks with spin network function
\begin{align}\psi^{\{y_{e}\}_{e\in\mathcal{E}_{\gamma}}}_{\mathrm{sup}}(\{g_{e}\}_{e\in\mathcal{E}_{\gamma}}):=\sum_{j\in(\mathbb{N}_{0}/2)^{\mathcal{E}_{\gamma}}}\omega(\{j_{e}\}_{e\in\mathcal{E}_{\gamma}})\bigg(\prod_{e\in\mathcal{E}_{\gamma}}Y_{e}^{2j_{e}}\bigg)\psi_{\{j_{e}\}_{e\in\mathcal{E}_{\gamma}}}(\{g_{e}\}_{e\in\mathcal{E}_{\gamma}}),\end{align}
where $Y_{e}:=\mathrm{tanh}(y_{e})$ as above. This is a perfect example of quasi-local holography: The theory describing quantum gravity of a $3$-dimensional bounded region of spacetime, namely the closed $3$-ball in this case, is for a specific choice of boundary state dual to a theory living purely on the boundary $2$-sphere, which is given by two copies of the Ising model, a well-known model of statistical physics, in this case.

\subsection{Ponzano-Regge Model on the Solid Torus and Holography}
After studying the case of a spherical boundary, the next more complicated boundary topology is given by the $2$-torus $T^{2}=S^{1}\times S^{1}$. A detailed study of holographic dualities in the Ponzano-Regge spin foam model of the solid torus $\overline{T}^{2}:=D^{2}\times S^{1}$, where $D^{2}$ denotes the disk (=closed $2$-ball), has been done in the series of papers \cite{G1,G2,G3,G4}, summarized in the PhD Thesis \cite{GoellerThesis}, for various different choices of boundary states. To start with, let us consider the following general cellular decomposition of the solid cylinder\footnote{Note that the solid clyinder is a \textit{topological manifold with corners}, i.e. a second countable Hausdorff space, which is locally homeomorphic to $\mathbb{R}_{\geq 0}^{d}=[0,\infty)^{d}$. Topologically, manifolds with corners are one and the same as manifolds with boundary, however, not in the differentiable category. \cite[p.415ff.]{LeeSmoothManifolds}} only consisting of prisms:

\begin{figure}[H]
\centering
\includegraphics[scale=1]{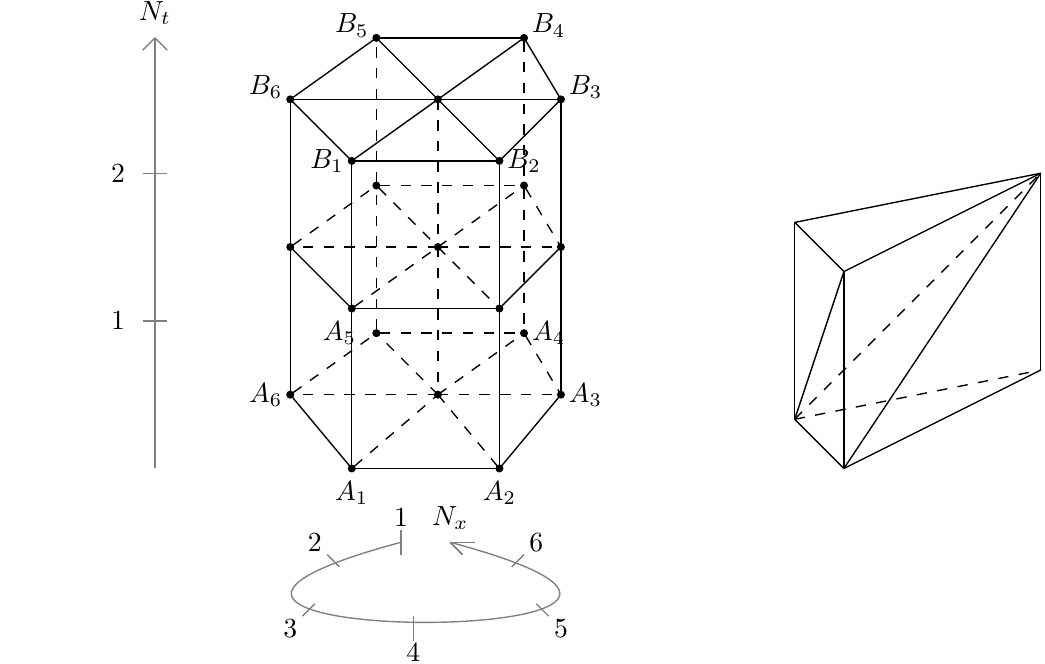}
\caption{A cellular decomposition of the solid cylinder consisting of prisms characterized by the number of horizontal and vertical layers.\label{TrianSolidTorus}}
\end{figure}

The cellular complex drawn above is characterized by two numbers: The number of vertical layers of prisms, which we denote by $N_{t}\in\mathbb{N}$, as well as the number of horizontal layers, i.e. the number of prisms in each horizontal slice, which we denote by $N_{x}\in\mathbb{N}$. In order to obtain a cellular decomposition of the solid torus, we have to identify the top and bottom of the cylinder. Note that there is some freedom in doing this, since the gluing can be done in several ways. Hence, we introduce the ``\textit{twist parameter}'' $N_{\gamma}$ defined by the equation
\begin{align}A_{i}\doteq B_{i+N_\gamma}\hspace{1cm}\forall i\in\{1,\dots,N_{x}\},\end{align}
where the indices in this equation are cyclic, e.g. $N_{x}+1=1$ and so on, and where the ``\textit{twist angle}'' $\gamma$, corresponding to a discrete Dehn twist, is defined by 
\begin{align}\gamma:=2\pi\frac{N_{\gamma}}{N_{x}},\end{align}
which tells us how we identify the outer vertices of the first slice with the outer vertices of the top slice. To sum up, we have constructed general cellular decompositions of the solid torus characterised by the three numbers $N_{x},N_{t}\in\mathbb{N}$ and $N_{\gamma}\in\{0,\dots,N_{x}-1\}$. The discretization above is of course not a simplicial one. However, one can easily obtain a triangulation by decomposing each prism by three tetrahedra as drawn on the right-hand side above. However, since we are dealing with the Ponzano-Regge model in this section, we do not have to work with a triangulation and hence, we take the cellular decomposition in terms of prisms as above. Now, in \cite{G3}, \cite[Ch.5]{GoellerThesis}, the starting point was to choose a particular spin network state living on the rectangular boundary dual $1$-skeleton of the torus discretizations of the class drawn above. More precisely, the spin network states are build out of coherent intertwiners, also called ``LS-intertwiners'' \cite{LS}, which in turn are build out of $\mathrm{SU}(2)$-coherent states à la Gilmore–Perelomov \cite{Perelomov}. To the edges, we assign the spins $T$ in time direction and $L$ in space direction. For more details about the detailed definition of the boundary state, see \cite[Sec.4.3.]{GoellerThesis}. Hence, the boundary state is peaked at a given geometry in the semi-classical limit and has a good behaviour in the asymptotic limit. Using a saddle point approximation, one obtains the following result:
\begin{align}\langle\mathcal{Z}_{\mathrm{PR}}^{\mathrm{gauge-fixed}}\vert\Psi_{\mathrm{coh.}}\rangle\approx\sum_{n=1}^{N_{x}-1}(-1)^{2TN_{t}n}\mathcal{A}(n)(2-2\mathrm{cos}(\gamma n))\prod_{k=1}^{\frac{N_{x}-1}{2}}\frac{1}{2-2\cos(\gamma k)},\end{align}
where $n$ can be understood as the winding number around the torus and where $\mathcal{A}(n)$ is some factor, whose dominant contribution in the asymptotic limit is given by $n=1$. The product in the formula above exactly reproduces the characters of the $3$-dimensional BMS-group in the asymptotic limit. As an important remark, let us stress that although this result was obtained by performing a saddle point approximation, the theory is still solved exactly in the bulk. The approximation is only on the level of the boundary and needed purely for computational reasons. Yet another result obtained in this context is the calculation of the exact amplitude when choosing a superposition of coherent spin network states in a similar fashion as in the case of the Ponzano-Regge model on the ball discussed previously, i.e. by using the generating functional of spin network states \cite{G4}, \cite[Ch.6]{GoellerThesis}.

\section{Coloured Graphs Representing the Solid Torus and their Transition Amplitudes}
The goal of this section is to construct some explicit colourable and bipartite triangulations of the solid torus $\overline{T}^{2}:=D^{2}\times S^{1}$. The general strategy is as follows: We start with the general cellular decomposition of the solid torus introduced in the previous chapter and try to find a possible way to triangulate each cell in this decomposition in such way that we obtain a proper face coloured and bipartite triangulation with the property that all boundary triangles have the same colour. Afterwards, we define a family of open $(3+1)$-coloured graphs dual to these triangulations, calculate the corresponding Ponzano-Regge transition amplitudes and discuss the role of the contractible cycle of the solid torus.
 
\subsection{Colourable and Bipartite Triangulations of the Solid Torus}
In Figure \ref{TrianSolidTorus}, we have introduced a cellular decomposition of the solid cylinder labelled by three numbers $N_{x},N_{t}\in\mathbb{N}$ and $N_{\gamma}\in\{0,\dots,N_{x}-1\}$. The right-hand side of this figure shows a way to triangulate each prisms by three tetrahedra without adding new vertices to the cellular complex. Now, it clear that we cannot just triangulate each prism in the complex in precisely this way, since the resulting simplicial complex is neither bipartite nor proper face-colourable. In general, a $d$-dimensional simplicial complex admits a coloured graph in $\mathfrak{G}_{d}$ representing it if and only if it is bipartite in the sense that there are two different types of $d$-simplices and only different types are allowed to share a common $(d-1)$-face, if it admits a $(d+1)$-vertex-colouring, which is injective on every $d$-simplex, and if none of the vertices on the boundary has colour $0$. A closer analysis reveals that we need at least two vertical layers and at least two horizontal layers. In other words, a colourable and bipartite simplicial complex of the type introduced above consists of basic building blocks with four tetrahedra. Such a basic building block is drawn in the following figure together with its bipartite structure and colouring:

\begin{figure}[H]
\centering
\includegraphics[scale=0.9]{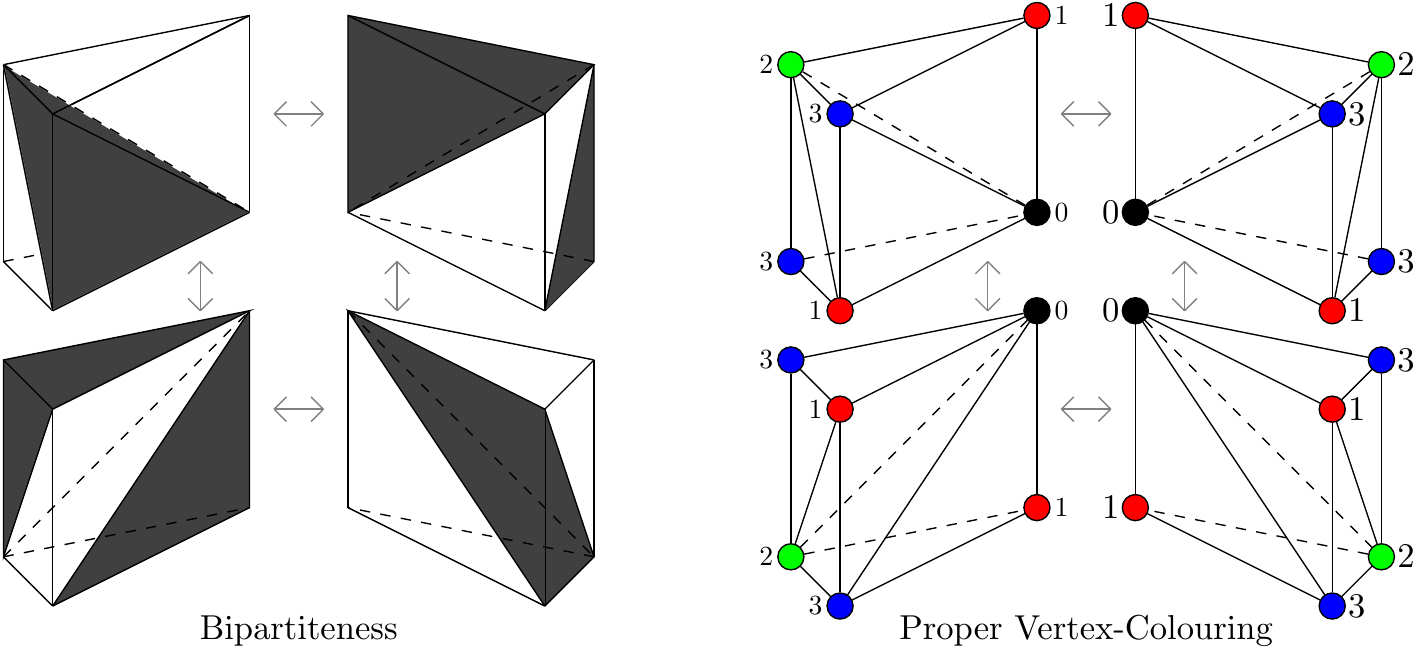}
\caption{Basic building block of a coloured and bipartite simplicial complex of the solid cylinder consisting of four prisms. The gluing of prisms is indicated by the grey arrows.}
\end{figure}

In the figure above, we have drawn a vertex colouring instead of a face colouring. Recall that this two things are one and the same and a proper face colouring can be obtained by assigning to each face the colour of the vertex on the opposite. Using this triangulation, we finally arrive at general proper face-coloured and bipartite triangulations of the solid torus characterized by three numbers $N_{x},N_{t}\in 2\mathbb{N}$ and $N_{\gamma}\in\{0,2,4,\dots,N_{x}-2\}$. Note that, due to the colouring, only even twists are possible, since we are only allowed to glue basic building block consisting of two vertical and two horizontal layers together. Using the figure above, it is straightforward to draw the coloured graph corresponding to a basic building block:

\begin{figure}[H]
\centering
\includegraphics[scale=0.85]{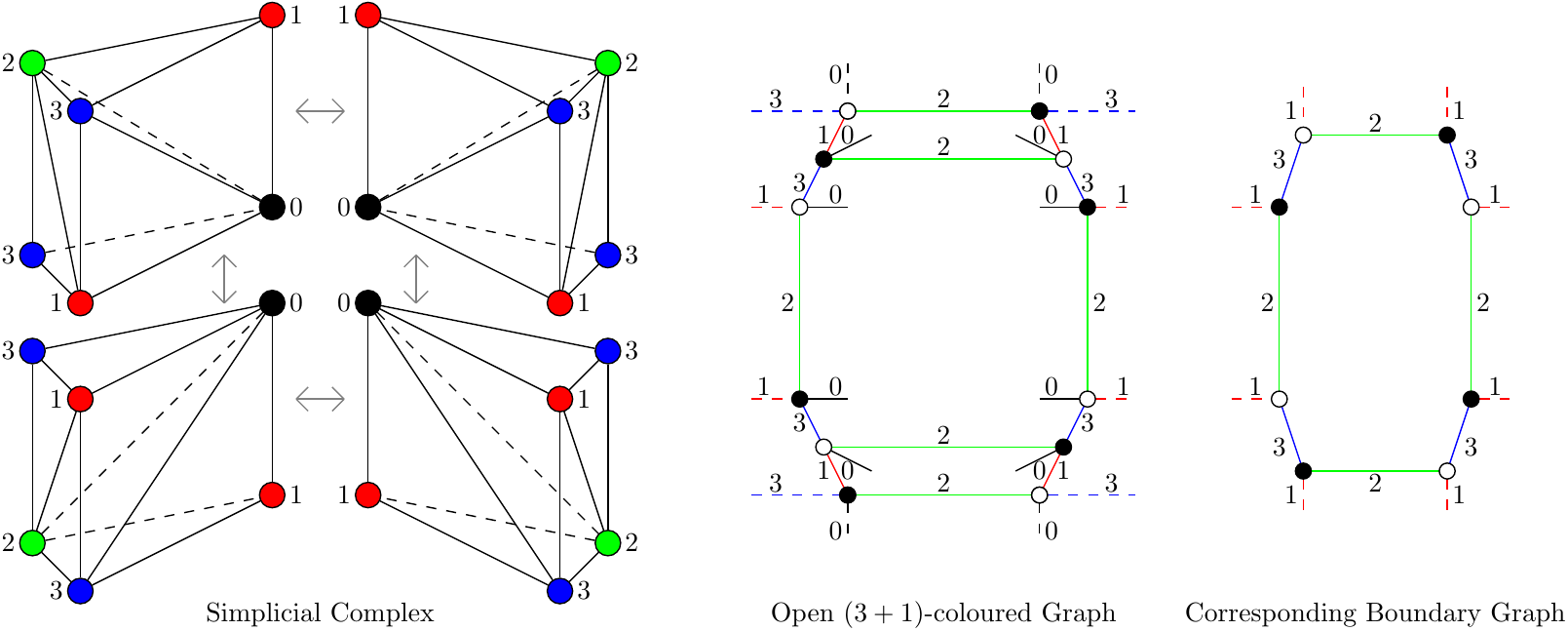}
\caption{Basic building block as a simplicial complex and as an open $(3+1)$-coloured graph with corresponding boundary graph.}
\end{figure}

The dotted lines in the figure above are those edges, to which we glue further building blocks. Hence, we have constructed a family of open $(3+1)$-coloured graphs which are dual to the solid torus $\overline{T}^{2}$. Such a graph is labelled and uniquely determined by the three parameters $N_{x},N_{t}\in 2\mathbb{N}$ and $N_{\gamma}\in\{0,2,4,\dots,N_{x}-2\}$ and has the following general form:

\begin{figure}[H]
\centering
\includegraphics[scale=1]{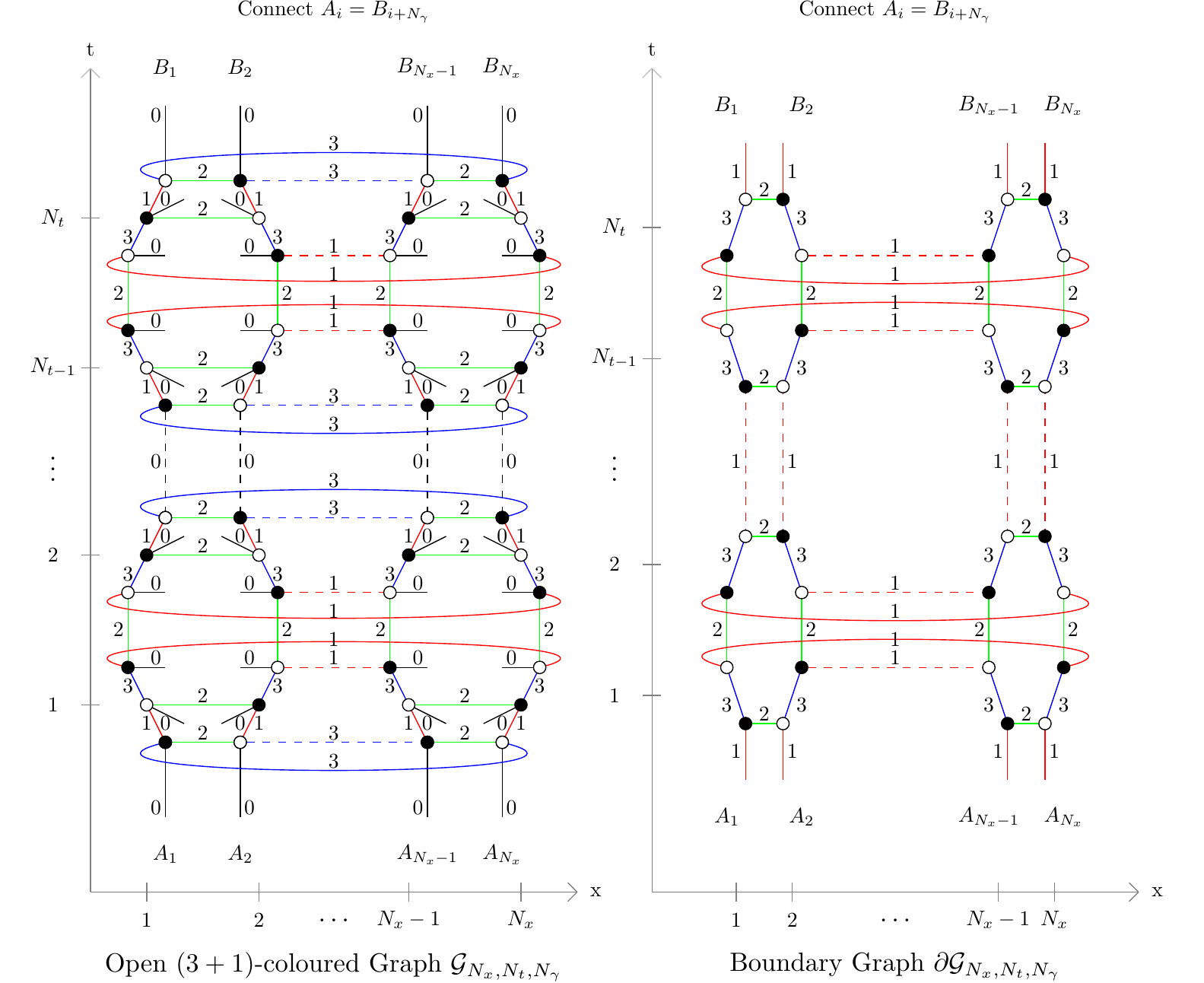}
\caption{A family of open $(3+1)$-coloured graphs $\mathcal{G}_{N_{x},N_{t},N_{\gamma}}\in\mathfrak{G}_{3}$ labelled by three parameters $N_{x},N_{t}\in 2\mathbb{N}$ and $N_{\gamma}\in\{0,2,4,\dots,N_{x}-2\}$, each representing the solid torus, as well as their boundary graphs $\partial\mathcal{G}_{N_{x},N_{t},N_{\gamma}}\in\overline{\mathfrak{G}}_{2}$.}
\end{figure}

It is straightforward to count the number of $k$-bubbles of these graphs for $k\in\{0,1,2,3\}$, or equivalently, the number of $(3-k)$-simplices of the corresponding complex:

\begin{itemize}
\item The number of $0$-bubbles, i.e. internal vertices of the graph $\mathcal{G}_{N_{x},N_{t},N_{\gamma}}$, or equivalently, the number of tetrahedra of the complex $\Delta_{\mathcal{G}_{N_{x},N_{t},N_{\gamma}}}$, is given by $\mathcal{B}^{[0]}=3N_{x}N_{t}$.
\item The number of $1$-bubbles, i.e. edges of the graph $\mathcal{G}_{N_{x},N_{t},N_{\gamma}}$, or equivalently, the number of triangles of the complex $\Delta_{\mathcal{G}_{N_{x},N_{t},N_{\gamma}}}$, is given by $\mathcal{B}^{[1]}=7N_{x}N_{t}$ from which $2N_{x}N_{t}$ are external legs, i.e. triangles living purely on the boundary of the complex.
\item The number of $2$-bubbles, i.e. faces of the graph $\mathcal{G}_{N_{x},N_{t},N_{\gamma}}$, or equivalently, the number of edges of the complex $\Delta_{\mathcal{G}_{N_{x},N_{t},N_{\gamma}}}$, is given by $\mathcal{B}^{[2]}=N_{t}+5N_{t}N_{x}$ from which $3N_{t}N_{x}$ are non-cyclic faces, i.e. edges living purely on the boundary of the complex.
\item The number of $3$-bubbles of the graph $\mathcal{G}_{N_{x},N_{t},N_{\gamma}}$, or equivalently, the number of vertices of the complex $\Delta_{\mathcal{G}_{N_{x},N_{t},N_{\gamma}}}$, is given by $\mathcal{B}^{[3]}=N_{t}+N_{t}N_{x}$ from which $N_{t}N_{x}$ are open $3$-bubbles, i.e. vertices living on the boundary of the complex.
\end{itemize}

As a quick consistency check, let calculate the Euler characteristic of the simplicial complex dual to the open graph $\mathcal{G}_{N_{x},N_{t},N_{\gamma}}$ as well as of the boundary complex, which gives
\begin{align}\chi(\Delta_{\mathcal{G}_{N_{x},N_{t},N_{\gamma}}})&=N_{t}+N_{t}N_{x}-(N_{t}+5N_{t}N_{x})+7N_{t}N_{x}-3N_{t}N_{x}=0\\\chi(\partial\Delta_{\mathcal{G}_{N_{x},N_{t},N_{\gamma}}})&=\chi(\Delta_{\partial\mathcal{G}_{N_{x},N_{t},N_{\gamma}}})=N_{t}N_{x}-3N_{t}N_{x}+2N_{t}N_{x}=0\end{align}
as it should.

\subsection{Transition Amplitudes of the Solid Torus Graphs}
Let us calculate the general form of the transition amplitudes corresponding to the solid torus graphs $\mathcal{G}_{N_{x},N_{t},N_{\gamma}}$. For this, we have to label the boundary edges by group elements. Let us introduce the following notation for each basic building block:

\begin{figure}[H]
\centering
\includegraphics[scale=1.4]{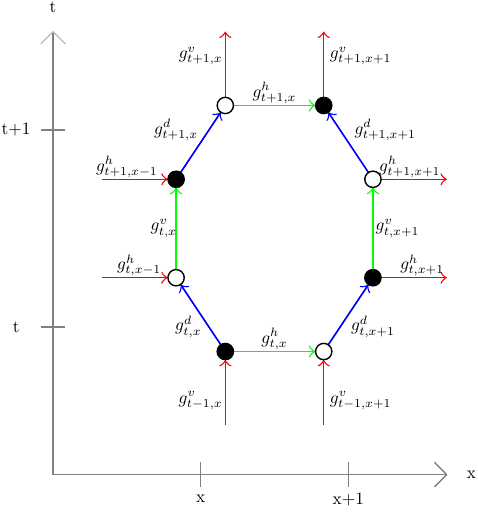}
\caption{Labelling of edges of the boundary graph $\partial\mathcal{G}_{N_{x},N_{t},N_{\gamma}}$ with group elements as well as chosen orientation of boundary edges.\label{TorusBoundaryLabel}}
\end{figure}

Each prism in the simplicial complex is labelled by a pair $(t,x)$ and has two triangles on the boundary and hence is represented by a pair of vertices of different type in the boundary graph. For each such pair, we label the connecting diagonal edge of colour $3$ (=blue) by $g_{t,x}^{d}$. Furthermore, each pair at position $(t,x)$ has a unique horizontal edge, which is connected to the pair at $(t,x+1)$ and has either colour $1$ (=red) or $2$ (=green). We label this edge by a group element $g_{t,x}^{h}$. Last but not least, each pair located at $(t,x)$ has a unique vertical edge, which is connected to the pair $(t+1,x)$ and has colour $1$ (=red). This edge is labelled by the group element $g_{t,x}^{v}$. Furthermore, note that we choose the orientation of boundary edges as drawn above and not the canonical orientation for coloured graphs discussed earlier, for convenience. The amplitudes of the coloured Boulatov model have the following form:

\begin{Proposition}\label{AmplitudesTorus} (Transition Amplitudes of the Solid Torus)\newline
Let $N_{x},N_{t}\in 2\mathbb{N}$ and $N_{\gamma}\in\{0,2,\dots,N_{x}-2\}$ and $\Psi$ be a spin network living on the boundary graph $\partial\mathcal{G}_{N_{x},N_{t},N_{\gamma}}$ with corresponding spin network function $\psi\in L^{2}(\mathrm{SU}(2)^{3N_{t}N_{x}}/\mathrm{SU}(2)^{2N_{t}N_{x}})$. Then the $\mathrm{SU}(2)$-gauge fixed Ponzano-Regge partition function of the corresponding complex is given by
\begin{align*}\langle\mathcal{Z}_{\mathrm{PR}}^{\mathrm{gauge-fixed}}\vert\Psi\rangle=\delta^{N}(\mathds{1})^{N_{t}}\int_{\mathrm{SU}(2)}\mathrm{d}g\,\psi(g^{d}_{t,x}=g^{h}_{t,x}=g^{v}_{t\neq N_{t},x}=\mathds{1},g^{v}_{N_{t},x}=g).\end{align*}
\end{Proposition}

In other words, there is exactly one remaining integration, which is due to the existence of a non-contractible cycle. As a direct consequence, we see that the term appearing in the transition amplitudes of the coloured Boulatov model with respect to this boundary state is given by
\begin{equation}\begin{aligned}\langle\mathcal{A}_{\mathcal{G}_{N_{x},N_{t},N_{\gamma}}}^{\lambda,\mathrm{gauge-fixed}}\vert\Psi\rangle=&\frac{(\lambda\overline{\lambda})^{\frac{3N_{x}N_{t}}{2}}}{\delta^{N}(\mathds{1})^{\frac{N_{t}(3N_{x}-2)}{2}}}\times\\&\times\int_{\mathrm{SU}(2)}\mathrm{d}g\,\psi(g^{d}_{t,x}=g^{h}_{t,x}=g^{v}_{t\neq 0,x}=\mathds{1},g^{v}_{N_{t},x}=g).\end{aligned}\end{equation}
The power of $\delta^{N}(\mathds{1})$ in the denominator comes from the fact that there are $3N_{x}N_{t}$ internal vertices in the graph and hence we get in total a power of $N_{t}-3N_{x}N_{t}/2=-N_{t}(3N_{x}-2)/2$.

\begin{proof}In order to compute the gauge-fixed amplitudes, we have to choose a maximal tree in the graph $\mathcal{G}_{_{N_{x},N_{t},N_{\gamma}}}$. A possible choice is drawn in orange on the left-hand side of the figure below.

\begin{figure}[H]
\centering
\includegraphics[scale=1]{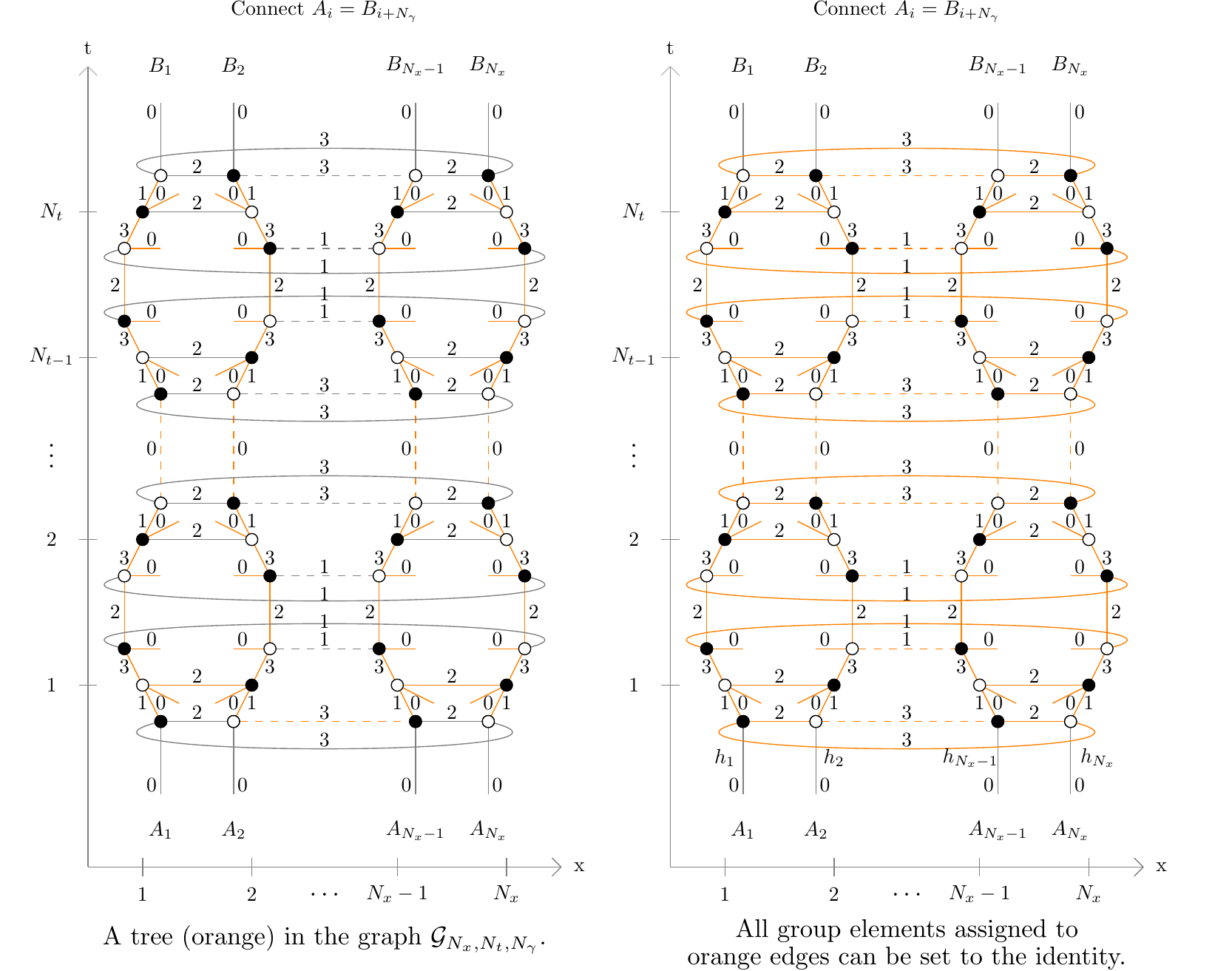}
\caption{A maximal tree in the graph $\mathcal{G}_{N_{x},N_{t},N_{\gamma}}$ (l.h.s.) drawn in orange. The right-hand side shows which additional edges can immediately be set to the identity by using the closed faces of the graph. The only remaining edges are the edges connecting the top and bottom and we denote the corresponding group elements by $h_{x}$ for $x\in\{1,\dots,N_{x}\}$.}
\end{figure}

Using all the closed faces of the graph $\mathcal{G}_{_{N_{x},N_{t},N_{\gamma}}}$ we can set almost all group elements assigned to edges to the identity. The only remaining ones are the group elements assigned to the edges of colour $0$ connecting the top and bottom. We label these elements by group elements $h_{x}$ with $x\in\{1,\dots,N_{x}\}$, as drawn in the right-hand side above. Hence, we also see that all the boundary group elements up to $\{g^{v}_{N_{t},x}\}_{x\in\{1,\dots,N_{x}\}}$ are set to the identity. Let us now restrict to the case with $N_{\gamma}=0$. The case $N_{\gamma}\neq 0$ can be done analogously, by slightly changing the notation. The relevant parts of the Ponzano-Regge partition function have the following form

\begin{equation}\begin{aligned}\int_{\mathrm{SU}(2)^{N_{x}}}\,\bigg (\prod_{x=1}^{N_{x}}\,\mathrm{d}h_{x}\bigg )\,\bigg (&\prod_{x=1}^{\frac{N_{x}}{2}}\delta^{N}(h_{2x-1}h_{2x})\bigg )\bigg (\delta^{N}(h_{1}h_{N_{x}})\prod_{x=1}^{\frac{N_{x}-2}{2}}\delta^{N}(h_{2x}h_{2x+1})\bigg )\times\\&\times\bigg (\prod_{x=1}^{\frac{N_{x}}{2}}\delta^{N}(g_{N_{t},2x-1}h_{2x-1}^{-1})\bigg )\bigg (\prod_{x=1}^{\frac{N_{x}}{2}}\delta^{N}(g_{N_{t},2x}h_{2x})\bigg )\end{aligned}\end{equation}

The first bracket of delta-functions contains all the faces of colour $02$ containing $h_{x}$, the second bracket contains all faces of colour $03$ containing $h_{x}$ and the last line contains all the faces of colour $01$ containing $h_{x}$ and the boundary group elements $g_{N_{t},x}^{v}$. The first bracket tells us that all the pairs starting at an odd number are equivalent with permuting orientation, i.e.
\begin{align}h_{1}=h_{2}^{-1}, h_{3}=h_{4}^{-1}, h_{5}=h_{6}^{-1},\dots\end{align}
whereas the second bracket tells us that the same is true for pairs starting at an even index, i.e.
\begin{align}h_{2}=h_{3}^{-1}, h_{4}=h_{5}^{-1}, h_{6}=h_{7}^{-1},\dots\end{align}
together with $h_{1}=h_{N_{x}}^{-1}$. Combining these two statements, we see that
\begin{align}h_{1}=h_{2}^{-1}=h_{3}=h_{4}^{-1}=h_{5}=\dots=h_{N_{x}}^{-1}.\end{align}
Now, the last line of delta-function exactly identifies all the boundary elements with one of the $h_{x}$'s and hence, respecting the orientation of the boundary group elements, we get
\begin{align}g^{v}_{N_{t},1}=g^{v}_{N_{t},2}=\dots=g^{v}_{N_{t},N_{x}-1}=g^{v}_{N_{t},N_{x}},\end{align}
from which the claim follows. The claimed prefactor $\delta^{N}(\mathds{1})$ follows from the fact that there are in total $N_{t}$ internal $3$-bubbles in the graph $\mathcal{G}_{N_{x},N_{t},N_{\gamma}}$ and hence $N_{t}$ internal $3$-cells not touching the boundary in the corresponding simplicial complex.
\end{proof}

\subsection{Different Contributions Depending on Choice of Contractible Cycle}\label{Cycle}
In the last section, we have constructed a family of open $(3+1)$-coloured graphs representing solid tori and calculated their $\mathrm{SU}(2)$ gauge fixed amplitudes. Now, the boundary of the solid torus is the $2$-torus $T^{2}$, which has two uncontractible cycles, since it is the Cartesian product of two circles. In other words, its fundamental group is $\pi_{1}(T^{2})\cong\mathbb{Z}^{2}$. However, the solid torus, whose fundamental group is $\pi_{1}(\overline{T}^{2})\cong\mathbb{Z}$, has only one uncontractible cycle, since one of the uncontractible cycles of its boundary becomes contractible through the bulk. Now, it turns out that this information is encoded in the Ponzano-Regge partition function of some general triangulation of the solid torus. Since we only fix the boundary topology in the Boulatov model with some fixed labelling of boundary group elements, we hence get two different contributions depending on how the bulk of the solid torus is glued to the boundary complex. Let us discuss this point in more details.\\
\\
As an example of an open $(3+1)$-coloured graph representing the solid torus, we take the simplest example of the general class introduced previously, i.e. the graph $\mathcal{G}_{N_{x},N_{t},N_{\gamma}}$ with $N_{x}=N_{t}=2$ and $N_{\gamma}=0$, which we denote simply by $\mathcal{G}$ in the following. The graph together with its boundary graph is drawn in the figure below.

\begin{figure}[H]
\centering
\includegraphics[scale=0.9]{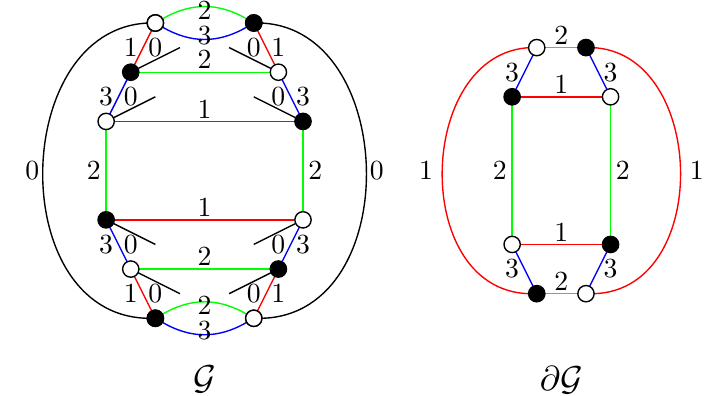}
\caption{The solid torus graph $\mathcal{G}$ with $N_{x}=N_{t}=2$ and $N_{\gamma}=0$ together with its boundary graph $\partial\mathcal{G}$.\label{SimpleTorusGraph}}
\end{figure}

In order to analyse the geometric interpretation of the transition amplitude of this graph, it is useful to consider the \textit{un-gauge-fixed} amplitude. In order to compute the transition amplitudes, let us label the boundary graph $\gamma$ by group elements. We will use a different notation as introduced in Figure \ref{TorusBoundaryLabel} for the family of solid torus graphs, since this notation would be a little bit too cumbersome for this simple example. The boundary graph together with its labellings with group elements and its dual simplicial complex is drawn in the figure below:

\begin{figure}[H]
\centering
\includegraphics[clip,trim=0 1.8cm 1.5cm 2cm,scale=0.8]{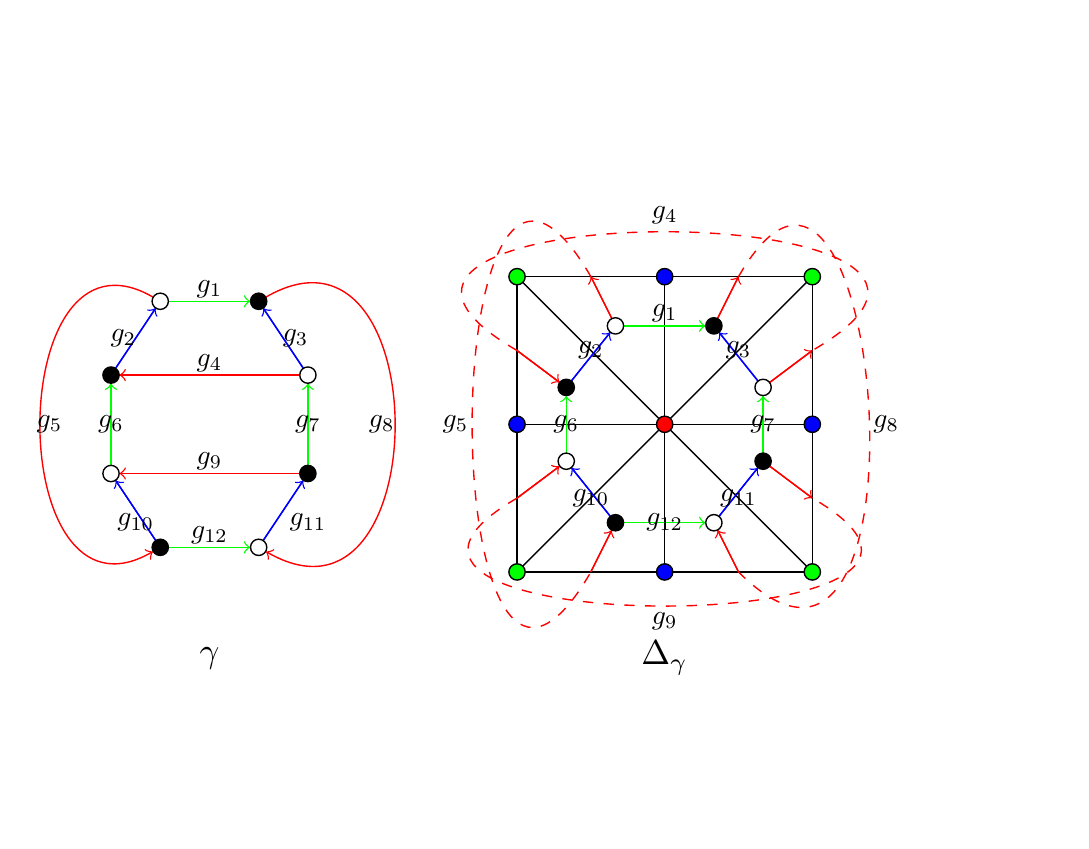}
\caption{The boundary graph $\gamma$ with group elements assigned to its edges (l.h.s.) and the corresponding simplicial complex (r.h.s.).}
\end{figure}

After a straightforward calculation, one finds the following Ponzano-Regge partition function corresponding to the graph $\mathcal{G}$ together with the labelling of boundary edges as above:
\begin{equation}\begin{aligned}\mathcal{Z}_{\mathrm{PR}}^{\mathcal{G}}[\{g_{1},\dots,g_{12}\}]=&\delta^{N}(\mathds{1})^{2}\delta^{N}(g_{7}g_{3}g_{1}^{-1}g_{2}^{-1}g_{6}^{-1}g_{9}^{-1})\delta^{N}_{(12)}(g_{4}g_{6}^{-1}g_{9}^{-1}g_{7})\times\\&\hspace*{1.2cm}\times\delta^{N}_{(12)}(g_{1}g_{8}g_{12}^{-1}g_{5}^{-1})\delta^{N}_{(13)}(g_{2}g_{5}g_{10}g_{9}^{-1}g_{11}^{-1}g_{8}^{-1}g_{3}^{-1}g_{4})\end{aligned}\end{equation}
We get a factor of $\delta^{N}(\mathds{1})^{2}$ as expected, since the graph $\mathcal{G}$ has two internal $3$-bubbles. Three of the delta functions above exactly correspond to three of the four faces of the boundary graph. However, the first delta function does not correspond to any face. In order to see that we indeed recover flatness of the boundary, let us rewrite this expression a little bit. For this, we rewrite one of the factors $\delta^{N}(\mathds{1})$ as $\delta^{N}(g_{7}g_{3}g_{1}^{-1}g_{2}^{-1}g_{6}^{-1}g_{9}^{-1})$, which is allowed, since this delta function already appears in the amplitude. Afterwards, we use some of the other delta functions to rewrite this delta function as
\begin{align}\delta^{N}(g_{7}g_{3}g_{1}^{-1}g_{2}^{-1}g_{6}^{-1}g_{9}^{-1})\to\delta^{N}(g_{9}g_{10}^{-1}g_{12}g_{11}).\end{align}
Hence, we see that we can rewrite the amplitude as 
\begin{equation}\begin{aligned}\label{Ampl1}\mathcal{Z}_{\mathrm{PR}}^{\mathcal{G}}[\{g_{1},\dots,g_{12}\}]=&\delta^{N}(\mathds{1})\delta^{N}(g_{9}g_{10}^{-1}g_{12}g_{11})\Delta^{N}_{\mathrm{flat}},\end{aligned}\end{equation}
where $\Delta^{N}_{\mathrm{flat}}$ denotes the product of all the delta functions corresponding to the faces of the boundary graph, i.e.
\begin{equation}\begin{aligned}\Delta^{N}_{\mathrm{flat}}=&\delta_{(23)}^{N}(g_{1}g_{3}^{-1}g_{7}^{-1}g_{11}^{-1}g_{12}^{-1}g_{10}g_{6}g_{2})\delta^{N}_{(12)}(g_{4}g_{6}^{-1}g_{9}^{-1}g_{7})\times\\&\times\delta^{N}_{(12)}(g_{1}g_{8}g_{12}^{-1}g_{5}^{-1})\delta^{N}_{(13)}(g_{2}g_{5}g_{10}g_{9}^{-1}g_{11}^{-1}g_{8}^{-1}g_{3}^{-1}g_{4}).\end{aligned}\end{equation}
To sum up, we see that the Ponzano-Regge partition function of the solid torus graph $\mathcal{G}$ encodes a theory of flat boundary connections, as it should be, together with an additional delta function. Now, it turns out that this additional delta function exactly encodes the information about which of the two uncontractible cycles of the boundary becomes contractible through the bulk, or in other words, how the interior is ``glued'' to the boundary. Before discussing this point, let us draw another graph representing the solid torus. More precisely, we look at the same simplicial complex, but this time we glue the bulk and boundary along a different cycle. Hence, let us look at the following graph, which we call $\mathcal{G}^{\prime}$:

\begin{figure}[H]
\centering
\includegraphics[scale=0.8]{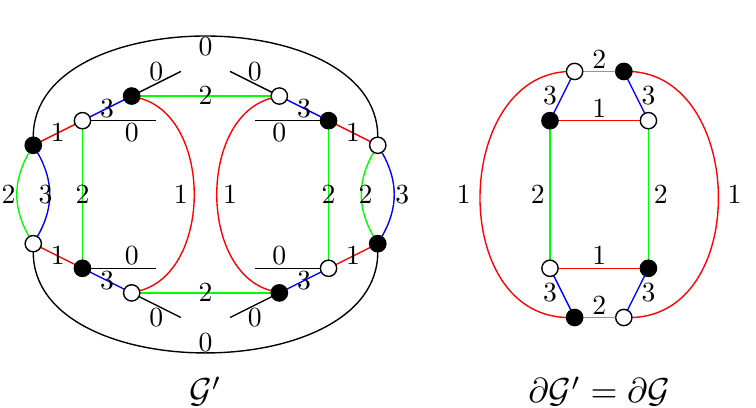}
\caption{An open $(3+1)$-coloured graph $\mathcal{G}^{\prime}$ obtained by rotating the graph $\mathcal{G}$ by 90 degrees.\label{SimpleTorusGraph2}}
\end{figure}

Note that the graph $\mathcal{G}^{\prime}$ is obtained by rotating the graph $\mathcal{G}$ by 90 degrees, however, since the boundary graph is symmetric under these rotations, we keep it with the same labelling as for $\mathcal{G}_{1}$. A straightforward calculation yields the following Ponzano-Regge partition function:
\begin{equation}\begin{aligned}\label{Ampl2}\mathcal{Z}_{\mathrm{PR}}^{\mathcal{G}^{\prime}}[\{g_{1},\dots,g_{12}\}]=&\delta^{N}(\mathds{1})\delta^{N}(g_{3}g_{8}g_{11}g_{7})\Delta^{N}_{\mathrm{flat}}\end{aligned}\end{equation}
The following figure shows the two graphs $\mathcal{G}$ and $\mathcal{G}^{\prime}$ as well as their simplicial complexes:

\begin{figure}[H]
\centering
\includegraphics[clip,trim=0 1.8cm 2cm 1.8cm,scale=0.62]{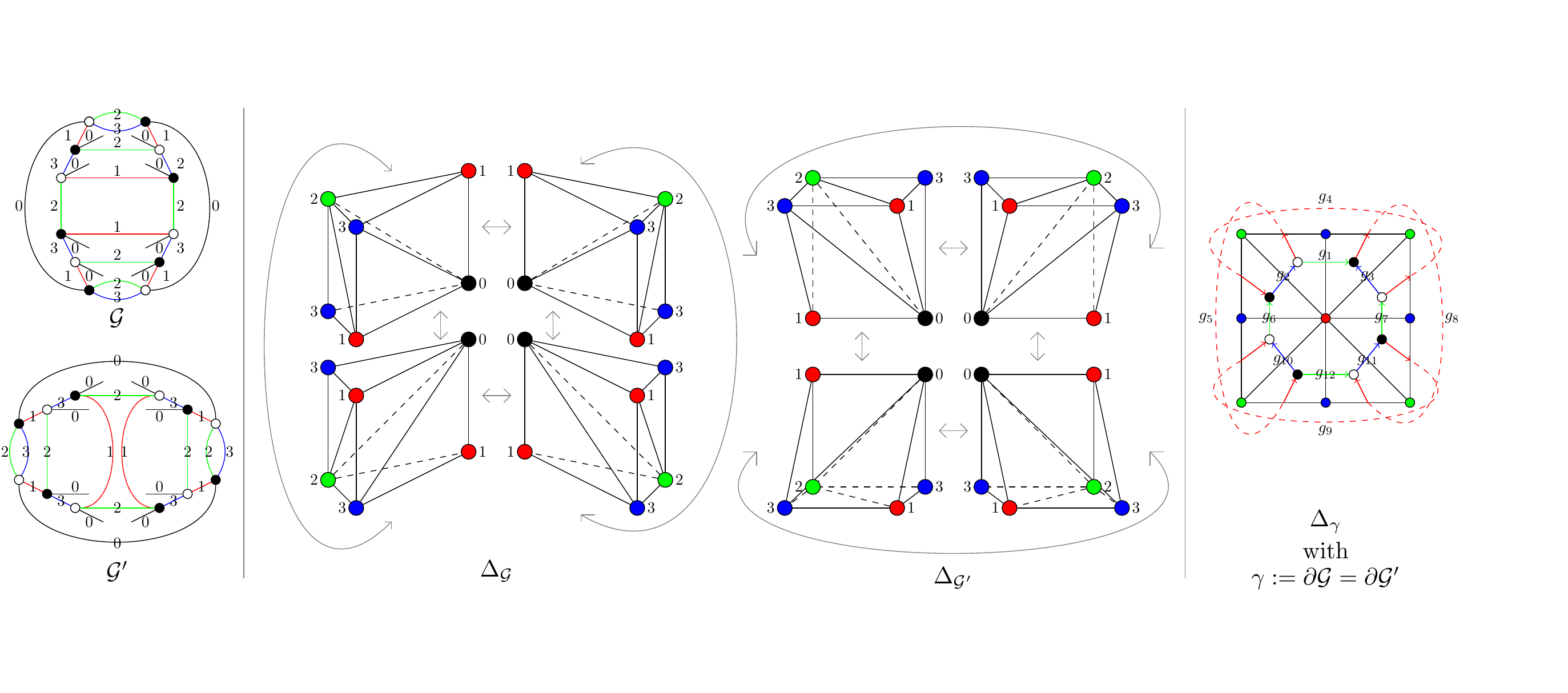}
\caption{The graphs $\mathcal{G}$ and $\mathcal{G}^{\prime}$ together with their simplicial complexes and the corresponding boundaries.}
\end{figure}

The graphs $\mathcal{G}_{1}$ and $\mathcal{G}_{2}$ are of course colour-isomorphic and their complexes are simplicial isomorphic. However, the amplitudes are different, since, the choice of which cycle of the labelled boundary graph $\gamma$ becomes contractible is different. Now, as already mentioned, the additional delta function $\delta^{N}(g_{9}g_{10}^{-1}g_{12}g_{11})$ in the amplitude of $\mathcal{G}$ (Equation (\ref{Ampl1})) and the delta function $\delta^{N}(g_{3}g_{8}g_{11}g_{7})$ in the amplitude of the graph $\mathcal{G}^{\prime}$ (Equation (\ref{Ampl2})) exactly tells us which cycle becomes contractible. This can be seen in by drawing this additional constraint on the boundary complex explicitly:

\begin{figure}[H]
\centering
\includegraphics[clip,trim=0 1.95cm 0cm 2.1cm,scale=0.9]{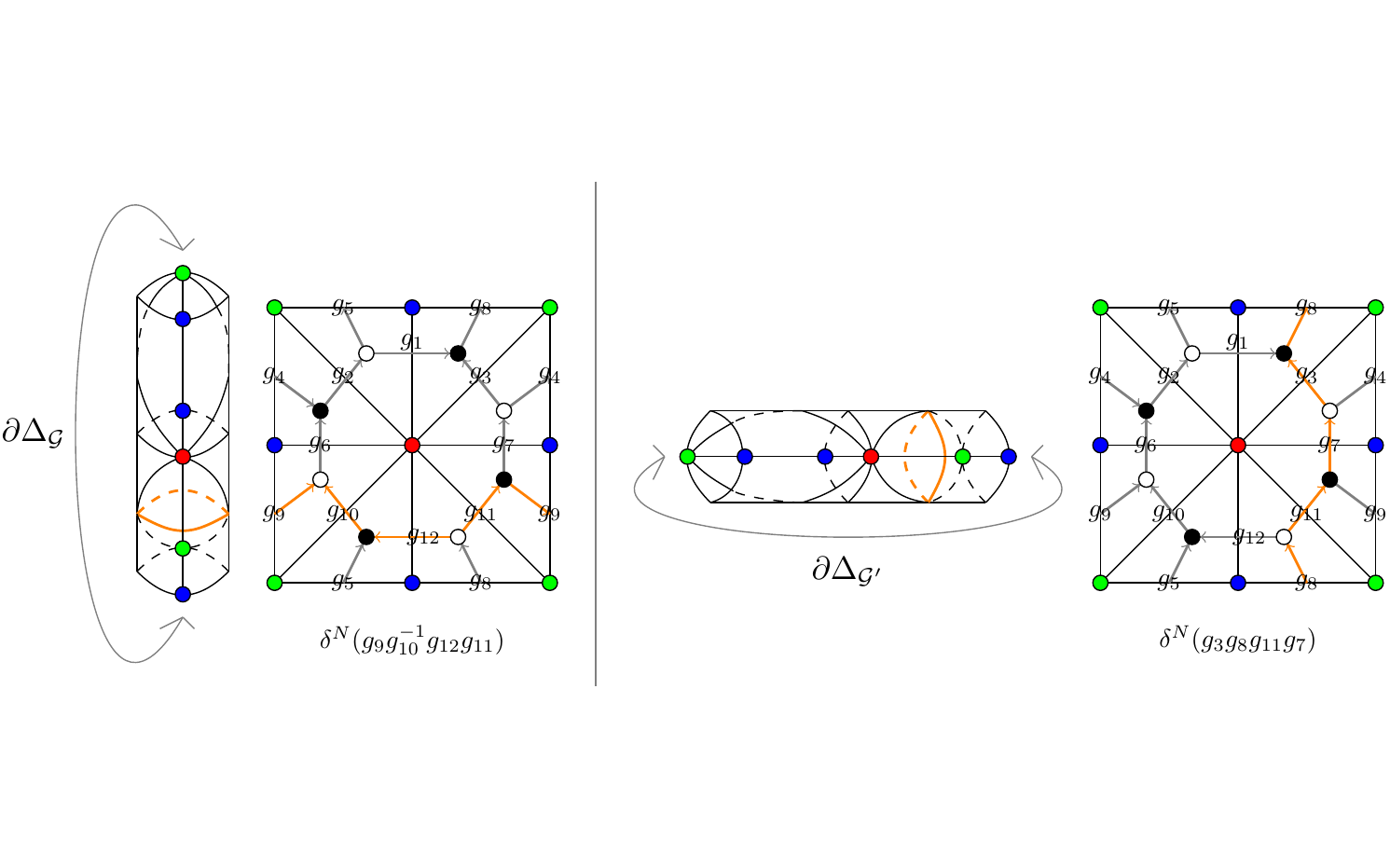}
\caption{Geometric interpretation of the additional constraint.}
\end{figure}

In other words, we get two different contributions from the solid torus in the transition amplitude of the Boulatov model, which differ by the choice of contractible cycle. In the case of the Ponzano-Regge model, these two transition amplitudes are the same up to relabelling, however, in the Boulatov model they are different, since we have fixed a \textit{labelled} boundary graph at the beginning. This is an interesting observation with respect to holography, since the information about which cycle becomes the contractible one through the bulk is not there at the boundary complex, however, the transition amplitude of the Boulatov model factorized into these two contributions and hence contains the information that there are two different choices. A conjecture at this point is that we get a similar result for more complicated boundary topologies: The transition amplitude factorized into different contributions, which differ by how the bulk is glued to the boundary. To sum up, let us conclude that the labelling of the boundary graph in the Boulatov model becomes relevant as opposed to the Ponzano-Regge model.

\section{Other Manifolds with Toroidal Boundary}
One of the main conceptual differences between the transition amplitudes of the Ponzano-Regge model and of the Boulatov model is that the latter contains generically a sum over bulk topologies, whereas in the former case, we have fixed the bulk to be a solid torus $\overline{T}^{2}$. Furthermore, by Theorem \ref{Pezzana}, we know that in fact all $3$-manifolds $\mathcal{M}$ with $\partial\mathcal{M}\cong T^{2}$ will appear in the transition amplitude of the Boulatov model. Hence, it is now natural to discuss how to construct other manifold with torus boundary, which are different to the solid torus. 

\subsection{Internal Connected Sums with Solid Torus}
A simple way to build new manifolds with boundary given by the $2$-torus $T^{2}=S^{1}\times S^{1}$, which are not homeomorphic to the solid torus $\overline{T}^{2}$, is by performing the (internal) connected sum of some closed $3$-dimensional manifold, which is not equal to the $3$-sphere $S^{3}$, with the solid torus. Let us discuss some examples. Similar as in our discussion in Section \ref{SimpleExample}, let us choose the following three closed $(3+1)$-coloured graphs representing $S^{2}\times S^{1}$, $\mathbb{R}P^{3}$ and $L(3,1)$:

\begin{figure}[H]
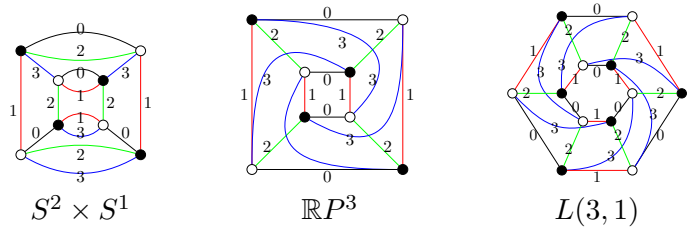

\captionsetup[subfigure]{labelformat=empty}
\centering
\subfloat[$S^{2}\times S^{1}$]{\includegraphics[width=0.12\textwidth]{SimpleExample3a.pdf}}\hspace{1cm}
\subfloat[$\mathbb{R}P^{3}$]{\includegraphics[width=0.15\textwidth]{SimpleExample3b.pdf}}\hspace{1cm}
\subfloat[$L(3,1)$]{\includegraphics[width=0.15\textwidth]{SimpleExample3c.pdf}}
\caption{Three closed $(3+1)$-coloured graphs representing manifolds.\label{ClosedManifolds}}
\end{figure}

As an example of a open $(3+1)$-coloured graph representing the solid torus, we take again the simplest example of the general class introduced in the last section, i.e. the graph $\mathcal{G}_{N_{x},N_{t},N_{\gamma}}$ with $N_{x}=N_{t}=2$ and $N_{\gamma}=0$, as drawn in the Figure \ref{SimpleTorusGraph}. Note that this graph is neither a core graph nor a crystallization, as it has two $3$-bubbles of colour $023$, from which one is a closed graph, which represents a $2$-sphere. Hence, it can easily be made into a core graph simply by applying a proper internal $1$-dipole move of colour $1$. However, in this section we are not interested in core graphs and hence, we use the graph as drawn above. In order to get open $(3+1)$-coloured graphs representing the connected sum of the solid torus with the graphs representing closed $3$-manifolds drawn above, we have to apply Corollary \ref{CorConSum}(1). In other words, we have to perform the graph-connected sum using a vertex not touching the boundary of the solid torus graph $\mathcal{G}$ and a vertex of the graphs in Figure \ref{ClosedManifolds} of different type (black vs. white). The graphs obtained by this procedure are drawn below.

\begin{figure}[H]
\captionsetup[subfigure]{labelformat=empty}
\centering
\subfloat[$\overline{T}^{2}\#(S^{2}\times S^{1})$]{\includegraphics[width=0.23\textwidth]{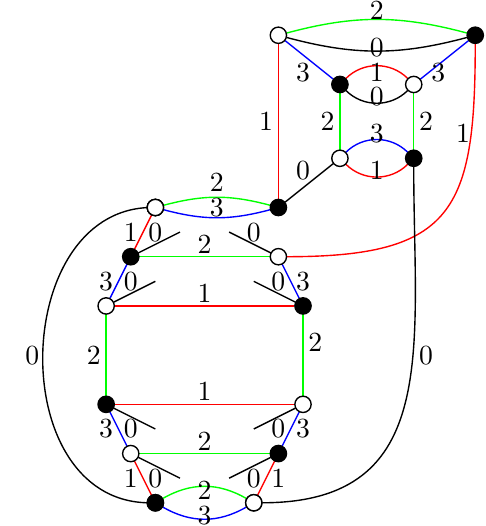}}\hspace{1cm}
\subfloat[$\overline{T}^{2}\#\mathbb{R}P^{3}$]{\includegraphics[width=0.23\textwidth]{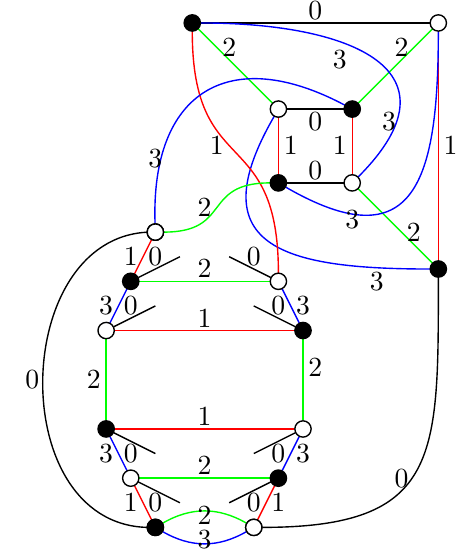}}\hspace{1cm}
\subfloat[$\overline{T}^{2}\# L(3,1)$]{\includegraphics[width=0.23\textwidth]{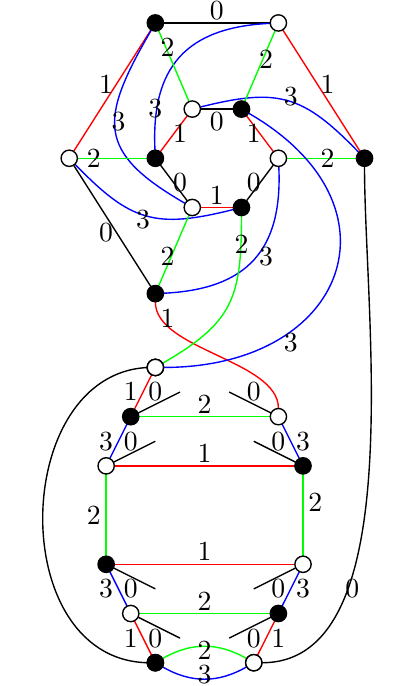}}
\caption{Three open $(3+1)$-coloured graphs representing manifolds with torus boundary obtained by performing internal connected sums.\label{OtherManifolds}}
\end{figure}

Note that all these graphs have the same boundary graph $\gamma:=\partial\mathcal{G}$, as they should. Now, in order to compute the transition amplitudes, let us label the boundary graph $\gamma$ by group elements. We will again use the same labelling of boundary edges as in the previous section:

\begin{figure}[H]
\centering
\includegraphics[trim=0 2cm 2cm 2cm,scale=1]{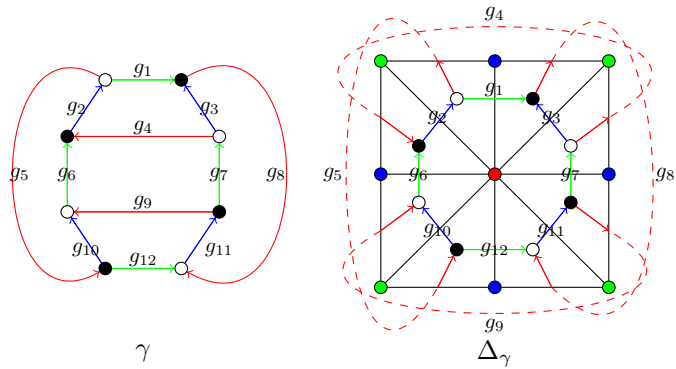}
\caption{The boundary graph $\gamma$ with group elements assigned to its edges (l.h.s.) and the corresponding simplicial complex (r.h.s.).\label{BG}}
\end{figure}

Using Proposition \ref{AmplitudesTorus}, we see that the $\mathrm{SU}(2)$-gauge fixed amplitudes of the solid torus graph $\mathcal{G}$ is given by
\begin{align}\mathcal{A}_{\mathcal{G}}^{\lambda,\mathrm{gauge-fixed}}[\{g_{1},\dots,g_{12}\}]=\frac{(\lambda\overline{\lambda})^{6}}{\delta^{N}(\mathds{1})^{4}}\bigg(\prod_{k=1,k\neq 5,8}^{12}\delta^{N}(g_{k})\bigg)\delta^{N}(g_{5}g_{8}^{-1}).\end{align}

If $\Psi$ is some spin network living on the boundary graph $\gamma$ and $\psi\in L^{2}(\mathrm{SU}(2)^{12}/\mathrm{SU}(2)^{8})$ the corresponding spin network function, then the term appearing in the transition amplitude is hence given by
\begin{align}\langle\mathcal{A}^{\lambda}_{\mathcal{G}}\vert\Psi\rangle=\frac{(\lambda\overline{\lambda})^{6}}{\delta^{N}(\mathds{1})^{4}}\int_{\mathrm{SU}(2)}\,\mathrm{d}g\,\psi(\{g_{k}=\mathds{1}\}_{k=1,\dots,12;k\neq 5,8},g_{5}=g_{8}=g).\end{align}
Choosing maximal trees in the three graphs drawn in Figure \ref{OtherManifolds}, it is straightforward to calculate the $\mathrm{SU}(2)$ gauge fixed amplitudes. The results one obtains are as follows:
\begin{align}
\mathcal{A}_{\overline{T}^{2}\#(S^{2}\times S^{1})}^{\lambda,\mathrm{gauge-fixed}}[\{g_{1},\dots,g_{12}\}]&=\frac{(\lambda\overline{\lambda})^{9}}{\delta^{N}(\mathds{1})^{6}}\bigg(\prod_{k=1,k\neq 5,8}^{12}\delta^{N}(g_{k})\bigg)\delta^{N}(g_{5}g_{8}^{-1})\\
\mathcal{A}_{\overline{T}^{2}\#\mathbb{R}P^{3}}^{\lambda,\mathrm{gauge-fixed}}[\{g_{1},\dots,g_{12}\}]&=\frac{(\lambda\overline{\lambda})^{9}}{\delta^{N}(\mathds{1})^{7}}\bigg(\prod_{k=1,k\neq 5,8}^{12}\delta^{N}(g_{k})\bigg)\delta^{N}(g_{5}g_{8}^{-1})\bigg(\int_{\mathrm{SU}(2)}\,\mathrm{d}h\,\delta^{N}(h^{2})\bigg)\\
\mathcal{A}_{\overline{T}^{2}\# L(3,1)}^{\lambda,\mathrm{gauge-fixed}}[\{g_{1},\dots,g_{12}\}]&=\frac{(\lambda\overline{\lambda})^{11}}{\delta^{N}(\mathds{1})^{9}}\bigg(\prod_{k=1,k\neq 5,8}^{12}\delta^{N}(g_{k})\bigg)\delta^{N}(g_{5}g_{8}^{-1})\bigg(\int_{\mathrm{SU}(2)}\,\mathrm{d}h\,\delta^{N}(h^{3})\bigg)
\end{align}

Hence, we see that in all three cases we get the same contribution coming from the boundary as for the solid torus. The remaining integrals over bulk group elements are coming from non-contractible cycles in the bulk. In other words, if we choose a spin network $\Psi$ as above, then we get in all three cases the following proportionality:
\begin{align}\langle\mathcal{A}^{\lambda}\vert\Psi\rangle\propto\int_{\mathrm{SU}(2)}\,\mathrm{d}g\,\psi(\{g_{k}=\mathds{1}\}_{k=1,\dots,12;k\neq 5,8},g_{5}=g_{8}=g)\end{align}

In fact, it turns out that this is true more generally, as the following proposition shows:

\begin{Proposition} (Amplitudes of Connected Sums with Solid Torus)\newline
Let $\mathcal{G}$ be the solid torus graph as above and $\mathcal{G}^{\prime}\in\overline{\mathfrak{G}}_{3}$ some closed $(3+1)$-coloured graph representing a manifold $\mathcal{N}$. Then the graph connected sum $\mathcal{G}\#_{\{v,w\}}\mathcal{G}^{\prime}$, where $v$ is some vertex of $\mathcal{G}$, which does not admit an adjacent external leg, and where $w$ is some vertex of $\mathcal{G}^{\prime}$, represents the manifold $\overline{T}^{2}\#\mathcal{N}$ and its transition amplitude is proportional to 
\begin{align*}\langle\mathcal{A}^{\lambda}\vert\Psi\rangle\propto\int_{\mathrm{SU}(2)}\,\mathrm{d}g\,\psi(\{g_{k}=\mathds{1}\}_{k=1,\dots,12;k\neq 5,8},g_{5}=g_{8}=g)\end{align*}
\end{Proposition}

\begin{proof}The fact that $\mathcal{G}\#_{\{v,w\}}\mathcal{G}^{\prime}$ represents the manifold $\overline{T}^{2}\#\mathcal{N}$ is a consequence of Corollary \ref{CorConSum}(1). The solid torus graph $\mathcal{G}$ has in total four vertices which do not admit adjacent external legs and these four vertices are distributed symmetrically, so it does not matter which one we choose. The general situation is sketched in the following figure:
\begin{figure}[H]
\centering
\includegraphics[scale=0.75]{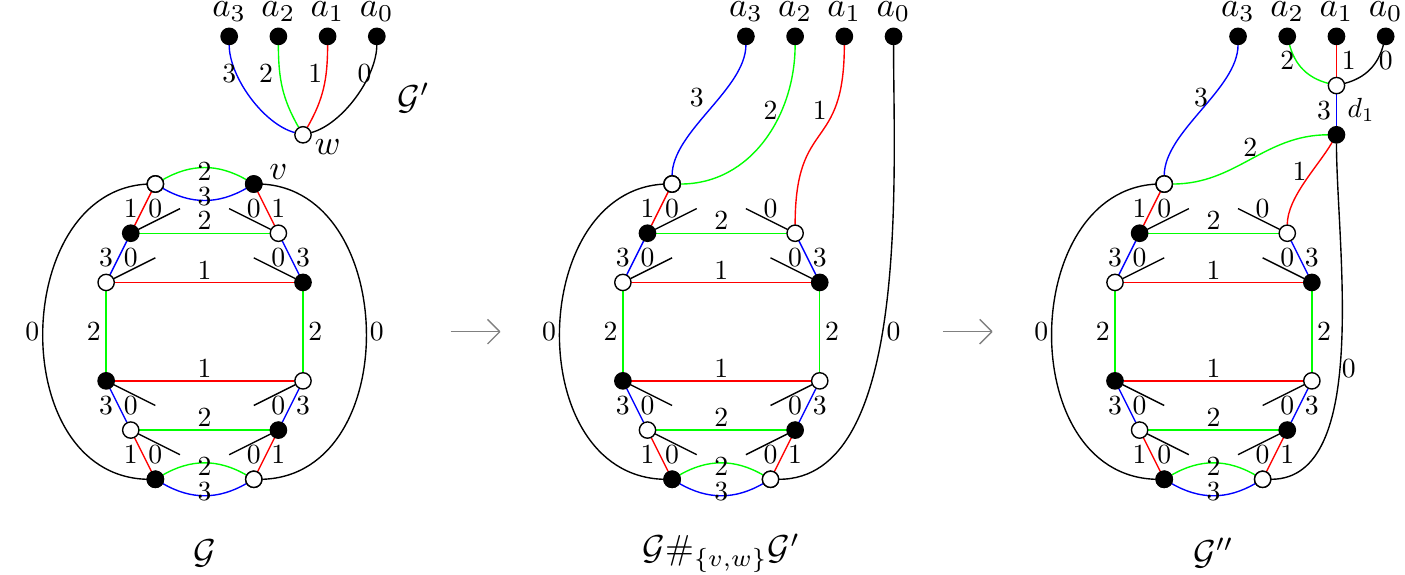}
\caption{The graph $\mathcal{G}\#_{\{v,w\}}\mathcal{G}^{\prime}$ and the graph $\mathcal{G}^{\prime\prime}$ obtained by performing a proper internal $1$-dipole move.}
\end{figure}
The left hand side shows the two graphs separated. The vertices $a_{i}$ belonging to the graph of $\mathcal{N}$ must of course not be distinct. In order to form the connected sum, we delete the vertices $v$ and $w$ and glue the produced half-edges together according their colouring. The resulting graph $\mathcal{G}\#_{\{v,w\}}\mathcal{G}^{\prime}$ is drawn in the middle. In order to obtain the result, it is more convenient to bundle the edges of colours $0,1,2$ together and this can be done by creating a 1-dipole $d_{1}$, as shown on the right-hand side. First of all, it is clear that $d_{1}$ is really a dipole, since the two $123$-bubbles separated by $d_{1}$ are clearly distinct. Furthermore, this dipole is proper and internal, since the $012$-bubble separated by the dipole $d_{1}$ containing the vertices $a_{2},a_{1}$ and $a_{0}$ represents a $2$-sphere, since, by assumption, $\mathcal{N}$ is a manifold. Hence, the amplitudes of the graph $\mathcal{G}^{\prime\prime}$ on the right-hand side and the graph drawn in the middle are equivalent up to factor of $\lambda\overline{\lambda}$. Now, let us choose a maximal tree in the graph on the right-hand side in order to fix the $\mathrm{SU}(2)$ gauge symmetry. A choice is drawn in the left-hand side of the following figure:
\begin{figure}[H]
\centering
\includegraphics[scale=0.75]{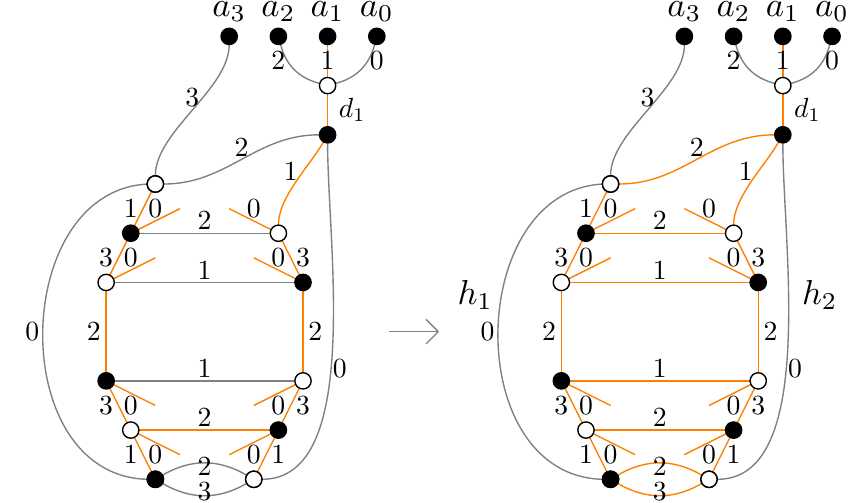}
\caption{A maximal tree in $\mathcal{G}^{\prime\prime}$ in orange (l.h.s.). The right-hand side shows which other edges can immediately be set to the identity using some of the closed faces of the graph.}
\end{figure}
Using all the closed faces of $\mathcal{G}^{\prime\prime}$ we set almost all group elements assigned to edges belonging to the graph $\mathcal{G}$ to the identity, up to the two edges of colour $0$ on the right- and left-hand side. We denote the group elements living on these two edges by $h_{1}$ and $h_{2}$ as drawn on the right-hand side above. Furthermore, we see that all the boundary group elements up to $g_{5}$ and $g_{8}$ are set to the identity. The contribution of the edges labelled by $h_{1}$ and $h_{2}$ to the amplitude is
\begin{align}\int_{\mathrm{SU}(2)^{2}}\,\mathrm{d}h_{1}\mathrm{d}h_{2}\,\delta^{N}(h_{1}h_{2})\delta^{N}(g_{5}h_{1})\delta^{N}(g_{8}h_{2}^{-1})\delta^{N}(h_{2}h_{1}H^{03})=\delta^{N}(g_{5}g_{8}^{-1})\delta^{N}(H^{03}),\end{align}
where $H^{03}$ denotes the product of group elements assigned to the path of colour $0$ and $3$ connecting the edges equipped with $h_{1}$ and $h_{2}$. Integrating over $h_{1}$ and $h_{2}$ leads to the delta-function $\delta^{N}(g_{5}g_{8}^{-1})$, which proves the claim.
\end{proof}

Of course, as before, we could also take the graph $\mathcal{G}^{\prime}$ drawn in Figure \ref{SimpleTorusGraph2}. In this case, we would get a different result, which differs by the choice of which cycle becomes contractible though the bulk. To sum up, for every manifold of the form $\overline{T}^{2}\#\mathcal{N}$, where $\mathcal{N}$ denotes some closed, connected and oriented $3$-manifold, we get two different contributions depending on how we glue the interior to the boundary, which are exactly the same contributions as the two different amplitudes of the solid torus discussed in Section \ref{Cycle}. Hence, when choosing the boundary graph $\gamma$ drawn in Figure \ref{BG}, we again see that the transition amplitude restricted to all manifolds of the form $\overline{T}^{2}\#\mathcal{N}$ factorized into two contributions, which differ by the cycle which becomes contractible through the bulk. Of course, there are also other manifolds with torus boundary, which we will briefly discuss below. Another open question is again how pseudomanifolds contribute to the transition amplitude of the coloured Boulatov model.

\subsection{Prime and Boundary-Prime Manifolds with Torus Boundary}
In the last section, we have seen that we get the same amplitudes for a large class of manifolds with torus boundary, namely for all the manifolds obtained by performing the internal connected sum of some closed $3$-manifold with the solid torus. However, there are many more manifolds with torus boundary. In the following, a ``$3$-manifold'' is always meant to be a compact, connected and orientable $3$-dimensional manifold possibly with non-empty connected boundary. Such a $3$-manifold $\mathcal{M}$ is called a ``\textit{prime manifold}'', if it is different from the $3$-sphere $S^{3}$ and if a decomposition of the form $\mathcal{M}=Q_{1}\# Q_{2}$ implies that either $Q_{1}\cong S^{3}$ or $Q_{2}\cong S^{3}$. In other words, $\mathcal{M}$ cannot be written as a non-trivial connected sum. Let us also remark that prime-manifolds are central in the classification of $3$-manifolds, since according to the famous ``\textit{Theorem of Kneser-Milnor}'' \cite{MK1a,MK1b}, every $3$-manifold $\mathcal{M}$ can be decomposed as
\begin{align}\mathcal{M}\cong P_{1}\#\dots\# P_{k},\end{align}
where all the $P_{i}$'s are prime manifolds. Furthermore, this decomposition is unique up to ordering and up to homeomorphism of its prime factors\footnote{Recall that the connected sum of two $3$-manifolds is in general not unique and depends on the chosen orientations. In order to avoid this ambiguity, we should in the following fix an orientation from the start.}. It is immediate from this definition that manifolds of the form discussed in the previous section, i.e. manifolds of the type $\overline{T}^{2}\#\mathcal{N}$, where $\mathcal{N}$ denotes a $3$-manifold without boundary, fall into the class of non-prime manifolds. However, there are of course also prime manifolds with torus boundary. Furthermore, even a non-prime manifold is not necessarily of the form $\overline{T}^{2}\#\mathcal{N}$, since there could also be a different splitting. If a $3$-manifold $\mathcal{M}$ has non-empty boundary, then there is also the notion of $\partial$-prime manifolds, which are $3$-manifolds with non-empty connected boundaries, which are different from the closed $3$-ball $B^{3}$, and for which a decomposition of the form $\mathcal{M}=Q_{1}\#_{\partial}\,Q_{2}$ implies that either $Q_{1}\cong B^{3}$ or $Q_{2}\cong B^{3}$. In fact, one can also proof an analogues theorem to the Theorem of Kneser-Milnor mentioned above \cite{MK2,MK4}: Every $3$-manifold with non-empty boundary can be written as \begin{align}\mathcal{M}\cong P_{1}\#_{\partial}\,\dots\#_{\partial}\,P_{k},\end{align}
where the $P_{i}$'s are $\partial$-prime manifolds. Again, this decomposition is unique up to ordering and up to homeomorphism of its $\partial$-prime factors\footnote{There is also a generalization for manifolds with several connected boundary components \cite{MK3}, where one extends the notion of the boundary connected sum to manifolds with several boundary connected components.}. Examples of non $\partial$-prime manifolds with torus boundary can be obtained by performing the boundary connected sum of the solid torus, or of any other manifold with torus boundary, with some manifold with spherical boundary, which is different from the closed $3$-ball. In the specific case where $\mathcal{M}$ is a $3$-manifold with boundary homeomorphic to a $2$-torus, it is actually not too hard to see that the notion of $\partial$-prime manifolds and prime manifolds is one and the same. We will not delve into a more detailed discussion. To sum up, manifolds of the form $\overline{T}^{2}\#\mathcal{N}$ discussed previously are only a small part of the class of $3$-manifolds with torus boundary, since there are many more such manifolds, e.g. manifolds which are ($\partial$-)prime. 
\phantomsection
\chapter*{Summary and Outlook}
\addtocontents{toc}{\vspace*{2ex}}
\addcontentsline{toc}{section}{\hspace{-16pt}\textbf{Summary and Outlook}}
\markboth{Summary and Outlook}{Summary and Outlook}
This thesis was devoted to the study of $3$-dimensional quantum gravity, whereby we focused mainly on the case of Riemannian gravity with cosmological constant $\Lambda=0$. In the first chapter, we have explored several aspects of Einstein's theory of general relativity in three dimensions by discussing its physical degrees of freedom, its Newtonian limit, the Riemann and Weyl curvature tensors as well as the solutions of Einstein's field equations in vacuum. All of these considerations have led us to the same conclusion: $3$-dimensional gravity has \textit{no local degrees of freedom} and is \textit{not propagating}, i.e. there are no gravitational waves, or formulated from a particle physical point of view, there are no ``gravitons''. Furthermore, $3$-dimensional gravity is a \textit{theory of constant curvature} and the solutions of Einstein's field equations in vacuum are locally flat (if $\Lambda=0$), locally de Sitter space (if $\Lambda>0$) or locally anti-de Sitter space (if $\Lambda<0$). In other words, Einstein's general theory of relativity in three dimensions is locally trivial and hence, it is a particular instance of a \textit{topological field theory}. We have also seen that $3$-dimensional gravity, formulated in its triadic Palatini formalism, is related to other famous and well-studied topological field theories of Schwarz type, namely to topological $BF$-theory and Chern-Simons theory. Although the dynamics of $3$-dimensional quantum gravity is very different from the $4$-dimensional case, due to the absence of local degrees of freedom, there are still many interesting global phenomena, like infinite-dimensional asymptotic symmetry groups, black hole solutions and local defects, which can be studied in the $3$-dimensional case in a non-perturbative way. Hence, $3$-dimensional gravity is often used as a simple toy model for studying phenomena, which we expect to be there also in the $4$-dimensional case.
\newline\hspace*{2cm}
In the second part of this thesis, we have discussed a particular model for $3$-dimensional quantum gravity, namely the \textit{Ponzano-Regge spin foam model}, which is in fact the first model for quantum gravity in three dimensions ever proposed. We have seen that the Ponzano-Regge model can be understood as the discretization of the quantum partition function of general relativity formulated as a $BF$-theory and that the result is a well-defined topological invariant, i.e. it does not depend on the chosen triangulation of some spacetime manifold. The boundary states of the Ponzano-Regge model are described by \textit{spin network states}, which encode the corresponding quantum geometric data and which provide a basis of the boundary Hilbert space. Besides the Ponzano-Regge model, we have also discussed other related spin foam models for $3$-dimensional quantum gravity, like the Lorentzian Ponzano-Regge model introduced by L. Freidel, as well as the Turaev-Viro model for $3$-dimensional quantum gravity with a positive cosmological constant.
\newline\hspace*{2cm}
After discussing the spin foam approach to quantum gravity, we have introduced yet another model for Riemannian quantum gravity in three dimensions and without a cosmological constant, namely the \textit{Boulatov model}, which is a particular example of a (simplicial) \textit{group field theory}. By definition, this model is a non-local field theory defined on three copies of the Lie group $\mathrm{SU}(2)$, whose fields are living on the space of possible geometries of a triangle and whose interaction term has the combinatorial structure encoding a tetrahedron. The Feynman diagrams of this model are cellular complexes, which are dual to $3$-dimensional simplicial complexes, and the corresponding Feynman amplitudes are exactly the Ponzano-Regge partition functions. From the physical point of view, the group field theory corresponding to some spin foam model can be viewed as a completion of the theory in the sense that it gives us some prescription for organizing the spin foam amplitudes corresponding to different simplicial complexes. An additional advantage of the group field theory formalism is that it is by definition a field theory and hence, we can apply all the techniques developed in (quantum) field theory. In this light, we saw that the Boulatov model generically incorporates a sum over all topologies, where each topology is weighted by its corresponding Ponzano-Regge functional.
\newline\hspace*{2cm}
We then introduced a slightly modified version of the Boulatov model by introducing a \textit{colouring} of its fields. As a consequence, we have seen that the Feynman graphs of this model can be viewed as bipartite and proper edge-coloured graphs, which are always dual do orientable and normal pseudomanifolds, i.e. we do not produce any other types of topological singularities. Interestingly, it turns out that the topology of these graphs is not only studied in the context of quantum gravity, but also in pure mathematics in a branch of geometric topology, which is nowadays known as \textit{crystallization theory}, and which goes back to the pioneering work of the school around P. Bandieri, A. Cavicchioli, M. Ferri, C. Gagliardi, M. Pezzana and many others. Many results have been obtained in this field and we reviewed many important definitions and results, whereby we focused on the general case of coloured graphs with non-empty boundary, representing not necessarily manifolds. Coloured graphs are particularly useful to study low-dimensional piecewise-linear manifolds, since one can directly read of many important objects of algebraic topology, like the homotopy and homology groups, as well as combinatorial quantities, like the Euler characteristic. Furthermore, one can represent many operations from topology, like the (boundary) connected sum and handle decompositions, in terms of coloured graphs. The main advantage is of course that one does not have to draw highly non-trivial and possibly high-dimensional simplicial complexes, but just the corresponding graphs representing them. We have only explored some of the many results obtained in crystallization theory, which were particularly important for our discussion of transition amplitudes in the context of the $3$-dimensional Boulatov model. However, there is much more to explore in the classical crystallization theory literature and there are many results, which could be of interest for the study of coloured tensor models and group field theories in general.
\newline\hspace*{2cm}
As a next step, we used these techniques in order to define suitable boundary observables, which we applied in order to rigorously define transition amplitudes of the coloured Boulatov model. For a fixed boundary spin network state living on some closed coloured graph representing a fixed boundary topology, these transition amplitudes include generically a sum over all possible $3$-manifolds, whose boundary is PL-homeomorphic to the given boundary topology. In order to write this sum in a more systematic way, we hence introduced a \textit{bubble rooting procedure} similar in spirit to the rooting procedure established in the seminal work of R. Gurau regarding the large $N$-limit of coloured tensor models. As a result, we were able to write the transition amplitudes as topological expansions in the sense that each summand in the expansion describes some fixed topology. \newline\hspace*{2cm}
Afterwards, the goal was to apply these techniques in order to analyse the transition amplitudes for some explicit choices of boundary graphs. As a first example, we have considered the simplest possible boundary graph with trivial topology, i.e. the \textit{elementary melonic $2$-sphere}. For this particular choice of boundary state we have obtained a remarkably simple result: Each manifold appearing in the transition amplitudes yields the same contribution including the boundary spin network state and hence, the transition amplitude factorized into a part containing all prefactors and remaining integrals from the bulk of various topologies and a fixed part containing the boundary state. However, there are still many question which remain open: First of all, we haven't discussed pseudomanifolds in more details. Furthermore, we did not discuss the question which topology appearing in the transition amplitude is the most dominant one. Another possible direction for some further work would be to try calculating the transition amplitude in this case explicitly. Of course, the factorized prefactor will in general be divergent, but it would be interesting to prove some Borel summability results, at least for a certain subclass of diagrams appearing in the topological expansion, for example by adding additional terms to the Boulatov action.
\newline\hspace*{2cm}The elementary melonic $2$-sphere is of course a very simple choice of boundary graph. As a next step, we have discussed a more complicated boundary topology, namely the $2$-torus. This choice is particularly interesting, since it appears naturally in the discussion of the asymptotic limit of AdS space as well as in the discussion of BTZ black holes in $3$-dimensional gravity. Furthermore, much insight has been obtained in the work by B. Dittrich, C. Goeller, E. R. Livine and A. Riello, who studied quasi-local holographic dualities in the context of Ponzano-Regge model on the solid torus. In particular, they were able to recover the BMS characters, which are the characters of the asymptotic symmetry group of asymptotically flat gravity, together with some quantum corrections, when choosing a specific boundary state with a clear geometric interpretation in the semi-classical limit. An open problem in this context is how this result changes when considering a sum over topologies, i.e. in the case in which we only fix the boundary graph, equipped with the same boundary state, but sum over all manifold with torus boundary. The last chapter of this thesis was devoted for a brief discussion of this generalization using the formalism and techniques developed in this thesis. In particular, we have constructed a family of triangulations of the solid torus, which admit a coloured graph representing them, and which have a similar structure as the solid torus triangulations used in the work in the Ponzano-Regge model mentioned above. We then have seen that all manifolds obtained by performing the internal connected sum of some closed $3$-manifold with the solid torus yield the same contribution and hence, we get again a factorization of the transition amplitude. An open problem is of course that there are also other manifolds with torus boundary, which cannot be decomposed as a connected sum. Such manifolds are called \textit{prime manifolds} and it is unclear if also these topologies yield the same contribution to the transition amplitude, although we would conjecture they do, because, by the discrete Bianchi identity, we always should recover a theory of flat boundary connections together with possibly additional constraints coming from uncontractible cycles of the boundary graph, which become contractible through the bulk. Furthermore, an obvious next step would be to explicitly calculate the transition amplitude and sum over diagrams of the coloured Boulatov model, by choosing the same kind of boundary state used in the work regarding the Ponzano-Regge model.
\newline\hspace*{2cm}
Yet another open problem in this context, which would be very interesting to look at in some future work, is the question of how to define the asymptotic limit in the context of the (coloured) Boulatov model and in GFT in general. A possible way to approach this question is to study non-internal proper dipole moves in more details. As we have seen in our discussion of crystallization theory, these type of moves leave the topology invariant, but explicitly change the number of boundary triangles by two. Although not discussed in this thesis explicitly, it is in fact not too hard to prove an analogues statement as Proposition \ref{AmplChangeInt} for non-internal dipole, when defining a dipole contraction of spin-networks in a suitable way. A closer analysis of the change of amplitudes under non-internal proper dipole moves could us provide with some first insight in the problem mentioned above. Of course, an open problem which still remains afterwards is how to define the asymptotic limit on the level of the Boulatov fields. The final goal in this context would be to calculate the full transition amplitude of the coloured Boulatov model with respect to the same boundary state as chosen in the work regarding the Ponzano-Regge model and to take the corresponding asymptotic limit, in order to relate the results to other works done in this context and in order to study the holographic dualities in this context.
\newline\hspace*{2cm}
Another interesting result regarding the case of a torus boundary is the observation that we get two different contributions depending on which of the two uncontractible cycles of the $2$-torus becomes contractible through the bulk. When working with the Ponzano-Regge model, which is a triangulation-independent theory, we fix both the boundary and bulk complex and hence, taking two triangulations of the solid torus, which only differ by how they are glued to the boundary and by which cycle becomes the contractible one through the bulk, the corresponding transition functions are the same up to relabelling of boundary edges. However, in the Boulatov model, in which we only fix the boundary complex with a fixed labelling, we will get both contributions and whenever the corresponding spin network state is not symmetric under these changes, we will get in fact two different contributions. In other words, the labelling of the boundary graph becomes important in the Boulatov model. The most interesting point in this observation is the holographic character: We a priori only fix the boundary graph, which does not include any information about how the bulk topologies are glued to it, but the corresponding transition amplitude encodes the information that there are in fact two different results regarding on which cycle becomes the contractible one, which is a pure bulk information. Of course, there is a lot more work to do in this direction: First of all, it is interesting to look at more complicated graphs representing the boundary torus, i.e. graphs involving a twist. Furthermore, it would be interesting to look at more complicated boundary topologies, like the genus $g=2$ surface. In fact, we have already provided the basics needed in order to start analysing these situations in this thesis: Simple boundary graphs representing surfaces of arbitrary genus were presented in Section \ref{GeneralConsiderations} and manifolds with boundary given by such surfaces can be obtained by performing the boundary connected sum of manifolds with torus boundary. A conjecture, which we would like to make at this point, is that the transition amplitude of the coloured Boulatov model always factorizes into several terms, where each term encodes flatness of the boundary together with an additional constraint, which appears in the form of a delta function containing a tricoloured path on the boundary graph, which is independent from the other delta functions, which encode flatness of the boundary connection. Of course, proving this claim rigorously requires much more work, but our results regarding the case of a boundary torus seem to support this conjecture. 
\newline\hspace*{2cm}
A completely different direction for some further work is to study the situation of Riemannian quantum gravity with a cosmological constant. In the second chapter of this thesis, we have introduced the Turaev-Viro spin foam model, which is defined using the quantum group $U(\mathfrak{su}(2))_{q}$ for some root of unity $q$, and which describes $3$-dimensional Riemannian quantum gravity with a positive cosmological constant $\Lambda\propto r^{-2}$, where $r\geq 3$ is defined via $q=\mathrm{exp}(\pi i/r)$. Now, it is actually not too hard to define a corresponding group field theory, which produces the Turaev-Viro model as its Feynman amplitudes. In fact, the corresponding model was already introduced in the original article introducing the Boulatov model \cite{BoulatovModel}. Of course, since it is a priori not clear how to integrate over a quantum group, we cannot use the group representation, as for the Boulatov model. However, we can define the Turaev-Viro group field theory using the spin-representation. An important feature of the Turaev-Viro model is that its corresponding partition functions are finite, as opposed to the Ponzano-Regge model. It would be interesting to define a coloured version of this theory and to analyse the transition amplitudes in this case. Furthermore, it would be interesting to study the relation of this theory with the Boulatov model. We have already discussed that the limit $q\to 1$ cannot always be taken and hence, the relation between the Ponzano-Regge model and Turaev-Viro model is not completely well-understood. Hence, it would be interesting to study the relation in more details on the level of group field theories.
\renewcommand{\theHchapter}{A\arabic{chapter}}
\appendix
\chapter{Appendix}
\fancyhead[LE]{\nouppercase{\rightmark}}

\section{Mathematical Gauge Theory and Topological BF-Theory}\label{gaugetheory}
In the first chapter of this appendix, we review some important concepts of mathematical gauge theory, which is the branch of mathematics studying the geometry of principal fibre bundles and their associated vector bundles. In physics, gauge theories are used to describe field theories, which are invariant under local transformations, but the corresponding formalism has also many important applications in pure mathematics, like Seiberg-Witten and Donaldson theory. In the end of this chapter, we discuss some aspects of $BF$-theory, which we have discussed in Section \ref{BF}, like its equations of motion and its gauge symmetries, in more details. The main references for the following discussion are the excellent textbooks \cite{Baum,Hamilton,RudolphSchmidt1,RudolphSchmidt2}. For applications in pure mathematics see for example the monographs  \cite{Donaldson,WildWorld}.

\subsection{A Glossary of Mathematical Gauge Theory}
In the first part of this chapter, we review some important concepts of mathematical gauge theory. We will not delve into a detailed discussion and the following part should be understood as a glossary, which introduces the most important definitions in order to fix the terminology which we have used in Chapter \ref{Chap1} of this thesis as well as in our discussion of $BF$-theory in Section \ref{GaugeSymmetriesBF} below. We will omit citations throughout this section and for more details the reader is guided to the literature cited above.\\
\\
\fbox{\textbf{Principal Fibre Bundles}:}\\
\\
The central objects in gauge theory are ``principal fibre bundles'', which combine the notion of fibre bundles from differential topology with notions from group theory. To start with, recall that a ``\textit{(smooth) fibre bundle}'' is a quadruple $(E,\mathcal{M},\pi,F)$ consisting of a smooth manifold $\mathcal{M}$, called the ``\textit{base space}'', a smooth manifold $E$, called ``\textit{total space}'', a smooth manifold $F$, called ``\textit{general fibre}'', as well as a smooth and surjective map $\pi:E\to\mathcal{M}$, such that $E$ is ``\textit{locally trivial}'', which means that $\forall x\in M$, there is an open neighbourhood $U_{x}\subset M$ of $x$ and a diffeomorphism $\Phi_{x}:\pi^{-1}(U_{x})\to U_{x}\times F$ such that the following diagram commutes:
\begin{equation*}\begin{tikzcd}
\pi^{-1}(U_{x}) \arrow[d,"\pi",swap] \arrow[rr,"\Phi_{x}"]& & U_{x}\times F \arrow[dll,"\operatorname{pr}_{1}"] \\ U_{x}
\end{tikzcd}\end{equation*}
The embedded submanifold $E_{x}:=\pi^{-1}(\{x\})$, which is diffeomorphic to $F$, is called ``\textit{fibre over $x$}''. Furthermore, we call a pair $(U_{x},\Phi_{x})$ ``\textit{bundle chart} around $x$'' and a collection of bundle charts, which cover $\mathcal{M}$, ``\textit{bundle atlas}''. Using this, a ``\textit{principal fibre bundle}'' is then defined to be a smooth fibre bundle $(P,\mathcal{M},\pi,G)$, where $G$ is some Lie group, called the ``\textit{structure group}'', together with a smooth right action $\mu:P\times G\to P$ with the following extra properties:
\begin{itemize}
\item[(1)]The action $\mu$ preserves the fibres of $\pi$, i.e. $\pi(\mu(p,g))=x$ for all $x\in\mathcal{M}$, $p\in P_{x}=\pi^{-1}(\{x\})$ and $g\in G$. Furthermore, the restricted action $\mu\vert_{P_{x}\times G}:P_{x}\times G\to P_{x}$ is simply transitive for all $x\in \mathcal{M}$. In the following, we will write $p\cdot g:=\mu(p,g)$.
\item[(2)]There exists a ``\textit{principal bundle atlas}'', i.e. a bundle atlas $\{(U_{i},\Phi_{i})\}_{i\in I}$, where $I$ denotes some index set, consisting only of ``\textit{$G$-equivariant maps}'', i.e. $\Phi_{i}(p\cdot g)=\Phi_{i}(p)\cdot g$ for all $i\in I$, $p\in P_{U_{i}}:=\pi^{-1}(U_{i})$ and $g\in G$, where the action on the right-hand side is the obvious one defined by $(x,g)\cdot h:=(x,g\cdot h)$ for all $x\in U_{i}$ and $g,h\in G$. 
\end{itemize}

\fbox{\textbf{Connections on Principal Bundles}:}\\
\\
Another central object in gauge theory are ``connection $1$-forms'', which correspond to the gauge fields in physics terminology. Let $G\to P\stackrel{\pi}{\to}\mathcal{M}$ be a principal fibre bundle with structure group $G$. A ``\textit{connection $1$-form}'', or ``\textit{gauge field}'', on this bundle is a $\mathfrak{g}$-valued $1$-form $A\in\Omega^{1}(P,\mathfrak{g})$ satisfying the following two conditions:
\begin{itemize}
\item[(1)]$A$ is of type $\mathrm{Ad}$, i.e. $(r_{g}^{\ast}A)_{p}=\mathrm{Ad}_{g^{-1}}\circ A_{p}$ for all $g\in G$ and $p\in P$, where $r_{g}:P\to G, p\mapsto p\cdot g$ denotes the right translation of the group action $\cdot:P\times G\to P$ and where $\mathrm{Ad}:G\to\mathrm{Aut}(\mathfrak{g})$ denotes the adjoint representation of $G$.
\item[(2)]$A(\widetilde{X})=X$ for all $X\in\mathfrak{g}$, where $\widetilde{X}$ denotes the ``\textit{fundamental vector field}'' corresponding to $X$, i.e. the vector field $\widetilde{X}\in\mathfrak{X}(P)$ defined by $\widetilde{X}_{p}:=\frac{\mathrm{d}}{\mathrm{d}t}\big\vert_{t=0}(p\cdot\mathrm{exp}_{G}(tX))$ for all $p\in P$.
\end{itemize} 
If $s\in\Gamma^{\infty}(U,P)$ is a ``\textit{local gauge}'', i.e. a local section of $P$, where $U\subset\mathcal{M}$ is some open set, then we can define the form $A_{s}:=s^{\ast}A\in\Omega^{1}(U,\mathfrak{g})$, which is now a $1$-form living on $\mathcal{M}$. Using a basis $\{T_{a}\}_{a}$ of the Lie algebra $\mathfrak{g}$, this form can be decomposed as $A_{s}=\sum_{a}A_{s}^{a}T_{a}$, where $A_{s}^{a}\in\Omega^{1}(U)$ are $\mathbb{R}$-valued forms. Furthermore, if $(U,\varphi=\{x^{\mu}\})$ is a local manifold chart, we can introduce the maps $A_{\mu}^{a}:=A_{s}^{a}(\partial_{\mu})\in C^{\infty}(U)$ and hence we can write $A^{a}_{s}=\sum_{\mu}A_{\mu}^{a}\mathrm{d}x^{\mu}$ for every $a$. The fields $\{A_{\mu}^{a}\}_{\mu,a}$ are then called ``\textit{(local) gauge boson fields}''.\\
\\
If $A\in\Omega^{1}(P,\mathfrak{g})$ is a connection $1$-form on the principal bundle $P$, then we define the vector space $H_{p}:=\mathrm{ker}(A_{p})\subset T_{p}P$. This defines an ``\textit{Ehresmann connection}'', i.e. a vector subbundle $H\subset TP$ (a ``\textit{distribution on $P$}'') such that $T_{p}P=H_{p}\oplus V_{p}$ for all $p\in P$, where $V_{p}:=T_{p}(P_{\pi(p)})$ denotes the ``\textit{vertical tangent space at $p$}'', and such that $r_{g\ast}(H_{p})=H_{p\cdot g}$ for all $p\in P$ and $g\in G$. We call a tangent vector contained in $H_{p}\subset T_{p}P$ ``\textit{horizontal}''.\\
\\
\fbox{\textbf{Curvature}:}\\
\\
The ``\textit{curvature}'' of a connection $1$-form $A\in\Omega^{1}(P,\mathfrak{g})$ is the $2$-form $F^{A}\in\Omega^{2}(P,\mathfrak{g})$ defined via the ``\textit{structure equation}''
\begin{align}F^{A}:=\mathrm{d}A+\frac{1}{2}[A\wedge A],\end{align}
where $[\cdot\wedge\cdot]:\Omega^{k}(P,\mathfrak{g})\times\Omega^{l}(P,\mathfrak{g})\to\Omega^{k+l}(P,\mathfrak{g})$ is the wedge-product defined via the Lie-bracket of $\mathfrak{g}$, i.e.
\begin{align}[\alpha\wedge\beta]:=\sum_{a}(\alpha^{a}\wedge\beta^{b})[T_{a},T_{b}],\end{align}
for all $\alpha=\alpha^{a}T_{a}\in\Omega^{k}(P,\mathfrak{g})$ and $\beta=\beta^{a}T_{a}\in\Omega^{l}(P,\mathfrak{g})$, where $\alpha^{a}$ and $\beta^{b}$ are real-valued forms. Again, using a local gauge $s\in\Gamma^{\infty}(U,P)$, this can locally be written as $F_{s}^{A}:=s^{\ast}F^{A}\in\Omega^{2}(U,\mathfrak{g})$. Using a basis of the Lie algebra $\mathfrak{g}$, we decompose this as $F^{A}_{s}=\sum_{a}F^{a}_{s}T_{a}$ with $F_{s}^{a}\in\Omega^{2}(U)$ and using a local chart $(U,\varphi=\{x^{\mu}\})$, we can further write $F_{s}^{a}=\sum_{\mu,\nu}F_{\mu\nu}^{a}\mathrm{d}x^{\mu}\wedge\mathrm{d}x^{\nu}$. The collection $\{F_{\mu\nu}^{a}\}_{\mu,\nu,a}\in C^{\infty}(U)$ is called ``\textit{(local) field strength} of the gauge boson field $A$''.\\
\\
\fbox{\textbf{Associated Vector Bundles}:}\\
\\
Let $(V,\rho)$ be some finite-dimensional real (resp. complex) representation of a Lie group $G$ and $G\to P\stackrel{\pi}{\to}\mathcal{M}$ be a principal $G$-bundle. Then we define a real (resp. complex) vector bundle, called the ``\textit{associated vector bundle}'', as follows: First of all, we define a free right group action $(P\times V)\times G\to (P\times V)$ of $G$ on $P\times V$ via
\begin{align}(p,v)\cdot g:=(p\cdot g,\rho(g^{-1})v)\end{align}
for all $(p,v)\in P\times V$, $g\in G$. One can show that the quotient $E:=(P\times V)/G$ is a well-defined smooth manifold and that the manifold $E$ is a well-defined vector bundle with projection $\pi_{E}:E\to\mathcal{M}$ given by $\pi_{E}([p,v]):=\pi(p)$ for all $[p,v]\in E$. The fibres of this bundle are given by $E_{x}:=(P_{x}\times V)/G$ for all $x\in\mathcal{M}$. This bundle is usually denoted by
\begin{align}P\times_{\rho}V:=E.\end{align}
Sections in the associated vector bundle can be described by sections of the principal bundle $P$. More precisely, if $s\in\Gamma^{\infty}(U,P)$ with $U\subset\mathcal{M}$ open is a local gauge, then for every smooth map $f\in C^{\infty}(U,V)$ there is a section $\tau\in\Gamma^{\infty}(U,E)$ defined via
\begin{align}\tau(x)=[s(x),f(x)]\end{align}
for all $x\in U$ and vice versa. Last but not least, if $\langle\cdot,\cdot\rangle_{V}$ is a $G$-invariant, non-degenerate and symmetric (resp. hermitian) bilinear (resp. sesquilinear) form on $V$, then one can define a vector bundle (pseudo-)metric $\langle\cdot,\cdot\rangle_{E}\in\Gamma^{\infty}(E^{\ast}\otimes E^{\ast})$ of $E$ via
\begin{align}\langle [p,v],[p,w]\rangle_{E}:=\langle v,w\rangle_{V}\end{align}
for all $[p,v],[p,w]\in E$. One can easily verify that this is indeed a bundle metric and that the definition is independent of the choice of representatives of $[p,v],[p,w]$. This metric is called the ``\textit{bundle metric induced by $\langle\cdot,\cdot\rangle_{V}$}''. Let us remark that associated vector bundles are central in the standard model of particle physics, since matter fields are defined to be sections of associated bundles (possibly twisted by spinor bundles for fermions) for some given representations.\\
\\
\fbox{\textbf{Exterior Covariant Derivative}:}\\
\\
Let $G\to P\stackrel{\pi}{\to}\mathcal{M}$ be a principal $G$-bundle with a connection $1$-form $A\in\Omega^{1}(P,\mathfrak{g})$ and $(V,\rho)$ be a representation of $G$. Then, we define a differential operator on $V$-valued differential forms, i.e. a family of maps $D_{A}:\Omega^{k}(P,V)\to\Omega^{k+1}(P,V)$ for $k\in\mathbb{N}_{0}$, via
\begin{align}(D_{A}\omega)_{p}(v_{1},\dots,v_{k}):=(\mathrm{d}\omega)_{p}(\mathrm{pr}_{H}v_{1},\dots,\mathrm{pr}_{H}v_{k})\end{align}
for all $\omega\in\Omega^{k}(P,V)$, $v_{1},\dots,v_{k}\in T_{p}P$ and $p\in P$, where $\mathrm{pr}_{H}:T_{p}P\to H_{p}$ denotes the obvious projection to the horizontal tangent space $H_{p}:=\mathrm{ker}(A_{p})$. Note that this differential operator can be used to define the curvature, since
\begin{align}D_{A}A=F^{A}\in\Omega^{2}(P,\mathfrak{g}).\end{align}
In order to define a covariant derivative on the base manifold $\mathcal{M}$, let us introduce the following terminology: We define the vector space $\Omega^{k}_{\mathrm{hor}}(P,V)^{\rho}$ of ``\textit{horizontal $k$-forms of type $\rho$}'' consisting of $k$-forms $\omega\in\Omega^{k}(P,V)$ with the following two properties:
\begin{itemize}
\item[(1)]$\omega_{p}(v_{1},\dots,v_{k})=0$ whenever at least one of the vectors $v_{1},\dots,v_{k}\in T_{p}P$ is vertical $\forall p\in P$.
\item[(2)]$(r_{g}^{\ast}\omega)_{p}=\rho(g^{-1})\circ\omega_{p}$
for all $p\in P$ and for all $g\in G$.
\end{itemize}
One can easily verify that the curvature $2$-form corresponding to some connection $1$-form $A\in\Omega^{1}(P,\mathfrak{g})$ is horizontal and of type $\mathrm{Ad}$, i.e. $F^{A}\in\Omega_{\mathrm{hor}}^{2}(P,\mathfrak{g})^{\mathrm{Ad}}$. Now, it turns out that there is the following isomorphism:
\begin{align}\Omega^{k}_{\mathrm{hor}}(P,V)^{\rho}\cong\Omega^{k}(\mathcal{M},E),\end{align}
where $E:=P\times_{\rho}V$ denotes the associated vector bundle. In particular, this means that we can view the curvature $F^{A}$ as an element $F^{A}_{\mathcal{M}}$ of $\Omega^{2}(\mathcal{M},\mathrm{Ad}(P))$, where $\mathrm{Ad}(P):=P\times_{\mathrm{Ad}}\mathfrak{g}$ denotes the ``\textit{adjoint bundle}''. The differential operator $D_{A}$ maps forms of $\Omega^{k}_{\mathrm{hor}}(P,V)^{\rho}$ into forms of $\Omega^{k+1}_{\mathrm{hor}}(P,V)^{\rho}$ and hence we get a well defined family of operators
\begin{align}\mathrm{d}_{A}:\Omega^{k}(\mathcal{M},E)\to\Omega^{k+1}(\mathcal{M},E).\end{align}
If we restrict this map to $\Omega^{0}(\mathcal{M},E):=\Gamma^{\infty}(E)$, then we get a connection $\nabla^{A}:=\mathrm{d}_{A}\vert_{\Gamma^{\infty}(E)}:\Gamma^{\infty}(E)\to\Gamma^{\infty}(T^{\ast}\mathcal{M}\otimes E)$ of the vector bundle $E$. This connection is usually called ``\textit{covariant derivative}''. Now, it turns out that the exterior covariant derivative $\mathrm{d}_{\nabla^{A}}$ induced by this connection is exactly $\mathrm{d}_{A}$, i.e. it holds that $\mathrm{d}_{A}=\mathrm{d}_{\nabla^{A}}$. As a consequence, we can compute $\mathrm{d}_{A}\omega$ for some $\omega\in\Omega^{k}(\mathcal{M},E)$ in some local frame $\{e_{a}\}\subset\Gamma^{\infty}(U,E)$, where $U\subset\mathcal{M}$ open, via
\begin{align}(\mathrm{d}_{A}\omega)\vert_{U}=\sum_{a}(\mathrm{d}\omega^{a}\otimes e_{a}+(-1)^{k}\omega^{a}\wedge\nabla_{A}e_{a}),\end{align} where $\omega^{a}$ are the real-valued coordinate forms of $\omega$, i.e. $\omega\vert_{U}=\sum_{a}\omega^{a}\otimes e_{a}$. As a last remark, let us note that if $\langle\cdot,\cdot\rangle_{V}$ is some $G$-invariant, non-degenerate and symmetric bilinear form on $V$, then the covariant derivative $\nabla^{A}$ is compatible with the induced bundle metric $\langle\cdot,\cdot\rangle_{E}$, i.e.
\begin{align}X(\langle s,t\rangle_{E})=\langle\nabla^{A}_{X}s,t\rangle_{E}+\langle s,\nabla^{A}_{X}t\rangle_{E}\end{align}
for all sections $s,t\in\Gamma^{\infty}(E)$ and vector fields  $X\in\mathfrak{X}(\mathcal{M})$.\\
\\
\fbox{\textbf{The Bianchi-Identity}:}\\
\\
Let $G\to P\stackrel{\pi}{\to}\mathcal{M}$ be a principal $G$-bundle with a connection $1$-form $A\in\Omega^{1}(P,\mathfrak{g})$. Then, as discussed above, we can define the corresponding curvature $F^{A}\in\Omega^{2}(P,\mathfrak{g})$, which can equivalently be viewed as $2$-form $F^{A}_{\mathcal{M}}\in\Omega^{2}(\mathcal{M},\mathrm{Ad}(P))$. The ``\textit{Bianchi identity}'' is the following formula
\begin{align}\mathrm{d}F^{A}+[A\wedge F^{A}]=0,\end{align}
which follows from a straightforward calculation. Using the exterior covariant derivative, this can also be written as $D_{A}F^{A}=0$, or equivalently, as $\mathrm{d}_{A}F^{A}_{\mathcal{M}}=0$.\\
\\
\fbox{\textbf{Gauge Transformations}:}\\
\\
A ``\textit{(global) gauge transformation}'' of some principal $G$-bundle $G\to P\stackrel{\pi}{\to}\mathcal{M}$ is defined to be a principal bundle automorphism $f$ of $P$, i.e. a smooth diffeomorphism $f:P\to P$ satisfying the following two properties:
\begin{itemize}
\item[(1)]$f$ is a fibre bundle automorphism, i.e. $\pi\circ f=\pi$.
\item[(2)]$f$ is $G$-equivariant, i.e. $f(p\cdot g)=f(p)\cdot g$ for all $p\in P$, $g\in G$.
\end{itemize}
We will denote the set of all gauge transformations by $\mathcal{G}(P)$. Once can show that there is the following group isomorphism:
\begin{align}\mathcal{G}(P)\to C^{\infty}(P,G)^{G},f\mapsto\sigma_{f},\end{align}
where $C^{\infty}(P,G)^{G}$ is the subset of $C^{\infty}(P,G)$ consisting of smooth functions $\sigma:P\to G$ satisfying $\sigma(p\cdot g)=g^{-1}\sigma(p)g$ for all $p\in P$ and $g\in G$ and where $\sigma_{f}$ is the map defined by $f(p)=p\cdot\sigma_{f}(p)$ for all $p\in P$. Note that the gauge group $C^{\infty}(P,G)^{G}$ can be given the structure of an infinite-dimensional Hilbert Lie group \cite[p.7]{Schmid}. In physics, one usually defines gauge transformations to be maps living on the spacetime manifolds $\mathcal{M}$. This can be related to the mathematical notion of gauge transformations as follows: For every local gauge $s\in\Gamma^{\infty}(U,P)$ with $U\subset\mathcal{M}$ open, there is an isomorphism of groups $C^{\infty}(P_{U},G)^{G}\cong C^{\infty}(U,G)$, where $P_{U}:=\pi^{-1}(U)\subset P$. If $G$ is abelian, then one can show that there is also a global isomorphism, i.e. we can choose $U=\mathcal{M}$.\\
\\
Let us discuss the action of the gauge group $\mathcal{G}(P)$ on various different objects. The proof of the claimed equalities is straightforward.
\begin{itemize}
\item[(1)]If $A\in\Omega^{1}(P,\mathfrak{g})$ is a connection $1$-form, then $f\in\mathcal{G}(P)$ acts on $A$ as
\begin{align}f^{\ast}A=\mathrm{Ad}_{\sigma_{f}^{-1}}\circ A+\sigma_{f}^{\ast}\mu_{G},\end{align}
where $\mu_{G}\in\Omega^{1}(G,\mathfrak{g})$ denotes the Maurer-Cartan form on $G$, i.e. the $1$-form defined by $(\mu_{G})_{g}(v):=\mathrm{d}_{g}L_{g^{-1}}(v)\in T_{e}G\cong\mathfrak{g}$ for all $g\in G$, $v\in T_{g}G$ with $L_{g}:G\to G,h\mapsto gh$.
\item[(2)]If $(V,\rho)$ is some representation of $G$ and $E:=P\times_{\rho}V$ the corresponding associated vector bundle, then $f\in\mathcal{G}(P)$ acts on $[p,v]\in E$ as 
\begin{align}f\cdot [p,v]:=[f(p),v].\end{align}
One can easily verify that this is independent on the choice of representative of $[p,v]$.
\item[(3)]Let $(V,\rho)$ and $E$ as in (2). Then $f\in\mathcal{G}(P)$ acts on $\omega\in\Omega^{k}_{\mathrm{hor}}(P,V)^{\rho}$ as
\begin{align}f^{\ast}\omega=\rho(\sigma_{f}^{-1})\circ\omega.\end{align}
As an application, we see that the curvature $F^{A}\in\Omega^{2}_{\mathrm{hor}}(P,\mathfrak{g})^{\mathrm{Ad}}$ of some connection $1$-form $A\in\Omega^{1}(P,\mathfrak{g})$ transforms as
\begin{align}F^{f\ast A}=f^{\ast}F^{A}=\mathrm{Ad}_{\sigma_{f}^{-1}}\circ F^{A}.\end{align}
Now, recall that there is an isomorphism $I:\Omega^{k}_{\mathrm{hor}}(P,V)^{\rho}\to\Omega^{k}(\mathcal{M},E)$. Then, one can easily show that under this isomorphism $I(f^{\ast}\omega)=f\cdot I(\omega)$, where the action on the right-hand side is the action as defined in (2) applied point-wise to forms in $\Omega^{k}(\mathcal{M},E)$.
\end{itemize}

\subsection{Application to Topological BF-Theory}\label{GaugeSymmetriesBF}
As an application of the formalism discussed previously, let discuss topological $BF$-theory in more details. More precisely, we will discuss rigorously the gauge symmetries of $BF$-theory as well as a precise mathematical derivation of the equations of motion. Let us fix the following date:

\begin{itemize}
\item[(1)] A $d$-dimensional smooth, compact and oriented manifold $\mathcal{M}$ without boundary and a principal $G$-bundle $G\to P\stackrel{\pi}{\to}\mathcal{M}$, where $G$ is some finite-dimensional Lie group with Lie algebra $\mathfrak{g}:=\mathrm{Lie}(G)$.
\item[(2)]An $\mathrm{Ad}$-invariant symmetric and non-degenerate inner product $\langle\cdot,\cdot\rangle_{\mathfrak{g}}$ on $\mathfrak{g}$. Recall that this inner product induces a bundle pseudo-metric 
$\langle\cdot,\cdot\rangle_{\mathrm{Ad}(P)}\in\Gamma^{\infty}(\mathcal{M},\mathrm{Ad}(P)^{\ast}\otimes\mathrm{Ad}(P)^{\ast})$ on the adjoint bundle $\mathrm{Ad}(P)=P\times_{\mathrm{Ad}}\mathfrak{g}$, where $\mathrm{Ad}:G\to\mathrm{Aut}(\mathfrak{g})$ denotes the adjoint representation of $G$.
\end{itemize}

Recall that the bundle metric $\langle\cdot,\cdot\rangle_{\mathrm{Ad}(P)}$ induced by the inner product $\langle\cdot,\cdot\rangle_{\mathfrak{g}}$ induces a wedge product on $\mathrm{Ad}$-valued forms, i.e. a family of maps 
\begin{align}\mathrm{tr}(\cdot\wedge\cdot):\Omega^{k}(\mathcal{M},\mathrm{Ad}(P))\times\Omega^{l}(\mathcal{M},\mathrm{Ad}(P))\to\Omega^{k+l}(\mathcal{M})\end{align}
for all $k,l\in\mathbb{N}_{0}$. More precisely, we set for all $\alpha\in\Omega^{k}(\mathcal{M},\mathrm{Ad}(P))$ and $\beta\in\Omega^{l}(\mathcal{M},\mathrm{Ad}(P))$
\begin{equation}\begin{aligned}\mathrm{tr}(\alpha\wedge\beta)_{x}&(v_{1},\dots,v_{k+l}):=\\&\frac{1}{k!l!}\sum_{\sigma\in\mathfrak{S}^{k+l}}\mathrm{sign}(\sigma)\langle\alpha_{x}(v_{\sigma(1)},\dots,v_{\sigma(k)}),\beta_{x}(v_{\sigma(k+1)},\dots,v_{\sigma(k+l)})\rangle_{\mathrm{Ad}(P)_{x}}\end{aligned}\end{equation}
for all $v_{1},\dots,v_{k+l}\in T_{p}\mathcal{M}$ and $x\in\mathcal{M}$. If $G$ is compact and simple and $\langle\cdot,\cdot\rangle_{\mathfrak{g}}$ positive definite, then $\langle\cdot,\cdot\rangle_{\mathfrak{g}}$ is necessarily a (negative) multiple of the Killing form, which is the reason for the notation ``$\mathrm{tr}$'' used above. Using this, we define ``\textit{$BF$-theory}'' via the action functional
\begin{align}\mathcal{S}_{\mathrm{BF}}:\Omega^{d-2}(\mathcal{M},\mathrm{Ad}(P))\times\mathcal{C}(P)\to\mathbb{R},\end{align}
where $\mathcal{C}(P)\subset \Omega^{1}(P,\mathfrak{g})$ denotes the set of all connection $1$-forms on $P$, which is an infinite-dimensional affine space over $\Omega^{1}_{\mathrm{hor}}(P,\mathfrak{g})^{\mathrm{Ad}}$, defined for all $B\in\Omega^{d-2}(\mathcal{M},\mathrm{Ad}(P))$ and for all connection $1$-forms $A\in \mathcal{C}(P)$ by
\begin{align}\mathcal{S}_{\mathrm{BF}}[B,A]:=\int_{\mathcal{M}}\! \operatorname{tr}(B\wedge F_{\mathcal{M}}^{A}).\end{align}
First of all, let us proof that the action of $BF$-theory is invariant under gauge transformations:

\begin{Proposition} (Gauge-Invariance of the $BF$-Action)\newline
Let $f\in\mathcal{G}(P)$ be a global gauge transformation. Then
\begin{align*}\mathcal{S}_{\mathrm{BF}}[f\cdot B,f^{\ast}A]=\mathcal{S}_{\mathrm{BF}}[B,A]\end{align*}
for all $B\in\Omega^{d-2}(\mathcal{M},\mathrm{Ad}(P))$ and for all $A\in\mathcal{C}(P)$, where $f\cdot B$ denotes the action of gauge transformations on forms in $\Omega^{d-2}(\mathcal{M},\mathrm{Ad}(P))$, as defined previously
\end{Proposition}

\begin{proof}
The claim is rather obvious and follows directly from $\mathrm{Ad}$-invariance of $\langle\cdot,\cdot\rangle_{\mathfrak{g}}$. First of all, let $\sigma_{f}\in C^{\infty}(P,G)^{G}$ be the map defined by $f(p)=p\cdot\sigma_{f}(p)$ for all $p\in P$. Then, recall that 
\begin{align}F^{f^{\ast}A}=f^{\ast}F^{A}=\mathrm{Ad}_{\sigma_{f}^{-1}}\circ F^{A}.\end{align}
Furthermore, recall that if we view the curvature as an element in $\Omega^{2}(\mathcal{M},\mathrm{Ad}(P))$, then we have equivalently
\begin{align}F_{\mathcal{M}}^{f^{\ast}A}=f\cdot F_{\mathcal{M}}^{A}.\end{align}
Now, using the definition of the Lagrangian of $BF$-theory, i.e. 
\begin{equation}\begin{aligned}\operatorname{tr}(B&\wedge F_{\mathcal{M}}^{A})_{x}(v_{1},\dots,v_{d})=\\&=\frac{1}{2!(d-2)!}\sum_{\sigma\in\mathfrak{S}^{d}}\mathrm{sign}(\sigma)\langle B_{x}(v_{\sigma(1)},\dots,v_{\sigma(d-2)}),F_{\mathcal{M},x}^{A}(v_{\sigma(d-1)},v_{\sigma(d)})\rangle_{\mathrm{Ad}(P)_{x}}\end{aligned}\end{equation}
for all $x\in\mathcal{M}$ and for all $v_{1},\dots,v_{d}\in T_{x}\mathcal{M}$, the claim follows from the fact that the $\mathrm{Ad}(P)$-bundle metric is gauge-invariant, i.e.
\begin{align}\langle f\cdot [p,v],f\cdot [p,w]\rangle_{\mathrm{Ad}(P)_{x}}=\langle [p,v],[p,w]\rangle_{\mathrm{Ad}(P)_{x}}\end{align}
for all $x\in\mathcal{M}$, for all $p\in P_{x}$ and for all $[p,v],[p,w]\in\mathrm{Ad}(P)$ as well as the fact that the action of $f$ on $\mathrm{Ad}(P)$-valued forms is defined point-wise.
\end{proof}

Besides the gauge-symmetry, it turns out that $BF$-theory has also another type of symmetry thanks to the Bianchi-identity, which can be interpreted as some kind of translational invariance. For this, we need the following preliminary Lemma:

\begin{Lemma}\label{LeibnizBundleForms} (Leibniz Rule for Bundle-Valued Forms)\newline
Let $(V,\rho)$ be a finite-dimensional real representation and $E:=P\times_{\rho} V$ be the associated vector bundle. Furthermore, let $\langle\cdot,\cdot\rangle_{E}$ denote the bundle metric on $E$ induced by a $V$-invariant symmetric and non-degenerate inner product on $V$. Then 
\begin{align*}\mathrm{d}(\mathrm{tr}(\alpha\wedge\beta))=\mathrm{tr}(\mathrm{d}_{A}\alpha\wedge\beta)+(-1)^{k}\mathrm{tr}(\alpha\wedge\mathrm{d}_{A}\beta)\end{align*}
for all $\alpha\in\Omega^{k}(\mathcal{M},E)$ and for all $\beta\in\Omega^{l}(\mathcal{M},E)$.\end{Lemma}

\begin{proof}We prove the claim locally: Let $\{e_{a}\}_{a}\subset\Gamma(U,E)$ be a local frame of $E$ defined on some open subset $U\subset\mathcal{M}$. Then we can write $\alpha\in\Omega^{k}(\mathcal{M},E)$ and $\beta\in\Omega^{l}(\mathcal{M},E)$ in this frame as $\alpha\vert_{U}=\sum_{a}\alpha^{a}\otimes e_{a}$ and $\beta\vert_{U}=\sum_{a}\beta^{a}\otimes e_{a}$ for real-valued coordinate forms $\alpha^{a}\in\Omega^{k}(U)$ and $\beta^{a}\in\Omega^{l}(U)$. Hence, we can write the left-hand side of the equation we would like to prove as
\begin{equation}\label{BFlhs}\begin{aligned}\mathrm{d}&(\mathrm{tr}(\alpha \wedge\beta))\vert_{U}=\mathrm{d}\bigg(\sum_{a,b}(\alpha^{a}\wedge\beta^{b})\langle e_{a},e_{b}\rangle_{E}\bigg )=\\&=\sum_{a,b=1}(\mathrm{d}\alpha^{a}\wedge\beta^{b}+(-1)^{k}\mathrm{d}\alpha^{a}\wedge\mathrm{d}\beta^{b})\langle e_{a},e_{b}\rangle_{E}+(-1)^{k+l}\sum_{a,b=1}\alpha^{a}\wedge\beta^{b}\wedge\mathrm{d}\langle e_{a},e_{b}\rangle_{E},\end{aligned}\end{equation}
where we used the Leibniz rule for real-valued forms and the fact that $\langle e_{a},e_{b}\rangle\in C^{\infty}(U,\mathbb{R})=\Omega^{0}(U)$. In order to write down the left-hand side, recall that by definition of $\mathrm{d}_{A}$, we have that 
\begin{equation}\begin{aligned}\mathrm{d}_{A}\alpha\vert_{U}:&=\sum_{a=1}(\mathrm{d}\alpha^{a}\otimes e_{a}+(-1)^{k}\alpha^{a}\wedge\nabla^{A}e_{a})=\\&=\sum_{a=1}(\mathrm{d}\alpha^{a}+(-1)^{k}\sum_{c=1}(\alpha^{c}\wedge{\omega^{a}}_{c}))\otimes e_{a},\end{aligned}\end{equation}
where we introduced the local connection forms ${\omega^{a}}_{b}\in\Omega^{1}(U)$ corresponding to the connection $\nabla^{A}$, i.e. we write $\nabla^{A}e_{a}=\sum_{b=1}{\omega^{b}}_{a}\otimes e_{b}$. We get a similar expression for $\mathrm{d}_{A}\beta$. Using this, we can write the right-hand side as 
\begin{equation}\label{BFrhs}\begin{aligned}(\mathrm{tr}(\mathrm{d}_{A}\alpha &\wedge\beta)+(-1)^{k}\mathrm{tr}(\alpha\wedge\mathrm{d}_{A}\beta))\vert_{U}=\\=&\sum_{a,b=1}(\mathrm{d}\alpha^{a}\wedge\beta)\langle e_{a},e_{b}\rangle_{E}+(-1)^{k}\sum_{a,b,c=1}(\alpha^{c}\wedge{\omega^{a}}_{c}\wedge\beta^{b})\langle e_{a},e_{b}\rangle_{E}+\\ &+(-1)^{k}\sum_{a,b=1}(\alpha^{a}\wedge\mathrm{d}\beta^{b})\langle e_{a},e_{b}\rangle_{E}+(-1)^{k+l}\sum_{a,b,c=1}(\alpha^{a}\wedge\beta^{c}\wedge{\omega^{b}}_{c})\langle e_{a},e_{b}\rangle_{E}.\end{aligned}\end{equation} 
Comparing (\ref{BFlhs}) and (\ref{BFrhs}), we see that the claimed equality is fulfilled, if and only if 
\begin{align}\label{iff}\sum_{a,b=1}\alpha^{a}\wedge\beta^{b}\wedge\mathrm{d}\langle e_{a},e_{b}\rangle_{E}\stackrel{!}{=}\sum_{a,b,c=1}(\alpha^{c}\wedge\beta^{b}\wedge{\omega^{a}}_{c}+\alpha^{a}\wedge\beta^{c}\wedge{\omega^{b}}_{c}))\langle e_{a},e_{b}\rangle_{E}.\end{align}
Now, recall that the covariant derivative $\nabla^{A}$ is compatible with the induced bundle metric, i.e.
\begin{align}\mathrm{d}(\langle e_{a},e_{b}\rangle_{E})=\langle\nabla^{A}e_{a},e_{b}\rangle_{E}+\langle e_{a},\nabla^{A}e_{b}\rangle_{E}.\end{align}
Using the connection forms ${\omega^{a}}_{b}$ of $\nabla^{A}$, this can be written as
\begin{align}\mathrm{d}(\langle e_{a},e_{b}\rangle_{E})=\sum_{c=1}{\omega^{c}}_{a}\langle e_{c},e_{b}\rangle_{E}+\sum_{c=1}{\omega^{c}}_{b}\langle e_{a},e_{c}\rangle_{E}.\end{align}
Plugging this into the left-hand side of Equation (\ref{iff}), we get the right-hand side, as required.
\end{proof}

\begin{Proposition} (Translational Invariance of $BF$-Theory)\newline
For all $B\in\Omega^{d-2}(\mathcal{M},\mathrm{Ad}(P))$, for all $A\in\mathcal{C}(P)$ and for all $\eta\in\Omega^{d-3}(\mathcal{M},\mathrm{Ad}(P))$ it holds that
\begin{align*}\mathcal{S}_{\mathrm{BF}}[B+\mathrm{d}_{A}\eta,A]=\mathcal{S}_{\mathrm{BF}}[B,A].\end{align*}
\end{Proposition}

\begin{proof}Using Lemma \ref{LeibnizBundleForms} as well as the Bianchi identity $\mathrm{d}_{A}F^{A}_{\mathcal{M}}=0$, we have that
\begin{equation}\begin{aligned}\mathcal{S}_{\mathrm{BF}}[B+\mathrm{d}_{A}\eta,A]&=\int_{\mathcal{M}}\! \operatorname{tr}((B+\mathrm{d}_{A}\eta)\wedge F_{\mathcal{M}}^{A})=\int_{\mathcal{M}}\! \operatorname{tr}(B\wedge F_{\mathcal{M}}^{A})+\int_{\mathcal{M}}\! \operatorname{tr}(\mathrm{d}_{A}\eta\wedge F_{\mathcal{M}}^{A})=\\&=\int_{\mathcal{M}}\! \operatorname{tr}(B\wedge F_{\mathcal{M}}^{A})+\int_{\mathcal{M}}\! \mathrm{d}(\mathrm{tr}(\eta\wedge F_{\mathcal{M}}^{A}))+(-1)^{d}\int_{\mathcal{M}}\! \mathrm{tr}(\eta\wedge \underbrace{\mathrm{d}_{A}F_{\mathcal{M}}^{A}}_{=0})=\\&=\int_{\mathcal{M}}\! \operatorname{tr}(B\wedge F_{\mathcal{M}}^{A})=\mathcal{S}_{\mathrm{BF}}[B,A],\end{aligned}\end{equation}
where we used Stoke's Theorem in the last to last step as well as the fact that $\partial\mathcal{M}=\emptyset$.
\end{proof}

To end the discussion, let us derive the equations of motion of topological $BF$-theory:

\begin{Proposition} (Equation of Motions of $BF$-Theory)\newline
The equation of motions corresponding to the $BF$-action are given by
\begin{align*}F_{\mathcal{M}}^{A}=0\hspace{1cm}\text{and}\hspace{1cm}\mathrm{d}_{A}B=0\end{align*}
for $B\in\Omega^{d-2}(\mathcal{M},\mathrm{Ad}(P))$ and $A\in\mathcal{C}(P)$.
\end{Proposition}

\begin{proof}
Let us first calculate the critical points of the $BF$-action with respect to the $B$-field. Let $\alpha\in\Omega^{d-2}(\mathcal{M},\mathrm{Ad}(P))$, then
\begin{align}\frac{\mathrm{d}}{\mathrm{d}t}\bigg\vert_{t=0}\mathcal{S}_{\mathrm{BF}}[B+t\alpha,A]=\frac{\mathrm{d}}{\mathrm{d}t}\bigg\vert_{t=0}\int_{\mathcal{M}}\! t\cdot\operatorname{tr}(\alpha\wedge F_{\mathcal{M}}^{A})=\int_{\mathcal{M}}\! \operatorname{tr}(\alpha\wedge F_{\mathcal{M}}^{A})\stackrel{!}{=}0,\end{align}
which is fulfilled for all $\alpha$ if and only if $F_{\mathcal{M}}^{A}=0$, by non-degeneracy. To derive the second equation of motion, we use the fact that
\begin{align}F_{\mathcal{M}}^{A+t\alpha}=F_{\mathcal{M}}^{A}+t(\mathrm{d}_{A}\alpha)+\mathcal{O}(t^{2}).\end{align}
for all $\alpha\in\Omega^{2}(\mathcal{M},\mathrm{Ad}(P))$ \cite[p.418]{Hamilton}. As a consequence, using Lemma \ref{LeibnizBundleForms} as well as Stoke's Theorem, we have that 
\begin{align}\frac{\mathrm{d}}{\mathrm{d}t}\bigg\vert_{t=0}\mathcal{S}_{\mathrm{BF}}[B,A+t\alpha]=\int_{\mathcal{M}}\! \operatorname{tr}(B\wedge \mathrm{d}_{A}\alpha)=-\int_{\mathcal{M}}\! \operatorname{tr}(\mathrm{d}_{A}B\wedge \alpha)\stackrel{!}{=}0,\end{align}
which is satisfies for all $\alpha\in\Omega^{1}(\mathcal{M},\mathrm{Ad}(P))$ if and only if $\mathrm{d}_{A}B=0$.
\end{proof}
\newpage
\section{Cellular Complexes, Triangulations and Pseudomanifolds}\label{SimplicalComplexes}
In the second part of this appendix, we give a brief overview of some important mathematical tools for discretization, which is very important for background-independent approaches of quantum gravity, but also in many other areas of physics and mathematics. Some useful references for the following chapter are \cite[Ch.5]{LeeTopologicalManifolds} and \cite[Ch.1]{MunkresBook}.

\subsection{Cellular Decompositions and CW-Complexes}
Let us start by briefly discussing the general concept of cellular decompositions and CW-complexes. First of all, let us introduce the following terminology: An ``\textit{open $k$-cell}'' is any topological space homeomorphic to the open unit ball
\begin{align}B^{k}:=B_{1}(0):=\{x\in\mathbb{R}^{k}\mid\Vert x\Vert<1\}\subset\mathbb{R}^{k}.\end{align} equipped with the standard topology. A closed $k$-cell is any topological space homeomorphic to the closed unit ball $\overline{B^{k}}$. The number $k$ is also called the ``\textit{degree}'' of the cell. Of course, every open ball is an open cell, but there are many more examples: For example, every compact and convex subset with non-empty interior of $\mathbb{R}^{k}$ is a closed $k$-cell and its interior is an open $k$-cell. In particular, every closed interval is a closed $1$-cell and every compact region in the plane bounded by a regular polygon is a closed 2-cell. The idea of a ``cellular decomposition'' of some topological space is that the space can be ``build up'' by gluing cells along their boundaries in order to obtain cells of increasing dimensions. In general, a cellular decomposition of some topological space is a collection of disjoint cells, such their union is equivalent to the given space. However, we need some more restrictions, because an arbitrary partition will tell us nothing about the topology of our given space. Therefore, we assume that the boundary of each cell is attached in some reasonable way to lower-dimensional cells:

\begin{Definition} (Cellular Decompositions and Complexes)\newline
Let $(X,\mathcal{T})$ be a topological Hausdorff space. A cellular decomposition of $X$ is a partition (=open cover consisting of non-empty disjoint subsets) $(c_{i})_{i\in I}$ consisting of open cells of different dimensions, such that for every cell $c_{i}$ of dimension $k\geq 1$, there is a continuous map $f_{i}:\overline{B^{k}}\to X$ such that $f_{i}\vert_{B^{k}}$ is a homeomorphism between $B^{k}$ and $c_{i}$ and such that $f_{i}$ maps the boundary $\partial \overline{B^{k}}$ in a union of cells in $(c_{i})_{i\in I}$ of dimension $<k$.
\end{Definition}

A topological Hausdorff space together with a cellular decomposition is called ``\textit{cellular complex}''. Such a complex is called ``\textit{finite}'', if there are only finitely many cells and ``\textit{locally finite}'', if the collection of open cells is locally
finite. A very general and important type of cellular decompositions are CW-complexes, introduced by J. H. C. Whitehead \cite{CWComplexes}, which are central in algebraic topology. They are defined as follows:

\begin{Definition} (CW-Complex)\newline
Let $(X,\mathcal{T})$ be a topological Hausdorff space with a cellular decomposition $(c_{i})_{i\in I}$. We call it a ``CW-complex'', if the following additional conditions are satisfied:
\begin{itemize}
\item[(1)]For every cell $c_{i}$ of dimension $k\geq 1$, there is a continuous map $f_{i}:\overline{B^{k}}\to X$ such that $f_{i}\vert_{B^{k}}$ is a homeomorphism between $B^{k}$ and $c_{i}$ and such that $f_{i}$ maps the boundary $\partial \overline{B^{k}}$ into a \textit{finite} union of cells in $(c_{i})_{i\in I}$ of dimension $<k$.
\item[(2)]A subset $M\subset X$ is closed if and only if $M\cap f_{i}(\overline{B^{k}})$ is closed for all $i\in I$.
\end{itemize}
\end{Definition}

The name ``CW-complex'' comes from ``\textit{closure-finiteness}'' (condition (1)) and ``\textit{weak topology}'' (condition (2)). Note that every locally finite cellular complex is in particular a CW-complex, as proven in \cite[p.132f.]{LeeTopologicalManifolds}. 

\subsection{Abstract Simplicial Complexes and Their Geometric Realization}
Simplicial complexes are a particular class of CW-complexes consisting only of ``\textit{simplices}'', which are higher-dimensional analogue of triangles and tetrahedra. These complexes can either be defined abstractly or embedded in Euclidean space. Let us start with the abstract definition:

\begin{Definition}\label{SimplComp} (Abstract Simplicial Complex)\newline
Let $\mathcal{V}$ be a finite set. Furthermore, let $\Delta\subset\mathcal{P}(\mathcal{V})\textbackslash\emptyset$ be a collection of non-empty finite subsets of $\mathcal{V}$ satisfying the following two properties:
\begin{itemize}
\item[(1)]$\Delta$ contains all singletons, i.e. $\{v\}\in\Delta$ for all $v\in\mathcal{V}$.
\item[(2)]For any non-empty $\tau\subset\sigma$ for $\sigma\in\Delta$ it holds that $\tau\in\Delta$.
\end{itemize}
We call the pair $(\mathcal{V},\Delta)$ ``(abstract) simplicial complex with vertex set $\mathcal{V}$''.
\end{Definition}

If it is clear from the context, a simplicial complex $(\mathcal{V},\Delta)$ is often simply denoted by $\Delta$. In addition to the above definition, let us fix the following terminology:
\begin{itemize}
\item An element $v\in\mathcal{V}$ is called ``\textit{vertex}'' and an element $\sigma\in\Delta$ is called ``\textit{simplex}''. Any non-empty subset $\tau\subset\sigma$ is called ``\textit{face} of $\sigma$''.
\item The ``\textit{dimension}'' of a simplex $\sigma\in\Delta$ is the number $d\in\mathbb{N}_{0}$ defined by $d:=\vert\sigma\vert-1$. A $d$-dimensional simplex is also called $d$-simplex. A $k$-dimensional face of some $d$-simplex $\sigma$ is called ``\textit{$k$-face of $\sigma$}''. Let us denote the set of all $d$-simplices by $\Delta_{d}$. Vertices are by definition $0$-simplices, i.e. $\mathcal{V}=\Delta_{0}$. The ``\textit{dimension of a simplicial complex} $\Delta$'' is the maximal number $d\in\mathbb{N}_{0}$ such that $\Delta_{d}\neq\emptyset$.
\item The collection of all simplices with dimension smaller equal to some $k\in\{0,\dots,d\}$ is called the ``\textit{$k$-skeleton}'' of the complex $\Delta$.
\item A ``\textit{subcomplex} of $\Delta$'' is a subset $\mathcal{S}\subset\Delta$ such that $(\bigcup\mathcal{S},\mathcal{S})$ is again an abstract simplicial complex. 
\item A simplicial complex $\Delta$ is called ``\textit{finite}'', if $\Delta$ is a finite set. More generally, we say that $\Delta$ is ``\textit{locally finite}'', if every vertex in $\Delta$ only belongs to finitely many simplices.
\end{itemize}

Let now $S\subset\Delta$ be some subset of the abstract simplicial complex $\Delta$, which is not necessarily a subcomplex, i.e. it does not have to be an abstract simplicial complex by itself. Then, one often uses the following terminology:

\begin{itemize}
\item[(1)]The ``\textit{closure} $\mathrm{Cl}_{\Delta}(S)$ of $S$'' is the smallest subcomplex of $\Delta$ containing $S$, i.e. 
\begin{align}\mathrm{Cl}_{\Delta}(S):=\{\sigma\in\Delta\mid \exists\tau\in S:\sigma\subset\tau\}.\end{align}
As an example, the closure of a single $k$-simplex $\sigma$ is the set consisting of the simplex itself together with all its faces. Note that $S$ is a subcomplex of $\Delta$ if and only if $\mathrm{Cl}_{\Delta}(S)=S$.
\item[(2)]The ``\textit{star}'' of a single simplex $\sigma\in\Delta$ is defined to be set of all simplices in $\Delta$ having $\sigma$ as face. More explicitly:
\begin{align}\mathrm{St}_{\Delta}(\sigma):=\{\tau\in\Delta\mid\sigma\subset\tau\}.\end{align}
The ``star of $S$'' is then the union of the stars of all its simplices. Note that the star is in general just a collection of sets and not a subcomplex. Hence, one often defines the ``\textit{closed star}'', which is the subcomplex $\mathrm{Cl}_{\Delta}(\mathrm{St}_{\Delta}(S))$. Note also that some authors define the star directly in this way.
\item[(3)]The ``\textit{link} of $S$'' is defined to be $\mathrm{Lk}_{\Delta}(S):=\mathrm{Cl}_{\Delta}(\mathrm{St}_{\Delta}(S))\textbackslash\mathrm{St}_{\Delta}(\mathrm{Cl}_{\Delta}(S))$. If $\sigma\in\Delta$ is some single simplex, then its links is hence given by
\begin{align}\mathrm{Lk}_{\Delta}(\sigma)=\{\tau\in\mathrm{Cl}_{\Delta}(\mathrm{St}_{\Delta}(\sigma))\mid \tau\cap\sigma=\emptyset\}=\{\tau\in\Delta\mid \tau\cup\sigma\in\Delta\text{ and }\tau\cap\sigma=\emptyset\}.\end{align}
The link of some subset $\mathcal{S}$ is again a subcomplex of $\Delta$.
\end{itemize}

The following figure shows a $2$-dimensional abstract simplicial complex $\Delta$ as well as the star, closed star and link of a vertex $v$:

\begin{figure}[H]
\centering
\includegraphics[scale=0.8]{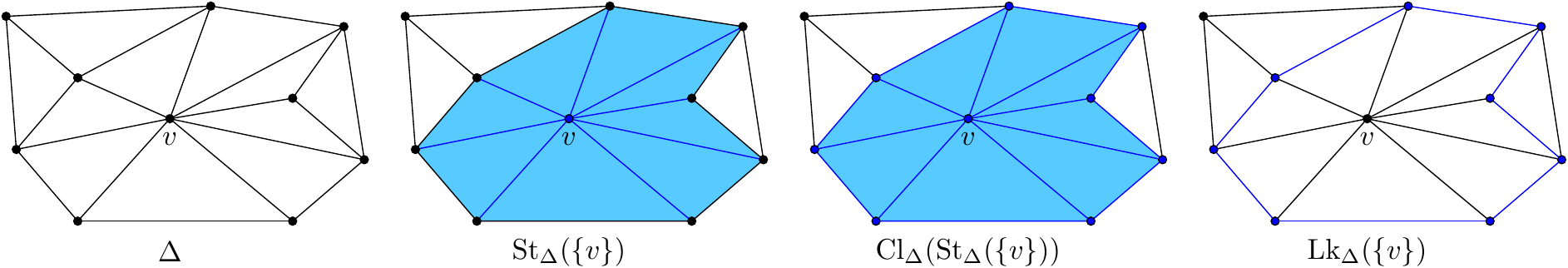}
\caption{A $2$-dimensional simplicial complex and the star $\mathrm{St}_{\Delta}(\{v\})$, closed star $\mathrm{Cl}_{\Delta}(\mathrm{St}_{\Delta}(\{v\}))$ and link $\mathrm{Lk}_{\Delta}(\{v\})$ of a vertex $v$ in $\Delta$ drawn in blue.}
\end{figure}

Abstract simplicial complexes are in general purely combinatorial objects. In order to equip them with a topology, we need to notion of a ``geometric realization''. Let us discuss this in more details: Let $v_{0},\dots,v_{k}\in\mathbb{R}^{n}$ be vectors. They are called ``\textit{affinely independent}'', if they are not contained in any $<k$-dimensional affine subspace of $\mathbb{R}^{n}$, or in other words, if $v_{1}-v_{0},\dots,v_{k}-v_{0}$ are linearly independent vectors.

\begin{Definition} (Euclidean Simplices)\newline
Let $v_{0},\dots,v_{k}\in\mathbb{R}^{n}$ be affinely-independent vectors. The ``Euclidean $k$-simplex spanned by $\{v_{0},\dots,v_{k}\}$'' is the convex hull of these vectors, i.e.
\begin{align*}\Delta^{k}:=\bigg\{\sum_{i=0}^{k}t_{i}v_{i}\,\bigg\vert\, \sum_{i=0}^{k}t_{i}=1\text{ and } 0\leq t_{i}\leq 1\bigg\}.\end{align*}
\end{Definition}

Note that every Euclidean $k$-simplex is a closed $k$-cell, i.e. a topological space (with its subspace topology induced from $\mathbb{R}^{n}$) homeomorphic to the closed unit ball, as defined in the previous section. The vectors $v_{0},\dots,v_{k}\in\mathbb{R}^{n}$ of a simplex $\Delta^{k}$ are called ``\textit{vertices}''. A ``\textit{face}'' of some $k$-simplex $\sigma$ is a simplex spanned by a non-empty subset of vertices of $\sigma$. Next, let us define Euclidean simplicial complexes in the following way \cite[p.149]{LeeTopologicalManifolds}

\begin{Definition} (Euclidean Simplicial Complex)\newline
A Euclidean simplicial complex is a collection $\Delta$ of Euclidean simplices in $\mathbb{R}^{n}$ such that 
\begin{itemize}
\item[(1)]Every face of a simplex in $\Delta$ is again in $\Delta$.
\item[(2)]The intersection of two simplices is either empty or the face of both simplexes.
\item[(3)]$\Delta$ is locally finite.
\end{itemize}
The set $\mathcal{P}:=\bigcup_{\Delta\in\mathcal{K}}\Delta$ is called polyhedron with simplicial decomposition $\Delta$. 
\end{Definition}

Obviously, every Euclidean simplicial complex can in particular be understood as an abstract simplicial complex, where an abstract $k$-simplex $\sigma$ corresponding to a Euclidean $k$-simplex spanned by $v_{0},\dots,v_{k}\in\mathbb{R}^{n}$ is just the set $\sigma=\{v_{0},\dots,v_{k}\}$. The vertex set of this abstract simplicial complex is hence given by the union of all the vertices of the Euclidean $k$-simplices. Furthermore, every Euclidean simplicial complex is a CW-complex of its polyhedron consisting of all the interiors of its simplices. In order to relate an abstract simplicial complex to a Euclidean one, we need the following notion of equivalence:

\begin{Definition} (Isomorphism of Abstract Simplicial Complexes)\newline
Let $(\mathcal{V}_{1},\Delta_{1})$ and $(\mathcal{V}_{2},\Delta_{2})$ be two abstract simplicial complexes. A pair of maps $(\varphi:\Delta_{1}\to\Delta_{2},\varphi_{0}:\mathcal{V}_{1}\to\mathcal{V}_{2})$ is called ``simplicial map'', if it is of the form $\varphi(\{v_{0},\dots,v_{k}\})=\{\varphi_{0}(v_{0}),\dots,\varphi_{0}(v_{k})\}$, i.e. $\varphi$ maps simplices of $\Delta_{1}$ into simplices of $\Delta_{2}$. If both $\varphi$ and $\varphi_{0}$ are bijective, then $(\varphi^{-1},\varphi_{0}^{-1})$ is also a simplicial map and we call $(\varphi,\varphi_{0})$ ``isomorphism''.\end{Definition}

If it clear from the context, we usually just call $\varphi$ simplicial map. Let $\Delta$ be some abstract simplicial complex. We call some Euclidean simplicial complex, whose underlying abstract simplicial complex is isomorphic to $\Delta$, ``\textit{geometric realization of $\Delta$}''. We denote the corresponding polyhedron by $\vert\Delta\vert$. Note that $\vert\Delta\vert$ has a topology, namely the topology induced from $\mathbb{R}^{n}$. We also say that $\Delta$ is a ``\textit{triangulation of the topological space $\vert\Delta\vert$}''. It turns out that such a geometric realization always exists:

\begin{Definition and Proposition}Every finite abstract simplicial complex $\Delta$ has a geometric realization $\vert\Delta\vert$ embedded in $\mathbb{R}^{n}$ for sufficient high $n\in\mathbb{N}$ (in fact, one can choose $n=2\mathrm{dim}(\Delta)+1$), i.e. a Euclidean simplicial complex, which is isomorphic to it.\end{Definition and Proposition}

\begin{proof}See for example Proposition 5.41 in \cite[p.154]{LeeTopologicalManifolds}.\end{proof}

\begin{Remarks}\begin{itemize}\item[]
\item[(a)]For two Euclidean simplicial complexes $\Delta_{1}$ and $\Delta_{2}$, it turns out that a map $\varphi:\Delta_{1}\to\Delta_{2}$ is a simplicial isomorphism if and only if $\varphi$ induces a homeomorphism between $\vert\Delta_{1}\vert$ and $\vert\Delta_{2}\vert$ (see Lemma 2.8. in \cite{MunkresBook}). Hence, the geometric realization of an abstract simplicial complex is unique up to homeomorphism.
\item[(b)]The result can be extended to finite-dimensional, countable and locally finite simplicial complexes. There also more general constructions for arbitrary abstract simplicial complexes, however, in this case we have to extend the meaning of geometric realization.
\end{itemize}\end{Remarks}

As mentioned in the Remark above, if the underlying abstract simplicial complexes of two Euclidean complexes are simplicial isomorphic, then there corresponding polyhedra are homeomorphic. Since every abstract simplicial complex has a geometric realization, we can state the following:

\begin{Proposition} 
If two finite abstract simplicial complexes $\Delta_{1}$ and $\Delta_{2}$ are simplicial isomorphic, then their geometric realizations are homeomorphic.
\end{Proposition}

There is also a stronger notion of isomorphism for abstract simplicial complexes, namely the concept of ``piecewise-linear maps''. In order to state the definition, we need the concept of a ``subdivision'': Let $\Delta_{1}$ and $\Delta_{2}$ be two abstract simplicial complexes. Then $\Delta_{1}$ is called a ``\textit{(simplicial) subdivision}'' of $\Delta_{2}$, if $\vert\Delta_{1}\vert$ and $\vert\Delta_{2}\vert$ are homeomorphic and if every simplex in $\Delta_{1}$ is contained in a simplex of $\Delta_{2}$. Using this, PL-homeomorphisms are defined as follows:

\begin{Definition} (PL-Homeomorphism)\newline
Let $\Delta_{1}$ and $\Delta_{2}$ be two abstract simplicial complexes. A map $\varphi:\Delta_{1}\to\Delta_{2}$ is called ``piecewise-linear'' if there exits subdivisions $\Delta^{\prime}_{1}$ and $\Delta^{\prime}_{2}$, such that $\varphi:\Delta^{\prime}_{1}\to\Delta^{\prime}_{2}$ is a simplicial map. If this induced map is in addition a simplicial isomorphism, then we say that $\varphi$ is ``PL-homeomorphism'' and we call $\Delta_{1}$ and $\Delta_{2}$ ``PL-homeomorphic''.\end{Definition}

In other words, two abstract simplicial complexes are PL-homeomorphic if and only if there exist (simplicial) subdivisions of them, which are simplicial isomorphic. Of course, every simplicial map is a PL-map, but not vice versa.

\begin{Remark}Two PL-homeomorphic abstract simplicial complexes do always have homeomorphic geometric realizations. However, the reverse is in general not true, i.e. homeomorphic realizations does not imply that the corresponding complexes are PL-homeomorphic. As we will see below, even if the geometric realization is a manifold it turns out that in sufficient high dimensions there a many inequivalent PL-structures.\end{Remark}

Let us briefly discuss the notion of a ``\textit{boundary of a simplicial complex}'', which will be used in the following when talking about a triangulation of a manifold with boundary. The notion of a boundary only makes sense for ``pure'' simplicial complexes: A simplicial complex is called ``\textit{pure}'', if every simplex $\sigma\in\Delta$ of dimension $<d$ is the face of some $d$-simplex. Using this, the boundary is defined as follows:

\begin{Definition} (Boundary of a Pure Simplicial Complex)\newline
Let $\Delta$ be a pure abstract simplicial complex. The ``boundary of $\Delta$'' is then defined to be
\begin{align*}\partial\Delta:=\mathrm{Cl}_{\Delta}(\{\tau\in\Delta_{k}\mid k\leq d-1\land \exists ! \sigma\in\Delta_{d}:\tau\subset\sigma\}).\end{align*}
In other words, it is the smallest subcomplex consisting of all $k$-simplices belonging to exactly one $d$-simplex.
\end{Definition}

The boundary is by construction a $(d-1)$-dimensional simplicial complex. Note that the boundary complex does not have to be connected, even in the case $\Delta$ is connected.\\
\\
If $\sigma$ is a $d$-simplex, then we can assign an ``\textit{orientation}'' to it by ordering its vertices (up to even permutations). In general, there are always two possible orientations of a $d$-simplex. A choice of orientation induces an orientation of its faces. As an example, let $[v_{0},\dots,v_{d}]$ be the orientation of a $d$-simplex, i.e. the equivalence class of the ordering of its vertices $v_{0},\dots,v_{d}$ up to even permutation. Then the $(d-1)$-face $\sigma^{\hat{i}}$ with vertices $v_{0},\dots,\hat{v_{i}},\dots,v_{d}$ gets the orientation
\begin{align}(-1)^{i}[v_{0},\dots,\hat{v_{i}},\dots,v_{d}],\end{align}
where a factor of $(-1)$ in front of an equivalence class means to take the opposite orientation. Using this, we can define ``oriented simplicial complexes'' for ``\textit{non-branching}'' simplicial complexes, in which every $(d-1)$-simplex is the face of exactly one or two $d$-simplices.

\begin{Definition} (Orientation of a Non-Branching Simplicial Complex)\newline
Let $\Delta$ be a non-branching abstract simplicial complex. An ``orientation of $\Delta$'' is a choice of an orientation for each $d$-simplex $\sigma\in\Delta_{d}$, such that each $(d-1)$-simplex, which is the face of exactly two $d$-simplices, gets different induced orientations from these two $d$-simplices. If such a choice exists, then we say that $\Delta$ is ``orientable'', otherwise, we say that $\Delta$ is ``non-orientable''.\end{Definition}

Note that for orientable simplicial complexes there are always two choices of an orientation. Last but not least, let us define the Euler characteristic of an abstract simplicial complex. This quantity is for example used to classify $2$-dimensional manifolds.

\begin{Definition} (Euler Characteristic)\newline
Let $\Delta$ be a $d$-dimensional abstract simplicial complex. Then the ``Euler characteristic of $\Delta$'' is the integer $\chi(\Delta)$ defined by
\begin{align*}\chi(\Delta)=\sum_{i=0}^{d}(-1)^{i}\vert\Delta_{i}\vert.\end{align*}
\end{Definition}

The Euler characteristic is a topological invariant, meaning that the Euler characteristic is the same for two abstract simplicial complexes with homeomorphic and in fact even homotopy equivalent geometric realizations. More generally, one can define the Euler characteristic for finite CW-complexes. Also in this case, it is a topological invariant \cite[p.373f.]{LeeTopologicalManifolds}.

\subsection{Triangulations and Piecewise-Linear Manifolds}\label{TriangTheorem}
After having introduced the notion of simplicial complexes, we are now in the position to define the concept of a ``triangulation'' of a manifold, which abstracts the idea of dividing a space into triangles. In full generality, a triangulation of a topological space is defined as follows:

\begin{Definition}\label{Trian} (Triangulation)\newline
Let $(X,\mathcal{T})$ be a topological Hausdorff space. A pair $(\Delta,\varphi)$ consisting of an abstract simplicial complex $\Delta$ and a homeomorphism $\varphi:\vert\Delta\vert\to X$ is called a ``triangulation of $X$''.\end{Definition}

\begin{Remark}A triangulation of a $d$-dimensional manifold $\mathcal{M}$ is always pure and non-branching. Hence, it makes sense to define the boundary complex and to talk about orientability of the corresponding simplicial complex of the triangulation. If $\mathcal{M}$ has a boundary, then the boundary complex triangulates the boundary $\partial\mathcal{M}$. Furthermore, if a triangulable manifold $\mathcal{M}$ is orientable then also its triangulation is and a choice of orientation of the corresponding complex results into a choice of orientation of $\mathcal{M}$.\end{Remark}

For manifolds, there is a slightly stronger notion of triangulations, which is usually called a ``piecewise-linear structure''. It can be defined either using a special type of atlas or using a special type of triangulation. Let us start with the first definition:

\begin{Definition} (Piecewise-Linear Manifolds)\newline
Let $\mathcal{M}$ be a $d$-dimensional topological manifold. A ``piecewise-linear structure on $\mathcal{M}$'', or short ``PL-structure'', is an atlas in which all transition functions are piecewise-linear maps.\end{Definition}

As already mentioned, the concept of a PL-structure can also be defined via a specific type of triangulation, which is usually called a ``combinatorial triangulation'':

\begin{Definition}(Combinatorial Triangulations)\newline
A triangulation $(\Delta,\varphi)$ of a topological manifold possibly with boundary is called ``combinatorial'', if the link of every vertex in $\Delta$ is PL-homeomorphic to the $(d-1)$-dimensional standard PL-sphere or PL-homeomorphic to the $(d-1)$-dimensional standard PL-ball.\end{Definition}

\newpage
\begin{Remarks}\begin{itemize}\item[]
\item[(a)]The ``\textit{standard $PL$-sphere} of dimension $d$ is defined to be the boundary of a $(d+1)$-simplex and the ``\textit{standard $PL$-ball}'' of dimension $d$ is defined to be a single $d$-simplex.
\item[(b)]Adding the detail that links of vertices are homeomorphic to the \textit{standard} PL-sphere or \textit{standard} PL-ball is actually only relevant in $d=4$, since for $d\neq 4$, one can show that the $d$-sphere has a unique PL-structure \cite{PoinConj}. The case $d=4$ is still an open problem known as ``\textit{smooth $4$-dimensional Poincaré conjecture}'' \cite{Exotic4Sphere}.
\end{itemize}\end{Remarks}

\begin{Proposition}\label{CombTrian} (Combinatorial Triangulations and PL-Manifolds)\newline
Every combinatorial triangulation of a topological manifold $\mathcal{M}$ gives rise to a PL-structure and vice versa.\end{Proposition}

\begin{proof}For the direction ``$\Rightarrow$'', see Corollary 1.16 in \cite[p.24f.]{Hudson}. For direction ``$\Leftarrow$'' see the Remark in \cite[p.26]{Hudson}. 
\end{proof}

\begin{Remarks}\begin{itemize}\item[]
\item[(a)]In order to distinguish combinatorial triangulations to the general concept defined in Definition \ref{Trian}, the latter one are sometimes called ``\textit{simplicial triangulations}''. 
\item[(b)]In general, it is also not to hard to see that the geometric realization of \textit{any} simplicial complex with the property that the links of all its vertices represents spheres or balls is a manifold and in fact, by the proposition above, a PL-manifold. Hence, such complexes are also often called ``\textit{combinatorial manifolds}''.
\end{itemize}\end{Remarks}

It turns out that we do not have to make a distinction between triangulation and combinatorial triangulations in sufficient low dimensions, as the following proposition shows:

\begin{Proposition}\label{Comb3} All triangulations of a $d$-dimensional manifold with $d\leq 4$ are combinatorial triangulations.\end{Proposition}

\begin{proof}For $d=2,3$, the claim follows from the triangulation theorems of Radó and Moise stated below. For $d=4$, one can show that the claim is equivalent to the famous ``\textit{Poincaré conjecture}'', which states that any $3$-dimensional closed and simply-connected $3$-manifold is homeomorphic to the $3$-sphere. This conjecture is now known to be true, since it follows from ``Thurston's geometrization conjecture'', which was proven by G. Perelman in 2002/3 \cite{Pere1,Pere2}.\end{proof}

\begin{Remark}
It is important to note that this result is in general not true in $d>4$. In other words, not every triangulation of a manifold with dimension larger then $4$ has the property that links are spheres. In fact, one can show, by using ``Connon's double suspension theorem'', that there are triangulations of $S^{5}$ admitting links, which are not even manifolds. \cite{Edwards}\end{Remark}

For a long time it was an unsolved problem whether all manifolds admit a piecewise-linear structure or a triangulation. Historically, it was conjectured by H. Poincaré \cite{ASComp1} (translated in modern language) that every smooth manifold admits a triangulation. In 1926, it was conjectured by H. Kneser that also all topological manifolds are triangulable \cite{Kneser}. Yet another famous conjecture in this context is the so-called ``\textit{Hauptvermutung}'' \cite{HV} (German for ``\textit{main conjecture}'') by E. Steinitz and H. F. F. Tietze from 1908, which states that any two triangulations of a triangulable topological space are PL-homeomorphic. Poincaré's conjecture turns out to be true, as stated in the theorem below. However, the answer to Kneser's conjecture as well as to the Hauptvermutung is now known to be negative. Nevertheless, it turns out that at least for low dimensions Kneser's conjecture is true:

\begin{Theorem} (Triangulation Theorems for Manifolds)
\begin{itemize}
\item Every $2$- and $3$-dimensional topological manifold admits a piecewise-linear structure, which is unique up to piecewise-linear equivalence. The result for $d=2$ was proven by T. Radó in 1924 \cite{Rado} and the result for $d=3$ by E. E. Moise in 1952 \cite{Moise}. An alternative proof for $d=3$ was shortly after given by R. H. Bing in 1959 \cite{Bing}.
\item Every smooth manifold (and in fact every $C^{1}$-manifold) of arbitrary dimension admits a unique piecewise-linear structure, which is unique up to piecewise-linear equivalence. This was proven by S. S. Cairns and J. H. C. Whitehead \cite{Cairns,Whitehead}.
\end{itemize}
\end{Theorem}

In higher dimensions, topological manifolds do not always admit piecewise-linear structures or triangulations. As an example, in 1982 M. H. Freedman showed that the $4$-dimensional compact and simply-connected manifold $E_{8}$, which is constructed using the root lattice of the simple Lie group with the same name, does not admit any piecewise-linear structure \cite{Fredmann}. In 1990, it was proven by S. Akbulut and J. D. McCarthy that this manifold does not even admit a triangulation \cite{AkbulutMcCarthy}. There are also more general results: First of all, there is a theorem by R. C. Kirby and L. C. Siebenmann that shows that for every $d\geq 5$, there is a $d$-dimensional topological manifold, which does not admit any piecewise-linear structure \cite{KirbySiebenmann}. Furthermore, as recently shown by C. Manolescu, for every $d\geq 5$ there is a $d$-dimensional topological manifold, which does not even admit a triangulation \cite{Manolescu}. For more details, see also \cite{ManolescuLN}. Last but not least, there are also manifolds, which admit triangulations, but not PL-structures. Examples include manifolds of the type $E_{8}\times T^{k}$, where $T^{k}$ denotes the $k$-torus with $k\geq 1$ \cite{Rudyak}.\\
\\
Another nice result concerning the topology of low-dimensional manifolds is the fact that every topological manifold of dimension $d\leq 3$ admits a smooth structure, which is unique up to diffeomorphism. Combining this with the triangulation theorems above shows that in $d\leq 3$, the topological, smooth and piecewise-linear category of manifolds are equivalent. Hence, we can work with arbitrary $3$-manifolds using their topological, smooth and piecewise-linear structure simultaneously. Again, this is in general not true in higher dimensions: It is a well-known fact that there are topological manifolds in dimension $d\geq 4$, which do not admit any smooth structure (e.g. $E_{8}$). Furthermore, it is also well known that there are topological manifolds of dimension $d=4$ (and in fact only in this dimension as shown by R. C. Kirby and L. C. Siebenmann \cite{KirbyBook}) with an infinite number of inequivalent smooth structures and hence also with an infinite number of inequivalent PL-structures. Examples include the ``$K3$-surface'' and, as more recently discovered, the manifold $S^{2}\times S^{2}$ \cite{S2S2}. Another subtlety in higher dimensions is the fact that a PL-structure does not have to come from a smooth structure. An explicit example for this observation is the $10$-dimensional ``Kervaire manifold'', which admits a PL-structure, but not a smooth structure \cite{Kervaire}. Again, this is only true in sufficient high dimensions, since one can show that every PL-manifold of dimension $d\leq 7$ has at least one compatible smooth structure and for $d<7$, this smooth structure is also unique up to diffeomorphism (``Munkres-Hirsch-Mazur obstruction theory'', \cite{Munkres,HirschMazur}). In other words, for $d\leq 6$, the smooth and piecewise-linear category are equivalent. Let us stress that the latter statement does not mean that any manifold of dimension $d\leq 6$ has a unique smooth structure, but rather that the number of inequivalent smooth structures is the same as the number of inequivalent PL-structures for $d\leq 6$. More details can be found in the first chapter of the review \cite{TrianReview}.

\subsection{Pseudomanifolds}\label{Pseudomanifolds}
In quantum gravity approaches including a sum over topologies in their perturbative expansion, like tensor models and group field theories, one often has to deal with more singular combinatorial objects, which are not manifolds. Hence, we have to discuss some generalizations. A very important concept in this context are pseudomanifolds, which are defined as follows:

\begin{Definition} (Pseudomanifolds)\newline
Let $\Delta$ be a finite abstract $d$-dimensional simplicial complex. We call its geometric realization $\vert\Delta\vert$ a ``$d$-dimensional pseudomanifold'', if the following conditions are fulfilled:
\begin{itemize}
\item[(1)]$\Delta$ is ``\textit{pure}'', i.e. every simplex $\sigma\in\Delta$ of dimension $<d$ is the face of some $d$-simplex.
\item[(2)]$\Delta$ is ``\textit{non-branching}'', i.e. every $(d-1)$-simplex is face of exactly one or two $d$-simplices.
\item[(3)]$\Delta$ is ``\textit{strongly-connected}'', i.e. for every two $d$-simplices $\sigma,\tau\in\Delta_{d}$, there is a sequence of $d$-simplices $\sigma=\sigma_{1},\sigma_{2},\dots,\sigma_{k}=\tau$ such that $\sigma_{l}\cap\sigma_{l+1}$ is a $(d-1)$-simplex $\forall l$.
\end{itemize}
\end{Definition}

As before, the boundary of a pure simplicial complex is the smallest subcomplex consisting of all the $k$-simplices with $k\leq d-1$, which belong to precisely one $d$-simplex. Combining this with property (2), we hence see that the boundary $\partial\Delta$ is exactly the subcomplex consisting of all the $(d-1)$-simplices, which are the face of only on $d$-simplex as well as all their lower-dimensional faces. We call the geometric realization of the subcomplex $\partial\Delta$ ``\textit{boundary of the pseudomanifold}''. If $\partial\Delta\neq\emptyset$, then we call $\vert\Delta\vert$ ``\textit{pseudomanifold with boundary}'', otherwise we call $\vert\Delta\vert$ ``\textit{pseudomanifold without boundary}''.\\
\\
One can easily show that every compact, connected and triangulable manifold is a pseudomanifold. However, the converse is in general not true. As an example, pseudomanifolds can contain some kind of singularities. An often cited example is the ``\textit{pinched torus}'', which is a $2$-dimensional pseudomanifold obtained by identifying two distinct points on the $2$-sphere. \\
\\
An important class of pseudomanifolds are ``normal pseudomanifolds'', which are defined as follows:

\begin{Definition} (Normal Pseudomanifolds)\newline
Let $\vert\Delta\vert$ be a $d$-dimensional pseudomanifold. We call it ``normal'' if the links of all simplices of dimension $k\in\{1,\dots,d-2\}$ define themselves pseudomanifolds.\end{Definition}

The crucial condition in the normality-property is strongly-connectedness: In general, every link of a pseudomanifold is pure and non-branching, but can fail to be strongly-connected. In fact, they can even fail to be connected at all. For a proof of this statement, see for example the appendix in \cite{GurauColouredGFTPseudo}. An example of a non-normal pseudomanifold is the pinched torus mentioned above, since the link of the vertex, which identifies the two distinct point on the $2$-sphere, consists of two distinct circles and hence is not connected. Furthermore, it is important to note that while the boundary of any pseudomanifold is in general a pseudomanifold without boundary, the boundary of a \textit{normal} pseudomanifold does not have to be \textit{normal} itself, even if the boundary is connected. An example is discussed in Section \ref{SecBubbles}.\\
\\
For a more detailed discussion of pseudomanifolds with or without boundary in general, see for example the classical textbook \cite[p.90ff.]{SeifertTopology}.
\newpage
\section{Useful Concepts of Group Theory and Harmonic Analysis}
In this chapter of the appendix, we discuss some useful definitions and concepts of group theory and harmonic analysis. In the first part, we briefly review the definition of the Haar measure on locally-compact Lie groups as well as the Theorem of Peter-Weyl. Furthermore, we explain the relation between invariant elements and intertwiners mentioned in our discussion of spin networks. Last but not least, we introduce the non-commutative group Fourier transform, which we have mentioned in our discussion of different representations of the Boulatov fields, and review some of its properties. 

\subsection{The Haar Measure of Compact Lie Groups}\label{HaarMeasureAppendix}
Let $G$ be a locally compact Lie group. Then there is a very natural choice of measure on $G$, which respects the group structure and is usually called the ``Haar measure''. Historically, this measure was introduced by A. Haar in 1933 \cite{Haar} and A. Weil \cite{Weil} proved a general existence and uniqueness theorem, which states the following\footnote{The original proof by A. Weil uses the axiom of choice in form of Tychonoff’s theorem. However, it was shown by H. Cartan that this can also be avoided \cite{Cartan}.}: 

\begin{Definition and Theorem} (Haar-Measure: Existence and Uniqueness)\newline
Let $G$ be a locally-compact Lie group. Then there exists a unique (up to a positive multiplication constant) non-trivial Radon measure $\mu:\mathcal{B}(G)\to [0,\infty]$ on the Borel-$\sigma$-algebra $\mathcal{B}(G)$, which is ``left-invariant'', i.e. $\mu(gM)=\mu(M)$ for all $g\in G$ and $M\in\mathcal{B}(G)$, where $gM$ denotes the set $gM:=\{gm\mid m\in M\}$. This measure is usually called the ``(left-invariant) Haar measure''.\end{Definition and Theorem}

The integration with respect to the Haar measure is usually simply denoted by 
\begin{align}\int_{G}\,f(g)\,\mathrm{d}g:=\int_{G}\,f(g)\,\mathrm{d}\mu(g).\end{align}
Note that left-invariance of the Haar measure implies on the level of integration that 
\begin{align}\int_{G}f(g)\,\mathrm{d}g=\int_{G}f(hg)\,\mathrm{d}g\end{align}
for all measurable functions $f:G\to\mathbb{C}$ and for all $h\in G$. Similarly, one can show the existence of a right-invariant Radon measure on locally-compact groups, i.e. a measure $\mu:\mathcal{B}(G)\to [0,\infty]$ satisfying $\mu(Mg)=\mu(M)$ for all $g\in G$ and $M\in\mathcal{B}(G)$, called the ``\textit{right-invariant Haar measure}''. If the left- and right-invariant Haar measures of some group $G$ agree, we get a well-defined ``\textit{bi-invariant}'' Haar measure. Groups with this property are usually called ``\textit{unimodular}''. For such groups, the Haar measure satisfies in addition $\mu(M)=\mu(M^{-1})$, where $M^{-1}:=\{m^{-1}\mid m\in M\}$ for all $M\in\mathcal{B}(G)$, or in other words, 
\begin{align}\int_{G}f(g)\,\mathrm{d}g=\int_{G}f(g^{-1})\,\mathrm{d}g\end{align}
for all functions $f:G\to\mathbb{C}$. This is obviously the case for abelian groups as well as for discrete groups, but it turns out that it is also the case for compact Lie groups. Another property of the Haar measure of compact groups is that it is finite, i.e. $\mu(G)<\infty$, and hence, we can normalize the Haar measure such that $\mu(G)=1$, or in other words, $\int_{G}\,\mathrm{d}g=1$.

\begin{Examples}\begin{itemize}\item[]
\item[(1)]If $G$ is a finite group, then the Haar measure is given by the counting measure. 
\item[(2)]The Lebesgue measure of $\mathbb{R}^{d}$ and $\mathbb{C}^{d}$ is the Haar measure on the additive groups $(\mathbb{R}^{d},+)$ and $(\mathbb{C}^{d},+)$. These two groups are only locally but not globally compact. However, they are abelian and hence unimodular.
\item[(3)]Recall that an element of $g$ of $\mathrm{SU}(2)$ can be parametrized as $g=\mathrm{exp}(i\theta \vec{n}\cdot\vec{\sigma})$
with $\theta\in [0,\pi]$, $\vec{n}\in S^{2}$ and where $\vec{\sigma}=(\sigma_{1},\sigma_{2},\sigma_{3})$ denote the Pauli matrices. Then the normalized Haar measure on $\mathrm{SU}(2)$ is given by
\begin{align}\mathrm{d}g=\frac{2}{\pi}\mathrm{d}\theta (\sin(\theta))^{2}\mathrm{d}^{2}\vec{n}\end{align}
where $\mathrm{d}^{2}\vec{n}$ denotes the normalized measure on the $2$-sphere $S^{2}$. \cite[Appendix B]{FreidelPonzanoRegge1}
\end{itemize}\end{Examples}

\subsection{Theorem of Peter-Weyl}\label{PeterWeylAppendix}
Consider a compact Lie group $G$ together with its bi-invariant Haar measure. Then we can define the Hilbert space $(L^{2}(G),\langle\cdot,\cdot\rangle_{L^{2}})$ consisting of (equivalence classes of almost everywhere equivalent) measurable functions $f:G\to\mathbb{C}$, which are square-integrable, i.e.
\begin{align}\int_{G}\,\vert f(g)\vert^{2}\,\mathrm{d}g<\infty,\end{align}
where the inner product $\langle\cdot,\cdot\rangle_{L^{2}}$ on this space is for two elements $f,k\in L^{2}(G)$ defined by
\begin{align}\langle f,k\rangle_{L^{2}}:=\int_{G}\,\overline{f(g)}k(g)\,\mathrm{d}g.\end{align}
In this section, we would like to briefly discuss the famous ``Theorem of Peter-Weyl'', which is central in the discussion of compact Lie groups and harmonic analysis. Let us start by recalling some basic definitions. A ``\textit{unitary representation of $G$} '' on some Hilbert space $(\mathcal{H},\langle\cdot,\cdot\rangle)$ is a Lie group homomorphism of the type $\rho:G\to\mathcal{U}(\mathcal{H})$, where $\mathcal{U}(\mathcal{H})\subset\mathrm{Aut}(\mathcal{H})$ denotes the set of unitary operators on $\mathcal{H}$. Two unitary representations $\rho_{1}:G\to\mathcal{U}(\mathcal{H}_{1})$ and $\rho_{2}:G\to\mathcal{U}(\mathcal{H}_{1})$ are called ``\textit{unitary equivalent}'', if there exists a unitary operator $U:\mathcal{H}_{1}\to\mathcal{H}_{2}$ such that $U\circ\rho_{1}(g)=\rho_{2}(g)\circ U(g)$ for all $g\in G$. Last but not least, a unitary representation $\rho:G\to\mathcal{U}(\mathcal{H})$ is called ``\textit{irreducible}'', if there exists no non-trivial invariant subspace. Now, let us denote by $\Lambda$ the set of equivalence classes of unitary and irreducible representations. Then, the Theorem of Peter-Weyl, proven by H. K. H. Weyl and F. Peter in 1927 \cite{PeterWeyl}, states the following \cite{Dieck}:

\begin{Theorem} (of Peter-Weyl)\newline
Let $G$ be a compact Lie group. Then the set $\{\sqrt{\operatorname{dim}(\rho)}\rho_{ij}\mid \rho\in\Lambda\}$
is an orthonormal basis of the Hilbert space $L^{2}(G)$, where $\rho_{ij}:G\to\mathbb{C}$ denote the ``matrix elements'' of a representation $\rho\in\Lambda$, which are defined via $\rho_{ij}(g):=\langle e_{i}\vert\rho(g)e_{j}\rangle$ for some orthonormal basis $\{e_{i}\}_{i\in I}$ of the corresponding Hilbert space $\mathcal{H}_{\rho}$.\end{Theorem}

In the context of spin networks, it is useful to state the above theorem in a more abstract form. For this, let us first of all define the ``\textit{right regular representation}'' $R:G\to\mathcal{U}(L^{2}(G))$ and the ``\textit{left-regular representation}'' $L:G\to\mathcal{U}(L^{2}(G))$ by
\begin{align}(R_{g}(f))(h):=f(hg)\hspace{1cm}\text{and}\hspace{1cm}(L_{g}(f))(h):=f(g^{-1}h)\end{align}
for all $f\in L^{2}(G)$ and for all $h,g\in G$. Note that both of these representations are unitary precisely because the normalized Haar measure of a compact Lie group is bi-invariant. Now, the important point is that these two representations can be used to decompose the Hilbert space $L^{2}(G)$ in terms of finite-dimensional subspaces, which are invariant under these representations. To see this, first of all observe the following: For some given $(\rho,\mathcal{H})\in\Lambda$, there are $d_{\rho}:=\operatorname{dim}_{\mathbb{C}}(\mathcal{H})$ finite-dimensional Hilbert spaces, namely the subspaces spanned by the rows of the representation, i.e.
\begin{align}\mathcal{H}_{\rho}^{i}:=\operatorname{span}\{\rho_{ij}\mid 1\leq j\leq d_{\rho}\}.\end{align}
These subspaces are clearly $R$-invariant and the restriction of $R$ to $\mathcal{H}_{i}^{\rho}$ coincides with $\rho$, because $R_{g}\vert_{\mathcal{H}_{i}^{\rho}}$ is realized by the same matrix as $\rho(g)$, as the following short calculation shows:
\begin{align}R_{h}\bigg (\sum_{j}c_{j}\rho_{ij}(g)\bigg )=\sum_{j}c_{j}\rho_{ij}(gh)=\sum_{j,k}c_{j}\rho_{ij}(g)\rho_{jk}(h)\end{align}
and hence $R_{h}:\mathcal{H}_{\rho}^{i}\to \mathcal{H}_{\rho}^{i}, v\mapsto \rho(h)(v)$. Now, the Theorem of Peter-Weyl is equivalent to say that there is the following isomorphism of representations:
\begin{align}L^{2}(G)\cong \bigoplus_{\rho\in\Lambda}\bigoplus_{i=1}^{d_{\rho}}\mathcal{H}_{\rho}^{i}.\end{align}
A similar construction can be done for the left-regular representation, i.e.
\begin{align}L^{2}(G)\cong \bigoplus_{\rho\in\Lambda}\bigoplus_{i=1}^{d_{\rho}}(\mathcal{H}_{\rho}^{i})^{\ast},\end{align}
where $(\mathcal{H}_{\rho}^{i})^{\ast}$ are spanned by the complex-conjugate rows of matrix elements. Furthermore, we can also define another representation, namely by regarding $L^{2}(G)$ as the representation space of a unitary representation of the group $G\times G$, i.e.
\begin{align}\tau:G\times G\to \mathcal{U}(L^{2}(G)), (g,h)\mapsto R_{g}L_{h}=L_{h}R_{g}\end{align}
which is called the ``\textit{bi-regular representation}''. Again, this representation is unitary by bi-invariance of the Haar measure on $G$. The Peter-Weyl theorem then states that
\begin{align}L^{2}(G)\cong\bigoplus_{\rho\in\Lambda}\mathcal{H}_{\rho}\otimes (\mathcal{H}_{\rho})^{\ast}.\end{align}
More details can be found in \cite{Dieck} and \cite[p.32ff.]{ProvenziThesis}.

\subsection{Invariant Elements and Intertwiners}\label{InvEleInter}
First of all, recall that the set of invariant elements of some given compact Lie group $G$ and some given representation $(\rho,\mathcal{H})$ of $G$ is defined by
\begin{align}\mathrm{Inv}_{\rho}(\mathcal{H}):=\{v\in \mathcal{H}\mid \rho(g)v=v\,\forall g\in G\}.\end{align}
Now, suppose that $(\rho_{1},\mathcal{H}_{1})$ and $(\rho_{2},\mathcal{H}_{2})$ are two unitary representations of some given compact Lie group $G$. We then can construct a representation of $G$ on the set of linear maps $\mathrm{Hom}(\mathcal{H}_{1},\mathcal{H}_{2})$
\begin{align}\eta: G&\to \mathrm{Aut}(\mathrm{Hom}(\mathcal{H}_{1},\mathcal{H}_{2}))\end{align}
by $\eta(g)(f):=\rho_{1}(g)\circ f\circ \rho_{2}(g^{-1})$ for all $f\in\mathrm{Hom}(\mathcal{H}_{1},\mathcal{H}_{2})$. Now, it turns out that the set of invariant elements of this representations is related to the set of ``\textit{intertwiners}'' from $\mathcal{H}_{1}$ to $\mathcal{H}_{2}$, i.e. the subset $\mathrm{Int}(\mathcal{H}_{1},\mathcal{H}_{2})\subset\mathrm{Hom}(\mathcal{H}_{1},\mathcal{H}_{2})$ of linear maps $f:\mathcal{H}_{1}\to\mathcal{H}_{2}$ satisfying
\begin{align}\rho_{2}(g)\circ f=f\circ\rho_{1}(g)\end{align}
for all $g\in G$. More precisely, let us proof the following claim:

\begin{Proposition} (Invariant Elements and Intertwiners)\newline
It holds that
\begin{align*}\label{eq}\mathrm{Inv}_{\eta}(\mathrm{Hom}(\mathcal{H}_{1},\mathcal{H}_{2}))=\mathrm{Int}(\mathcal{H}_{1},\mathcal{H}_{2}).\end{align*}
\end{Proposition}

\begin{proof}\item[]
\begin{itemize}
\item[``$\subset$'']Let $f\in\mathrm{Inv}_{\eta}(\mathrm{Hom}(\mathcal{H}_{1},\mathcal{H}_{2}))$, i.e. $f\in\mathrm{Hom}(\mathcal{H}_{1},\mathcal{H}_{2})$ is such that $\eta(g)(f)=f$ for all $g\in G$. In order to show that $f\in \mathrm{Int}(\mathcal{H}_{1},\mathcal{H}_{2})$, we have to verify the intertwining property, i.e.
\begin{align}\rho_{2}(g)\circ f=f\circ\rho_{1}(g). \end{align}
Since $f$ is an invariant element, it trivially satisfies the following equation:
\begin{align}f=\int_{G}\,\mathrm{d}g\,\rho_{2}(g)\circ f\circ\rho_{1}(g^{-1}),\end{align}
where $\mathrm{d}g$ denotes the normalized Haar measure of $G$. As a consequence, we have that 
\begin{equation}\begin{aligned}\rho_{2}(g)\circ f &=\rho_{2}(g)\circ \int_{G}\,\mathrm{d}h\,\rho_{2}(h)\circ f\circ\rho_{1}(h^{-1})=\int_{G}\,\mathrm{d}h\,\rho_{2}(gh)\circ f\circ\rho_{1}(h^{-1})=\\&=\int_{G}\,\mathrm{d}h\,\rho_{2}(gh)\circ f\circ\rho_{1}(h^{-1}g^{-1}g)=\int_{G}\,\mathrm{d}h\,\rho_{2}(h)\circ f\circ\rho_{1}(h^{-1}g)=\\&=\int_{G}\,\mathrm{d}h\,\rho_{2}(h)\circ f\circ\rho_{1}(h^{-1})\circ\rho(g)=f\circ\rho_{1}(g),\end{aligned}\end{equation}
where we used that $\mathrm{d}(gh)=\mathrm{d}h$, by bi-invariance of the Haar measure.
\item[``$\supset$'']Let $f\in\mathrm{Int}(\mathcal{H}_{1},\mathcal{H}_{2})$. Then we have that
\begin{align}\eta(g)(f)=\rho_{1}(g)\circ f\circ\rho_{2}(g^{-1})=f\end{align}
where we used the intertwining property. This shows that $f\in\mathrm{Inv}_{\eta}(\mathrm{Hom}(\mathcal{H}_{1},\mathcal{H}_{2}))$.
\end{itemize}\end{proof}
In the case when $\mathcal{H}_{1}$ and $\mathcal{H}_{2}$ are finite-dimensional, there is the canonical isomorphism $\mathcal{H}_{1}\otimes\mathcal{H}_{2}^{\ast}\cong\mathrm{Hom}(\mathcal{H}_{1},\mathcal{H}_{2})$. As consequence, we have the following immediate Corollary:

\begin{Corollary} There is the following isomorphism:
\begin{align*}\mathrm{Inv}_{\eta}(\mathcal{H}_{1}\otimes\mathcal{H}_{2}^{\ast})\cong\mathrm{Int}(\mathcal{H}_{1},\mathcal{H}_{2}).\end{align*}\end{Corollary}

This can easily be seen by the fact that we not only have the isomorphism $\mathcal{H}_{1}\otimes\mathcal{H}_{2}^{\ast}\cong\mathrm{Hom}(\mathcal{H}_{1},\mathcal{H}_{2})$ on the level of vector spaces, but there is in fact also a isomorphism of representations (i.e. a bijective intertwiner) between the representation $\eta$ on $\mathrm{Hom}(\mathcal{H}_{1},\mathcal{H}_{2})$ and the (inner) tensor product $\rho_{1}\otimes\rho_{2}^{\ast}$ on $\mathcal{H}_{1}\otimes\mathcal{H}_{2}^{\ast}$, which we hence also denote by $\eta$ and where $\rho_{2}^{\ast}$ denotes the dual representation of $\rho_{2}$.

\subsection{Non-Commutative Group Fourier Transform}\label{NCFourier}
The non-commutative group Fourier transform, firstly introduced in \cite{FreidelPonzanoRegge3} for the Lie group $\mathrm{SO}(3)$, is a natural choice of mapping between the configuration space and the corresponding momentum space of a Lie group. Furthermore, it establishes a correspondence between group field theories and non-commutative quantum field theory. For a review of the general notion of this type of Fourier transform see for example \cite{Raasakka1,Raasakka2}.\\
\\
Let $G$ be a Lie group, usually assumed to be compact, or at least locally compact, such that we can define a Haar measure on $G$. The cotangent bundle $T^{\ast}G$ is diffeomorphic to $G\times\mathfrak{g}^{\ast}$, where $\mathfrak{g}$ denotes the Lie algebra corresponding to $G$ and $\mathfrak{g}^{\ast}$ its dual, because the tangent bundle of a Lie group is a trivial vector bundle. The definition of the non-commutative Fourier transform is based on maps of the type
\begin{equation}\begin{aligned}e:T^{\ast}G\cong G\times\mathfrak{g}^{\ast}&\to \mathbb{C},\\ (g,X)&\mapsto e_{g}(X)\end{aligned}\end{equation}
called the ``\textit{plane-waves of $G$}''. These maps are defined in such a way that the delta function corresponding to the group $G$ can be written as 
\begin{align}\delta_{G}(g)=\int_{\mathfrak{g}^{\ast}}\frac{\mathrm{d}^{d}x}{(2\pi)^{d}}\,e_{g}(x).\end{align}
where $d=\mathrm{dim}(G)=\mathrm{dim}_{\mathbb{R}}(\mathfrak{g})$ and where $\mathrm{d}^{d}x$ denotes the Lebesgue-measure on $\mathfrak{g}^{\ast}$. Another very important object in the definition of the non-commutative group Fourier transform is the so-called ``\textit{star product}'', denoted by $\star$, which is a product respecting the group structure. For two plane waves it is defined such that
\begin{align}(e_{g}\star e_{h})(x):=e_{g}(x)\star e_{h}(x):=e_{gh}(x)\end{align}
for all $g,h\in G$ and for all $x\in\mathfrak{g}^{\ast}$. Note that this product is associative, but in general not commutative. Furthermore, a unit with respect to this product if given by $e_{e}(x)$, where the lower $e$ denotes the neutral element of $G$. Now let $\varphi\in L^{2}(G)$. Then we define the non-commutative group Fourier transform via
\begin{align}\label{Fourier}\mathcal{F}(\varphi)(x):=\hat{\varphi}(x):=\int_{G}\,\mathrm{d}g\,e_{g^{-1}}(x)\varphi(g).\end{align}
Let us denote the image of this transformation by $L^{2}_{\star}(\mathfrak{g}^{\ast}):=\mathcal{F}(L^{2}(G))$. Using the above definition, we can extend the $\star$-product to all of $L^{2}_{\star}(\mathfrak{g}^{\ast})$. Furthermore, we can equip the space $L^{2}_{\star}(\mathfrak{g}^{\ast})$ with an inner product defined via 
\begin{align}\langle\varphi,\psi\rangle_{L^{2}_{\star}(\mathfrak{g}^{\ast})}:=\int_{\mathfrak{g}^{\ast}}\,\frac{\mathrm{d^{d}x}}{(2\pi)^{d}}\,\overline{\varphi(x)}\star\psi(x)\end{align}
for all $\varphi,\psi\in L^{2}_{\star}(\mathfrak{g}^{\ast})$. The inverse transform is given by
\begin{align}\label{InverseFourier}\mathcal{F}^{-1}(\hat{\varphi})(g)=\varphi(g)=\int_{\mathfrak{g}^{\ast}}\,\frac{\mathrm{d^{d}}x}{(2\pi)^{d}}\,e_{g}(x)\star \varphi(x)\end{align}
as the following short calculation shows:
\begin{equation}\begin{aligned}\int_{\mathfrak{g}^{\ast}}\,\frac{\mathrm{d^{d}}x}{(2\pi)^{d}}\,e_{g}(x)\star \hat{\varphi}(x)&=\int_{G}\,\mathrm{d}h\,\int_{\mathfrak{g}^{\ast}}\,\frac{\mathrm{d^{d}}x}{(2\pi)^{d}}\,e_{g}(x)\star e_{h^{-1}}(x)\varphi(h)=\\&=\int_{G}\,\mathrm{d}h\,\int_{\mathfrak{g}^{\ast}}\,\frac{\mathrm{d^{d}}x}{(2\pi)^{d}}\,e_{gh^{-1}}(x)\varphi(h)=\\&=\int_{G}\,\mathrm{d}h\,\delta_{G}(gh^{-1})\varphi(g)=\varphi(g).\end{aligned}\end{equation}
If the plane-waves satisfy the condition $e_{g^{-1}}(x)=\overline{e_{g}(x)}$, then the non-commutative group Fourier transform is actually an isometry, i.e. it satisfies
\begin{align}\langle\varphi,\psi\rangle_{L^{2}(G)}=\langle\mathcal{F}(\varphi),\mathcal{F}(\psi)\rangle_{L^{2}_{\star}(\mathfrak{g}^{\ast})}=\langle\hat{\varphi},\hat{\psi}\rangle_{L^{2}_{\star}(\mathfrak{g}^{\ast})},\end{align}
which can easily be shown by a straightforward calculation:
\begin{equation}\begin{aligned}\langle\hat{\varphi},\hat{\psi}\rangle_{L^{2}_{\star}(\mathfrak{g}^{\ast})}&=\int_{\mathfrak{g}^{\ast}}\,\frac{\mathrm{d}^{d}x}{(2\pi)^{d}}\,\overline{\hat{\varphi}(x)}\star \hat{\psi}(x)=\\&=\int_{\mathfrak{g}^{\ast}}\,\frac{\mathrm{d}^{d}x}{(2\pi)^{d}}\,\int_{G}\,\mathrm{d}g\,\int_{G}\,\mathrm{d}h\,\overline{\varphi(g)}\psi(h)\,\overline{e_{g^{-1}}(x)}\ast e_{h^{-1}}(x)=\\&=\int_{\mathfrak{g}^{\ast}}\,\frac{\mathrm{d}^{d}x}{(2\pi)^{d}}\,\int_{G}\,\mathrm{d}g\,\int_{G}\,\mathrm{d}h\,\overline{\varphi(g)}\psi(h) e_{gh^{-1}}(x)
=\\&=\int_{G}\,\mathrm{d}g\,\int_{G}\,\mathrm{d}h\,\delta_{G}(gh^{-1})\overline{\varphi(g)}\psi(h)=\\&=\int_{G}\,\mathrm{d}g\,\overline{\varphi(g)}\psi(h)=\langle\varphi,\psi\rangle_{L^{2}(G)}\end{aligned}\end{equation}
In other words, the non-commutative Fourier transform is a unitary operator $\mathcal{F}:L^{2}(G)\to L^{2}_{\star}(\mathfrak{g}^{\ast})$, which also establishes the analogy with the standard Fourier transform from functional analysis. Note also that there are some slightly different conventions in the literature: As an example, some authors define the non-commutative group Fourier transform with $e_{g}(x)$ in Formula (\ref{Fourier}) and hence with $e_{g^{-1}}(x)$ in Formula (\ref{InverseFourier}), e.g. in \cite{OritiNonCom}. \\
\\
The explicit form of the plane-waves $e_{g}(X)$ depends on the specific group and can be highly non-trivial. Usually they take the form 
\begin{align}e_{g}(x)=\mathrm{exp}\bigg(i\sum_{i=1}^{d}Z_{i}(g)x^{i}\bigg),\end{align}
where $Z(g)$ denote some coordinates of the group manifold $G$. As an example, for the group $\mathrm{SO}(3)$ one could use $Z^{i}(g)=-\frac{i}{2}\operatorname{tr}(g\sigma^{i})$, where $\sigma^{i}$ denote the Pauli-matrices. The case of $\mathrm{SU}(2)$ is more subtle. One has to be careful in order to avoid the same problem, which we had in the derivation of the Ponzano-Regge model, namely that using plane-waves defined in a similar manner as for $\mathrm{SO}(3)$ would lead to the ambiguity $Z^{i}(g^{-1})=Z^{i}(-g)$ and hence, we would not get a bijective Fourier transform in this case. In order to avoid this issue, one has to multiply the above formula by a polarization vector keeping track of the hemisphere to which $g$ belongs. Also the star-product has to be modified accordingly. Alternatively, one can also work with $\mathrm{SO}(3)$ instead and multiply the exponent by a sign factor $\operatorname{sign}(\operatorname{tr}(g))$. In other words, we basically use the isomorphism $\mathrm{SO}(3)\cong\mathrm{SU}(2)/\mathbb{Z}_{2}$. For more details about the specific case of $G=\mathrm{SU}(2)$ see for example \cite{Freidel07,Livine09,DupuisSpinors}.\\
\\
Last but not least, the delta function\footnote{For general locally compact Lie groups, the delta function of $L^{2}_{\star}(\mathfrak{g}^{\ast})$ is a distribution, however, it turns out to be a regular function for compact Lie groups \cite[p.3]{Raasakka1}.} of the space $L^{2}_{\star}(\mathfrak{g}^{\ast})$ is given by
\begin{align}\delta(x):=\int_{G}\,\mathrm{d}g\,e_{g_{-1}}(0)e_{g}(X).\end{align}
This can easily be seen by the following short calculation:
\begin{equation}\begin{aligned}\int_{\mathfrak{g}^{\ast}}\frac{\mathrm{d}^{d}x}{(2\pi)^{d}}\,\delta(x)\star\hat{\varphi}(x)&=\int_{\mathfrak{g}^{\ast}}\frac{\mathrm{d}^{d}x}{(2\pi)^{d}}\,\int_{G}\,\mathrm{d}g\,e_{g^{-1}}(0)e_{g}(x)\star\hat{\varphi}(x)=\\&=\int_{G}\,\mathrm{d}g\,e_{g^{-1}}(0)\varphi(g)=\hat{\varphi}(0).\end{aligned}\end{equation}
The generalization of the non-commutative Fourier transform for field in several variables is straightforward: One uses a plane wave $e_{g}(x)$ for each variable in the definition of the non-commutative group Fourier transform above. The star-product is then defined component-wise.
\newpage
\section{Some Additional Remarks on Topological Equivalence of Coloured Graphs}\label{DipolesAppendix}
In the last chapter, we discuss some additional topics concerning dipole moves in coloured graph with boundary. First of all, we construct an explicit example of a dipole move, which does not fulfil the properties described in Theorem \ref{DipoleProper}, but which still turns out to be proper. Secondly, we discuss some effects non-proper dipole moves can have. Furthermore, we introduce a second set of moves, which can be used to describe topological equivalence of coloured graph with non-empty boundaries. These moves are called ``wounds'' and can be replaced by a set of four proper dipole moves. We end this discussion be briefly mentioning a class of pseudomanifolds, which is stable under non-proper wound moves and non-proper dipole contractions.

\subsection{Non-Proper Dipole Moves and ``Accidental'' Proper Moves}
The Theorem of Ferri-Gagliardi (Theorem \ref{DipoleProper}) gives two conditions for general dipoles in open graphs to be proper. More precisely, we have seen that a $k$-dipole $d_{k}$ in some open $(d+1)$-coloured graph $\mathcal{G}\in\mathfrak{G}_{d}$ is proper, whenever it fulfils one of the following two properties:
\begin{itemize}
\item[(A)]At least one of the $(d+1-k)$-bubbles separated by the dipole represents a $(d-k)$-sphere.
\item[(B)]Both vertices involved in $d_{k}$ are touching the boundary and at least one of the $(d+1-k)$-bubbles separated by the dipole represents a $(d-k)$-ball.
\end{itemize}
Reversing this statement, we also see that whenever a $k$-dipole is non-proper, then it cannot have one of the properties (A) and (B). However, note that the Theorem of Ferri-Gagliardi is not an ``if-and-only-if-statement'', i.e. there are also dipoles, which do not fulfil (A) and (B), but which still turn out to be proper. However, one should note that these ``accidental proper dipole moves'' can always be replaced by a finite sequence of dipole moves of type (A) and (B), since by the Theorem of Casali (Theorem \ref{Casali}), moves of type (A) and (B) are enough to describe topological equivalence of coloured graphs. As an example of such an accidental proper dipole, consider the following dipole move in an open $(3+1)$-coloured graph $\mathcal{G}\in\mathfrak{G}_{3}$:

\begin{figure}[H]
\centering
\includegraphics[scale=1]{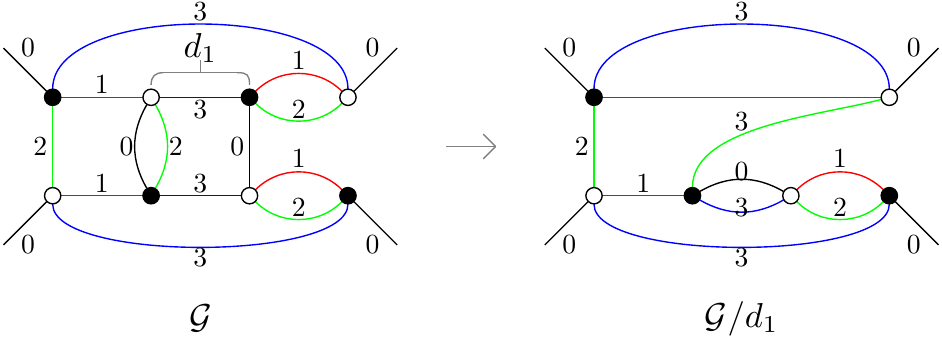}
\caption{An open $(3+1)$-coloured graph $\mathcal{G}$ and a proper dipole move not satisfying properties (A) and (B).}
\end{figure}

The edge of colour $3$ clearly defines a dipole $d_{1}$ in $\mathcal{G}$, since it separates two $3$-bubbles of colours $012$. Furthermore, note that both separated $3$-bubbles are open graphs and represent $2$-balls (=disks). However, the two vertices involved in the dipole are both not touching the boundary. Hence, this dipole does neither fulfil (A) nor (B). The boundary graphs of $\mathcal{G}$ and $\mathcal{G}/d_{1}$ are actually colour-isomorphic, as one can easily see: Both of them are the ``pillow graph'', but with different colouring. As a consequence, the two complexes dual to $\mathcal{G}$ and $\mathcal{G}/d_{1}$ have the same boundaries. Now, it turns out that they do not only have the same boundaries, but they also have the same bulk topology, since both the graphs $\mathcal{G}$ and $\mathcal{G}/d_{1}$ represent $3$-balls. This follows from the fact that both of them are related to the elementary melonic $3$-ball by a finite sequence of proper moves of type (A) and (B), as can be seen below. Hence, $d_{1}$ is a proper dipole.

\begin{figure}[H]
\centering
\includegraphics[scale=1]{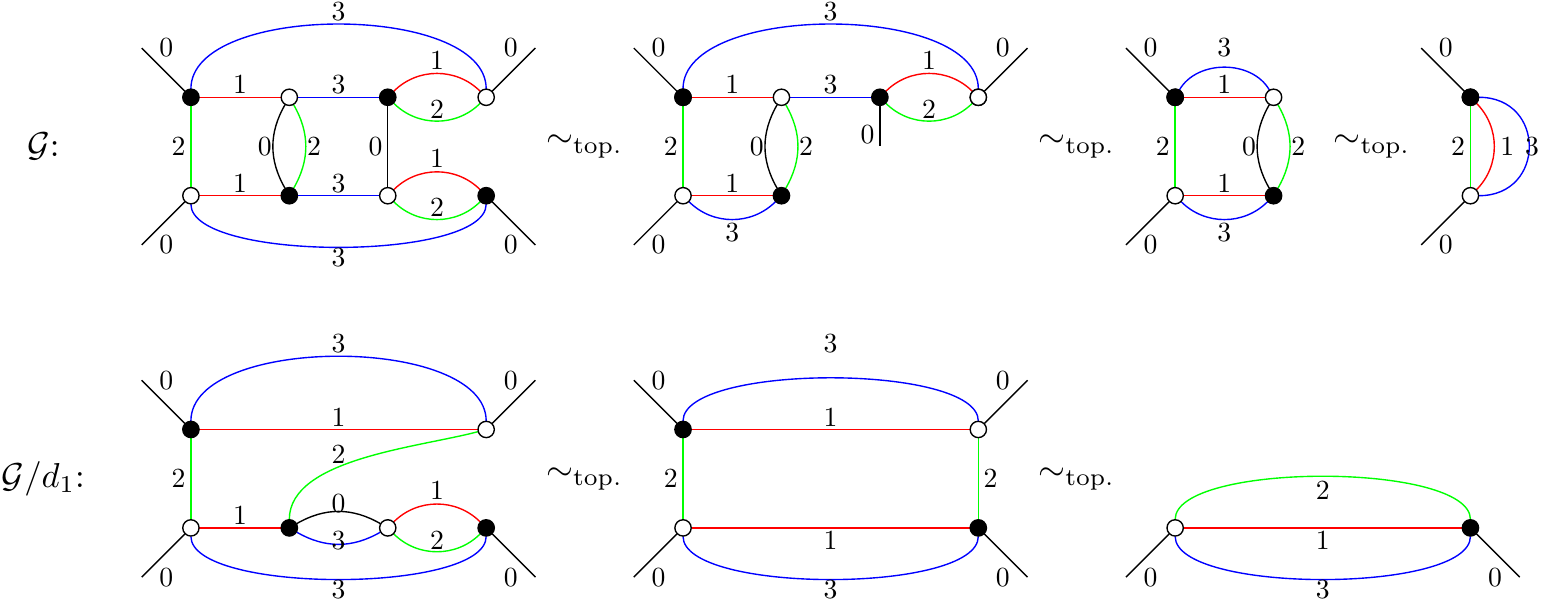}
\caption{Both the graphs $\mathcal{G}$ and $\mathcal{G}/d_{1}$ represent $3$-balls and hence $d_{1}$ is a proper dipole.}
\end{figure}

Next, let us discuss some effects of non-proper dipole moves. As already mentioned, non-proper dipoles cannot have properties (A) and (B). Let $\mathcal{G}\in\mathfrak{G}_{d}$ be an open $(d+1)$-coloured graph and $d_{k}=\{v,w\}$ a dipole within $\mathcal{G}$. Then, let us distinguish between the following types of dipoles
\begin{itemize}
\item[(1)]Both of the $(d+1-k)$-bubbles separated by the dipole $d_{k}$ are neither spheres nor balls.
\item[(2)]Either both $(d+1-k)$-bubbles separated by the dipole are $3$-balls, or one of them is a $3$-ball and the other one is neither a ball nor a sphere, however, at least one of the two vertices $v$ and $w$ is not touching the boundary.
\end{itemize}
Every dipole, which does not fulfil condition (A) and (B), does fall into one (and only one) of the two classes written above. However, it can happen that dipoles in these types turn out to be proper, as the accidental proper dipole move in the figure above shows. Dipoles, which fall into class (1), are always non-proper. To see this, recall that contracting a dipole means to perform the (boundary) connected sum of two links in the simplicial complex. If both the bubbles separated by the dipole are neither spheres nor balls, then contracting the dipole does change the topology of some subcomplex within the complex dual to $\mathcal{G}$. However, note that these dipoles can only occur in graphs which represents pseudomanifolds, since for manifolds, all bubbles are dual to spheres or balls. Hence, the only non-proper dipoles for graphs representing manifolds are of type (2). Dipoles in this class can be proper, as the example from above shows. On the other hand, non-proper dipoles in class (2) can have several different effects. One thing, which quite often happens, is that such a dipole produces a disconnected graph. An example is drawn below.

\begin{figure}[H]
\centering
\includegraphics[scale=1.2]{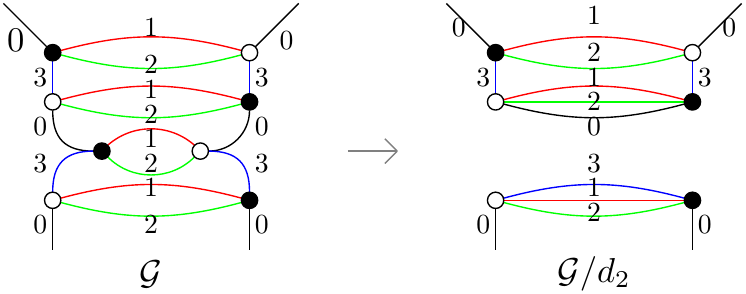}
\caption{A non-proper $2$-dipole of type (2), which produces a disconnected graph.}
\end{figure}

However, this must not necessarily be the case. There are also examples of non-proper dipole moves in class (2), which do not change the connectivity of the graph. As an example, let us look at the following $2$-dipole contraction:

\begin{figure}[H]
\centering
\includegraphics[scale=1.2]{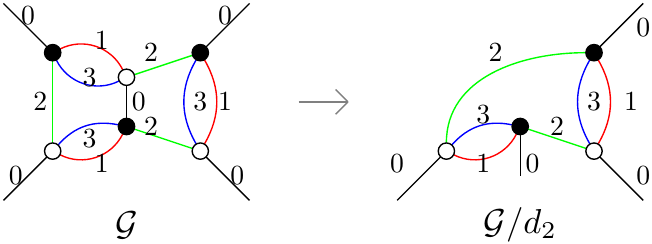}
\caption{A non-proper $2$-dipole contraction of type (2).}
\end{figure}

The $2$-dipole of colours 13 separates two open $2$-bubbles, which both represent $1$-balls (=closed intervals). However, one of the two vertices is not touching the boundary. In this case, one clearly sees that the graph $\mathcal{G}$ represents a pseudomanifold, since its unique $3$-bubbles of colours $012$ and $023$ both have two boundary components and represent cylinders. However, the graph $\mathcal{G}/d_{2}$ represents a manifolds (in fact, the $3$-ball), since the all its $3$-bubbles are either $2$-spheres or $2$-balls (=disks). Hence, this move is an example of a dipole move creating a manifold out of a pseudomanifold. As a consequence, this dipole is of course non-proper.

\subsection{Other Type of Topological Moves: Wounds}
For coloured graphs with non-empty boundary, one can also define another set of combinatorial and topological moves, which were introduced in \cite{Gagliardi87} and are called ``wounds''. As we will discuss below, these moves are just a combination of proper dipole moves and hence do not offer any new insights in topological equivalence of coloured graphs. Nevertheless, they still could be useful at some point, since they give another possibility to explicitly change the boundary graph of some given open graph. They are defined as follows:

\begin{Definition} (Wounds)\newline
Let $\mathcal{G}\in\mathfrak{G}_{d}$ be an open $(d+1)$-coloured graph with $\partial\mathcal{G}\neq\emptyset$ and $k\in\{0,\dots,d\}$. Furthermore, let $v,w$ be two internal vertices, which both admit an adjacent external leg. Then, we call a subgraph $\mathcal{W}_{k}$ consisting of all faces containing $v$ and $w$ ``$k$-wound of colours $i_{1},\dots,i_{k}$'', if the following two conditions are fulfilled:
\begin{itemize}
\item[(1)]The vertices $v$ and $w$ are contained in the same faces of colours $0i_{j}$ for all $j\in\{1,\dots,k\}$.
\item[(2)]The $(d+1-k)$-bubbles of colours $\mathcal{C}_{d}\textbackslash\{i_{1},\dots,i_{k}\}$ containing $v$ and $w$ are distinct.
\end{itemize}
\end{Definition}

Note that every non-internal proper $k$-dipole is of course also a $k$-wound, since all the non-cyclic faces containing the external legs and the two vertices are trivially the same and consist only of three edges. If $\mathcal{W}_{k}$ is a $k$-wound of colours $i_{1},\dots,i_{k}$ defined via two vertices $v,w$, then the boundary graph $\partial\mathcal{W}_{k}$, which is a subgraph of $\partial\mathcal{G}$, is a $k$-dipole of colours $i_{1},\dots,i_{k}$ in $\partial\mathcal{G}$ whenever $k<d$. If $k=d$, then $\partial\mathcal{W}_{k}$ is just a connected component of $\partial\mathcal{G}$, given by the elementary melonic $d$-sphere. If there is a wound in $\mathcal{G}$, then we construct a new graph as follows:

\begin{Definition} (Saturating a Wound)\newline
Let $\mathcal{G}\in\mathfrak{G}_{d}$ and $\mathcal{W}_{k}$ be a $k$-wound in $\mathcal{G}$ defined by two vertices $v$ and $w$. Then we define the graph $\mathcal{G}_{\overline{\mathcal{W}_{k}}}$ by connecting the two external legs connected to $v$ and $w$. We say that $\mathcal{G}_{\overline{\mathcal{W}_{k}}}$ is obtained from $\mathcal{G}$ by ``saturating the wound $\mathcal{W}_{k}$''. The reverse process is called ``opening the wound''.\end{Definition}

Let us briefly mentioned the effect of saturating a wound on the boundary graph of $\mathcal{G}$: If $k=d$, then saturating a wound results into removing a spherical component of $\partial\mathcal{G}$. If $k<d$, then saturating a wound results into contracting a $k$-dipole of the boundary graph $\partial\mathcal{G}$. An example of a $2$-wound saturation together with the effect on its boundary graph is drawn in the following figure.

\begin{figure}[H]
\centering
\includegraphics[scale=1]{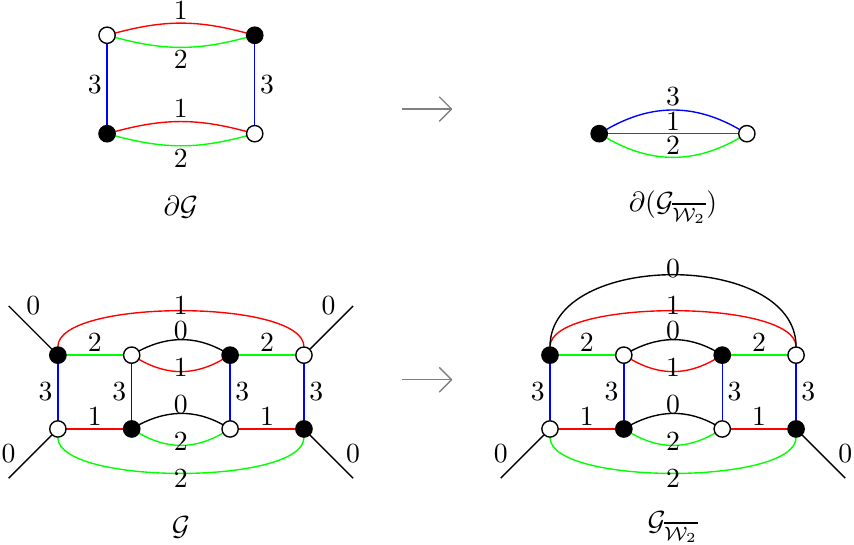}
\caption{An open $(3+1)$-coloured graph $\mathcal{G}\in\mathfrak{G}_{3}$ together with its boundary graph $\partial\mathcal{G}$ and the contraction of a $2$-wound $\mathcal{W}_{2}$ of colours $12$.\label{ExampleWound}}
\end{figure}

 As for dipole moves, we say that a wound saturation is ``proper'', if the (pseudo)manifolds dual to $\mathcal{G}$ and $\mathcal{G}_{\overline{\mathcal{W}_{k}}}$ are PL-homeomorphic. The $2$-wound drawn in Figure \ref{ExampleWound} above can easily be seen to be proper, since both the graphs $\mathcal{G}$ and $\mathcal{G}_{\overline{\mathcal{W}_{k}}}$ can be reduced to the elementary melonic $3$-ball by a sequence of proper dipole moves and hence represent $3$-balls themselves.\\
 \\
Of course, not every wound saturation is proper. As an example, consider an open graph with only two external legs. Then the collection of all non-cyclic faces form a $d$-wound, whose saturation leads to a closed graph. More generally, one can easily see that $d$-wounds are in general non-proper, even if there are more than two external legs, since they generically remove one (spherical) boundary component from the manifold. Furthermore, also for $k$-wounds with $k<d$ one can easily construct examples of non-proper wounds. In \cite[p.60]{Gagliardi87}, the authors found the following conditions for wounds to be proper:

\begin{Theorem} (Condition for Proper Wounds)\newline
Let $\mathcal{G}\in\mathfrak{G}_{d}$ and $\mathcal{W}_{k}$ be a $k$-wound with $k<d$ of colours $i_{1},\dots,i_{k}\in\mathcal{C}_{d}$ in $\mathcal{G}$ defined by two vertices $v$ and $w$. Furthermore, suppose that the following conditions are fulfilled:
\begin{itemize}
\item[(1)]At least one of the two $(d+1-k)$-bubbles of colours $\mathcal{C}_{d}\textbackslash\{i_{1},\dots,i_{k}\}$ separating the vertices $v$ and $w$ represents a $(d-k)$-ball.
\item[(2)]The $(k+1)$-bubble of colours $0,i_{1},\dots,i_{k}$ in $\mathcal{G}_{\overline{\mathcal{W}_{k}}}$ containing $v$ and $w$ represents a $k$-sphere.
\end{itemize}
Then $\mathcal{W}_{k}$ is equivalent to a sequence of four proper dipole moves and hence proper.
\end{Theorem}

\begin{proof}Consider a $k$-wound $\mathcal{W}_{k}$ in $\mathcal{G}$, whose colours are without loss of generality $1,\dots,k$. We denote the $(d+1-k)$-bubbles of colours $0,k+1,\dots,d$ separated by the wound by $\mathcal{B}_{v}^{0,k+1,\dots,d}$ and $\mathcal{B}_{w}^{0,k+1,\dots,d}$ respectively, and the spherical $k+1$-bubble in the saturated graph of colours $0,1,\dots,k$ containing the vertices $v$ and $w$ by $\mathcal{B}_{vw}^{0,1,\dots,k}$. Without loss of generality, we assume that the bubble $\mathcal{B}_{v}^{0,k+1,\dots,d}$ is the bubble separated by the wound, which represents a $(d-k)$-ball. Then, let us define a graph $\mathcal{G}^{\prime}$ by adding four new vertices to $\mathcal{G}$, which we denote by $v^{\prime},w^{\prime},v^{\prime\prime},w^{\prime\prime}$. We connect the vertex $v^{\prime}$ via an edge of colour $0$ to the external leg adjacent of $v$ and similarly the vertex $w^{\prime}$ to the external leg adjacent of $w$. Then, we connect the vertices $v^{\prime}$ and $v^{\prime\prime}$ by $d-k$ edges of colours $k+1,\dots,d$ and similarly, we connect $w^{\prime}$ and $w^{\prime\prime}$ by $d-k$ edges of colours $k+1,\dots,d$. Furthermore, we connect the vertices $v^{\prime}$ and $w^{\prime}$ by $k$ edges of colours $1,\dots,k$ and also the vertices $v^{\prime\prime}$ and $w^{\prime\prime}$ by $k$ edges of colours $1,\dots,k$. Last but not least, we add an external leg to $v^{\prime\prime}$ and one to $w^{\prime\prime}$. The resulting graph is a well-defined open $(d+1)$-coloured graph in $\mathfrak{G}_{d}$. The type of the vertices $v^{\prime},w^{\prime},v^{\prime\prime},w^{\prime\prime}$ (black vs. white) is such that $\mathcal{G}^{\prime}$ is bipartite. The structure of the graph $\mathcal{G}^{\prime}$ is sketched in the following figure:

\begin{figure}[H]
\centering
\includegraphics[scale=1.2]{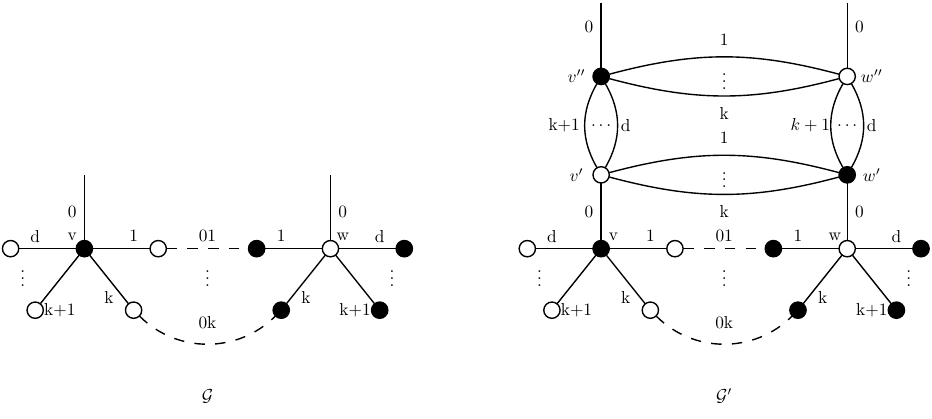}
\caption{The graph $\mathcal{G}$ and the graph $\mathcal{G}^{\prime}$ constructed above.}
\end{figure}

The dashed lines in the graph above sketch the bicoloured paths (=faces) connecting the corresponding vertices. Furthermore, note that the $d$ internal vertices connected to $v$ do not have to be distinct and similarly for $w$. By looking at the picture, it is clear that $\mathcal{G}^{\prime}$ has the same boundary graph as $\mathcal{G}$. In fact, $\mathcal{G}^{\prime}$ turns out to be equivalent to $\mathcal{G}$ up to proper dipoles. To see this, observe that $\{v^{\prime},v^{\prime\prime}\}$ together with the $d-k$ edges connecting them is an internal proper $(d-k)$-dipole, precisely because of assumption (2): The $(k+1)$-bubble of colours $0,1,\dots,k$ containing $v^{\prime}$, which is separated by this dipole, is just the graph obtained by adding an internal proper $k$-dipole between the vertices $v$ and $w$ in the graph $\mathcal{B}_{vw}^{0,1,\dots,k}$, which by assumption (2) represents a $k$-sphere. Contracting this dipole yields a graph which looks similar to $\mathcal{G}$ up to an internal proper $d$-dipole connected to $w$. Contracting this dipole finally yields $\mathcal{G}$. The procedure explained above is sketched in the following figure:

\begin{figure}[H]
\centering
\includegraphics[scale=0.95]{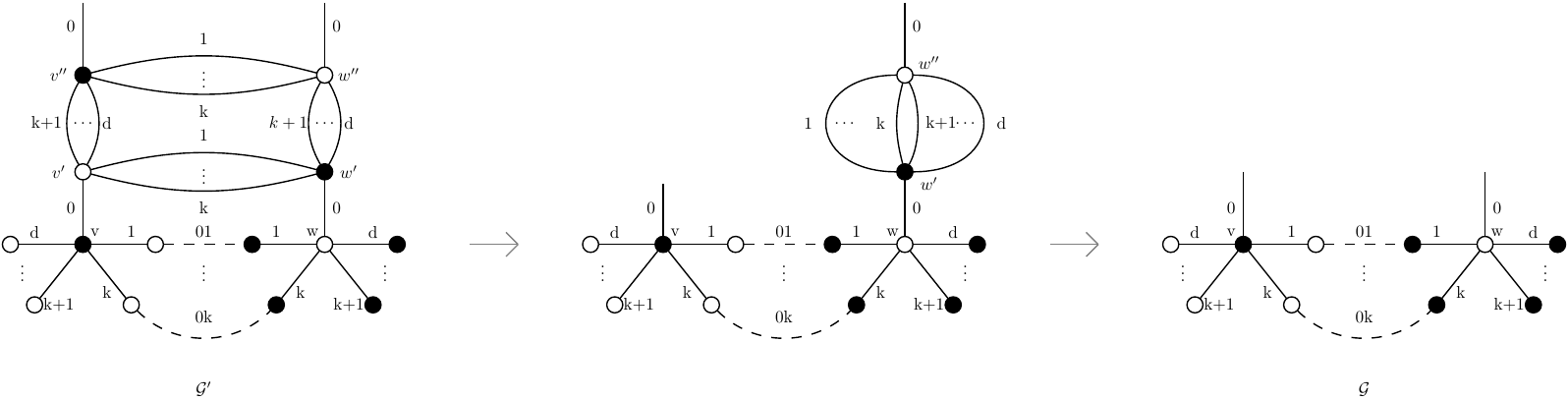}
\caption{The graph $\mathcal{G}^{\prime}$ can be reduced to $\mathcal{G}$ by performing first an internal proper $(d-k)$-dipole and then an internal proper $d$-dipole.}
\end{figure}

On the other hand, as a consequence of assumption (1), also $\{v^{\prime\prime},w^{\prime\prime}\}$ together with the $k$-edges connecting them is a proper dipole, namely a non-internal proper $k$-dipole. This follows from the fact that the $(d+1-k)$-bubble of colours $0,k+1,\dots,d$ containing the vertex $v^{\prime\prime}$ separated by this dipole is just the graph obtained by adding an internal proper $(d-k)$-dipole to the external leg adjacent to the vertex $v$ in the graph $\mathcal{B}_{v}^{0,k+1,\dots,d}$, which by assumption (1), represents a $(d+1-k)$-ball. Contracting this dipole yields a graph which looks like the saturated graph $\mathcal{G}_{\overline{\mathcal{W}_{k}}}$, in which we have added a $d$-dipole in-between the vertices $v$ and $w$. Contracting this dipole yields the graph $\mathcal{G}_{\overline{\mathcal{W}_{k}}}$. This is sketched in the figure below.

\begin{figure}[H]
\centering
\includegraphics[scale=0.95]{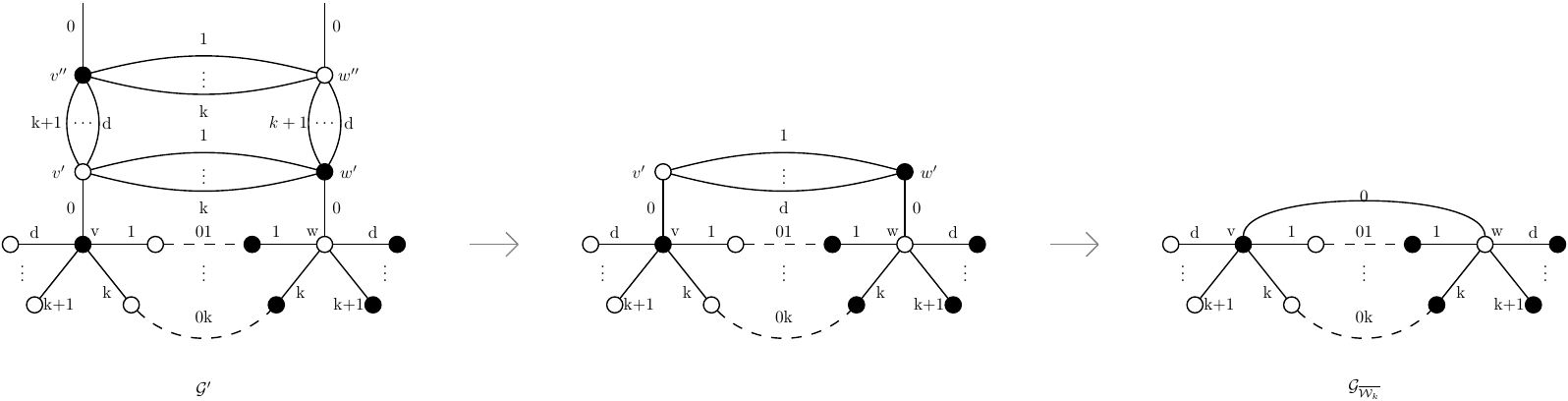}
\caption{The graph $\mathcal{G}^{\prime}$ can be reduced to $\mathcal{G}_{\overline{\mathcal{W}_{k}}}$ by performing first a non-internal proper $k$-dipole and then an internal proper $d$-dipole.}
\end{figure}

Hence, we have shown that $\mathcal{G}$ and $\mathcal{G}_{\overline{\mathcal{W}_{k}}}$ are related by a sequence of four proper dipole moves, which shows that $\mathcal{G}$ and $\mathcal{G}_{\overline{\mathcal{W}_{k}}}$ represent the same (pseudo)manifold. To sum up, saturating a wound means just to add two internal proper dipoles of types $d$ and $(d-k)$ to $\mathcal{G}$ followed by cancelling an non-internal proper $k$-dipole and an internal proper $d$-dipole.
\end{proof}

As an example, we can easily verify that the two assumption of the previous proposition are satisfied for the open $(3+1)$-coloured graph drawn in Figure \ref{ExampleWound} drawn above. Let us record the following immediate corollary:

\begin{Corollary} Let $\mathcal{G}\in\mathfrak{G}_{d}$ be a $(d+1)$-coloured graph representing a manifold. Then every $k$-wound in $\mathcal{G}$ is proper.\end{Corollary}

\begin{proof}Let $\mathcal{W}_{k}=\{v,w\}$ be a $k$-wound of colours $i_{1},\dots,i_{k}\in\mathcal{C}_{d}$ in $\mathcal{G}$. Since $\mathcal{G}$ represents a manifold, all its $k$-bubbles represent $(k-1)$-spheres or $(k-1)$-balls (Proposition \ref{ManifoldsGraphs}). Hence, assumption (1) of the proposition above is clearly satisfied. Let now $\mathcal{B}_{vw}$ be the $(k+1)$-bubble in $\mathcal{G}$ of colours $0,i_{1},\dots,i_{k}$ containing $v$ and $w$. By assumption, this bubble represents a $k$-ball, which only has two external legs. Connecting these two external legs hence produces a $k$-bubble, which represents a $k$-sphere.\end{proof}

In general, as explained in some previous parts, every non-internal proper $k$-dipole move in some open $(d+1)$-coloured graph induces a $k$-dipole in its boundary graph. However, the reverse of this statement is in general not true. What is true instead is that every proper $k$-dipole on the boundary graph corresponds to a $k$-wound in the corresponding open graph. This $k$-wound does not have to be proper, since manifolds with homeomorphic boundaries can obviously still have non-homeomorphic interiors.

\subsection{Pinched Manifolds and Non-Proper Dipole Moves}
In dimension $d=3$, there is a particular class of pseudomanifolds, which is stable under \textit{any} wound wove and dipole contraction (proper and non-proper). These type of topologies are called ``\textit{pinched manifolds}''. In order to define them, let us introduce the following terminology: The ``\textit{$\lambda$-punctured $2$-sphere}'' $S^{2}_{\lambda}$ for some $\lambda\geq 0$ is the surface obtained from the $2$-sphere $S^{2}$ by deleting the interior of $\lambda$ $2$-balls (=disks). In other words, $S^{2}_{\lambda}$ denotes the unique (up to homeomorphism) surface of genus $g=0$ with $\lambda$ boundary components. Its Euler characteristic is hence given by $\chi(S^{2}_{\lambda})=2-\lambda$. If $\lambda=0$, then $S^{2}_{\lambda=0}$ is just the $2$-sphere, if $\lambda=1$, then $S^{2}_{\lambda=1}$ is homeomorphic to the disk and if $\lambda=2$, $S^{2}_{\lambda=2}$ is homeomorphic to the cylinder. Using this, we define pinched manifolds as follows:

\begin{Definition} (Pinched Manifolds)\newline
A $3$-dimensional simplicial complex is called a ``combinatorial pinched manifolds'', if the link of every vertex represents a punctured $2$-sphere. The corresponding geometric realization is called a ``pinched manifold''.
\end{Definition}

Of course, every PL-manifold is also a pinched manifold, since $S^{2}_{\lambda=0}=S^{2}$ and $S^{2}_{\lambda=1}=D^{2}$. On the other hand, not every pinched manifold is a manifold. If it is pure, non-branching and strongly-connected, then it is, however, always a normal pseudomanifold, by definition. As proven in \cite{Gagliardi87}, if $\mathcal{G}$ is some open $(3+1)$-coloured graph and $\mathcal{W}_{k}$ is a $k\in\{1,2,3\}$ wound in $\mathcal{G}$, then $\mathcal{G}$ represents a pinched manifold if and only if $\mathcal{G}_{\overline{\mathcal{W}_{k}}}$ represents a pinched manifold. Moreover, if $d_{k}$ is a $k\in\{1,2,3\}$ dipole in $\mathcal{G}$, then $\mathcal{G}/d_{k}$ represents a pinched manifold whenever $\mathcal{G}$ represents a pinched manifold. The reverse of the latter statement is in general not true.\\
\\
As an example, consider again the following open $(3+1)$-coloured graph and the corresponding $2$-dipole contraction, which we have already discussed previously:

\begin{figure}[H]
\centering
\includegraphics[scale=1.2]{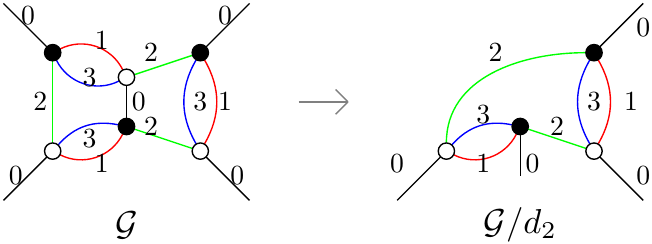}
\caption{A non-proper $2$-dipole contraction.}
\end{figure}

The graph $\mathcal{G}$ is clearly an open $(3+1)$-graph representing a pinched manifold, which is not a manifold: It has exactly five $3$-bubbles, from which one represents a $2$-sphere (the unique $3$-bubble with colours $123$), two represent the $2$-ball (the two $3$-bubbles with colours $013$) and two represent the $2$-punctured sphere $S^{2}_{2}$ (the unique $3$-bubbles with colours $012$ and $023$). Contracting the $2$-dipole drawn above, which is clearly a non-proper dipole, since both of the separated $2$-bubbles are open, but one of the vertices is internal, yields a manifold with different topology. This contracted manifold represented by $\mathcal{G}/d_{2}$ represents clearly the $3$-ball and hence is in particular again a pinched manifold.
%
%
%
%
%
%\chapter*{Directories}
%\addcontentsline{toc}{chapter}{Directories}
%
%
%
%
%
%\newpage
%\addcontentsline{toc}{section}{List of figures}
%{\hypersetup{linkcolor=black}
%\listoffigures
%
%
%
%
%
%Bibliography:
\newpage 
\addtocontents{toc}{\vspace*{2ex}}
\addcontentsline{toc}{section}{\hspace{-16pt}\textbf{Bibliography}}
\bibliographystyle{alpha} 
\bibliography{Bibliography}
\end{document}